\documentclass{article}

\usepackage{arxiv}

\usepackage[utf8]{inputenc} 
\usepackage[T1]{fontenc}    
\usepackage[colorlinks=true, linkcolor=blue, citecolor=blue, urlcolor=blue]{hyperref}       
\usepackage{url}            
\usepackage{booktabs}       
\usepackage{amsfonts}       
\usepackage{amsmath}
\usepackage{nicefrac}       
\usepackage{microtype}      
\usepackage{graphicx}
\usepackage{natbib}
\usepackage{doi}
\usepackage{subcaption}
\usepackage{xcolor}
\usepackage{CJKutf8}
\usepackage{float}
\usepackage{array}
\usepackage{paralist}
\usepackage{enumitem}
\usepackage{bbm}
\usepackage{multirow}
\usepackage{tabularx}
\usepackage{longtable}

\title{User Location Disclosure Fails to Deter Overseas Criticism but Amplifies Regional Divisions on Chinese Social Media}

\date{July 30, 2025}	

\author{ \href{https://orcid.org/0000-0002-1393-5417}{\includegraphics[scale=0.06]{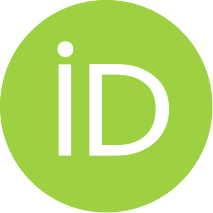}\hspace{1mm}Leo Yang Yang}\thanks{We thank Ting Chen, Yingjie Fan, Ruixue Jia, Yuan Liu, Jennifer Pan, Molly Roberts, Susan Shirk, Shaoda Wang, Zi Wang, Fei Zhou, Bolun Zhang, and participants at conferences and seminars at UCSD, Stanford, HKBU, SCNU, ZJU, and WHU for helpful comments and suggestions. We gratefully acknowledge support from the 21st Century China Center at UCSD, the King Center on Global Development at Stanford, and the Stanford Center on China’s Economy and Institutions. We also thank Yige Chen, Yujie Hu, Aiqi Li, Jiahui Liu, Yifei Lu, Jinwen Wu, Bin Xu, Zhuoluo Yang, and Ting Zeng for excellent research assistance. All errors are our own.} \\
	Department of Accountancy, Economics and Finance\\
	Hong Kong Baptist University\\
	Kowloon, Hong Kong SAR \\
	\texttt{leoyang@hkbu.edu.hk} \\
	\And
	\href{https://orcid.org/0000-0003-2041-6671}{\includegraphics[scale=0.06]{orcid.pdf}\hspace{1mm}Yiqing Xu} \\
	Department of Political Science\\
	Stanford University\\
	Stanford, CA, US, 92093 \\
	\texttt{yiqingxu@stanford.edu} \\
}

\renewcommand{\shorttitle}{User Location Disclosure on Chinese Social Media}

\hypersetup{
pdftitle={User Location Disclosure Fails to Deter Overseas Criticism but Amplifies Regional Divisions on Chinese Social Media},
pdfsubject={Econ.GN, cs.SI},
pdfauthor={Leo Yang Yang, Yiqing Xu},
pdfkeywords={Location Disclosure, Online Behavior, Social Divisions, Information Control, Authoritarianism},
}

\begin{document}
\begin{CJK}{UTF8}{gbsn}
\maketitle

\begin{abstract}
We examine the behavioral impact of a user location disclosure policy on Sina Weibo, China’s largest microblogging platform, using a unique high-frequency dataset of uncensored engagement—including tens of thousands of comments and replies—on prominent government and media accounts. The policy, publicly justified as a measure to curb misinformation and counter foreign influence, was abruptly rolled out on April 28, 2022. Using an interrupted time series design, we find no decline in participation by overseas users. Instead, it significantly reduced domestic engagement with local issues outside users’ home provinces, particularly among critical comments. Evidence indicates this decline was not driven by generalized fear or concerns about credibility, but by a surge in regionally discriminatory replies that raised the social cost of cross-provincial engagement. Our findings suggest that identity disclosure tools can have unintended consequences by activating existing social divisions in ways that reinforce state control without direct censorship.
\end{abstract}


\section{Introduction}

In October 2021, the Cyberspace Administration of China released draft regulations requiring platforms to display users’ IP-based locations, by province for domestic users and by country for overseas users. Officials framed the policy as a safeguard against misinformation and foreign interference, but critics viewed it as another step to constrain online speech \citep{reuters2022weibo, nytimes2022internet}. On March 17, 2022, less than a month after the Russian invasion of Ukraine, Sina Weibo, China’s largest microblogging site, began tagging the locations of users posting about the war. On April 28, it abruptly extended the policy to all posts and comments. Other major platforms soon followed, and by the end of 2022, location disclosure had become a standard feature across Chinese social media.

This paper examines the behavioral consequences of Weibo’s location disclosure policy. Authoritarian regimes have long relied on censorship \citep{king2013censorship, lorentzen2014strategic, roberts2018censored}, surveillance \citep{esarey2011digital, pan2023disguised}, and orchestrated engagement \citep{chen2017authoritarian, chen2017information, qin2017freer}. Increasingly, they also deploy identity disclosure tools, such as real-name registration or IP-based location tags, to shape behavior without direct repression. By making identity attributes visible, these tools can alter participation through reputational and social pressures. Prior work shows that regional cues can heighten antagonism and incivility \citep{peng2021amplification, chen2021science, yu2024topic, guo2023civilizing}, suggesting that disclosure may not only chill expression but also activate existing regional divisions. Yet the causal effects of such measures remain poorly understood.

\begin{figure}[!htbp]
    \centering
    \resizebox{1\textwidth}{!}{
    \begin{minipage}{\textwidth}
    \hspace{-1em}
    \begin{subfigure}[b]{0.55\textwidth}
        \includegraphics[width=0.9\textwidth]{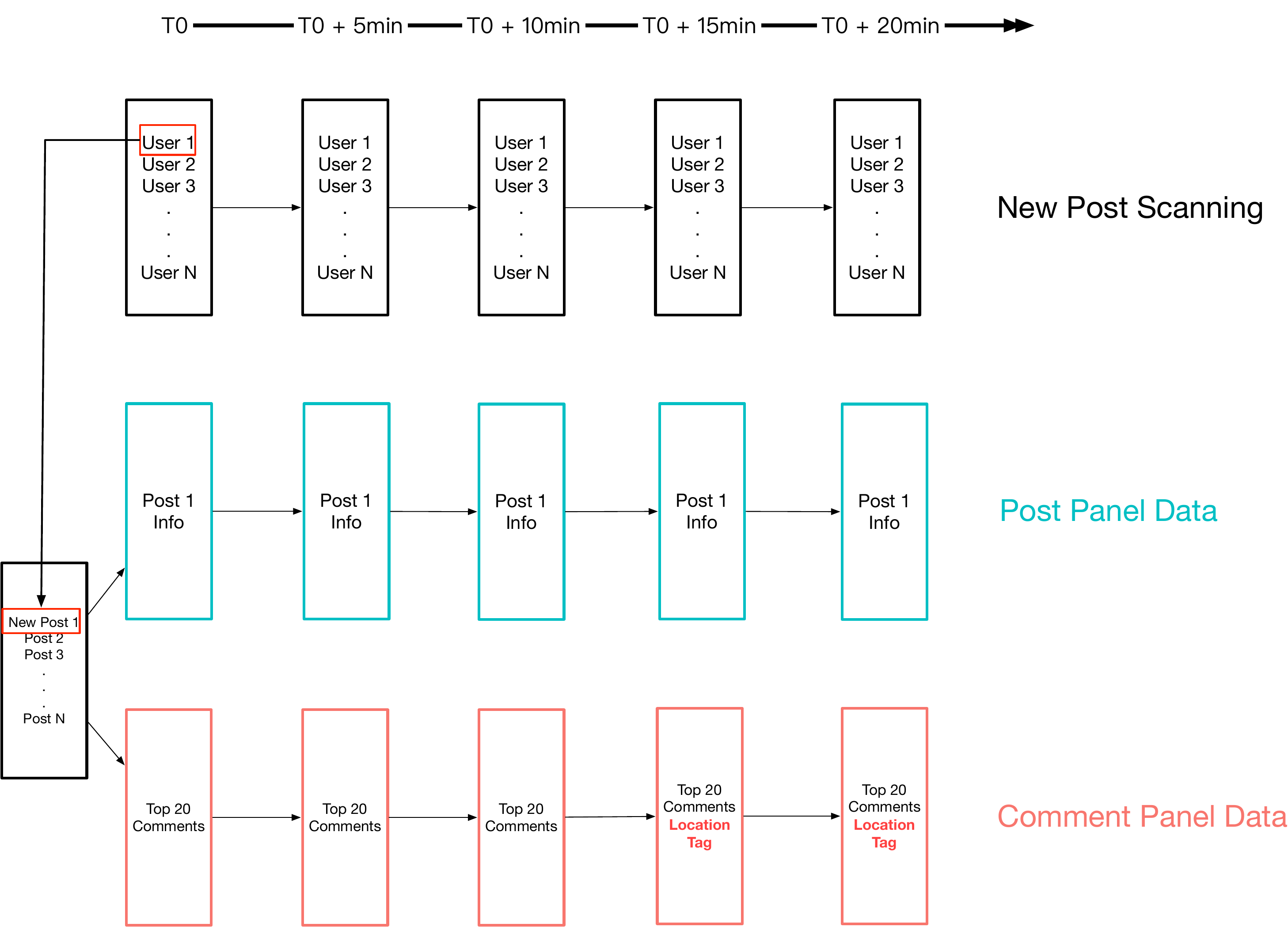}  
        \caption{Data Collection Pipeline}
        \label{fig:timeline}
    \end{subfigure}
    \hspace{-1em}
    \begin{subfigure}[b]{0.5\textwidth}
        \centering
        \includegraphics[width=\textwidth]{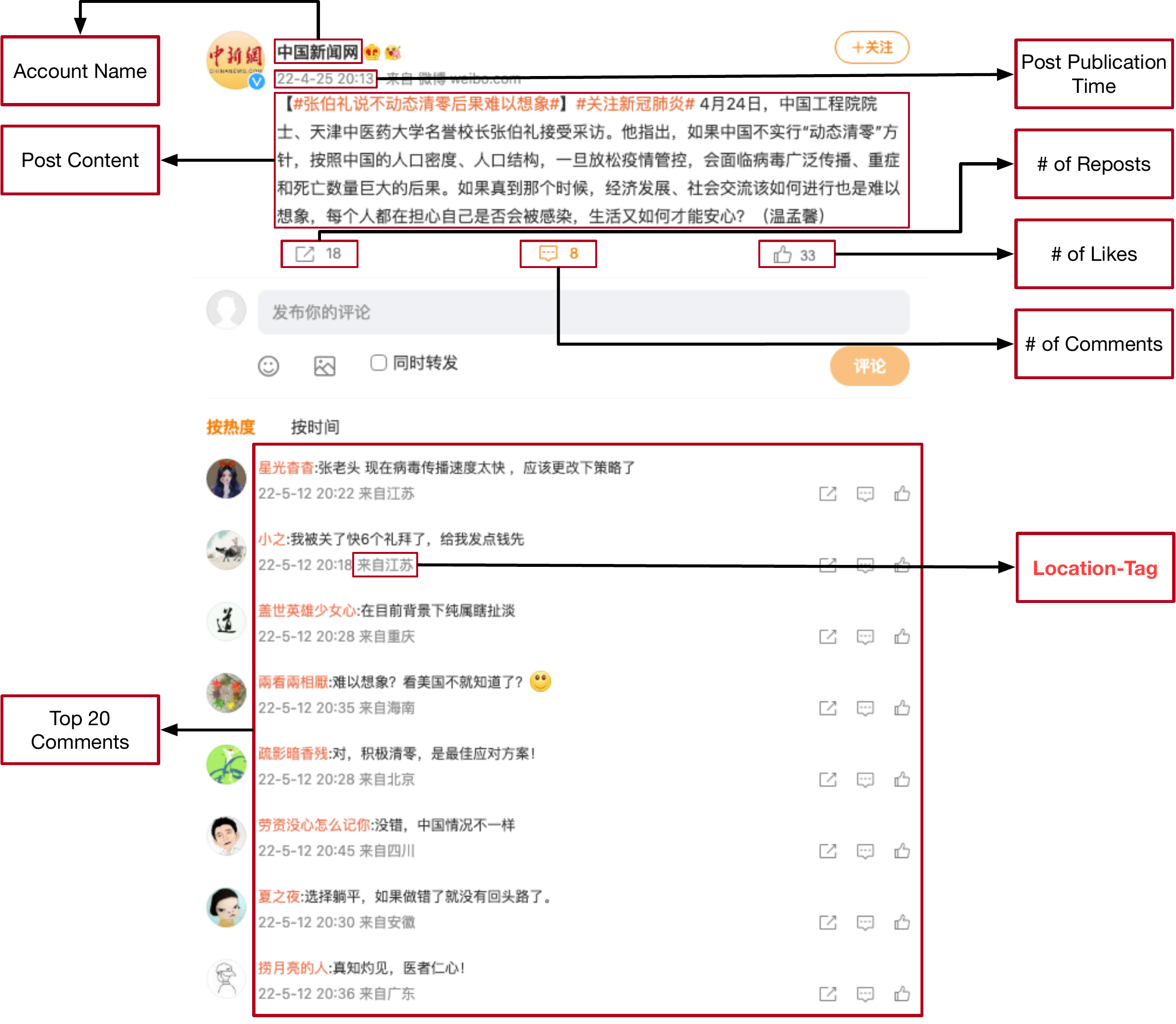}
        \caption{Variables Collected}
        \label{fig:variables}
    \end{subfigure}

    \vspace{1em}
    \hspace{-1em}
    \begin{subfigure}[b]{0.5\textwidth}
        \centering
        \begin{subfigure}[b]{\textwidth}
            \includegraphics[width=\textwidth]{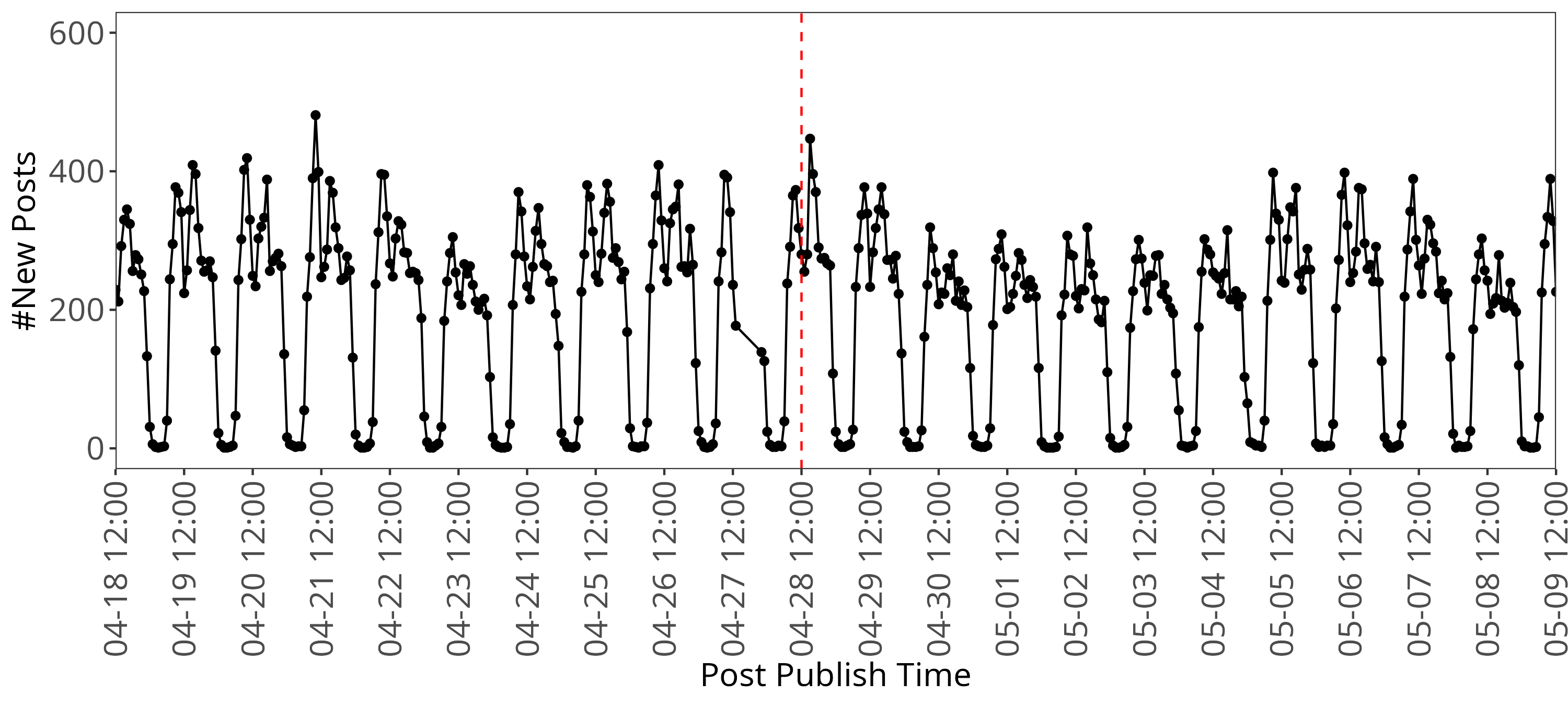}
            \caption{\#New Post Collected}
            \label{fig:data_new_posts}
        \end{subfigure}
        \vspace{1em}
        \begin{subfigure}[b]{\textwidth}
            \includegraphics[width=\textwidth]{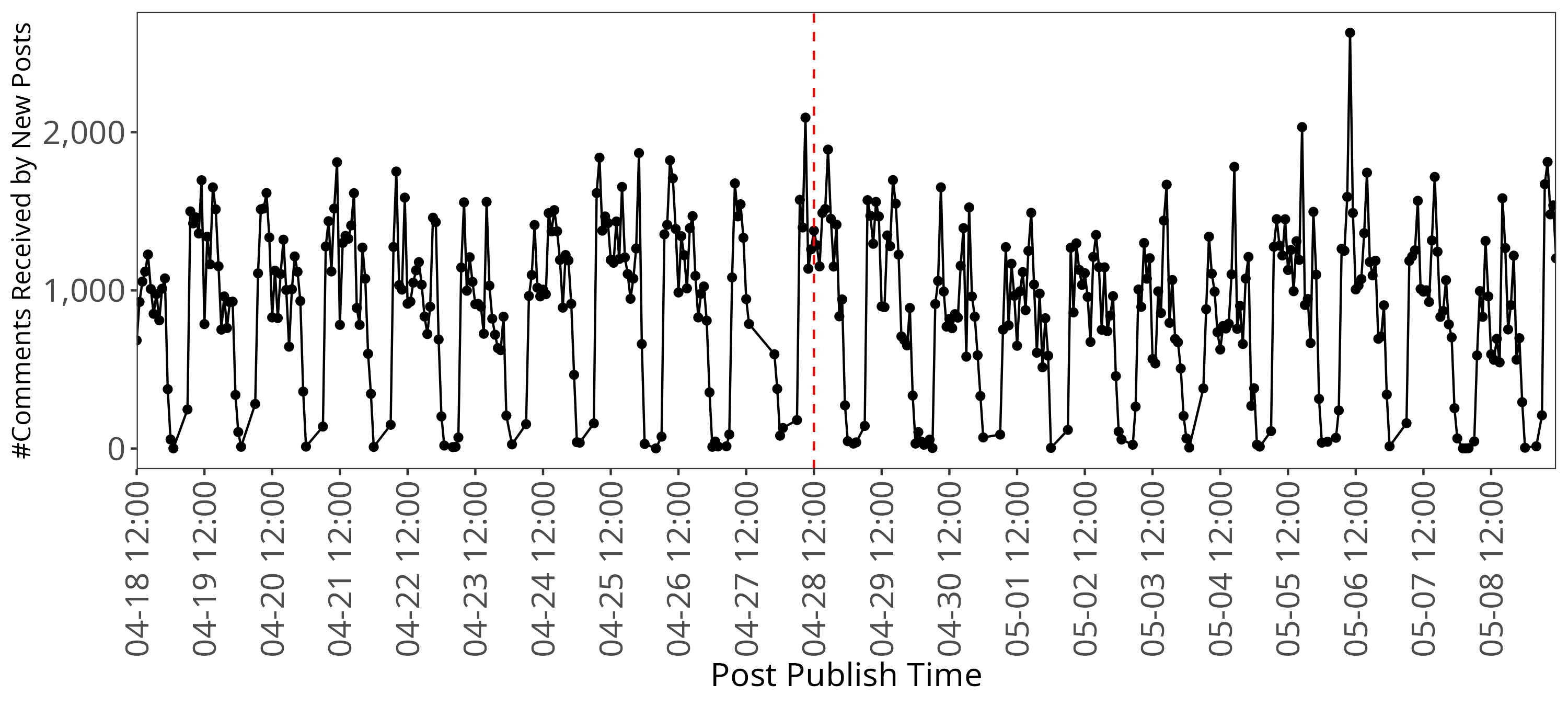}
            \caption{\#Comment Associated with New Posts}
            \label{fig:data_new_comments}
        \end{subfigure}
    \end{subfigure}
    \hspace{1em}
    \begin{subfigure}[b]{0.5\textwidth}
        \centering
        \begin{subfigure}[b]{\textwidth}
            \includegraphics[width=\textwidth]{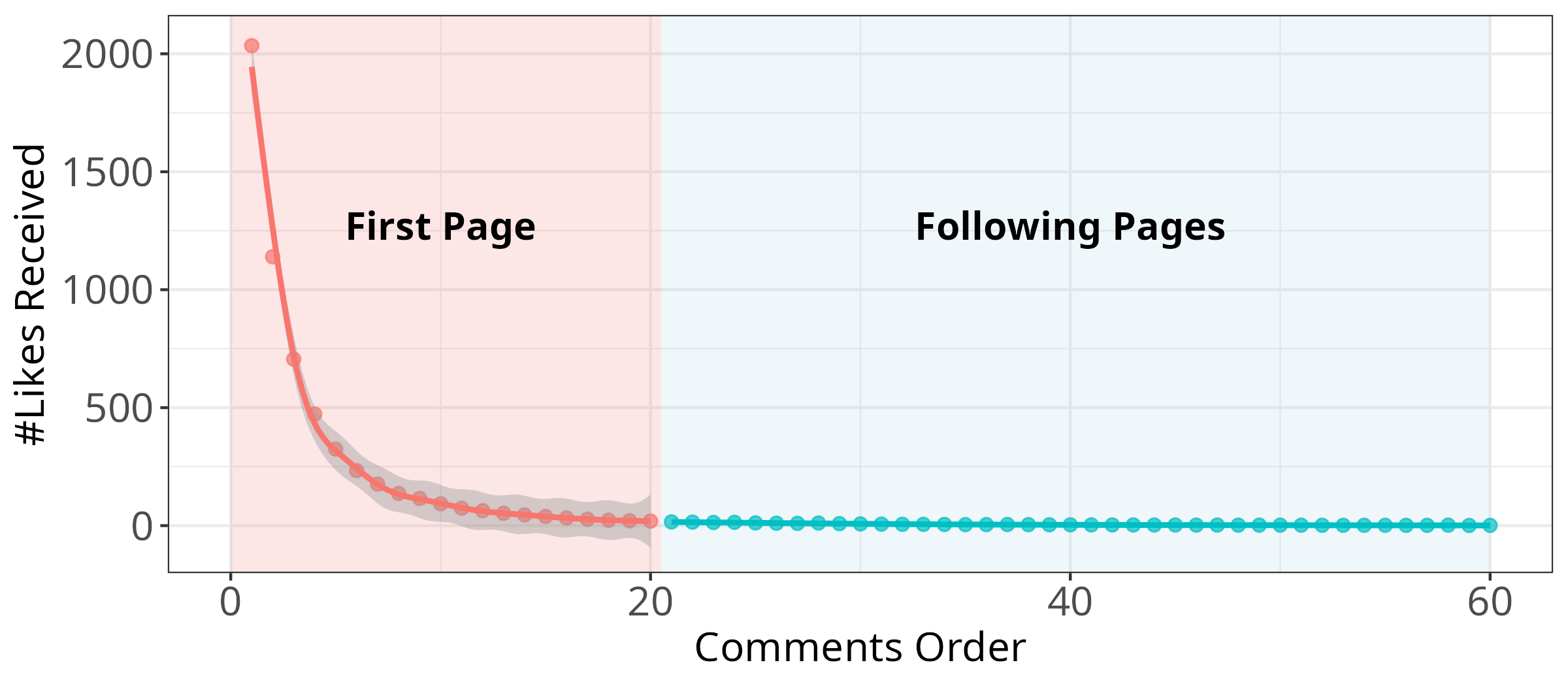}
            \caption{First-Page Engagement Concentration (Likes)}
            \label{fig:2nd_page_likes}
        \end{subfigure}
        \vspace{1em}
        \begin{subfigure}[b]{\textwidth}
            \includegraphics[width=\textwidth]{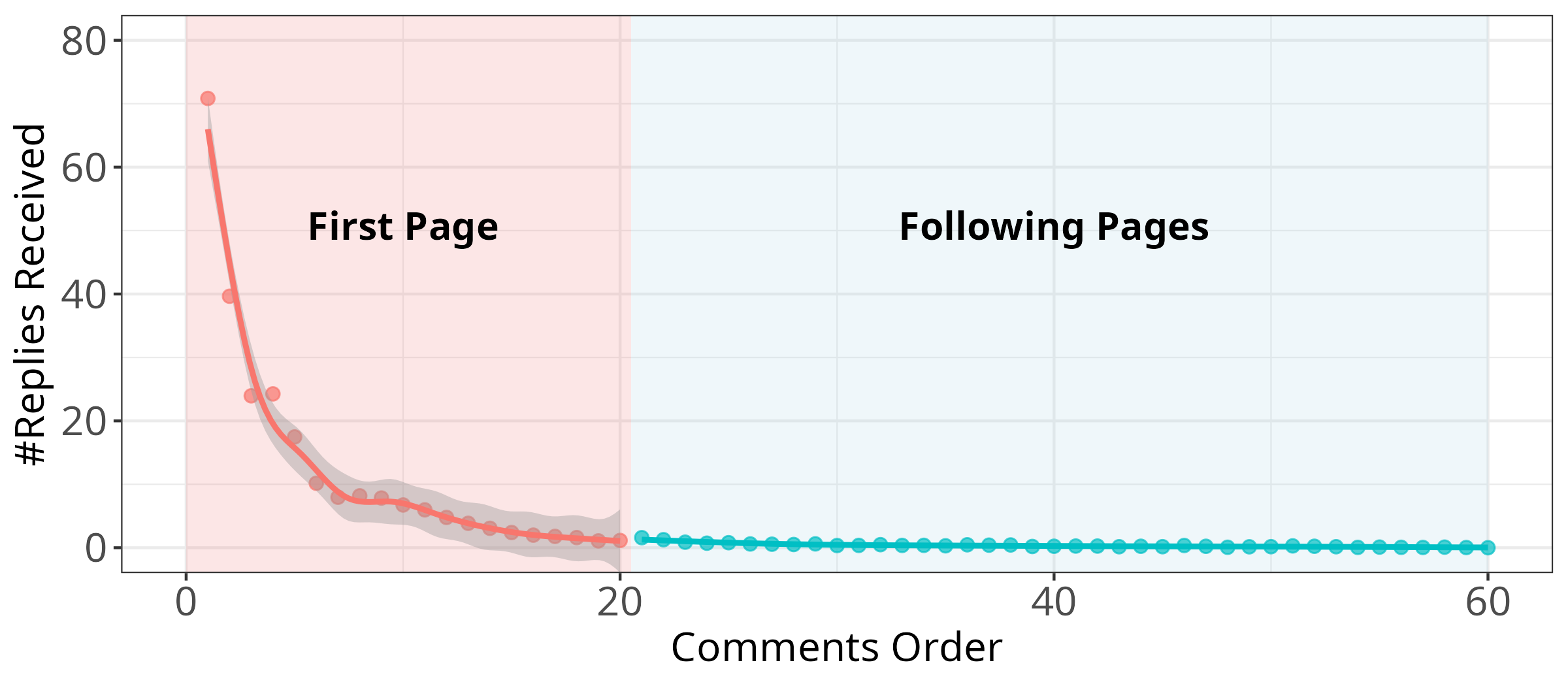}
            \caption{First-Page Engagement Concentration (Replies)}
            \label{fig:2nd_page_replies}
        \end{subfigure}
    \end{subfigure}
    \end{minipage}
    }
    \caption{\textbf{Data Collection} Figure~\ref{fig:timeline} illustrates our data collection process. We actively monitored the timelines of 165 government and media accounts. At time $T_0$, if a new post appeared, the system began tracking it and its top comments. Snapshots were taken every five minutes for the first 24 hours and every 24 hours for the following 10 days. Figure~\ref{fig:variables} provides an example of the information collected in each snapshot, including the publisher, publication time, post content, and engagement metrics, as well as the top 20 comments and their associated IP location tags. Figure~\ref{fig:data_new_posts} shows the number of new posts collected per day from April 18 to May 9, 2022. The vertical red line marks noon on April 28, when the user location disclosure policy was implemented. Figure~\ref{fig:data_new_comments} displays the daily distribution of new comments over the same period. Both figures show regular daily fluctuations, indicating consistent user activity. Figures~\ref{fig:2nd_page_likes} and~\ref{fig:2nd_page_replies} report average likes and replies by comment rank. Engagement drops sharply after the first page, and comments on following pages receive minimal interaction. Although we also estimate the causal effect on the total number of comments, our primary focus is on first-page comments, which account for the majority of meaningful engagement.}
   \label{fig:data_collection}
\end{figure}

Studying these effects poses serious empirical challenges. In tightly censored environments like Weibo, observed behavior often reflects censorship and deletion rather than genuine user response. Sensitive posts are removed quickly—3\% within 30 minutes and 90\% within 24 hours in some cases \citep{king_how_2013}—creating severe survivorship bias. Public datasets usually rely on post-censorship crawling, which overrepresents high-follower accounts and relies on narrow keyword lists. Causal inference is also difficult: randomized disclosure is ethically problematic, while observational studies based on deletion flags or self-reported geotags are vulnerable to selection bias.

We address these limitations with a unique high-frequency dataset that continuously monitored 165 prominent government and media accounts and their comment sections at approximately five-minute intervals. This approach captured tens of thousands of comments and replies before censorship or removal, substantially reducing survivorship bias and preserving authentic user responses. A rare Weibo implementation glitch further enabled us to recover pre-treatment geographic identities. Exploiting the abrupt and unannounced expansion of the policy as a natural intervention, we apply an interrupted time series design to estimate its causal effect. By comparing engagement immediately before and after the cutoff under plausible assumptions of no anticipation and smooth potential trends, we provide credible evidence on how identity disclosure reshaped online behavior. Figure~\ref{fig:data_collection} illustrates our data collection process, which we will explain in detail in the Methods section. 

We find that, contrary to the government’s stated aim, overseas users did not reduce their participation; if anything, they briefly increased their comments on international issues in a reactive backlash. By contrast, domestic users sharply curtailed their involvement in discussions of local issues outside their own provinces. This decline was not explained by shifts in post content but by the withdrawal of out-of-province participants, particularly those offering critical remarks. The new visibility of provincial identity increased the risk of backlash, deepened regional antagonism, and raised the social costs of cross-provincial engagement, thereby narrowing the scope of online discourse without direct censorship.

Beyond the authoritarian context, these findings contribute to a broader literature on how platform architecture shapes online behavior. Studies of credibility labels, state-affiliation tags, and feed-ranking algorithms show that design choices affect exposure, trust, and sharing even without changing beliefs \citep{flaxman2016filter, guess2023social, liang2022effects, papak2022algorithms, bradshaw2024look}. We extend this work by demonstrating that simple identity disclosure can silence participation by activating social cleavages, underscoring the broader importance of platform design in structuring the boundaries of online discourse in both authoritarian and democratic settings.

\section{Findings}
\label{sec:findings}

We now turn to the empirical findings. First, we examine whether the policy achieved its intended goal of deterring participation by overseas users. We then assess its effects on domestic engagement, distinguishing between local and non-local issues, and explore the mechanisms driving these patterns. 

\subsection{Revealing IP locations failed to deter overseas users but reduced domestic engagement}

The policy was designed to deter participation by overseas users with ``malicious intent,'' who were believed to interfere in domestic discourse by spreading misinformation and expressing views that challenge the official narrative. The expectation was twofold: that public location tags would help domestic audiences judge the credibility of online content, and that overseas users might withdraw out of concern about being discredited or attacked as part of a ``malicious foreign force'' (see Supplementary Information for background). To evaluate these claims, we distinguish between posts on international affairs, defined as those mentioning at least one foreign location, and non-international posts, which cover domestic issues or general topics without foreign reference.

\begin{figure}[!htbp]
    \centering
    \begin{subfigure}[b]{0.31\textwidth}
        \centering
        \includegraphics[width=\textwidth]{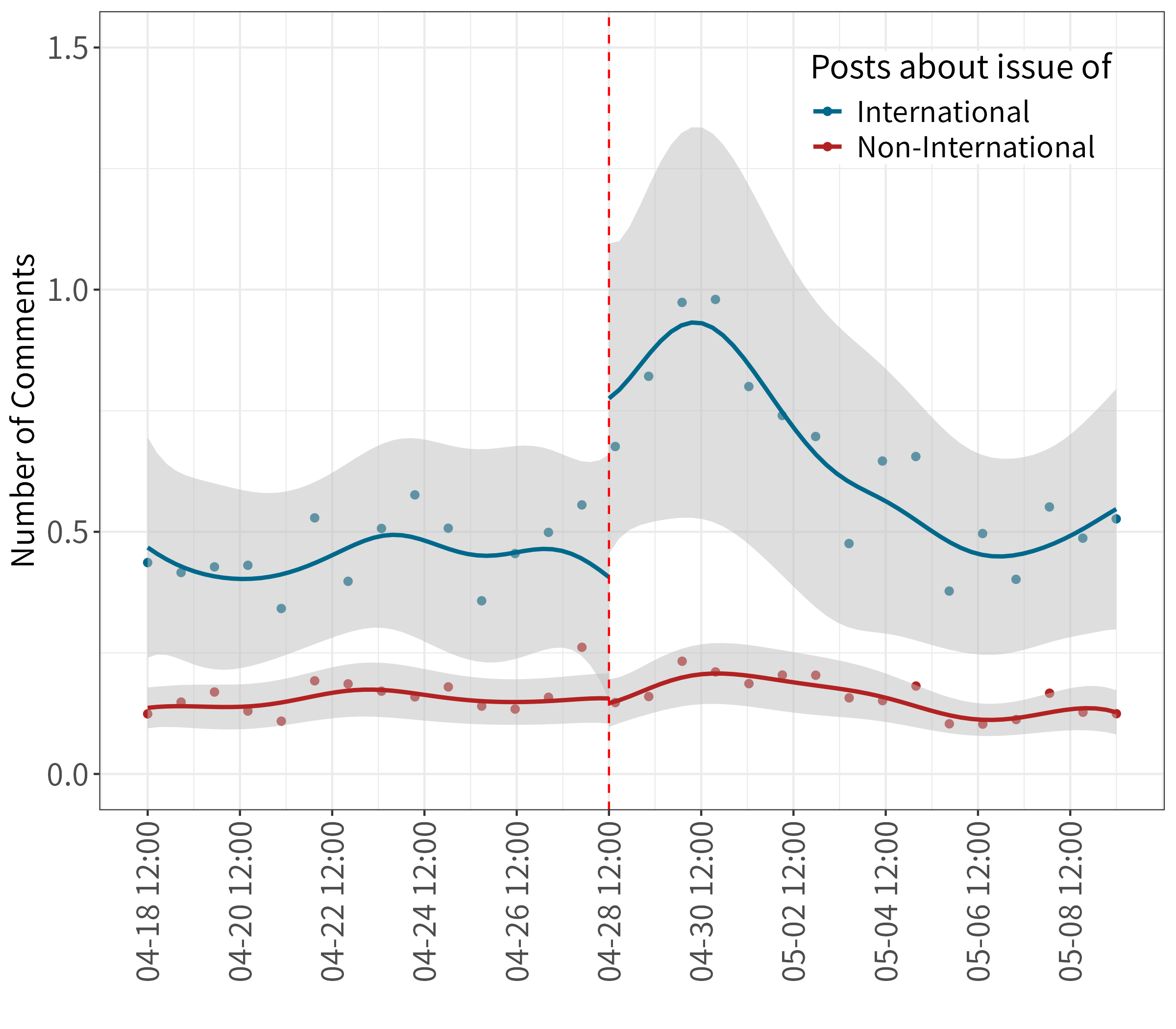}
        \caption{\footnotesize{\#Comments by Oversea Users}}
        \label{fig:overseas}
    \end{subfigure}
    \hfill
    \begin{subfigure}[b]{0.31\textwidth}
        \centering
        \includegraphics[width=\textwidth]{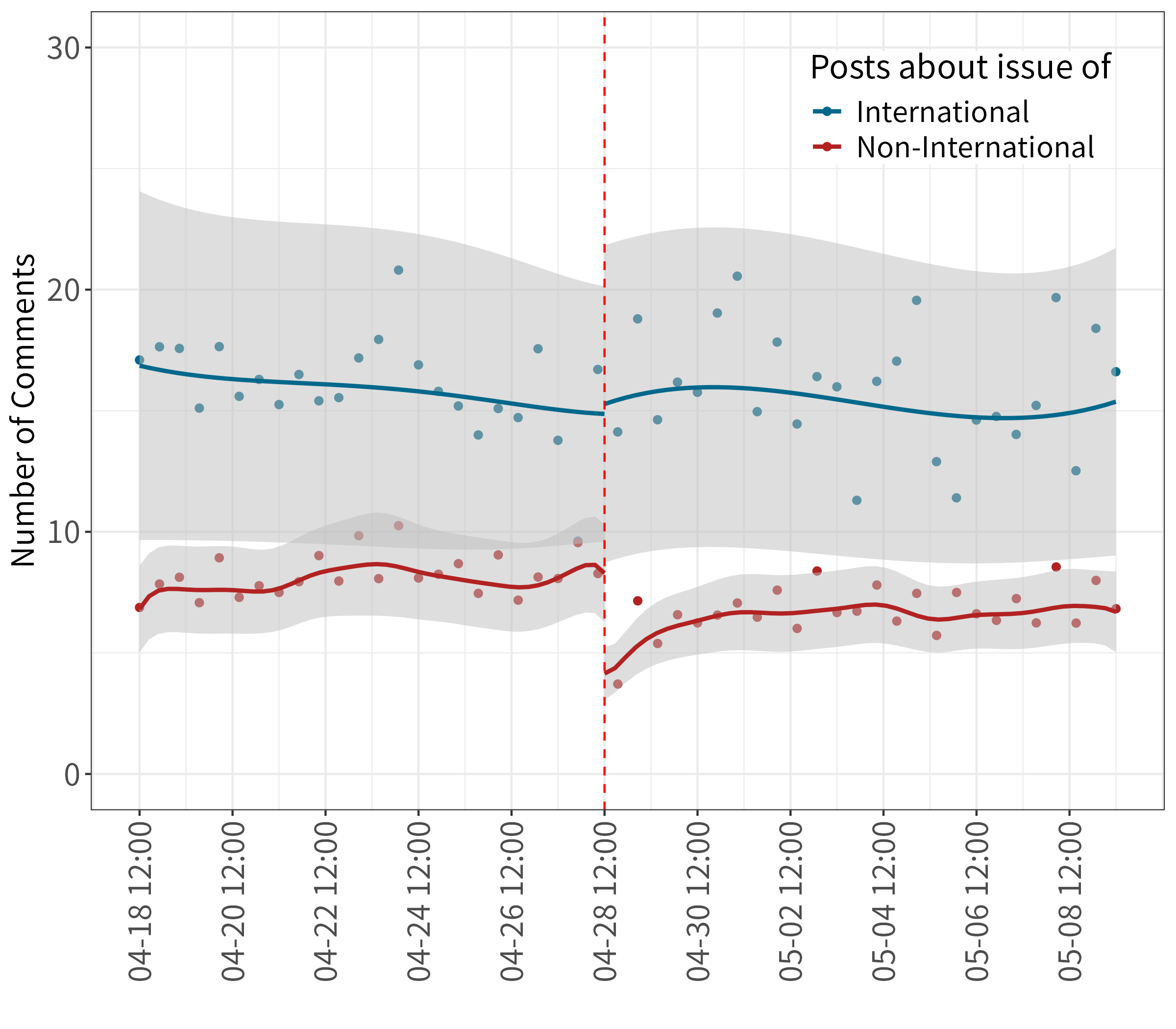}
        \caption{\footnotesize{\#Comments by Domestic Users}}
        \label{fig:domestic}
    \end{subfigure}
    \hfill
    \begin{subfigure}[b]{0.31\textwidth}
        \centering
        \includegraphics[width=\textwidth]{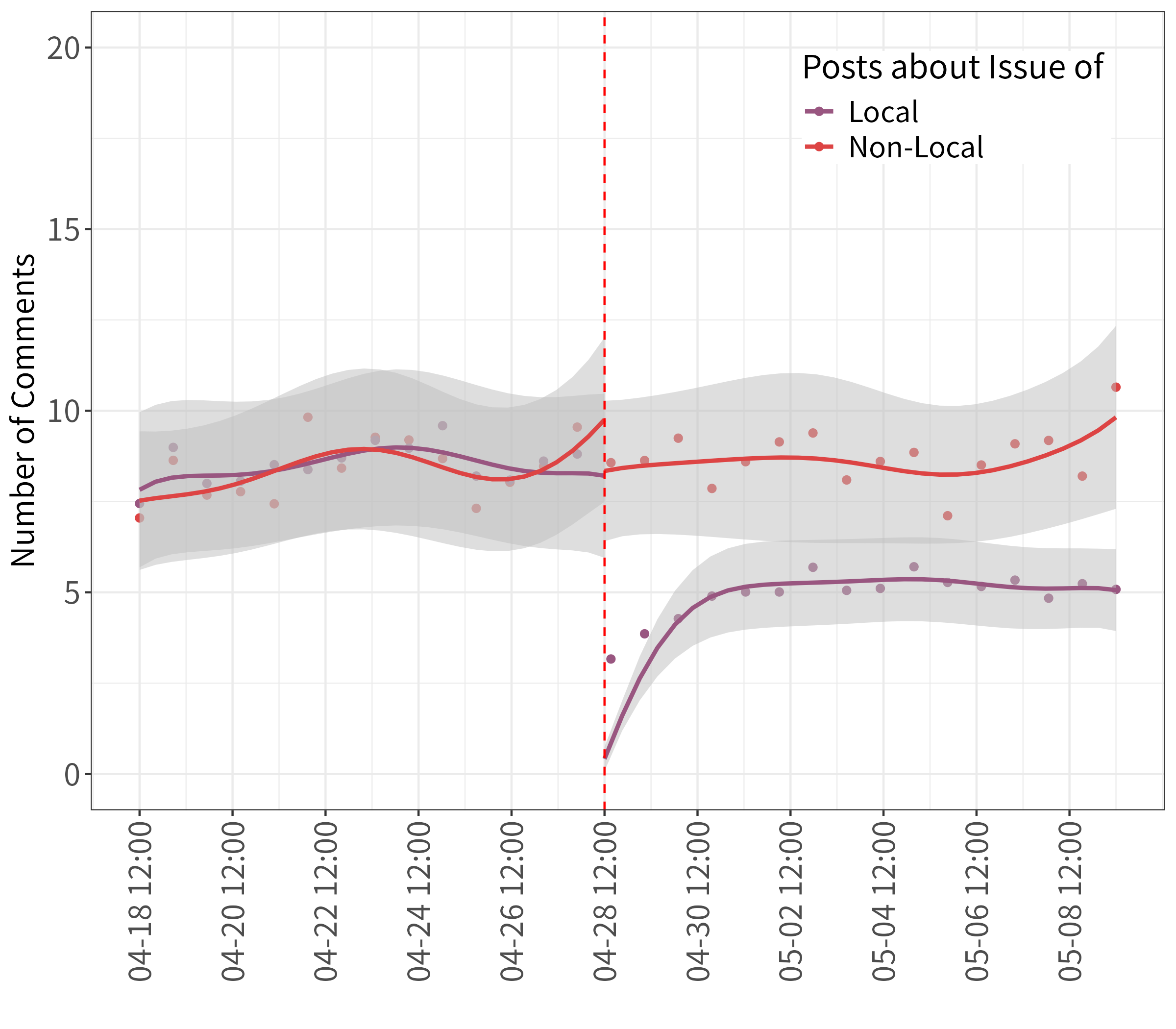}
        \caption{\footnotesize{\#Comments by Domestic Users\\on Non-International issues}}
        \label{fig:domestic-noninternational}
    \end{subfigure}

    \vspace{0.3cm}
    \begin{subfigure}[b]{0.31\textwidth}
        \centering
        \includegraphics[width=\textwidth]{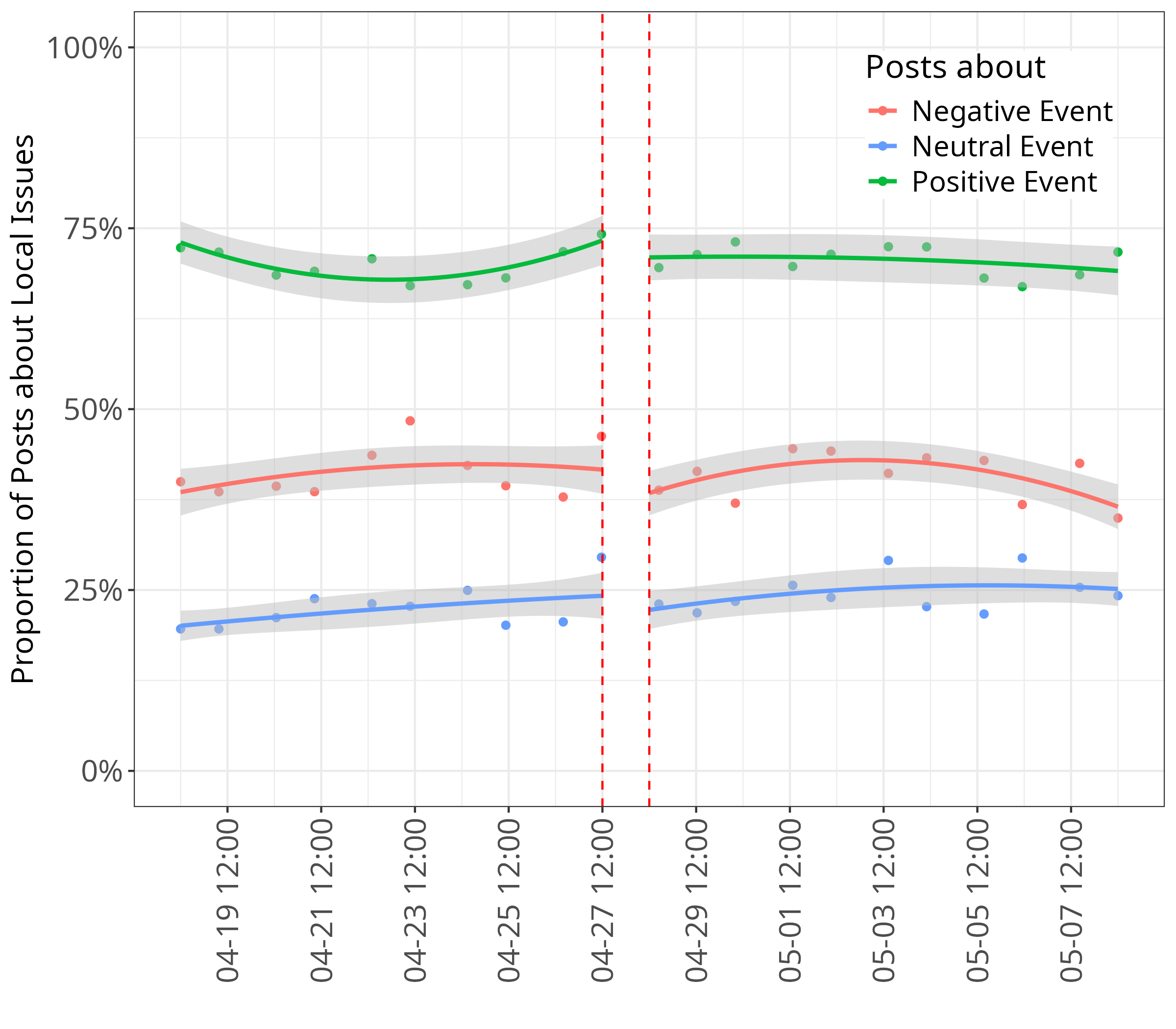}
        \caption{\footnotesize{Proportion of Local Posts Covering Positive \& Negative Events}}     \label{fig:content}
    \end{subfigure}
    \hfill
    \begin{subfigure}[b]{0.31\textwidth}
        \centering
        \includegraphics[width=\textwidth]{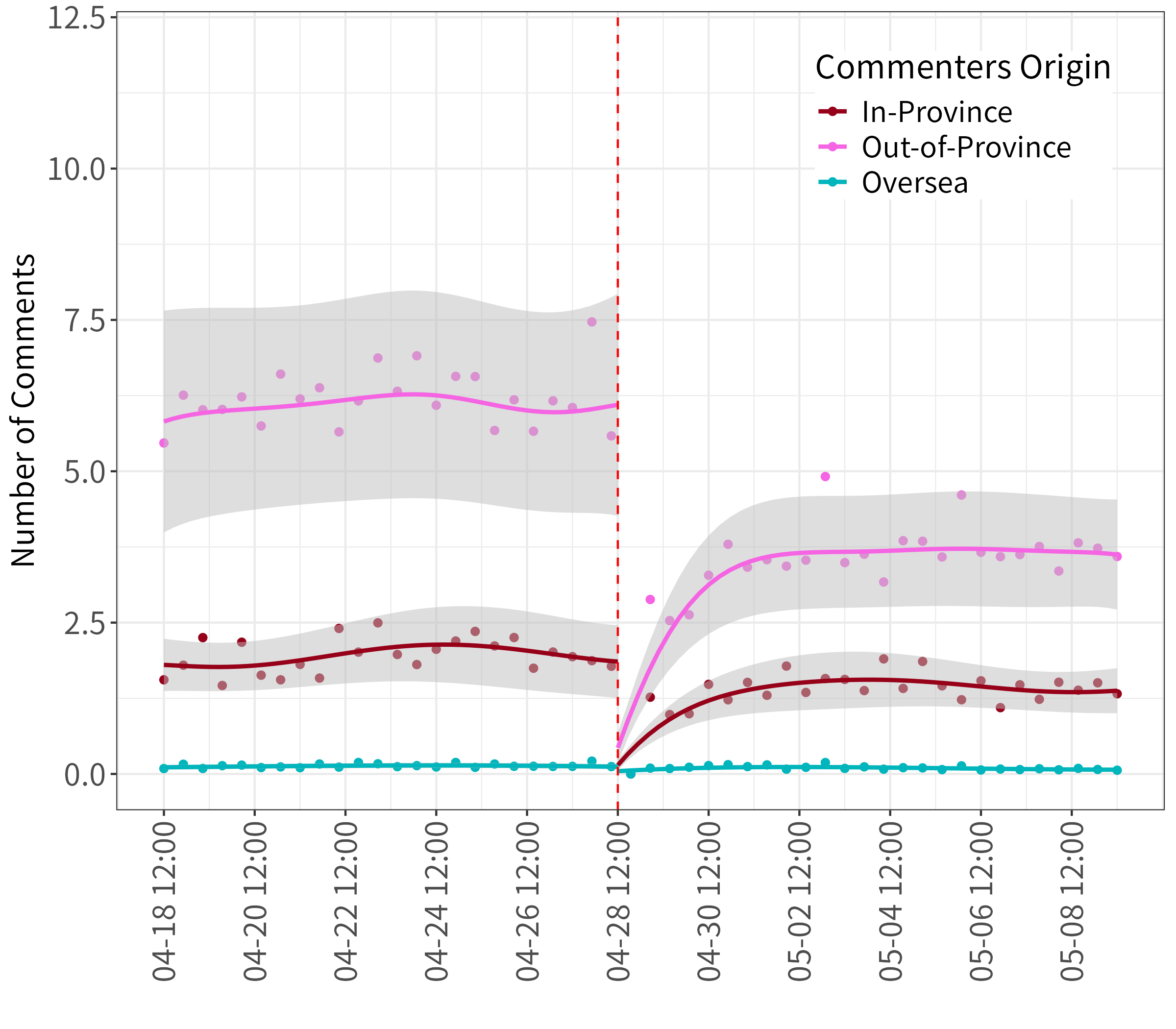}
        \caption{\footnotesize{\#Comments by Domestic Users: by User Origin}}
        \label{fig:local2}
    \end{subfigure}
    \hfill
    \begin{subfigure}[b]{0.31\textwidth}
        \centering
        \includegraphics[width=\textwidth]{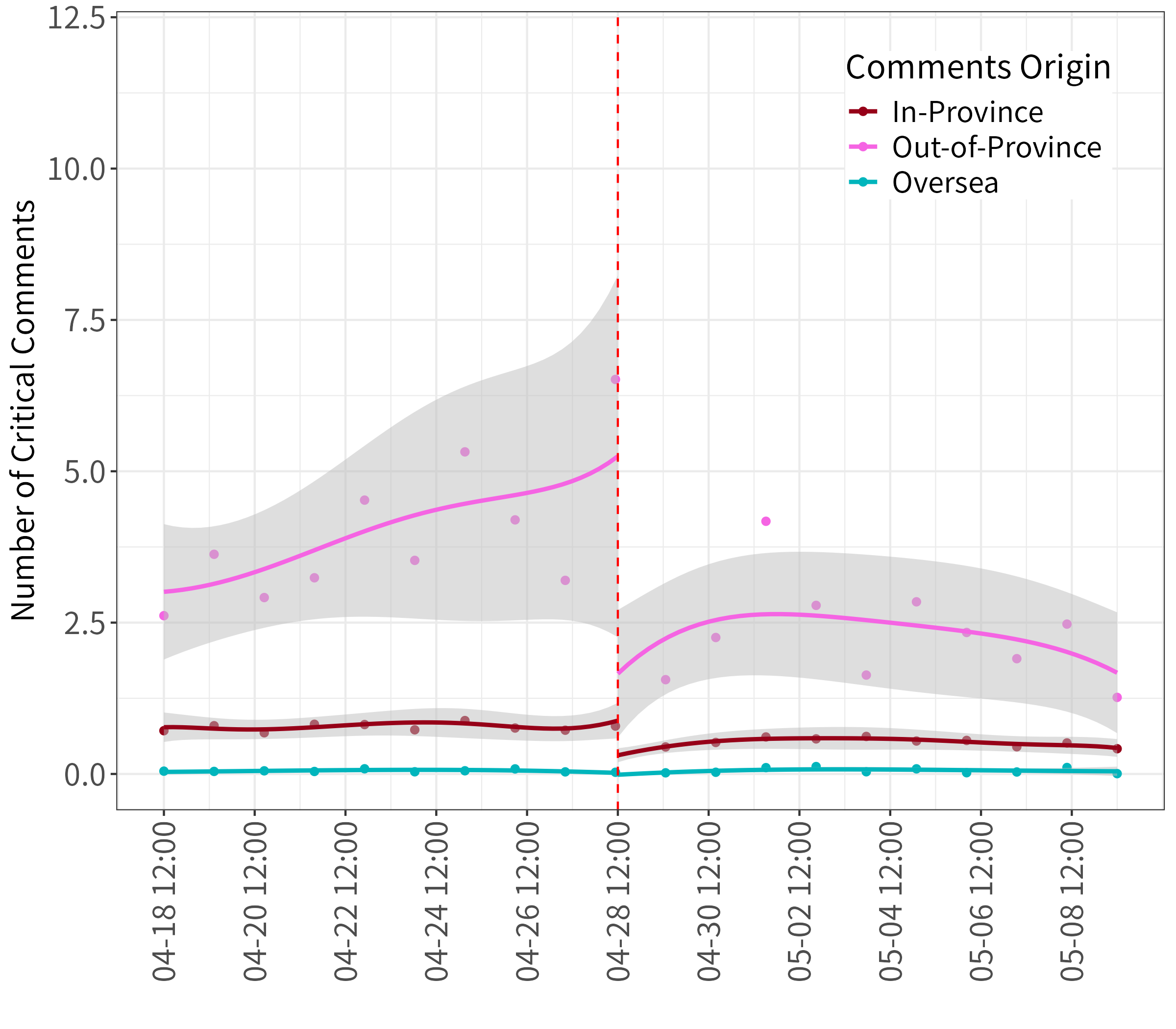}
        \caption{\footnotesize{\#Critical Comments by Domestic Users: by User Origin}}
        \label{fig:local3}
    \end{subfigure}

   \caption{\textbf{Engagement before and after User Location Disclosure.} Figures~\ref{fig:overseas} and~\ref{fig:domestic} show smoothed trends in the number of first-page comments from overseas and domestic users, respectively, on posts about international and non-international topics. Comment counts are aggregated over the first 48 hours after posting, and the vertical red line marks the implementation of the IP location disclosure policy at noon on April 28, 2022. Figure~\ref{fig:overseas} shows no decline in overseas engagement. Comments on international topics rose temporarily before returning to baseline, while comments on non-international topics remained stable. By contrast, Figure~\ref{fig:domestic} shows a sharp and sustained drop in domestic comments on non-international posts, with no change on international topics. Figure~\ref{fig:domestic-noninternational} confirms that this decline is driven by posts about local issues—defined as those mentioning a province name—while engagement with non-local topics remains stable. Figure~\ref{fig:content} shows no significant change in the distribution of local post content by sentiment, ruling out shifts in post type as an explanation. Figures~\ref{fig:local2} and~\ref{fig:local3} further show that the drop in engagement is driven primarily by out-of-province users and is accompanied by a reduction in dissenting comments. In-province engagement also declined but rebounded more quickly. These patterns suggest that user location disclosure discouraged nonlocal participation in local discussions. Confidence intervals are constructed based on cluster-robust standard errors clustered at the account level.}
    \label{fig:combined}
\end{figure}

Figure~\ref{fig:overseas} tests this deterrence hypothesis directly. The blue and red smoothed lines show the average number of comments from overseas users (based on platform-identified IP locations) on international and non-international posts, respectively, before and after the policy took effect. Contrary to expectations, there is no evidence of reduced overseas engagement. Comments on international topics rose sharply, from an average of 0.41 to 0.78 per post, before gradually returning to baseline. This short-lived surge suggests backlash rather than retreat. This result is unlikely to be driven by VPN masking because 99.1\% of user IP tags remained unchanged across the policy shift, and analysis in Section~G of the Supplementary Information shows that overseas users continued to comment on sensitive topics such as the Russian invasion of Ukraine, despite increased attacks from domestic commenters.  On non-international posts, overseas engagement remained stable, further weakening the deterrence hypothesis. Taken together, these results indicate that the policy failed in its stated goal of discouraging overseas participation. Note that the red line remains consistently lower than the blue line, but has much narrower confidence intervals. This is because overseas users are more likely to comment on posts related to international affairs; however, such posts are less frequently made by propaganda accounts compared to non-international posts, which explains the wider confidence intervals around the blue line.

Turning to domestic users, Figures~\ref{fig:domestic} shows a sharp and sustained decline in engagement. As in Figure~\ref{fig:overseas}, the blue and red smoothed lines represent the average number of comments on international and non-international posts, respectively—this time from domestic users, based on their platform-identified user locations. We see that comment volumes on international posts remained stable, but those on non-international posts fell immediately by about 50\%. Again, the wider confidence intervals around the blue line reflect the fact that government and media accounts post far fewer international than non-international topics, resulting in greater variability in the estimated trend. Disaggregating further, in Figure~\ref{fig:domestic-noninternational}, we find that this drop is concentrated in discussions of local issues, defined as posts mentioning a Chinese province. Engagement with posts concerning local affairs plummeted from an average of 8.21 to 0.42 comments per post after the policy (purple lines), while participation on non-local topics remained stable (red lines).\footnote{Consistent with this finding, Figure~G26 in the Supplementary Information shows that aggregate engagement metrics, including the total number of comments, likes, and reports, on posts about local issues also declined following the policy. However, using the aggregate metrics, we cannot distinguish whether this decrease is driven by domestic or international users.} 

In sum, while the policy did not deter overseas users, it sharply curtailed domestic engagement with local issues outside users’ home provinces. The intended targets were unaffected, but curiously, domestic discourse contracted in precisely the spaces where local identity cues were made most salient.

\subsection{Decline in discussions on local affairs were driven by out-of-province users}

Why did user engagement on posts about local issues fall so sharply after the policy took effect? We first test whether the decline reflects changes in what government and media accounts were posting. If these accounts shifted toward fewer negative or politically sensitive stories, reduced engagement could follow. Yet Figure~\ref{fig:content} shows that the distribution of local posts across positive, neutral, and negative events remained stable before and after implementation. These negative events include major accidents, scandals, and COVID-related lockdowns. Moreover, Figure~D5 in the Supplementary Information demonstrates that engagement fell for both positive and negative events alike. This rules out changes in post content as the main explanation.

If not content, then what may be the drivers? We consider two possibilities: a chilling effect, where users feared state repression once their locations became visible, or heightened social pressure, where visible location tags altered the dynamics of local discussions. To probe these possibilities, we compare ``in-province'' users—those who comment on posts about their own province—with ``out-of-province'' users—those who comment on posts about other regions. 

Figure~\ref{fig:local2} shows that the steepest decline came from out-of-province users. Their average comments per post fell from 6.10 to 0.42 and remained depressed in the days that followed. In-province users also reduced participation, but the drop was smaller (from 1.85 to 0.15 comments per post) and engagement rebounded quickly. This pattern indicates that the contraction of local discussions was driven mainly by the withdrawal of outsiders. The composition of comments also shifted. Figure~\ref{fig:local3} reveals that the share of critical or dissenting comments on local posts declined significantly once IP locations were displayed, largely because out-of-province users—who had previously contributed disproportionately to criticism—stopped commenting.

These findings make a chilling effect imposed by the state an unlikely primary explanation. If fear of repression were the main driver, we would expect users to be most cautious when commenting on affairs in their own province, where risks of being identified and penalized are highest. Instead, we observe the opposite pattern: out-of-province users withdrew, while in-province users continued to participate. Consistent with this interpretation, domestic users did not reduce their engagement on international or national (non-local) issues, including posts that drew critical comments (Figures~\ref{fig:domestic} and~\ref{fig:domestic-noninternational}).

In sum, the policy did little to deter overseas users but substantially dampened domestic discussion of local issues. The sharpest declines came from out-of-province users, who, once their locations became visible, appeared less willing to criticize regions where they did not reside. These results suggest that social pressure rooted in regional divisions, rather than fear of repression, may have played a central role. We turn next to the role of regional identity in shaping online discourse.

\subsection{Location disclosure deepened regional divisions and reduced participation on local issues} 

Why did domestic users—especially those commenting on other provinces—withdraw from local discussions after IP location disclosure? Beyond the chilling effect already considered, we evaluate two additional mechanisms.

The first possibility is concerns of loss of credibility. Location tags might have signaled that out-of-province commenters lacked local knowledge, reducing the perceived legitimacy of their contributions. This interpretation aligns with the policy’s stated rationale of curbing misinformation. If credibility concerns were the driving factor, however, we would expect similar declines among overseas users commenting on domestic issues and domestic users commenting on international issues. Yet neither group reduced engagement (Figure~\ref{fig:overseas}, red line; Figure~\ref{fig:domestic}, blue line). Therefore, credibility alone cannot explain the observed pattern.

\begin{figure}[!htbp]
    \centering
    \begin{subfigure}[b]{0.3\textwidth}
        \centering
        \includegraphics[width=\textwidth]{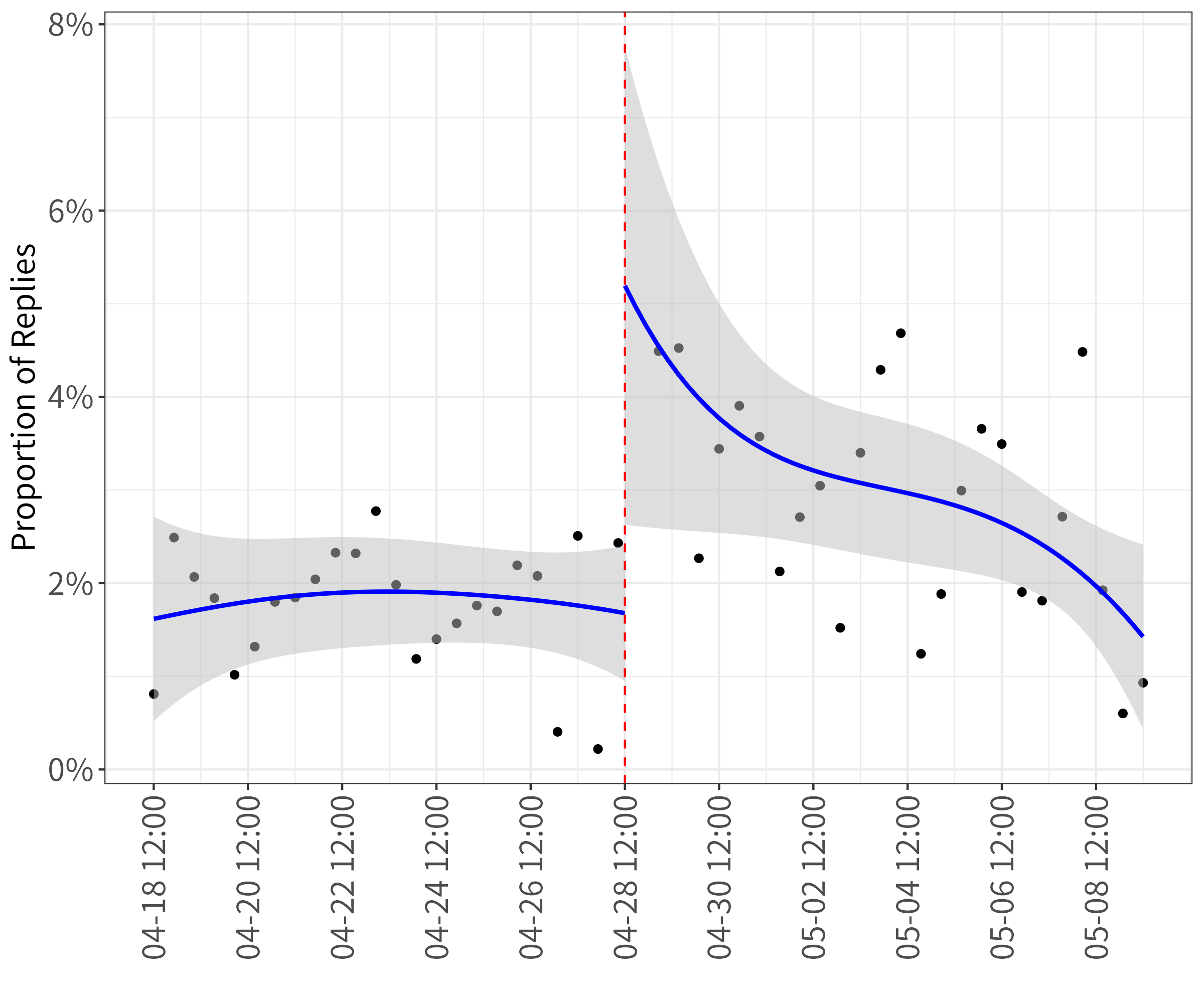}
        \caption{Intensified Regional Discrimination}
        \label{fig:division2}
    \end{subfigure}
    \begin{subfigure}[b]{0.3\textwidth}
        \centering
        \includegraphics[width=\textwidth]{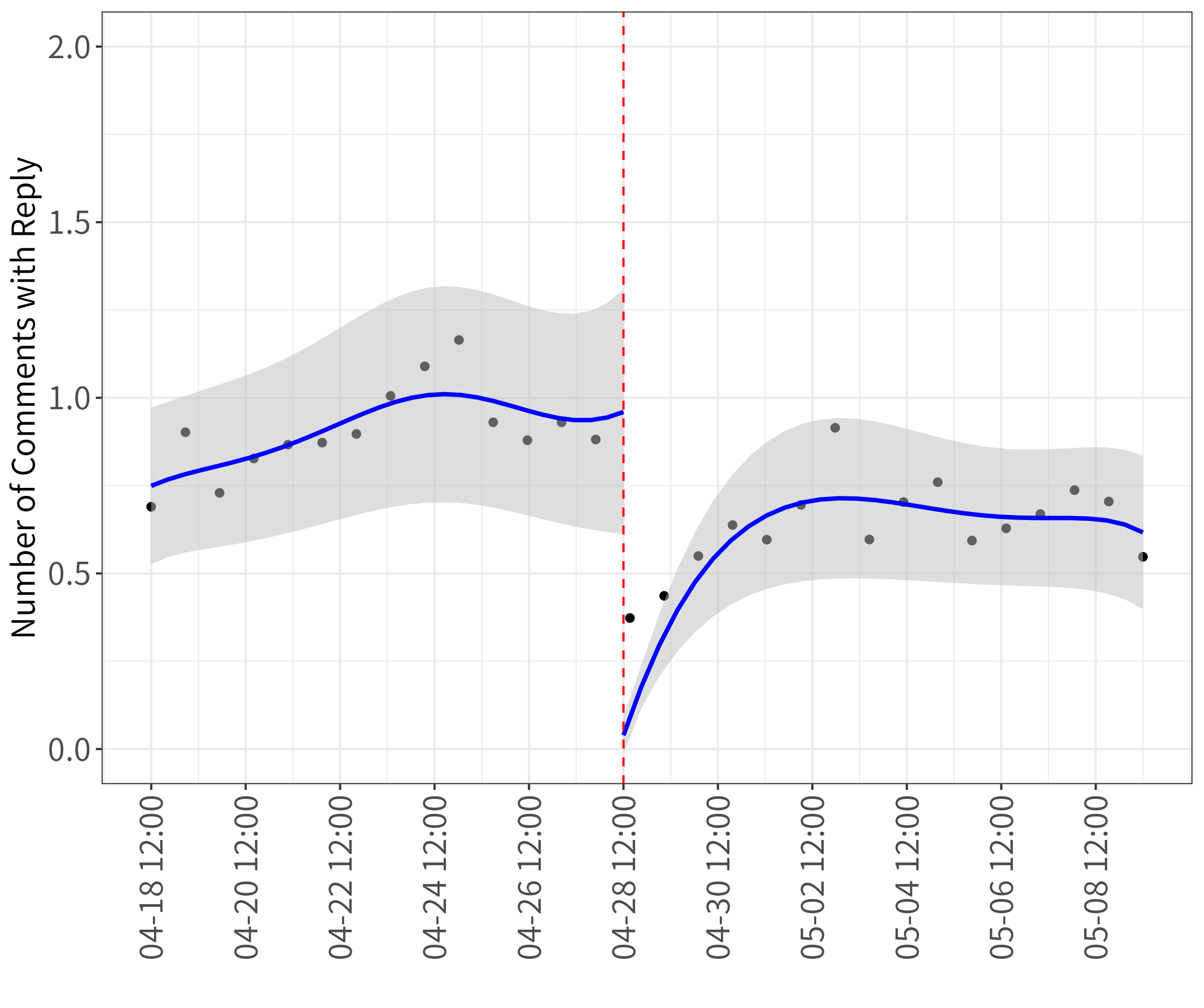}
        \caption{Less Comments with Reply}
        \label{fig:comment_reply}
    \end{subfigure}
    \hspace{0.2cm} 
    \begin{subfigure}[b]{0.3\textwidth}
        \centering
        \includegraphics[width=\textwidth]{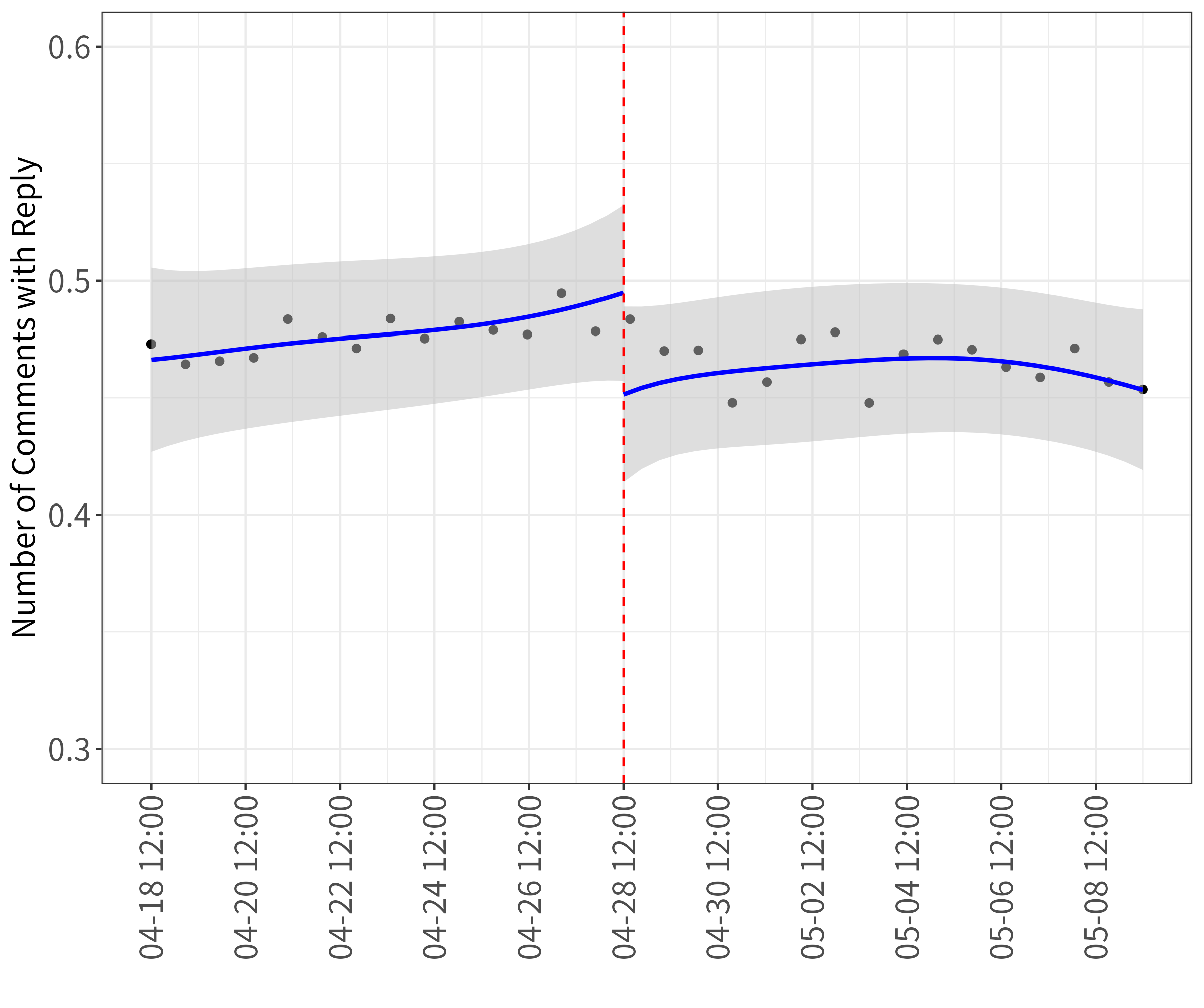}
        \caption{Reduced Coefficient of Variation}
        \label{fig:cov}
    \end{subfigure}
    \hspace{0.2cm} 
    \caption{\textbf{Change in Interactions in the Comment Sections} Figure~\ref{fig:division2} shows a sharp rise in regionally discriminatory replies following the introduction of IP location tags, particularly in response to controversial or critical comments. This suggests that the policy not only reduced cross-regional engagement but also intensified regional antagonism. Figure~\ref{fig:comment_reply} displays a decline in the proportion of comments receiving replies, indicating lower conversational depth and reduced willingness to interact. Figure~\ref{fig:cov} presents the coefficient of variation in comment floor numbers among first-page comments—a measure of comment section dynamism—and shows a noticeable drop after the policy took effect. The red dashed line in each figure marks the timing of the policy rollout. Taken together, the figures document a measurable shift in user behavior and interaction patterns following the introduction of IP-based location disclosure. Confidence intervals are constructed based on cluster-robust standard errors clustered at the account level.}
    \label{fig:pnas_decreased_activity}
\end{figure}

A more plausible explanation is fear of backlash from other users. Location tags made provincial origins visible, heightening the salience of regional identity and turning comment sections into sites of intergroup tension. Before the policy, criticism of local affairs often appeared anonymous or broadly directed. Afterward, even neutral observations could be read as attacks from outsiders. The case study in Section~H of the Supplementary Information illustrates this dynamic: a Fujian user commenting on population decline in the northeast invoked stereotypes about ``mafia'' in Heilongjiang, Liaoning, and Jilin, prompting retaliatory replies about fraud and scams in Fujian. Location tags thus transformed casual remarks into interprovincial confrontations.

To assess whether such exchanges were systemic, we applied an LLM to detect regionally discriminatory language in second-tier replies. Figure~\ref{fig:division2} shows a sharp increase in discriminatory replies after the rollout, driven almost entirely by cross-provincial interactions (Figure~G25, SI). Within-province exchanges remained stable. These results indicate that the policy deepened regional divisions and raised the reputational costs of participating in discussions about other regions.

This dynamic helps explain the broader decline in comment activity on local posts. As regional identity became more visible, users faced greater risks of confrontation—not just for critical remarks but for engaging at all. Even neutral comments were more likely to attract hostile replies, discouraging participation and reducing the likelihood of responses. Figure~\ref{fig:comment_reply} documents this shift: after disclosure, a smaller share of comments attracted replies, signaling thinner, more cautious engagement.

We further measure this contraction in discourse by examining the rotation of first-page comments using the coefficient of variation (CoV) of comment floor numbers, a proxy for dynamism in the comment section. Figure~\ref{fig:cov} shows a clear post-policy drop in CoV. With users more hesitant to comment—especially across provincial boundaries—top comments were displaced less often, producing a more static and less interactive hierarchy, hence the drop in the CoV.

In sum, location disclosure not only reduced overall engagement but also reshaped its structure. By making regional identity salient, it elevated the risks of cross-provincial participation, curtailed dissent, and produced a narrower, more fragmented conversation.

\section{Discussion}\label{sec:conclusion}

Our study examined the impact of Weibo’s location disclosure policy, focusing on how the abrupt introduction of IP-based location tags reshaped online engagement. The policy, framed as a tool to curb misinformation and foreign influence, did little to deter overseas users. Instead, it curtailed domestic participation in discussions of local issues, especially critical engagement by out-of-province users. Evidence suggests that this decline was not driven by fear of repression but by heightened exposure to regionally discriminatory backlash once users’ provincial origins became visible.

These findings highlight how authoritarian regimes can shape discourse without direct censorship. By embedding identity disclosure into platform design, authorities enable peer-based sanctions that suppress participation and narrow the scope of expression. Which such mechanisms may stabilize discourse in the short run, they risk muting early signals of discontent, with potential costs for the quality of governance \citep{chen2017information}.

Several limitations qualify our results. First, the analysis focuses on highly visible comment sections of government and media accounts, which may not reflect how users behave in smaller or less monitored spaces. Second, although our high-frequency collection reduces survivorship bias, we did not capture other forms of engagement such as reposting. In addition, IP tags provide only province- or country-level origins, limiting our ability to study how other identity dimensions—such as gender, class, or ethnicity—intersect with location in shaping discourse.

The scope of inference is also bounded by the sudden, unannounced rollout of the policy in an authoritarian setting. The causal effects we are able to identify are immediate behavioral shifts; longer-term adaptations may differ as users adjust or migrate to alternative venues. Future work could explore whether similar dynamics emerge in democratic contexts, where censorship risks are lower but identity cues may still intensify polarization, and extend analysis to other forms of identity disclosure beyond geography.

Nevertheless, our findings contribute to a growing literature on platform architecture and user behavior. Prior research shows that design features such as algorithmic curation, identity labeling, and credibility signals affect what users see, share, and trust \citep{guess2023social, liang2022effects, nassetta2020state, papak2022algorithms, bradshaw2024look}. We add to this work by showing that identity cues alone can alter participation even without removing content or correcting information. This underscores how design decisions—whether in democracies or authoritarian states—shape the boundaries of public discourse in subtle but consequential ways.

\section{Methods}

This study draws on a high-frequency dataset and a quasi-experimental design to assess the causal impact of Weibo’s abrupt rollout of IP-based location tags. The main challenges are capturing genuine user behavior in a tightly censored environment and separating the effect of the policy from other influences. We address these by constructing a real-time monitoring system that tracked posts and comments across the policy change and by applying an interrupted time series design that exploits the unanticipated timing of implementation.

\subsection{Data and Measurement}

We collaborated with industry experts to design a real-time data collection system that monitored 165 prominent Weibo accounts affiliated with the Chinese government. These include official accounts of government offices and state media outlets—such as newspapers, television stations, and major media outlets—identified in a series of reports on institutional influence in 2021.\footnote{These reports include multiple editions of Government Weibo Influence Reports (政务微博影响力报告), Sina Government Weibo Reports (新浪政务微博报告), and Weibo Trending Topics Data Reports (微博热搜榜数据报告). See Section B in the Supplementary Information for a full list of accounts we collected data from.} We focused on these accounts because, following the government's 2013 crackdown on key opinion leaders \citep{pan2025disguised}, the comment sections of these accounts—due to high traffic and visibility—have become important spaces for public expression and policy discussion. These accounts reach large audiences and maintain high engagement, making them well-suited for observing shifts in public sentiment. Our aim is to capture both the content posted by these accounts and the real-time public engagement their posts generate.

\paragraph{High-frequency Weibo Data} 

We monitored these 165 government and media accounts, all verified with a blue checkmark and ranked by follower count and user traffic. They include national and provincial government bodies (e.g., \emph{China Police Online}, \emph{Shanghai Release}) and major media outlets (e.g., \emph{Beijing Daily}, \emph{Global Times}). On average, each account had 12.7 million followers, and each post reached 419 thousand views. Our account selection spans both national and subnational levels, enabling us to observe patterns across a range of geographic and institutional contexts.

We conducted real-time monitoring of each account's timeline (i.e., their posts), each post's first-page comment section (displaying up to the top 20 comments), and associated engagement metrics at both the post and comment levels. We focus on the first-page comments because they were instantly available to readers when a post is clicked and draw the most user attention and interaction. As shown in Figure~\ref{fig:timeline}, our system scanned all monitored accounts, as well as recorded posts and comments, every five minutes. This allowed us to collect longitudinal data at both the account and post levels. When a new post appeared, the system initiated tracking and began recording data on the post and its top comments. Snapshots were captured every five minutes for the first 24 hours, and once every day for the following ten days. While most engagement occurred within the first few hours (see Figure~B3 in the Supplementary Information), the extended tracking window enabled us to capture delayed or atypical activity. Figure~\ref{fig:variables} illustrates the variables collected through scraping, including each post's content, publication time, and engagement metrics (number of reposts, likes, and comments), as well as the first-page of the comment section (top 20 comments). For each comment and their replies, we also recorded the associated IP location tag. Figure~\ref{fig:data_new_posts} and Figure~\ref{fig:data_new_comments} show the numbers of new posts, and new comments collected within a 48-hour window, respectively. Figures~\ref{fig:2nd_page_likes} and~\ref{fig:2nd_page_replies} show that, on average, comments appearing on and after the second page receive minimal engagement. 

\paragraph{Timeline and Location Tags} 
Our data span from April 18 to May 9, 2022. Although data collection was initially designed for a different research purpose, the sudden rollout of the user location disclosure policy at noon on April 28, 2022, created an unexpected but valuable opportunity to study its impact. Because monitoring began 11 days earlier, we have a rich set of pre-treatment observations. The policy was implemented just 16 minutes after its announcement, reducing the likelihood of anticipatory behavior and improving the credibility of causal identification.

The location tags reflect users' IP addresses at the time of commenting, although we do not observe or record the IPs themselves. Overseas users are labeled by country or region, while domestic users are tagged at the provincial level (e.g., Shanghai, Jiangsu, Shaanxi).\footnote{A potential concern is that users may mask their IP addresses using VPNs. in the Supplementary Information, we show that fewer than 1\% of users changed location tags during the study period. VPN usage among domestic users is rare and typically serves to switch between domestic and foreign IPs, not between provinces. To our knowledge, commercially available VPN services in China do not support intra-national IP masking, as all provinces operate within the same national firewall. These patterns suggest that provincial-level location tags are generally reliable. Excluding users whose location tags have changed does not alter on our findings.} 

A distinctive feature of our dataset is that it includes location tags for comments posted both before and after the policy was implemented, even though the regulation did not require platforms to apply these tags retroactively. For roughly 12 hours following the rollout, Weibo briefly displayed location tags on historical comments—an implementation glitch we captured in real time through continuous monitoring. This rare window enabled us to recover ``pre-treatment'' geographic information on commenters, which we analyze in later sections.

\paragraph{Key variables and Measurement}

The primary outcome of interest is the number of unique first-page comments (up to 20 top comments) per post during the observation window, used as a proxy for the vibrancy of the comment section. Since top comments are updated dynamically based on engagement, we aggregate all unique first-page comments captured across multiple snapshots over the eleven-day tracking window following each post. Focusing on first-page comments enables us to differentiate between user types—for example, overseas versus domestic users, and among domestic users, between in-province and out-of-province commenters. While we also examine aggregate metrics such as the total number of comments, we lack access to user location tags and comment content beyond the first page.

\begin{table}[htbp]
 \centering
  \caption{Measures / Variables Used in the Main Analyses}
  \label{tab:measures}\small
  \begin{tabular}{p{7.5cm}p{1.5cm}p{3.5cm}p{1.3cm}}
    \toprule
    Variable Definition & Level & Measurement Method & F1 Score \\
    \midrule
    Numbers of total comments \& unique top comments & Post & - & - \\[2pt]
    Numbers of unique top comments by overseas and domestic commenters  & Post & Based on commenters' IP location tags  & - \\[2pt]
    Numbers of unique top comments by in-province or out-of-province (excluding overseas)   & Post & Based on matches between post content and commenters' IP location tags  & - \\[2pt]
    Post content: international or non-international issue & Post & Dictionary method (123 countries/regions) & - \\[2pt]
    Post content: local or non-local issue & Post & Dictionary method (31 provinces) & - \\[2pt]
    Post content: positive, negative, or neutral event & Post & LLM‑assisted annotation + human validation & $0.721$ \\[2pt]
    Coefficient of variation (CoV) of top comments & Post & Formula & - \\[2pt]
    Comment stance: critical of the post or not & Comment  & Supervised BERT classifier + human validation &   $0.783$ \\[2pt]
    Comment receiving regionally discriminatory replies & Comment & LLM‑assisted annotation + human validation & $0.835$ \\[2pt]
    \bottomrule
  \end{tabular}
\end{table}

In addition to comment volume, we calculate the coefficient of variation (CoV) of the comments' floor numbers to capture the dynamism of the comment section. A floor number reflects the order in which a comment was posted, with smaller numbers indicating earlier submissions. We calculate the CoV as the standard deviation of the floor numbers of top comments divided by their mean:
$\text{CoV} = \sqrt{\frac{\sum_{i=1}^n (x_i - \mu)^2}{n}} \bigg/ \frac{\sum_{i=1}^n x_i}{n}$.
A higher CoV indicates more dynamic discussions, where later comments gain prominence through sustained engagement. In contrast, lower CoV values suggest static discussions dominated by early comments.

We also classify post types based on their content. International posts are those that mention one of 123 countries or regions outside China. We construct the list of 123 countries and regions (see Table~D4 in the Supplementary Information) by extracting all foreign country and region names that appeared in Weibo's post-treatment IP location tags. Local posts reference one of China's 31 provinces. We focus on provincial references, rather than prefectures or counties, because IP location tags are reported at the country level for overseas users and the provincial level for domestic users. Posts are further categorized as depicting a positive, negative, or neutral event (from the perspective of the Chinese government) using a large language model (LLM), followed by human validation.

To capture the substance of user engagement, we assess whether a comment is critical or supportive of the post's message using a supervised Bidirectional Encoder Representations from Transformers (BERT) classifier trained on 10,000 hand-labeled post-comment pairs. Classification takes into account the specific stance expressed in relation to the topic and the government's known position. This approach allows us to distinguish critical from supportive comments, even when the semantic tone is ambiguous (e.g., sarcastic praise or angry agreement).

To detect regional discrimination in replies, we use LLM-based annotation, validated by human coders, to identify whether a reply contains discriminatory content based on the commenter's location. We also track the frequency with which a user's IP location is mentioned in replies as an indirect measure of the salience of geographic identity in the discourse. Table~\ref{tab:measures} summarizes these variables and their corresponding measurement strategies. Table~C3 in the Supplementary Information provides the summary statistics.

\paragraph{Protecting Privacy} All data used in this study are publicly available. Nevertheless, we take additional steps to protect the privacy of commenters, many of whom are likely ordinary citizens in China. First, we do not collect any personally identifiable information, such as commenters' IP addresses. The location tags we use are limited to the provincial level and are too coarse to identify individuals. Second, we mask all user identifiers—including usernames, avatars, and URLs—by replacing them with randomly assigned IDs to ensure complete anonymity. Third, all data are stored on a secure server at the corresponding author's institution, with access restricted to authorized personnel only (see Table~B2 in the Supplementary Information for details.)

\subsection{Causal Identification}

Our identification strategy is an interrupted time series design. The unit of analysis is the post. Let $Y_{it}(1)$ and $Y_{it}(0)$ denote the potential outcomes for post $i$ at time $t$ under the treatment condition (the user location disclosure policy) and the control condition (absence of such a policy), respectively. Define $\mu_1(t) := \mathbb{E}[Y_{it}(1)]$ and $\mu_0(t) := \mathbb{E}[Y_{it}(0)]$ as the expected outcomes under each regime. Let~$c$ denote the time of policy implementation—noon on April 28, 2022. Our main estimand is the difference in expected outcomes at the time of the policy:
\vspace{-1em}\begin{center}
$\tau_{\text{SAITT}} := \mu_1(c) - \mu_0(c),$
\end{center}\vspace{-1em}
which captures the sample average instant treatment effect of the policy at the cutoff. Identification relies on two standard assumptions: (a) no anticipation, and (b) smoothness of $\mu_0(t)$ and $\mu_1(t)$ in a neighborhood around $c$ \citep{Linden2015, Bernal2017}. Given that the policy was not announced in advance and a large volume of posts appeared during the transition, both assumptions are plausible in our context.

Because an interrupted time series design is analogous to a regression discontinuity design with time as the forcing variable \citep{hausman2018regression}, we estimate $\tau_{\text{SAITT}}$ using local linear regression on either side of $c$, clustering standard errors at the account level. Bandwidths are selected using both the MSE-optimal procedure in \texttt{rdrobust} \citep{Calonico2014} and cross-validation. We show in the Supplementary Information Section~F that our results are robust to a wide range of bandwidth choices.

\paragraph{Robustness checks} 

We conduct a series of robustness checks, detailed in the Supplementary Information, to verify the stability of our main findings on the decline in comment volume due to the implementation of the user location disclosure policy. First, we vary the bandwidths used in the local smoothing, including those selected by the \texttt{rdrobust} package, bandwidths chosen via cross-validation, and a range of fixed windows (8, 24, 72, 96, and 120 hours). In all cases, the observed decline in comment volume remains consistent. Second, we re-estimate the models using a log-transformed outcome, $\log(\#comments + 1)$, to assess sensitivity to outliers. While the main analyses use the number of unique comments to accommodate posts with zero comments, the log transformation yields qualitatively similar results. Finally, we perform placebo tests by shifting the timing of the policy intervention to alternative dates. These placebo specifications show no discontinuities, reinforcing that the observed effects are indeed tied to the actual timing of the location disclosure policy.

\clearpage

\bibliographystyle{unsrtnat}
\bibliography{references}  

\clearpage
\thispagestyle{empty} 
\begin{center}
\vspace*{0.3\textheight} 
{\Huge \textbf{Supplementary Information}} 
\end{center}
\clearpage
\appendix
\renewcommand{\thepage}{S\arabic{page}} 
\setcounter{page}{1} 
\renewcommand{\thesection}{S\arabic{section}}
\renewcommand{\thesubsection}{S\arabic{section}.\arabic{subsection}}

\setcounter{section}{0}
\setcounter{figure}{0}
\setcounter{table}{0}

\renewcommand{\thefigure}{S\arabic{figure}}
\renewcommand{\thetable}{S\arabic{table}}

\tableofcontents
\clearpage

\section{Policy Background}

This section presents official announcements and explanations regarding Weibo’s IP location disclosure policy during both the initial trial and full rollout phases. The language used in these statements indicates that the policy was primarily intended to curb said misinformation and reduce the influence of overseas users on domestic discourse by making their IP locations visible to others on the platform. The timing and framing of the announcements align with the policy’s stated goal of deterring so-called “malicious” overseas activity, consistent with the account provided in the main text.

\subsection{Weibo policy announcement}

In March 2022, the user location disclosure policy was first introduced on a trial basis, targeting discussion threads related to COVID-19 and the Russo-Ukrainian War. Notably, the example image included in the official announcement highlights users with Ukrainian IP locations—suggesting that overseas commenters were a primary concern. This is illustrated at the bottom of Figure~\ref{fig:policy_announcement_0317} and its English translation in Figure~\ref{fig:policy_announcement_0317_eng}, with a close-up view provided in Figures~\ref{fig:policy_announcement_0317_pic} and~\ref{fig:policy_announcement_0317_pic_eng}.

About a month later, at 11:44 A.M. on April 28, Weibo announced the full rollout of the user location disclosure across all posts.\footnote{\url{https://weibo.com/1934183965/LqvYeCdBu}.} As stated in the announcement (Figures~\ref{fig:policy_announcement_0428} and~\ref{fig:policy_announcement_0428_eng}), the display of user location would be mandatory and ``cannot be turned on or off by users.'' Just 16 minutes later, at noon, the policy was implemented platform-wide.

\begin{figure}[p]  
    \centering
    \begin{subfigure}[b]{0.49\textwidth}
        \centering
        \includegraphics[width=\textwidth]{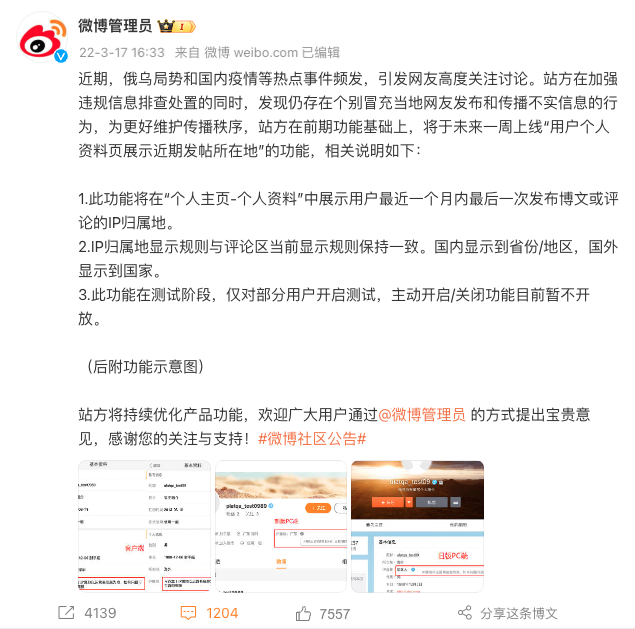}
        \caption{Announcement of Policy Trial}
        \label{fig:policy_announcement_0317}
    \end{subfigure}
    \hspace{0.05cm}
    \begin{subfigure}[b]{0.49\textwidth}
        \centering
        \includegraphics[width=\textwidth]{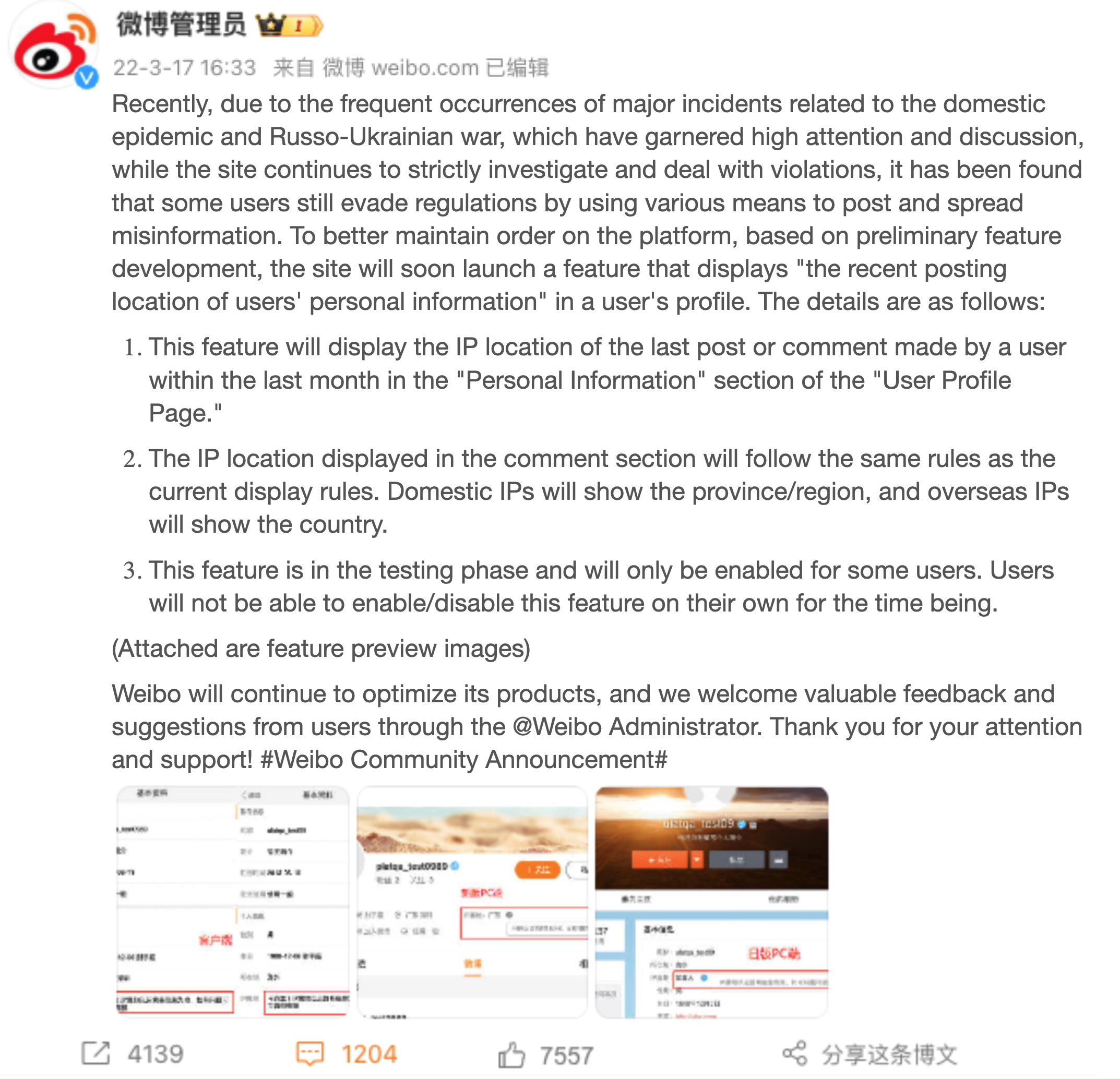}
        \caption{Announcement of Policy Trial (English)}
        \label{fig:policy_announcement_0317_eng}
    \end{subfigure}

    \vspace{0.4cm} 

    \begin{subfigure}[b]{0.48\textwidth}
        \centering
        \includegraphics[width=\textwidth]{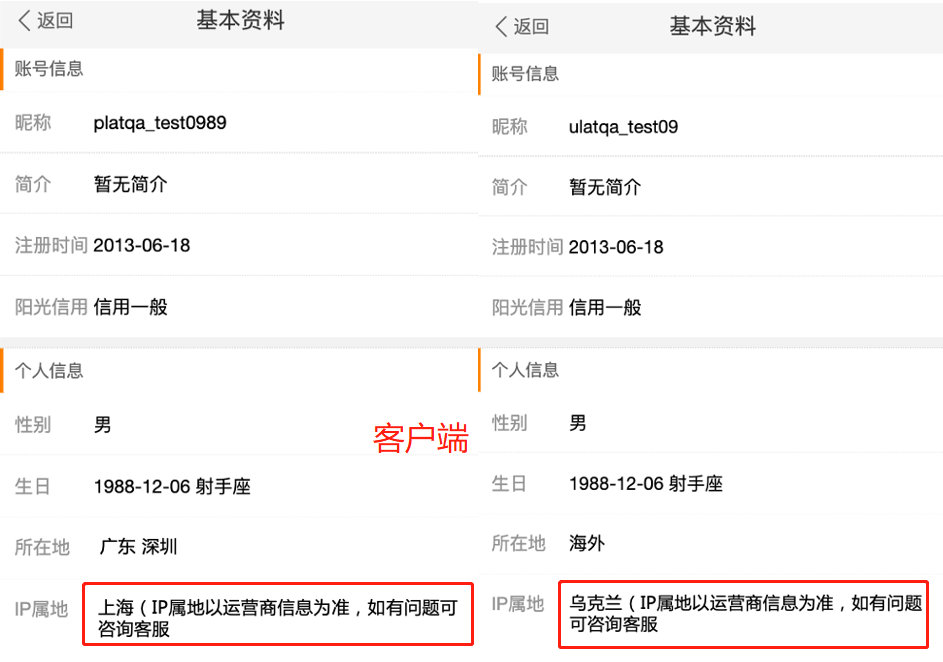}
        \caption{Emphasizing the Ukraine IP}
        \label{fig:policy_announcement_0317_pic}
    \end{subfigure}
    \hspace{0.3cm}
    \begin{subfigure}[b]{0.48\textwidth}
        \centering
        \includegraphics[width=\textwidth]{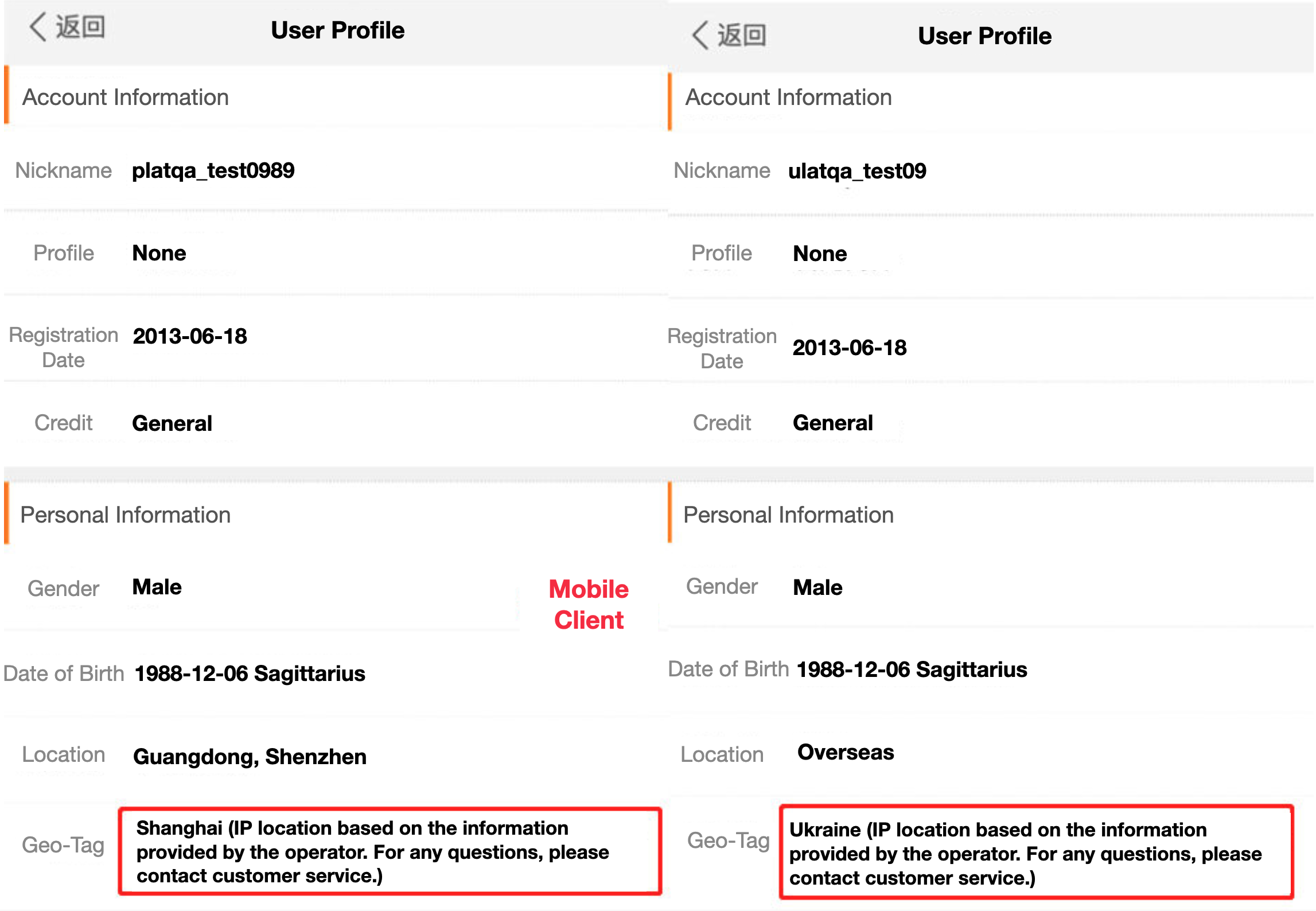}
        \caption{Emphasizing the Ukraine IP (English)}
        \label{fig:policy_announcement_0317_pic_eng}
    \end{subfigure}

    \caption{Policy Announcement (Part 1 of 2)}
\end{figure}

\begin{figure}[p]\ContinuedFloat 
    \centering
    \begin{subfigure}[b]{0.49\textwidth}
        \centering
        \includegraphics[width=\textwidth]{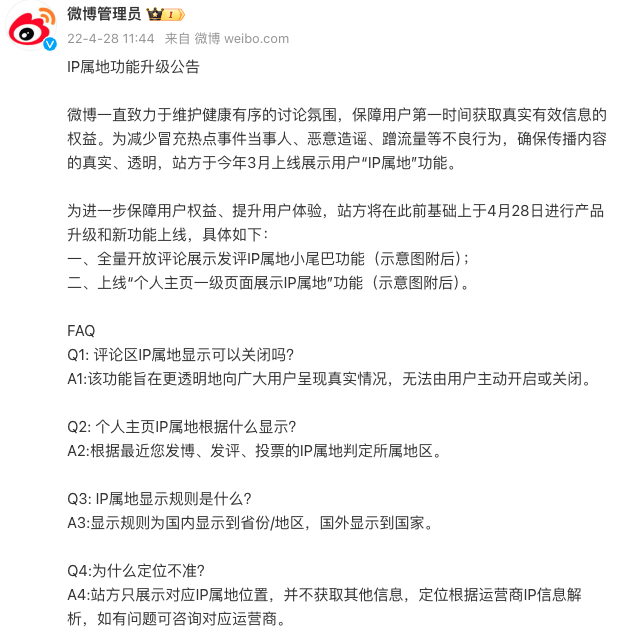}
        \caption{Announcement of Policy}
        \label{fig:policy_announcement_0428}
    \end{subfigure}
    \hspace{0.05cm}
    \begin{subfigure}[b]{0.49\textwidth}
        \centering
        \includegraphics[width=\textwidth]{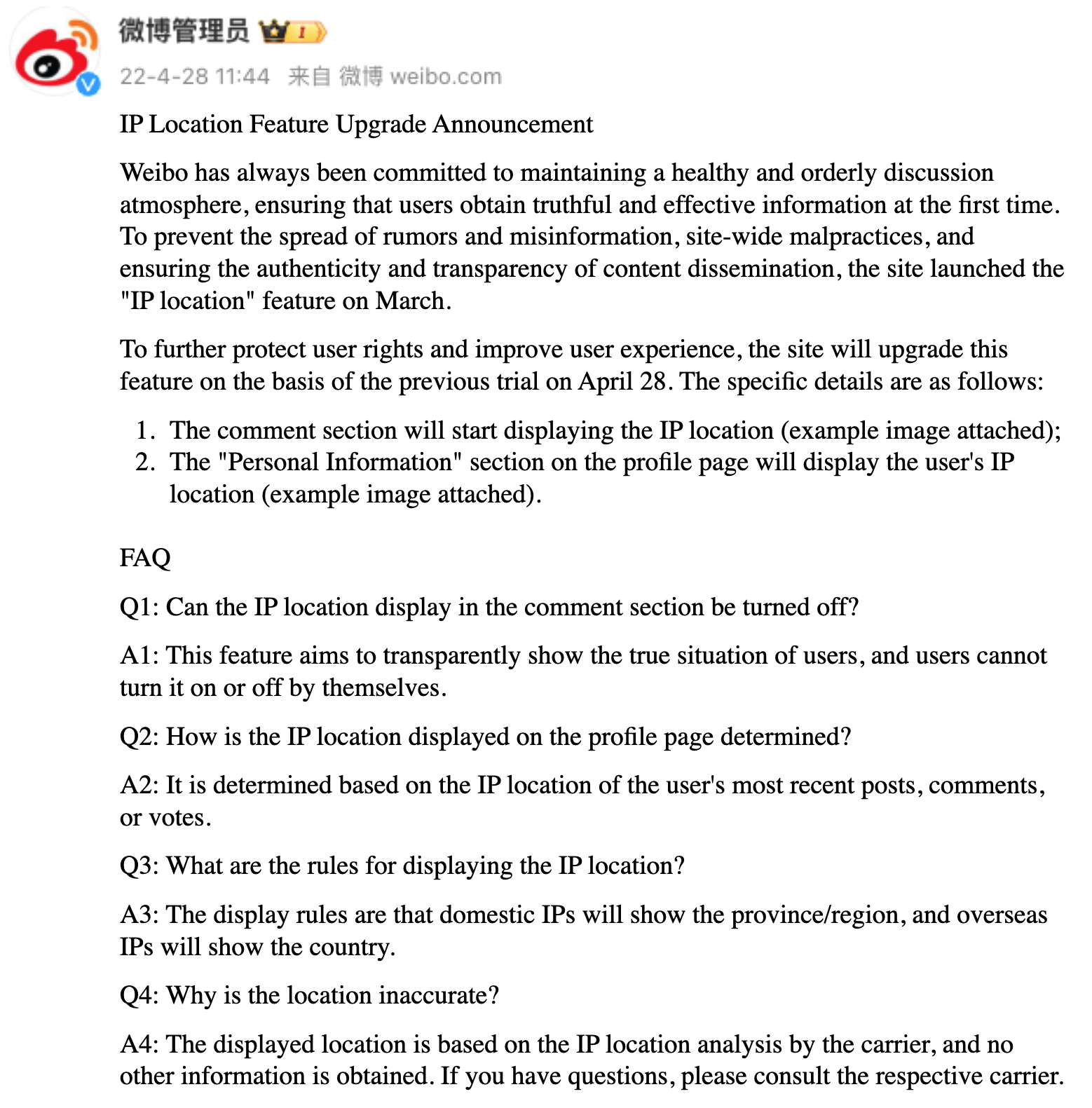}
        \caption{Announcement of Policy (English)}
        \label{fig:policy_announcement_0428_eng}
    \end{subfigure}

    \caption{Policy Announcement (Part 2 of 2)}
    \caption*{\textbf{Notes:} The figures above show official announcements regarding Weibo's user location disclosure. Figures~\ref{fig:policy_announcement_0317} and~\ref{fig:policy_announcement_0317_pic} illustrate the limited trial launched on March 17, 2022, which applied only to discussions about the Russo-Ukrainian War and COVID-19. In the sample image, Weibo explicitly highlighted users with Ukrainian IP addresses using a red rectangle. About a month later, at 11:44 A.M. on April 28, Weibo abruptly announced that the policy would be extended to all posts (Figure~\ref{fig:policy_announcement_0428}) (\url{https://weibo.com/1934183965/LqvYeCdBu}). Just 16 minutes after the announcement, at noon, the IP locations of all Weibo users were made publicly visible across the platform.}
\end{figure}

\subsection{Government press conference}

\begin{figure}[H]
    \centering
    \includegraphics[width=1\textwidth]{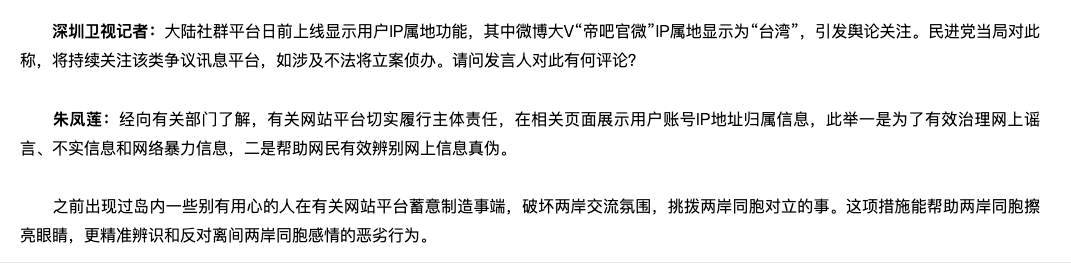}
    \caption{Records Related to User Location Disclosure in Press Conference }
    \label{fig:press_conference}
    \caption*{\textbf{Notes:} The excerpt above is from a May 11, 2022 interview with Zhu Fenglian, deputy director of the Information Bureau of the Taiwan Affairs Office. The translated script is as follows: \textbf{Shenzhen TV reporter}: Recently, a social platform on the mainland has launched a feature showing the IP locality of users. Among these, the IP locality for the Weibo account Di Bar Official'' is displayed as Taiwan,'' which has sparked discussions. The Democratic Progressive Party authorities have stated that they will continue to monitor such controversial information platforms and will initiate investigations if illegal activities are involved. What comments does the spokesperson have on this? \textbf{Zhu Fenglian}: After consulting with relevant departments, I learned that the website platforms are indeed fulfilling their responsibilities by displaying the IP address information of user accounts. This measure serves two purposes: first, to effectively manage online rumors, false information, and cyber violence; and second, to help netizens effectively distinguish the authenticity of online information. Previously, there have been instances where secessionists deliberately incited conflicts on these platforms to disrupt cross-strait exchanges and instigate opposition among compatriots across the strait. This policy will help compatriots from both sides to be more discerning and to recognize and oppose any inferior actions that harm the feelings between the compatriots on both sides of the strait.}
\end{figure}

When asked about the rationale for requiring all social media platforms to display user IP addresses, Zhu Fenglian, deputy director of the Information Bureau of the Taiwan Affairs Office (TAF) of the State Council, declined to give a direct answer during a press conference。\footnote{\url{http://www.gwytb.gov.cn/xwdt/xwfb/xwfbh/202205/t20220511_12435155.htm}.} Instead, she pointed to possible public benefits of the policy, stating that “some people with ulterior motives from Taiwan deliberately create trouble on relevant online platforms, disrupt the atmosphere of cross-strait exchanges, and provoke opposition among compatriots on both sides of the Strait. This measure can help compatriots sharpen their eyes, better identify, and oppose malicious actions that damage cross-strait relations.'' The full exchange is shown in Figure~\ref{fig:press_conference}.

\subsection{News coverage}

The user location disclosure on social media platforms like Weibo is designed to deter deceptive practices by overseas users. Social media companies are required to authenticate users' identities, and the display of IP-based location tags serves as a signal of external influence masquerading as domestic opinion. By publicly revealing user locations, the policy aims to curb the spread of misinformation and enhance transparency in online discourse. Numerous official commentaries have discussed the rationale and consequences of this policy. For example, one article notes, This functionality reveals the actual locations of users when they post or comment online, helping to unmask and debunk misinformation spread by foreign entities pretending to be local voices'' (Source: Huanqiu.com \url{https://m.huanqiu.com/article/47uUWruYUSR}). Another article from the \emph{People's Daily} states, The implementation of the IP display feature has uncovered that some accounts, which appeared to be local, were actually operated from abroad, falsifying their involvement in domestic issues.'' These reports underscore how the user location disclosure is framed as a tool to safeguard the integrity of online discourse by exposing foreign interference.

\clearpage

\section{Information on Data Collection}

Given the strict censorship on Chinese social media and our need to observe short-term changes in user behavior, we aimed to collect pre-censorship data in real time.

\subsection{List of monitored accounts}

We monitored the timelines of over 200 official Weibo accounts in real time, spanning various levels of Chinese government bodies and media outlets. Table~\ref{table:account_list} lists the 165 accounts included in this study, along with their corresponding English names.

{\tiny
\renewcommand{\arraystretch}{1.5} 
\begin{longtable}{|l|l|l|l|}
\caption{List of government-affiliated Weibo accounts being monitored} \label{table:account_list} \\
\hline
\textbf{Screen Name} & \textbf{English Name} & \textbf{Screen Name} & \textbf{English Name} \\ \hline 
\endhead
\multicolumn{4}{|c|}{\textbf{National Commercial Media}} \\ \hline
澎湃新闻 & The Paper & 封面新闻 & The Cover \\ \hline
每日经济新闻 & National Business Daily & 封面西洋镜 & Cover Xiyangjing \\ \hline
红星新闻 & Red Star News & 头条新闻 & Toutiao News \\ \hline
沸点视频 & Hot Video & 大米Video & Rice Video \\ \hline
荔枝新闻 & Litchi News & 海客新闻 & Haike News \\ \hline
新浪财经 & Sina Finance & 南方周末 & Southern Weekly \\ \hline
财经网 & caijing.com & 南方都市报 & Southern Metropolis Daily \\ \hline
凤凰网 & ifeng.com & 新京报 & The Beijing News \\ \hline
新浪新闻 & Sina News & 紧急呼叫 & Emergency Call \\ \hline
梨视频 & Pear Video & 新京报我们视频 & Beijing News Video \\ \hline

\multicolumn{4}{|c|}{\textbf{National Official Media}} \\ \hline
人民日报 & People's Daily & 中新视频 & China News Video \\ \hline
人民网 & People's Daily Online & 环球资讯 & Global Information \\ \hline
人民日报海外版-海外网 & People's Daily Overseas Edition & 观察者网 & Guancha.cn \\ \hline
新华社 & Xinhua News Agency & 观察者网微丢视频 & Guancha.cn Micro-drop Video \\ \hline
新华网 & Xinhuanet & 中国新闻周刊 & China Newsweek \\ \hline
央视新闻 & CCTV News & 财新网 & Caixin Online \\ \hline
央视网 & CCTV.com & 中国青年报 & China Youth Daily \\ \hline
央视网快看 & CCTV Quick Look & 检察日报 & Procuratorate Daily \\ \hline
央视财经 & CCTV Finance & 人民公安报 & People's Public Security Daily \\ \hline
央视频 & CCTV Video & 中国气象科普 & Meteorological Science \\ \hline
环球时报 & Global Times & 科技日报 & Science and Technology Daily \\ \hline
环球网 & huanqiu.com & 中国新闻网 & China News Service \\ \hline

\multicolumn{4}{|c|}{\textbf{National Government}} \\ \hline
中国警方在线 & China Police Online & 最高人民法院 & Supreme People's Court \\ \hline
共青团中央 & Central Committee of the CYL & 中国地震台网速报 & China Earthquake Networks \\ \hline
中国长安网 & China Chang'an Web & 中国警察网 & China Police Network \\ \hline
中国消防 & China Fire and Rescue & 中国妇女报 & China Women's News \\ \hline
中国反邪教 & China Anti-Cult Network & 中青报-温暖的BaoBao & China Youth Daily - Warm BaoBao \\ \hline
中国禁毒在线 & China Drug Control Online & 正义网 & jcrb.com \\ \hline
中国气象局 & China Meteorological Administration & 人民法院报 & People's Court Daily \\ \hline
最高人民检察院 & Supreme People's Procuratorate & 交通发布 & Transportation Release \\ \hline
中国交通 & China Transportation & 全国妇联女性之声 & All-China Women's Federation \\ \hline
国资小新 & SASAC Xiaoxin & 公安部刑侦局 & Criminal Investigation Bureau of MPS \\ \hline
战略安全与军控在线 & Strategic Security and Arms Control Online & 国家反诈中心 & National Anti-Fraud Center \\ \hline
公安部交通管理局 & Traffic Management Bureau of MPS & 应急管理部 & Ministry of Emergency Management \\ \hline
中国历史研究院 & Chinese Academy of History & 健康中国 & Healthy China \\ \hline

\multicolumn{4}{|c|}{\textbf{Regional Commercial Media}} \\ \hline
钱江晚报 & Qianjiang Evening News & 都市快报 & City Express \\ \hline
极目新闻 & Jimu News & 我苏特稿 & Wo Su Te Gao \\ \hline
成都商报 & Chengdu Economic Daily & 时间视频 & Time Video \\ \hline

\multicolumn{4}{|c|}{\textbf{Regional Official Media}} \\ \hline
广州日报 & Guangzhou Daily & 湖北日报 & Hubei Daily \\ \hline
北京日报 & Beijing Daily & 湖南日报 & Hunan Daily \\ \hline
四川观察 & Sichuan Observer & 吉林日报 & Jilin Daily \\ \hline
陕视新闻 & Shaanxi TV News & 新华日报 & Xinhua Daily \\ \hline
解放日报 & Jiefang Daily & 江西日报 & Jiangxi Daily \\ \hline
天津日报 & Tianjin Daily & 辽宁日报 & Liaoning Daily \\ \hline
重庆日报 & Chongqing Daily & 内蒙古日报 & Inner Mongolia Daily \\ \hline
安徽日报 & Anhui Daily & 青海日报 & Qinghai Daily \\ \hline
福建日报 & Fujian Daily & 大众日报 & Dazhong Daily \\ \hline
甘肃日报 & Gansu Daily & 山西日报 & Shanxi Daily \\ \hline
南方日报 & Nanfang Daily & 陕西日报 & Shaanxi Daily \\ \hline
贵州日报官微 & Guizhou Daily Official Weibo & 四川日报 & Sichuan Daily \\ \hline
海南日报 & Hainan Daily & 西藏日报 & Tibet Daily \\ \hline
河北日报 & Hebei Daily & 浙江日报 & Zhejiang Daily  \\ \hline
河南日报 & Henan Daily & 云南日报 & Yunnan Daily \\ \hline
黑龙江日报 & Heilongjiang Daily & 广西日报 & Guangxi Daily \\ \hline
宁夏日报 & Ningxia Daily & & \\ \hline

\multicolumn{4}{|c|}{\textbf{Regional Government}} \\ \hline
河北新闻网 & Hebei News Network & 重庆发布 & Chongqing Release \\ \hline
山西发布 & Shanxi Release & 四川共青团 & Sichuan Communist Youth League \\ \hline
辽宁发布 & Liaoning Release & 江苏共青团 & Jiangsu Communist Youth League \\ \hline
吉林发布 & Jilin Release & 青春山东 & Youth Shandong \\ \hline
黑龙江发布 & Heilongjiang Release & 浙江团省委 & Zhejiang Provincial Committee of the CYL \\ \hline
微博江苏 & Weibo Jiangsu & 安徽团省委 & Anhui Provincial Committee of the CYL \\ \hline
浙江发布 & Zhejiang Release & 广州共青团 & Guangzhou Communist Youth League \\ \hline
安徽发布 & Anhui Release & 三秦青年 & Sanqin Youth \\ \hline
福建发布 & Fujian Release & 河南共青团 & Henan Communist Youth League \\ \hline
江西发布 & Jiangxi Release & 青春上海 & Youth Shanghai \\ \hline
山东发布 & Shandong Release & 青春北京 & Youth Beijing \\ \hline
精彩河南 & Wonderful Henan & 甘肃共青团 & Gansu Communist Youth League \\ \hline
湖北发布 & Hubei Release & 青春江西 & Youth Jiangxi \\ \hline
这里是湖南 & This is Hunan & 云南共青团 & Yunnan Communist Youth League \\ \hline
广东发布 & Guangdong Release & 青年湖南 & Youth Hunan \\ \hline
四川发布 & Sichuan Release & 青春湖北 & Youth Hubei \\ \hline
这里是贵州 & This is Guizhou & 辽宁共青团 & Liaoning Communist Youth League \\ \hline
云南发布 & Yunnan Release & 黑龙江共青团 & Heilongjiang Communist Youth League \\ \hline
陕西发布 & Shaanxi Release & 广西共青团 & Guangxi Communist Youth League \\ \hline
甘肃发布 & Gansu Release & 津彩青春 & Colorful Youth Tianjin \\ \hline
青海发布 & Qinghai Release & 共青团福建省委 & Fujian Provincial Committee of the CYL \\ \hline
活力内蒙古 & Lively Inner Mongolia & 河北共青团 & Hebei Communist Youth League \\ \hline
贵州共青团 & Guizhou Communist Youth League & 山西共青团 & Shanxi Communist Youth League \\ \hline
西藏共青团 & Tibet Communist Youth League & 重庆共青团 & Chongqing Communist Youth League \\ \hline
新疆发布 & Xinjiang Release & 内蒙古团委 & Inner Mongolia League Committee \\ \hline
北京发布 & Beijing Release & 吉林共青团 & Jilin Communist Youth League \\ \hline
天津发布 & Tianjin Release & 宁夏共青团 & Ningxia Communist Youth League \\ \hline
上海发布 & Shanghai Release & 海南共青团 & Hainan Communist Youth League \\ \hline
\end{longtable}}

\clearpage

\subsection{Temporal patterns of engagement on Weibo}

\begin{figure}[h]
    \centering
    \caption{Change of Engagements Since Published}
    \label{fig:temporal}
    \begin{subfigure}[b]{0.4\textwidth}
        \centering
        \includegraphics[width=\textwidth]{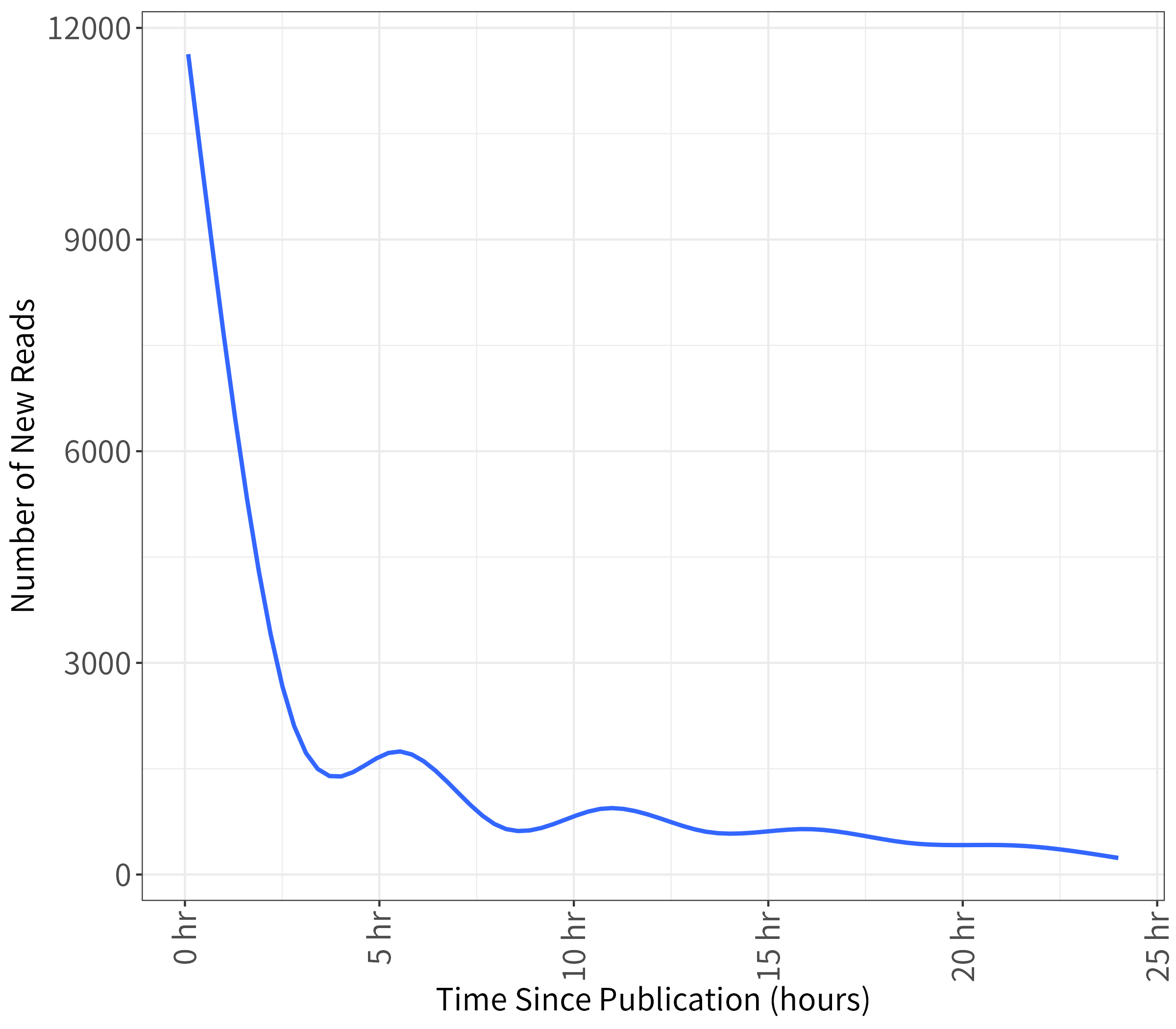}
        \caption{Number of New Reads}
        \label{fig:pnas_volume_reads_change}
        \end{subfigure}
    \hspace{0.1cm} 
        \begin{subfigure}[b]{0.4\textwidth}
        \centering
        \includegraphics[width=\textwidth]{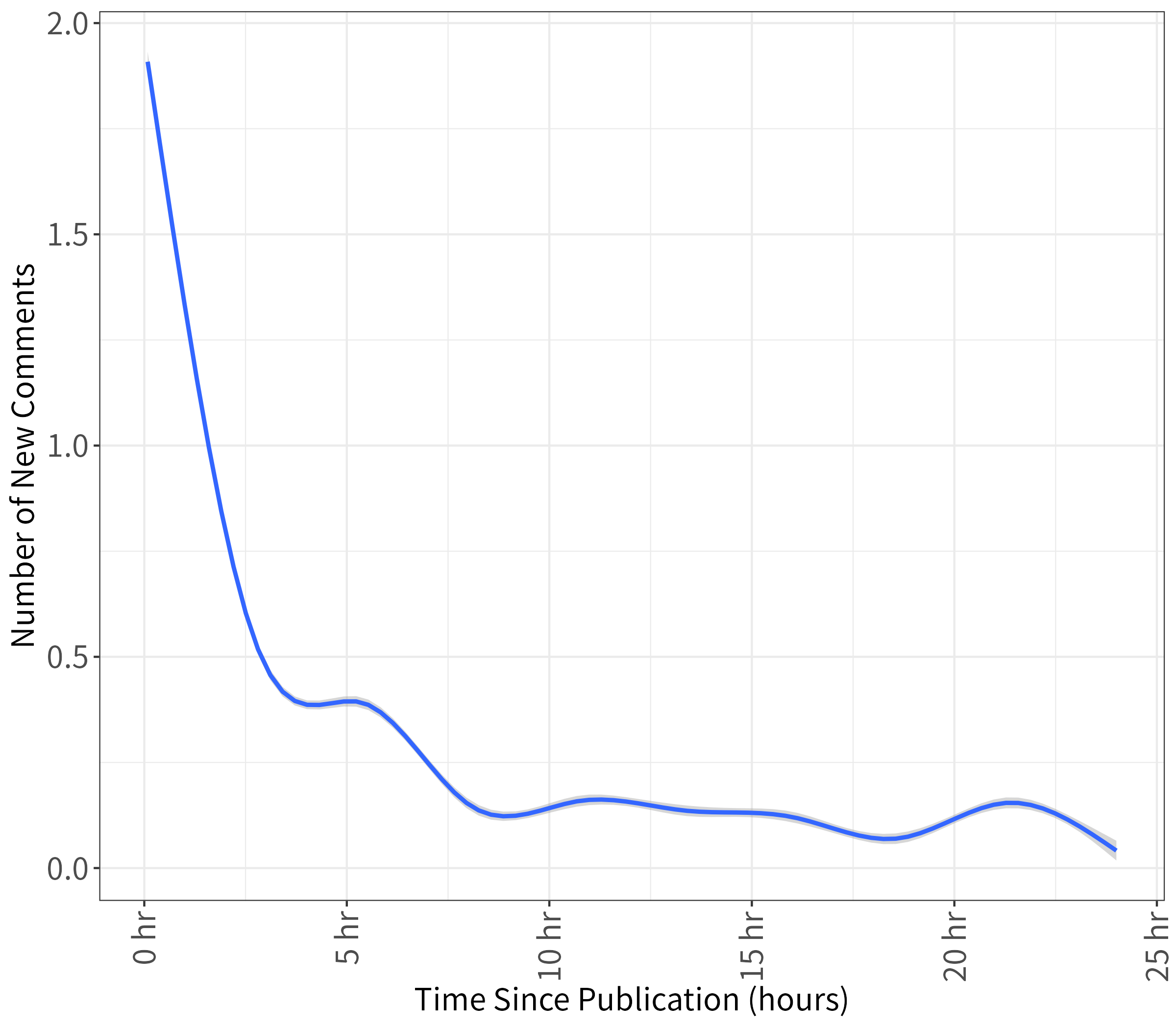}
        \caption{Number of New Comments}
    \label{fig:pnas_volume_comments_change}
        \end{subfigure}\\
        \begin{subfigure}[b]{0.4\textwidth}
        \centering
        \includegraphics[width=\textwidth]{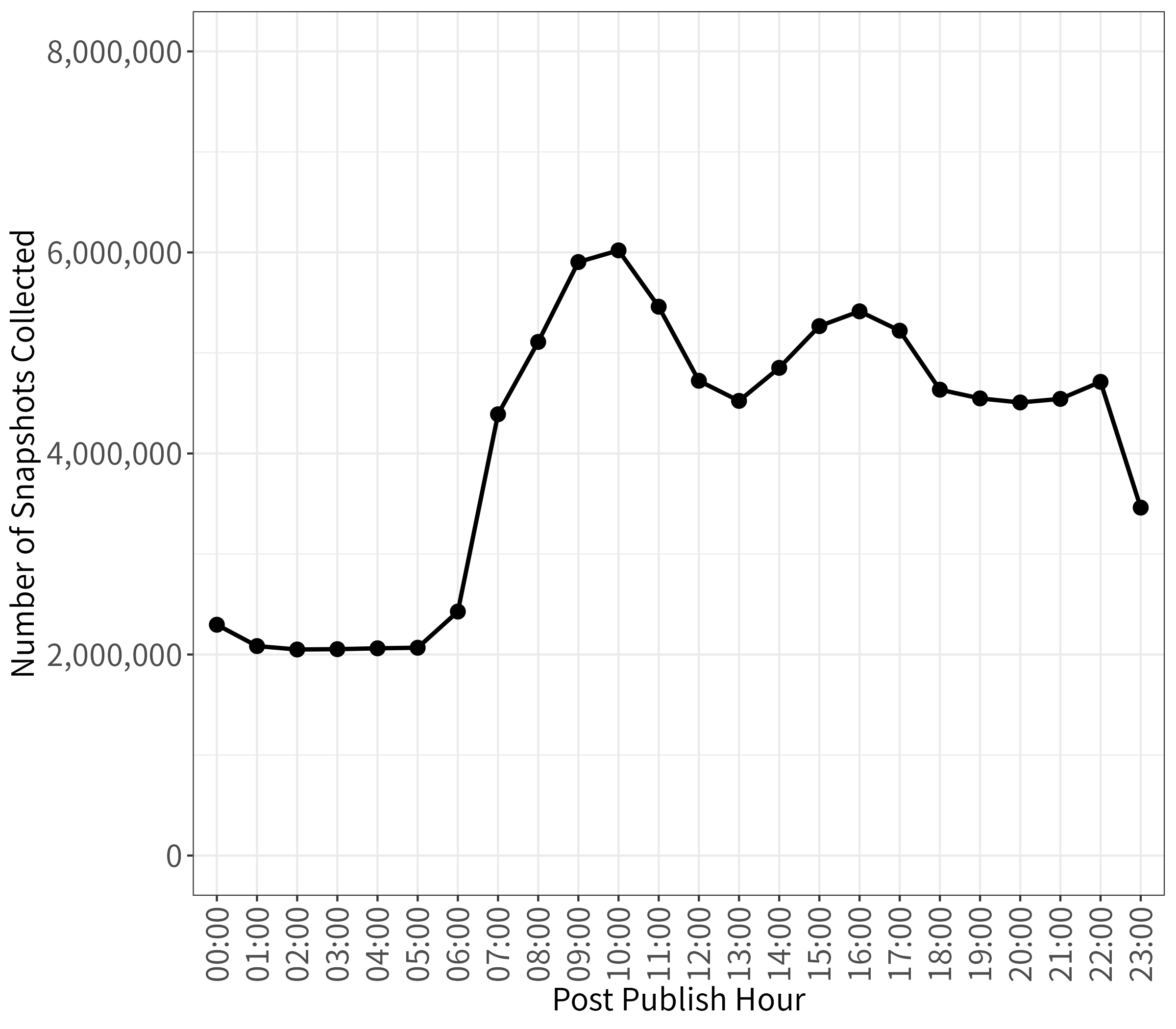}
        \caption{Number of Snapshots on Posts Collected by Hour}
        \label{fig:pnas_post_hour_snapshot}
        \end{subfigure}
            \hspace{0.1cm} 
        \begin{subfigure}[b]{0.4\textwidth}
        \centering
        \includegraphics[width=\textwidth]{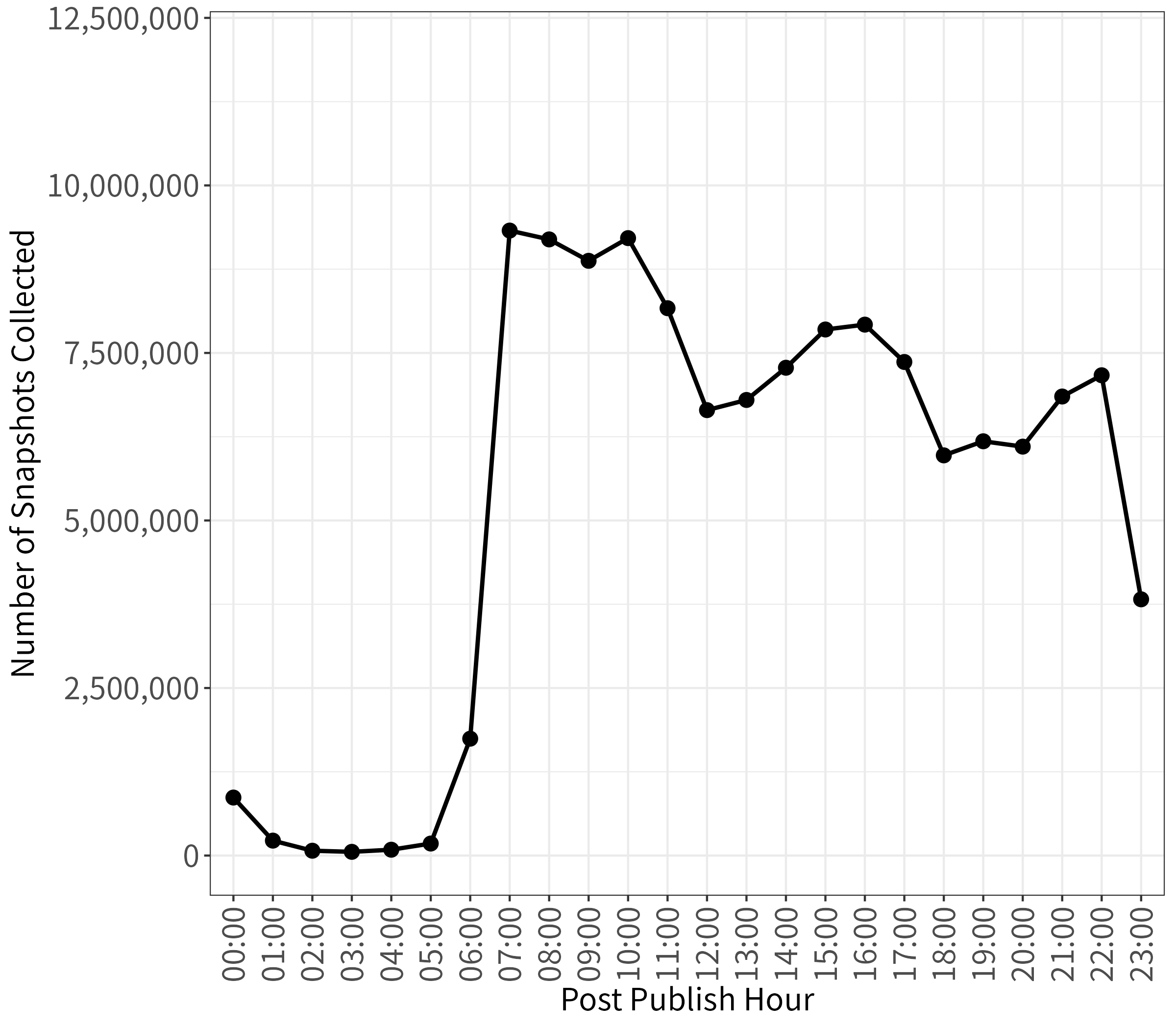}
        \caption{Number of Snapshots on Comments Collected by Hour}
        \label{fig:pnas_comment_hour_snapshot}
        \end{subfigure}
    \caption*{\textbf{Notes:} Panel~\ref{fig:pnas_volume_reads_change} shows the average number of new reads for a Weibo post tracked continuously for 24 hours after publication. Readership peaks immediately after posting, with an average of about 12,000 new reads, then drops sharply within the first four hours and stabilizes at a much lower level. This pattern is consistent across our full dataset of 94,009 posts. Panel~\ref{fig:pnas_volume_comments_change} presents the corresponding trend in comment activity. Comments also peak shortly after publication, though at a smaller scale, averaging about two at the peak, and decline rapidly to near zero within five hours. These patterns validate our monitoring strategy—five-minute intervals during the first 24 hours and daily snapshots thereafter—as sufficiently granular to capture the vast majority of user engagement.     
    Panel~\ref{fig:pnas_post_hour_snapshot} and Panel~\ref{fig:pnas_comment_hour_snapshot}  display the hourly distribution of post and comment snapshots aggregated across the observation period, with activity peaking between 08:00 and 12:00—typical morning posting hours. shows the hourly distribution of comment snapshots, with most activity occurring between 06:00 and 22:00, reflecting expected user engagement hours.}
\end{figure}

From April 18 to May 9, 2022, we continuously monitored all new posts and their first-page (top 20) comments from selected Weibo accounts. To ensure comprehensive coverage of user engagement, we recorded snapshots every five minutes during the first 24 hours after publication, followed by daily snapshots for the next ten days. Figure~\ref{fig:temporal} confirms that these intervals effectively captured the majority of user activity: most views and comments occurred within the first four hours after a post appeared, and overall engagement was concentrated between 6:00 a.m. and 11:00 p.m. Beijing Time.

\clearpage

\subsection{Protecting user privacy}

Since our data are publicly available, the study is exempt from institutional review board review. To ensure user privacy, we implemented the following anonymization procedures for all Sina Weibo data used in this study. These measures provide robust privacy protection and comply with ethical standards for research involving public online content.

\begin{table}[h]
  \centering
  \caption{Anonymization Protocol for \emph{Sina Weibo} Data}
  \label{tab:weibo_anonymization}
  \renewcommand{\arraystretch}{1.2} 
  \begin{tabularx}{\textwidth}{@{}p{5cm}X@{}}
    \toprule
    \textbf{Protocol Component} & \textbf{Anonymization Procedure} \\ \midrule
    Location Data Handling &
    Location tags are limited to the provincial level for mainland Chinese users and the country/region level for overseas users, as provided by Sina Weibo; no finer‑grained geographic information is collected or stored. \\[2pt]
    
    User Identifier Masking &
    Usernames, avatars, profile URLs, and other user‑specific references are removed and replaced with randomly generated unique IDs that cannot be reverse‑engineered. \\[2pt]
    
    Secure Storage and Access Control &
    Anonymized datasets are stored on encrypted servers at the corresponding author's institution; server access is restricted to authorized research personnel listed in the project documentation. \\[2pt]
    
    Data Sharing for Replication &
    Only aggregated data are shared externally for replication; individual‑level data remain confidential and are never disclosed. \\ 
    \bottomrule
  \end{tabularx}
\end{table}

\clearpage

\section{Descriptive Statistics}
Table~\ref{table:summary_of_account} provides descriptive statistics at both the account and post levels. Panel A summarizes the 165 accounts as of July 25, 2023. These accounts have an average of 12.7 million followers (median: 3.8 million), with the most followed account reaching approximately 152 million. They are highly active, posting an average of 21 times per day and producing over 87,000 posts in total. On average, each post receives 196 reposts, 81 comments, and 332 likes, indicating substantial user engagement.

Panel B presents post-level summary statistics for content published between April 17 and May 9, 2022. About 8\% of the posts concern international affairs; the remaining 92\% do not. Roughly half of the non-international posts address local issues by referencing one of China's 31 provincial-level divisions. Across all posts, the average totals over the 11-day observation window are 45 comments, 389 likes, 43 reposts, and 9.5 unique first-page comments.

Panel C focuses on posts about local issues. Among these, 62\% describe positive events, while 28\% depict negative events. These posts average 37 total comments, 351 likes, and 31 reposts. On average, each receives six unique first-page comments: 1.7 from in-province users, 5.6 from out-of-province users, and 0.13 from overseas users. The average coefficient of variation is 0.27.

\begin{table}[!htbp] 
\centering 
\caption{Summary Statistics of Account and Post} 
\label{table:summary_of_account} 
\resizebox{0.9\textwidth}{!}{%
\begin{tabular}{@{\extracolsep{5pt}}lcccccc} 
\\[-1.8ex]\hline 
\hline \\[-1.8ex] 
Statistic & \multicolumn{1}{c}{N} & \multicolumn{1}{c}{Mean} & 
           \multicolumn{1}{c}{Median} & \multicolumn{1}{c}{St. Dev.} & 
           \multicolumn{1}{c}{Min} & \multicolumn{1}{c}{Max} \\ 
\hline \\[-1.8ex] 
\multicolumn{7}{c}{Panel A: Account Level} \\
\hline
\#Followers & 165 & 12,765,524.00 & 3,807,360 & 24,491,766.00 & 25,604 & 151,991,693 \\ 
\#Posts & 165 & 87,019.05 & 71,714 & 60,339.40 & 2,835 & 330,482 \\ 
\#Posts Per Day & 165 & 21.38 & 19.06 & 12.58 & 1.21 & 65.07 \\ 
\#Reposts Per Post & 165 & 195.61 & 27.32 & 843.80 & 0.43 & 6,405.71 \\ 
\#Comments Per Post & 165 & 80.86 & 12.66 & 210.40 & 0.11 & 1,604.35 \\ 
\#Likes Per Post & 165 & 332.05 & 49.14 & 1,170.76 & 0.98 & 11,208.28 \\ 
\#Likes Per Comment & 165 & 3.32 & 2.03 & 3.68 & 0.00 & 23.89 \\ 
\hline 
\multicolumn{7}{c}{Panel B: Post Level (All Posts)} \\
\hline
International (0 or 1) & 94,067 & 0.08 & 0 & 0.28 & 0 & 1 \\  
Non-International (0 or 1)  & 94,067 & 0.92 & 1 & 0.28 & 0 & 1 \\ 
\hspace{1em} Local (0 or 1) & 86,285 & 0.51 & 1 & 0.50 & 0 & 1 \\ 
\hspace{1em} Non-Local (0 or 1) & 86,285 & 0.49 & 0 & 0.50 & 0 & 1 \\ 
\#Total Comments & 94,067 & 44.76 & 3 & 169.57 & 0 & 1,332 \\
\#Total Likes & 94,067 & 389.61 & 14 & 1,804.56 & 0 & 15,008 \\
\#Total Reposts & 94,067 & 43.19 & 4 & 168.13 & 0 & 1,335 \\
\#Unique First-Page Comments & 94,067 & 9.50 & 2 & 19.37 & 0 & 518 \\ 
\hline 
\multicolumn{7}{c}{Panel C: Post Level (Posts Concerning Local Issues)} \\
\hline 
Positive Event (0 or 1) & 44,392 & 0.62 & 1 & 0.49 & 0 & 1 \\
Negative Event (0 or 1) & 44,392 & 0.28 & 0 & 0.45 & 0 & 1 \\ 
Neutral Event (0 or 1) & 44,392 & 0.10 & 0 & 0.30 & 0 & 1 \\
\#Total Comments  & 44,392 & 36.70 & 3 & 136.85 & 0 & 1,069 \\
\#Total Likes & 44,392 & 351.32 & 12 & 1,627.92 & 0 & 13,429  \\
\#Total Reposts & 44,392 & 31.49 & 3 & 108.93 & 0 & 827 \\
\#Unique First-Page Comments & 44,392 & 6.54 & 3 & 11.41 & 0 & 185 \\ 
\#Unique First-Page Comments (In-Province) & 44,392 & 1.70 & 0 & 4.24 & 0 & 75 \\
\#Unique First-Page Comments (Out-of-Province) & 44,392 & 5.57 & 0 & 14.26 & 0 & 491 \\ 
\#Unique First-Page Comments (Overseas) & 44,392 & 0.13 & 0 & 0.64 & 0 & 28 \\ 
Coefficient of Variation & 44,392 & 0.27 & 0.00 & 0.32 & 0.00 & 2.88\\ 
\hline \\[-1.8ex] 
\end{tabular}}
\end{table}

\clearpage

\begin{figure}[H]
\centering
\includegraphics[width=0.7\textwidth]{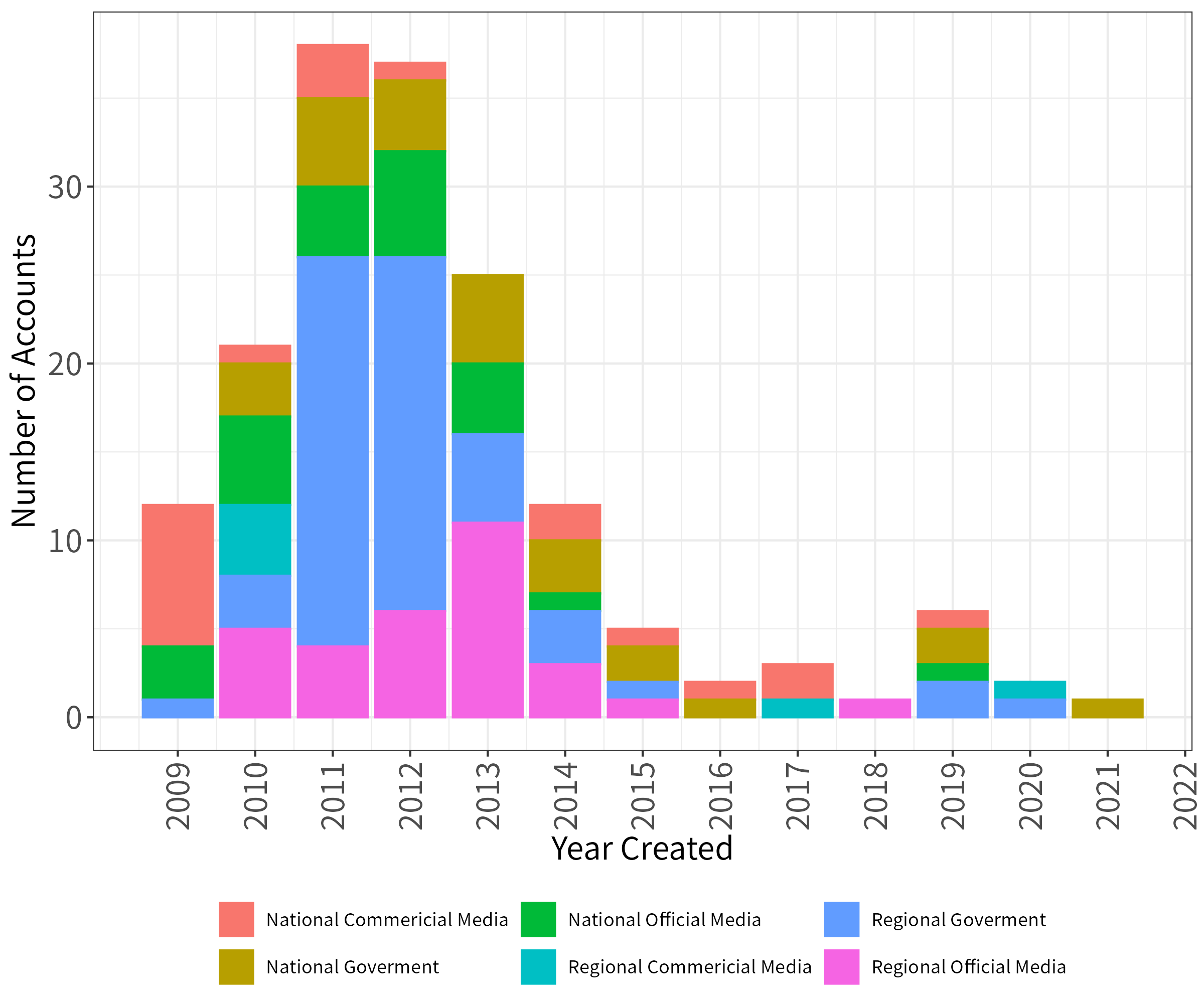}
        \caption{New Account Created by Year}
        \label{fig:pnas_account_created_year}
        \caption*{\textbf{Notes:} This figure presents a histogram of the 165 Weibo accounts analyzed in this study, categorized by creation year and account type. Most accounts were established between 2010 and 2013, with a peak in 2011, followed by a decline in new account registrations after 2014. Notably, all accounts were created prior to the implementation of the user location disclosure in 2022, with the most recent account dating to 2021. This distribution confirms that the observed changes in engagement are not attributable to newly created accounts, providing a stable foundation for analyzing the policy’s effects.}
\end{figure}

To further contextualize the scope of our data, Figure~\ref{fig:pnas_account_created_year} shows the distribution of account creation dates for the 165 Weibo accounts included in this study. The timing and frequency of account registration provide important context for interpreting engagement patterns and evaluating the potential influence of external interventions, such as the 2022 user location disclosure.

\clearpage

\section{Measurements}

This paper employs several measurements derived through dictionary methods, supervised learning, and large language models (LLMs). 

\subsection{Determining posts on international affairs}

We identify posts on international affairs using a dictionary-based approach. Specifically, we search for explicit mentions of at least one foreign location drawn from a keyword list constructed directly from Weibo’s post-treatment IP–location banners. After the implementation of the user location disclosure, Weibo began displaying each user’s IP region beneath their posts and comments. We recorded every foreign country or region that appeared in these banners during our study period, yielding a list of 123 unique locations (Table~\ref{table:classification_international}). This list corresponds precisely to the set of jurisdictions observable in our data and includes the general category “海外” (overseas), which Weibo uses when it cannot determine a more specific foreign location. To maintain specificity, we exclude any post that simultaneously references a Chinese province or major municipality.

\begin{table}[h!]
\caption{Keywords used to identify posts on international affairs}
\label{table:classification_international} 
\centering
\resizebox*{\textwidth}{!}{%
\begin{tabular}{|p{2.6cm}|p{3cm}||p{2.6cm}|p{3cm}||p{2.6cm}|p{3cm}|}
\hline
\textbf{Chinese Keyword} & \textbf{English Name} &
\textbf{Chinese Keyword} & \textbf{English Name} &
\textbf{Chinese Keyword} & \textbf{English Name} \\ \hline
阿富汗 & Afghanistan & 阿尔及利亚 & Algeria & 安哥拉 & Angola \\
阿根廷 & Argentina & 澳大利亚 & Australia & 奥地利 & Austria \\
孟加拉 & Bangladesh & 比利时 & Belgium & 巴西 & Brazil \\
文莱 & Brunei & 布基纳法索 & Burkina Faso & 布隆迪 & Burundi \\
柬埔寨 & Cambodia & 喀麦隆 & Cameroon & 加拿大 & Canada \\
乍得 & Chad & 智利 & Chile & 哥伦比亚 & Colombia \\
哥斯达黎加 & Costa Rica & 古巴 & Cuba & 捷克 & Czech Republic \\
科特迪瓦 & Côte d'Ivoire & 刚果民主共和国 & Democratic Republic of the Congo & 丹麦 & Denmark \\
多米尼加 & Dominican Republic & 厄瓜多尔 & Ecuador & 埃及 & Egypt \\
赤道几内亚 & Equatorial Guinea & 埃塞俄比亚 & Ethiopia & 芬兰 & Finland \\
法国 & France & 冈比亚 & Gambia & 格鲁吉亚 & Georgia \\
德国 & Germany & 加纳 & Ghana & 希腊 & Greece \\
格林纳达 & Grenada & 危地马拉 & Guatemala & 几内亚 & Guinea \\
圭亚那 & Guyana & 香港 & Hong Kong & 匈牙利 & Hungary \\
冰岛 & Iceland & 印度 & India & 伊朗 & Iran \\
伊拉克 & Iraq & 以色列 & Israel & 意大利 & Italy \\
日本 & Japan & 哈萨克斯坦 & Kazakhstan & 肯尼亚 & Kenya \\
吉尔吉斯斯坦 & Kyrgyzstan & 老挝 & Laos & 拉脱维亚 & Latvia \\
立陶宛 & Lithuania & 卢森堡 & Luxembourg & 澳门 & Macau \\
马达加斯加 & Madagascar & 马拉维 & Malawi & 马来西亚 & Malaysia \\
马里 & Mali & 马耳他 & Malta & 马绍尔群岛 & Marshall Islands \\
毛里塔尼亚 & Mauritania & 毛里求斯 & Mauritius & 墨西哥 & Mexico \\
蒙古 & Mongolia & 黑山 & Montenegro & 摩洛哥 & Morocco \\
莫桑比克 & Mozambique & 缅甸 & Myanmar & 尼泊尔 & Nepal \\
荷兰 & Netherlands & 新西兰 & New Zealand & 尼日尔 & Niger \\
尼日利亚 & Nigeria & 北马其顿 & North Macedonia & 挪威 & Norway \\
海外 & Overseas & 巴基斯坦 & Pakistan & 巴拿马 & Panama \\
巴布亚新几内亚 & Papua New Guinea & 秘鲁 & Peru & 菲律宾 & Philippines \\
波兰 & Poland & 葡萄牙 & Portugal & 卡塔尔 & Qatar \\
刚果共和国 & Republic of the Congo & 罗马尼亚 & Romania & 俄罗斯 & Russia \\
留尼汪岛 & Réunion & 沙特 & Saudi Arabia & 塞内加尔 & Senegal \\
塞尔维亚 & Serbia & 新加坡 & Singapore & 斯洛伐克 & Slovakia \\
斯洛文尼亚 & Slovenia & 所罗门群岛 & Solomon Islands & 南非 & South Africa \\
韩国 & South Korea & 南苏丹 & South Sudan & 西班牙 & Spain \\
斯里兰卡 & Sri Lanka & 苏丹 & Sudan & 瑞典 & Sweden \\
瑞士 & Switzerland & 台湾 & Taiwan & 塔吉克斯坦 & Tajikistan \\
坦桑尼亚 & Tanzania & 泰国 & Thailand & 东帝汶 & Timor-Leste \\
多哥 & Togo & 特立尼达和多巴哥 & Trinidad and Tobago & 土耳其 & Turkey \\
乌干达 & Uganda & 乌克兰 & Ukraine & 阿联酋 & United Arab Emirates \\
英国 & United Kingdom & 美国 & United States & 乌兹别克斯坦 & Uzbekistan \\
委内瑞拉 & Venezuela & 越南 & Vietnam & 赞比亚 & Zambia \\
津巴布韦 & Zimbabwe &  &  &  &  \\
\hline
\end{tabular}}
\end{table}

\subsection{Determining posts on local affairs}

We classify posts as concerning local affairs if they mention at least one of the 31 provincial-level administrative divisions in mainland China, including the four centrally administered municipalities. The corresponding keyword list used for this classification is provided in Table~\ref{table:classification_local}. Posts containing any of these location references are flagged as local, while those without such mentions are treated as non-local.

\begin{table}[h!]
\caption{Keywords used to identify posts on local affairs}
\label{table:classification_local} 
\centering
\resizebox{\textwidth}{!}{%
\begin{tabular}{|p{2.6cm}|p{3cm}||p{2.6cm}|p{3cm}||p{2.6cm}|p{3cm}|}
\hline
\textbf{Chinese Keyword} & \textbf{English Name} &
\textbf{Chinese Keyword} & \textbf{English Name} &
\textbf{Chinese Keyword} & \textbf{English Name} \\ \hline
安徽 & Anhui & 北京 & Beijing & 重庆 & Chongqing \\
福建 & Fujian & 甘肃 & Gansu & 广东 & Guangdong \\
广西 & Guangxi & 贵州 & Guizhou & 海南 & Hainan \\
河北 & Hebei & 黑龙江 & Heilongjiang & 河南 & Henan \\
湖北 & Hubei & 湖南 & Hunan & 内蒙古 & Inner~Mongolia \\
江苏 & Jiangsu & 江西 & Jiangxi & 吉林 & Jilin \\
辽宁 & Liaoning & 宁夏 & Ningxia & 青海 & Qinghai \\
陕西 & Shaanxi & 山东 & Shandong & 上海 & Shanghai \\
山西 & Shanxi & 四川 & Sichuan & 天津 & Tianjin \\
西藏 & Tibet & 新疆 & Xinjiang & 云南 & Yunnan \\
浙江 & Zhejiang &  &  &  &  \\ \hline
\end{tabular}}
\end{table}

\subsection{Classifying event sentiment in posts}

We measure the sentiment of events described in Weibo posts from the perspective of the Chinese government, classify commenters' stances toward public issues relative to official policies, and identify regional discrimination in replies. This section outlines the methodologies used to construct these metrics, including the prompts and parameters employed in LLMs, and describes how we validate automated classifications through cross-validation with hand-labeled data.

\begin{figure}[h]
\centering
\includegraphics[width=0.8\textwidth]{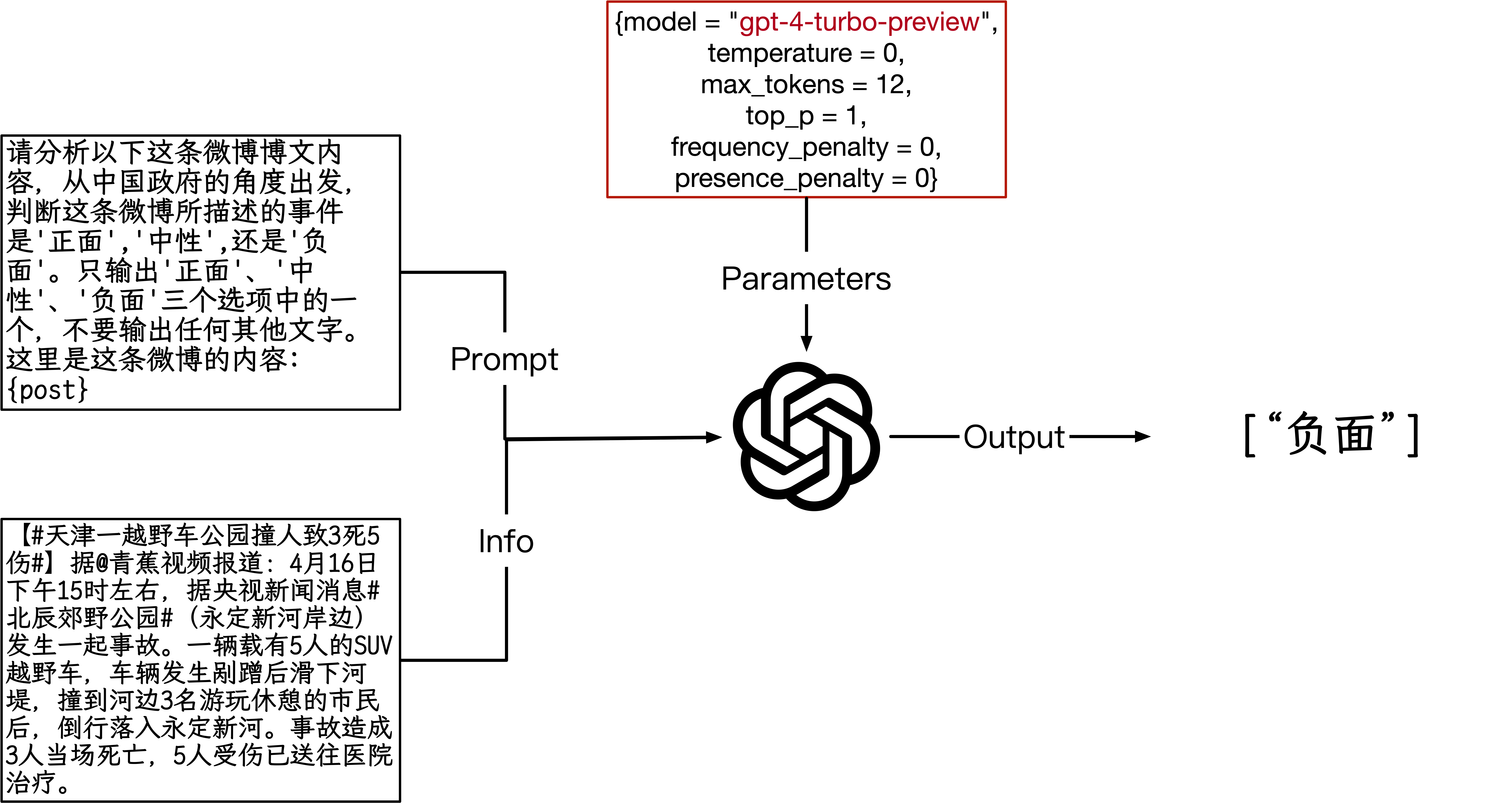}
\caption{Model for Event Sentiment Classification}
\label{fig:comment_sentiment_model}
\caption*{\textbf{Notes:} This figure illustrates the model used to classify the sentiment of events described in Weibo posts from the perspective of the Chinese government. Posts were processed using a structured prompt instructing the model to label each event as "positive," "neutral," or "negative" with no explanatory text. The model used was GPT-4-turbo-preview, with fixed parameters to ensure deterministic outputs. The example shown involves a Weibo post reporting a fatal accident, which the model correctly classifies as "negative." This approach allows for consistent and scalable sentiment classification across a large corpus of posts.

\textbf{Translation of Prompt:} "Please analyze the following Weibo content and determine, from the perspective of the Chinese government, whether the event described is 'positive,' 'neutral,' or 'negative.' Output only one of these three categories without additional text. Here is the content of the Weibo post: \{post\}."

\textbf{Translation of Info:} "[\#3 dead, 5 injured as SUV plunges into park lake in Tianjin\#] According to Qingjiao Video Report: Around 15:00 on April 16, according to CCTV news \#Beichen Yedi Park (Yongding Xinhe Riverside)\#, an accident occurred. An SUV with five people on board accidentally slid into the river embankment and collided with three citizens resting by the riverbank, then fell into the Yongding Xinhe. The accident caused 3 deaths on-site, and 5 injured were sent to the hospital for treatment."

\textbf{Translation of Output:} "Negative".}
\end{figure}

Figure~\ref{fig:comment_sentiment_model} illustrates the model used to classify the sentiment of events depicted in Weibo posts. Each post is systematically evaluated using a structured prompt that instructs the LLM to assign a sentiment label—positive, neutral, or negative—from the perspective of the Chinese government. This framing is essential, as the government's likely interpretation of a post shapes the tone and risk profile of user engagement. Notably, if the observed decline in critical comments were due to a reduction in negative post content, it would provide an alternative explanation for the engagement drop we document. We address this possibility directly in our analysis.

\clearpage

\subsection{Classifying post topics}

\begin{figure}[h!]
\centering
\includegraphics[width=0.8\textwidth]{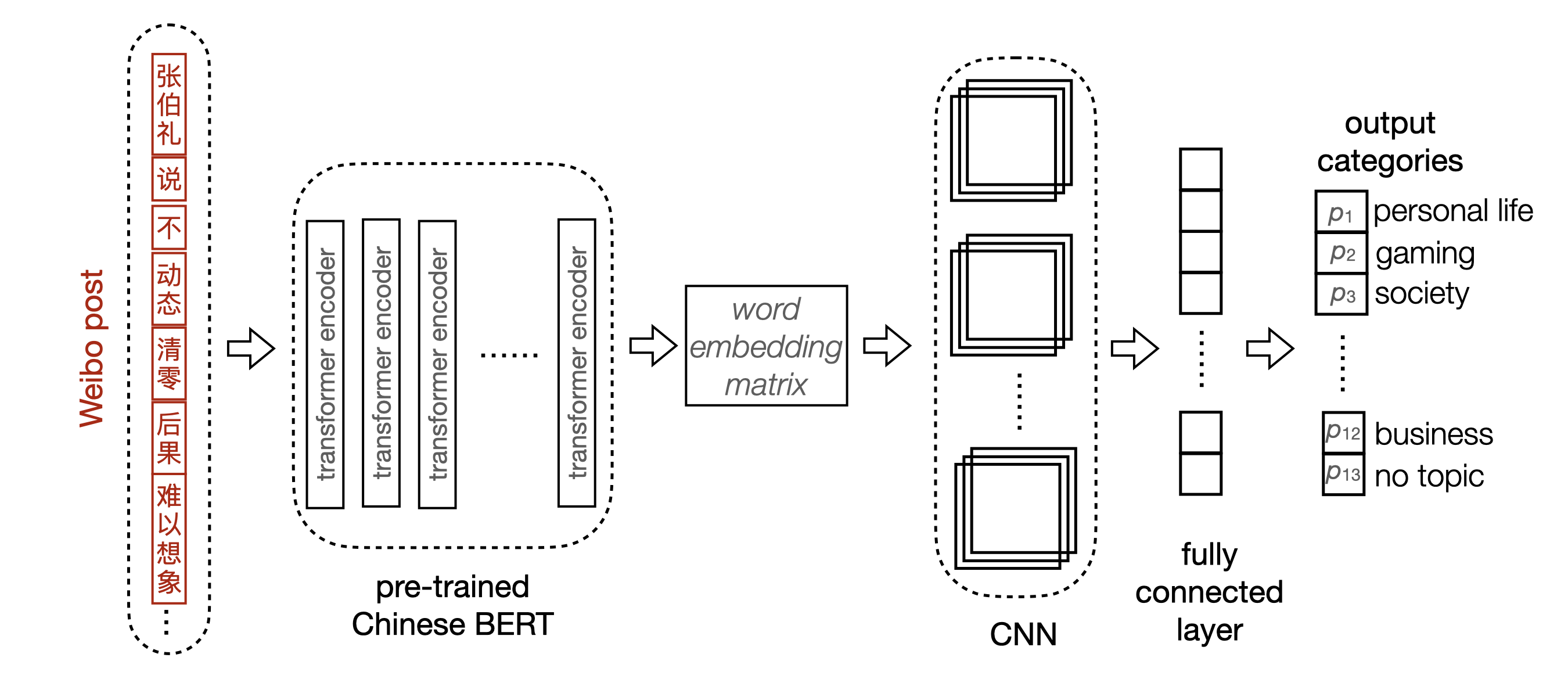}
\caption{Model for Post Topic Classification}
\label{fig:post_topic_model}
\caption*{\textbf{Notes:} This figure illustrates the workflow of our Weibo post topic classification model. Post texts are first tokenized and processed through a pre-trained Chinese BERT model to generate contextualized word embeddings. These embeddings are then passed through a convolutional neural network (CNN) to extract local semantic features relevant for topic distinction. A fully connected layer subsequently maps these features to output probabilities across predefined topic categories, enabling classification into specific content domains.}
\end{figure}

Figure~\ref{fig:post_topic_model} illustrates the architecture of our topic classification model for Weibo posts. We use a pre-trained Chinese BERT model to extract contextual embeddings from the post text, which are then passed through a convolutional neural network (CNN) to capture semantic features. A fully connected layer generates probability scores for predefined topic categories, including personal life, gaming, society, business, and others.

\clearpage

\subsection{Classifying comment stance}

\begin{figure}[h!]
\centering
\includegraphics[width=0.8\textwidth]{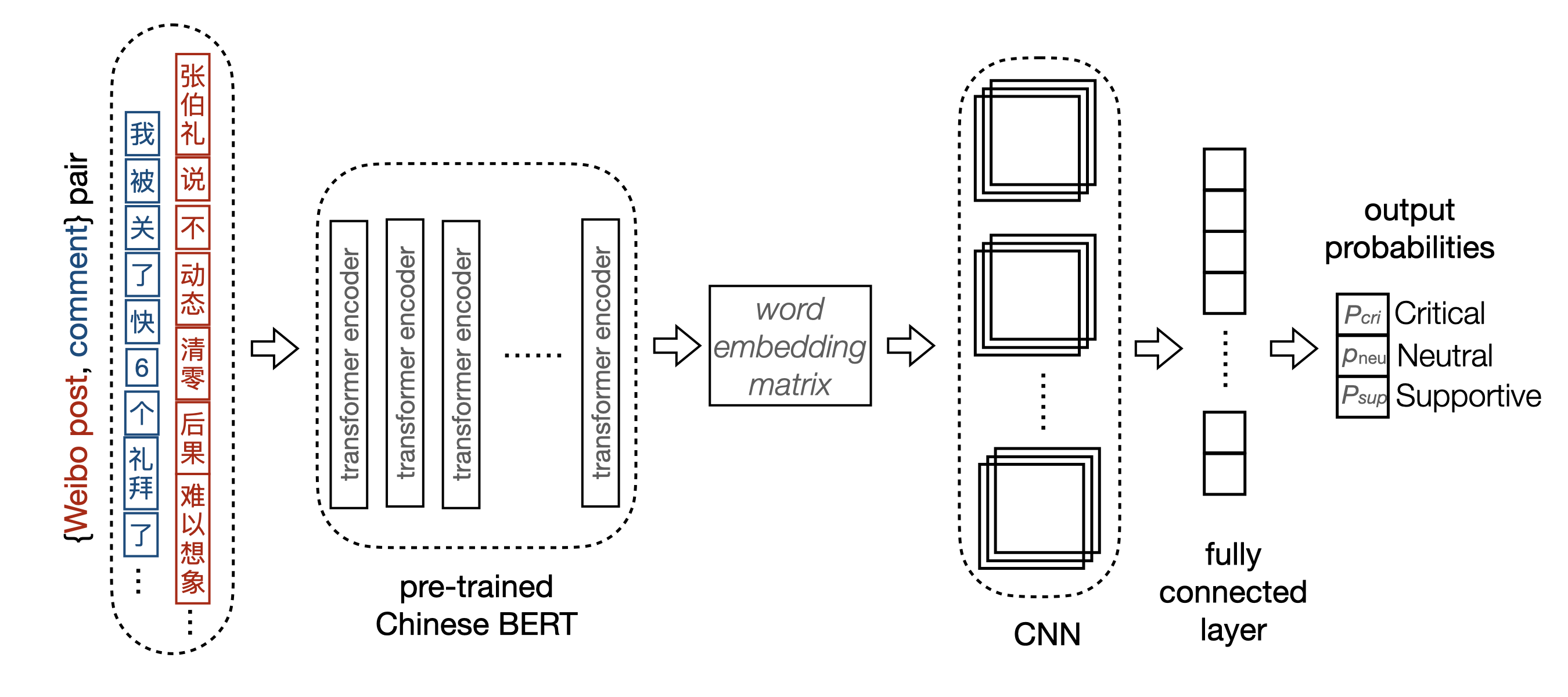}
\caption{Model for Comment Stance Classification}
\label{fig:comment_stance_model}
\caption*{\textbf{Notes:} This figure illustrates the architecture of the Comment Stance Classification model used in our analysis. Each Weibo post and its associated comment are jointly processed as input to a pre-trained Chinese BERT model, which generates contextual embeddings. These embeddings are then passed through a convolutional neural network (CNN) to extract localized semantic features relevant to stance detection. A series of fully connected layers then maps these features to one of three stance categories—supportive, neutral, or critical—by computing the corresponding output probabilities.}
\end{figure}

Figure~\ref{fig:comment_stance_model} illustrates the model used for classifying the stance of comments relative to government policy. The BERT+CNN model categorizes each comment as \textbf{supportive}, \textbf{neutral}, or \textbf{critical}, based on its alignment with the government's position at the time. Comments are labeled supportive if they express agreement with or approval of official policy, neutral if they are unrelated or express no clear position, and critical if they convey disagreement or disapproval. This classification enables us to assess public sentiment in relation to specific policy contexts and evaluate patterns of agreement or dissent across users and topics.

\clearpage

\subsection{Classifying regionally discriminatory comment replies}

\begin{figure}[h!]
\centering
\includegraphics[width=0.8\textwidth]{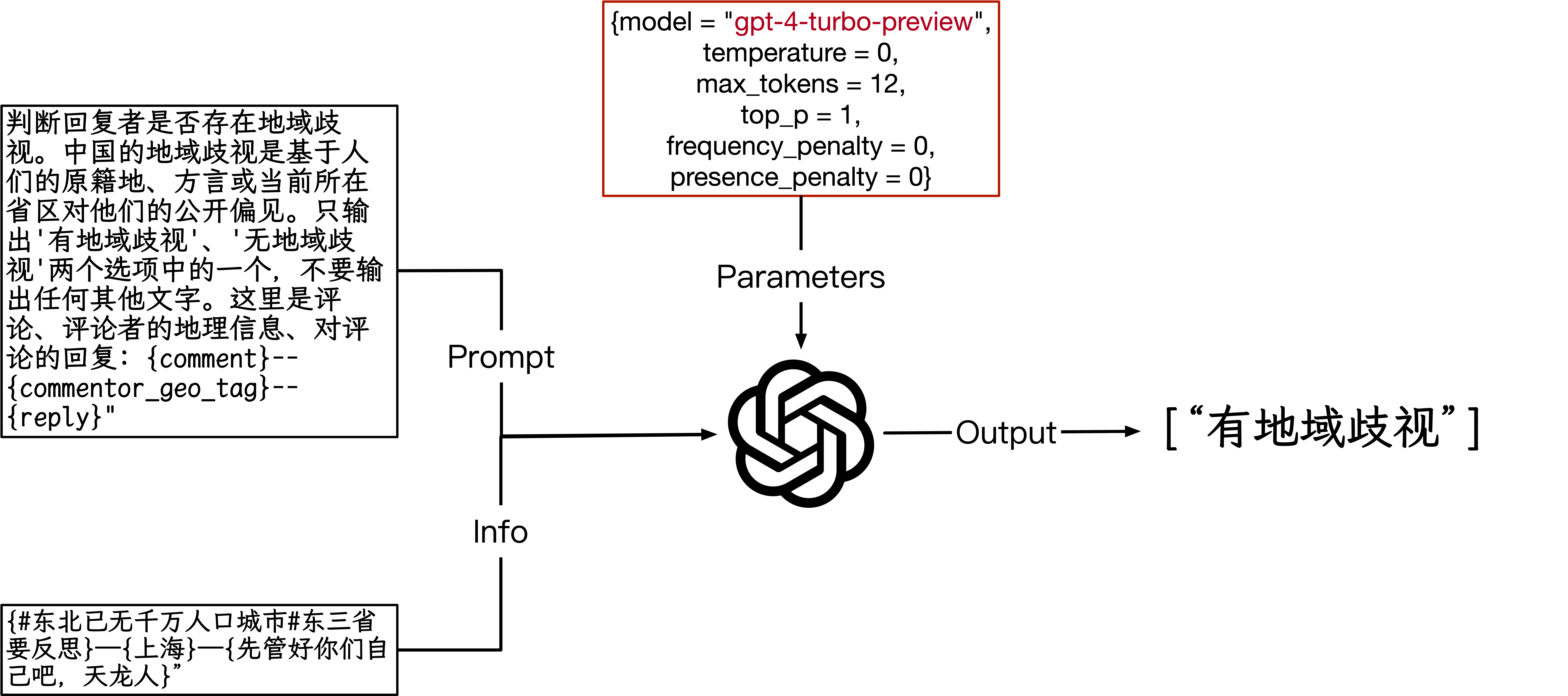}
\caption{Model for Regional Discrimination Classification}
\label{fig:comment_regional_discrimination}
\caption*{\textbf{Notes:} This figure illustrates the model used to identify regional discriminatory replies to comments. Comments were input into the model with a structured prompt instructing it to determine if a reply to the comment exhibited regional discrimination based on biases toward people's origin or their publicly perceived prejudices about specific provinces. The prompt explicitly instructed the model to output either "has regional discrimination" or "no regional discrimination" without additional text. Model parameters utilized GPT-4-turbo-preview with settings ensuring deterministic outcomes. The depicted example involves an exchange indicating regional discrimination, which the model accurately classifies as such.

\textbf{Translation of Prompt:} "Determine if the commenter exhibits regional discrimination. In China, regional discrimination is based on people's origin, language, or publicly expressed biases towards their current province. Output only 'has regional discrimination' or 'no regional discrimination' without additional text. Here is the comment, the commenter's geographic information, and the reply to the comment: \{comment\}--\{commentor\_geo\_tag\}--\{reply\}."

\textbf{Translation of Info:} "{\#Northeast China doesn't have a single city with over ten million people\#. The three northeastern provinces should reflect} — {Shanghai} — {Better focus on yourselves, Tianlong people}."

\textbf{Translation of Output:} "has regional discrimination"}
\end{figure}

Figure~\ref{fig:comment_regional_discrimination} presents the model used to detect regional discrimination in replies to Weibo comments. The model evaluates each comment–reply pair to identify biases associated with geographic origin or regional stereotypes. Such discrimination often appears in subtle or indirect forms—through innuendo, homonyms, or culturally specific references—posing challenges for accurate detection. Nonetheless, identifying these instances systematically is crucial for quantifying their prevalence and assessing whether regional hostility intensified following the implementation of the user location disclosure.

\clearpage

\subsection{Validating measurements}

Table~\ref{table:classification_validation} reports the validation results for our classification models based on human-labeled data. We randomly sampled 3,000 records from the dataset and tasked research assistants (RAs) with labeling them using the same prompts and information provided to the LLM and BERT models. These manually labeled records serve as the gold standard for evaluating model performance. The table presents key metrics—Precision, Recall, and F1 Score—for the Event Sentiment, Comment Stance, and Regional Discrimination tasks. Theses metrics indicate that the automated classifications are relatively accurate and reliable.

\begin{table}[!htbp] 
\centering 
\caption{Classification Validation} 
\label{table:classification_validation} 
\resizebox{0.7\textwidth}{!}{%
\begin{tabular}{@{\extracolsep{5pt}} l l ccc} 
\\[-1.8ex]\hline 
\hline \\[-1.8ex] 
Task & Class & Precision & Recall & F1 Score \\ 
\hline \\[-1.8ex] 
\multirow{5}{*}{\textbf{Event Sentiment}} 
  & Negative Event & 0.481 & 0.881 & 0.622 \\ 
  & Neutral Event & 0.750 & 0.610 & 0.673 \\ 
  & Positive Event & 0.932 & 0.761 & 0.838 \\ 
  & Macro-F1 & 0.721 & 0.751 & 0.711 \\ 
  & Micro-F1 & 0.721 & 0.721 & 0.721 \\ 
\hline \\[-1.8ex] 
\multirow{5}{*}{\textbf{Comment Stance}} 
  & Critical & 0.665 & 0.894 & 0.763 \\ 
  & Neutral & 0.793 & 0.641 & 0.709 \\ 
  & Supportive & 0.891 & 0.874 & 0.883 \\ 
  & Macro-F1 & 0.783 & 0.803 & 0.785 \\ 
  & Micro-F1 & 0.783 & 0.783 & 0.783 \\ 
\hline \\[-1.8ex] 
\multirow{4}{*}{\textbf{Regional Discrimination}} 
  & Discrimination & 0.771 & 0.883 & 0.823 \\ 
  & Non-Discrimination & 0.898 & 0.797 & 0.844 \\ 
  & Macro-F1 & 0.835 & 0.840 & 0.834 \\ 
  & Micro-F1 & 0.835 & 0.835 & 0.835 \\ 
\hline \\[-1.8ex] 
\end{tabular}}
\end{table}

\clearpage

\section{Estimation for the Interrupted Time Series Design}
\label{app:method:localpoly} 

Off-the-shelf \texttt{R} functions for local polynomial regression (\texttt{np}, \texttt{rdrobust}, \texttt{locfit}) do not support the simultaneous inclusion of kernel weights, cluster (account) fixed effects, and cluster-robust standard errors with confidence intervals. To address this, we developed a customized routine that: (a) applies Gaussian kernel weights; (b) incorporates account fixed effects; (c) estimates coefficients via weighted least squares with a cluster-robust sandwich variance estimator; and (d) selects bandwidths through cross-validation.

\subsection*{Setup and model}
Let $y_{gi}$ denote the outcome for observation $i$ in cluster (account) $g$, and let $t_{gi}$ be the corresponding timestamp measured in seconds. We are interested in estimating the instantaneous treatment effect at the cutoff $t_0 = \text{2022-04-28 12{:}00}$. Define the centered running variable as $\Delta_{gi} = t_{gi} - t_0$, and let the treatment indicator be $D_{gi} = \mathbbm{1}{\Delta_{gi} \ge 0}$.

For a polynomial order $p$ (we use $p=1$ and $p=2$), the model that underlies our estimator is
\begin{equation}
  y_{gi}
  \;=\;
  \underbrace{\alpha}_{\text{baseline}}
  \;+\;
  \underbrace{\gamma_g}_{\text{cluster FE}}
  \;+\;
  \underbrace{\tau}_{\substack{\text{treatment}\\\text{effect}}}D_{gi}
  \;+\;
  \sum_{j=1}^{p}
      \beta_j\,\Delta_{gi}^{\,j}
  \;+\;
  \sum_{j=1}^{p}
      \theta_j\, D_{gi}\,\Delta_{gi}^{\,j}
  \;+\;
  \varepsilon_{gi},
  \label{eq:model}
\end{equation}
where $\gamma_g$ captures any time-invariant heterogeneity at the account
level.  Model~\ref{eq:model} is estimated only with observations whose
$|\Delta_{gi}|$ lies in a bandwidth window $[-h,h]$; observations are
moreover re-weighted by a Gaussian kernel:
\begin{equation}
  K(u)\;=\;\frac{1}{\sqrt{2\pi}}\exp\bigl(-u^2/2\bigr),
  \qquad
  u=\frac{\Delta_{gi}}{h}.
  \label{eq:kernel}
\end{equation}
The resulting weights are $w_{gi}=K\!\bigl(\Delta_{gi}/h\bigr)$. Model~(\ref{eq:model}) is then estimated using ordinary least squares. 

\subsection*{Cluster-robust standard errors}
Let $G$ be the number of clusters and $\mathbf{X}_g, \mathbf{W}_g, \hat{\boldsymbol\varepsilon}_g$ the design matrix, weight matrix and residual vector for cluster $g$. Our estimator uses the covariate-adjusted cluster-robust-variance formula:
\begin{equation}
  \widehat{\mathrm{Var}}\!\bigl(\hat{\boldsymbol\beta}\bigr)
  = (\mathbf{X}^{\top}\mathbf{W}\mathbf{X})^{-1}
    \left(
      \sum_{g=1}^{G}
      \mathbf{X}_g^{\top}\mathbf{W}_g\,
      \hat{\boldsymbol\varepsilon}_g
      \hat{\boldsymbol\varepsilon}_g^{\top}\,
      \mathbf{W}_g\mathbf{X}_g
    \right)
    (\mathbf{X}^{\top}\mathbf{W}\mathbf{X})^{-1}.
  \label{eq:CRV}
\end{equation}
Taking the square root of the appropriate diagonal element yields the cluster-robust standard error $\hat{\tau}$, $\widehat{\mathrm{se}}(\hat{\tau})$. The confidence intervals are obtained through normal approximation. 

\subsection*{Bandwidth selection via cross-validation}
For a candidate bandwidth $h$ we perform $K$-fold block cross-validation: split the running variable into $K$ non-overlapping bins that are \emph{outside} an exclusion zone around $t_0$, re-estimate Model~(\ref{eq:model}) on $K-1$ bins, and compute the root-mean-squared prediction error (RMSE) on the hold-out bin. We then choose
\begin{equation}
  \hat{h}
  =\arg\min_{h\in\mathcal{H}}
   \frac{1}{K}
   \sum_{k=1}^{K}
     \operatorname{RMSE}_k(h).
  \label{eq:CV}
\end{equation}
The statistics reported in the “CV Bandwidth” column of Table \ref{table:treatment_effect} and in the figures employ the cross-validated bandwidth obtained this way.

\clearpage

\section{Additional Information on the Main Findings}

In this section, we present additional details supporting our main empirical findings. We include regression tables to report key causal estimates and conduct a series of robustness checks and placebo tests to test the robustness of the results reported in the main text.

\subsection{Causal estimates}

Table~\ref{table:treatment_effect} presents the estimated causal effects of user location disclosure on a range of outcomes across multiple bandwidths. These results extend the findings shown in Figures~1 and~2 of the main text. The CV Bandwidth'' column reports the optimal bandwidth selected using the \texttt{rdrobust()} function from the R package \texttt{rdrobust}. The RD Bandwidth'' and ``RD'' columns show the corresponding bandwidths and treatment effect estimates. To assess robustness, we also report estimates using alternative bandwidths of 8, 24, 72, 96, and 120 hours. The estimated effects are consistent across specifications, which shows that our main estimates are robust.

Overall, Table~\ref{table:treatment_effect} shows that user location disclosure reduced user engagement, with the most pronounced effects observed among non-local and out-of-province commenters, at least in the short term. The share of critical comments also declined significantly following the policy's implementation, suggesting a stronger impact on dissenting expression. These patterns hold across a range of bandwidths, reinforcing the conclusion that the observed drop in comment volume and critical engagement was directly driven by the user location disclosure policy.

\begin{table}[!htbp] \centering 
  \caption{Treatment Effect Estimates} 
  \label{table:treatment_effect} \vspace{0em}
\resizebox{1\textwidth}{!}{
\begin{tabular}{@{\extracolsep{5pt}} clccc|cc|ccccc} 
\\[-1.8ex]\hline 
\hline \\[-1.8ex] 
& & & & \multicolumn{6}{c}{Causal Estimates Using Different Bandwidth} \\
& & & & \multicolumn{6}{c}{(Standard Error)} \\
\cmidrule{4-12}
Figure & Group & Dependent Variable & CV Bandwidth & $\hat\tau$ & RD Bandwidth & $\hat\tau$ & 8 Hours & 24 Hours & 72 Hours & 96 Hours & 120 Hours\\ 
\hline \\[-1.8ex] 
Figure~1a & Topic: International & \#Comments & $32$ & 0.369 & 107 & 0.342 & 0.545 & 0.407 & 0.396 & 0.435 & 0.454 \\ 
 & & & & (0.146) & & (0.123) & (0.228) & (0.162) & (0.103) & (0.101) & (0.101)\\ 
Figure~1a & Topic: Non-International & \#Comments & $32$ & -0.011 & 93 & -0.010 & 0.214 & 0.005 & 0.004 & 0.019 & 0.027\\ 
& & & & (0.019) & & (0.022) & (0.050) & (0.019) & (0.015) & (0.014) & (0.013)\\ 
\hline \\[-1.8ex]
Figure~1b & Topic: International & \#Comments & $96$ & 0.407 & 122 & -2.078 & 0.378 & -1.783 & -0.191 & 0.407 & 0.661 \\ 
 & & & & (1.363) & & (1.858) & (3.038) & (1.780) & (1.511) & (1.363) & (1.276) \\ 
Figure~1b & Topic: Non-International & \#Comments & $16$ & -4.110 & 95 & -4.603 & -0.019 & -4.607 & -3.711 & -3.353 & -3.178\\ 
 & & & & (0.736) & & (0.429) & (0.827) & (0.799) & (0.634) & (0.585) & (0.565) \\ 
\hline \\[-1.8ex]
Figure~1c & Topic: Non-Local & \#Comments & $40$ & -1.399 & 89 & -1.473 & 1.201 & -1.511 & -0.631 & -0.194 & 0.030 \\ 
& & & & (0.785) & & (0.757) & (1.385) & (0.897) & (0.577) & (0.537) & (0.529)\\ 
Figure~1c & Topic: Local & \#Comments & $32$ & -7.794 & 99 & -7.793 & -0.887 & -7.599 & -6.748 & -6.392 & -6.224 \\ 
 & & & & (1.077) & & (0.496) & (0.803) & (1.047) & (0.969) & (0.928) & (0.910)\\ 
\hline \\[-1.8ex]
Figure~1e & User: In-Province & \#Comments & $48$ & -1.710 & 88 & -1.896 & -0.121 & -1.869 & -1.500 & -1.426 & -1.399\\ 
 & & & & (0.310) & & (0.171) & (0.229) & (0.316) & (0.307) & (0.307) & (0.307)\\ 
Figure~1e & User: Out-of-Province & \#Comments & $40$ & -5.665 & 103 & -5.702 & -0.801 & -5.559 & -5.008 & -4.735 & -4.600 \\
 & & & & (0.870) & & (0.395) & (0.650) & (0.858) & (0.763) & (0.725) &  (0.709)\\ 
Figure~1e & User: Oversea & \#Comments & $96$ & -0.073 & 96 & -0.108 & 0.060 & -0.100 & -0.086 & -0.073 & -0.066  \\ 
 & & & & (0.016) & & (0.017) & (0.036) & (0.021) & (0.017) & (0.016) & (0.015)\\ 
\hline \\[-1.8ex]
Figure~1f & User: In-Province & \#Critical Comments & $40$ & -0.572 & 88 & -0.615 & -0.436 & -0.596 & -0.410 & -0.370 & -0.355 \\ 
 & & & & (0.157) & & (0.092) & (0.338) & (0.175) & (0.130) & (0.120) & (0.116) \\ 
Figure~1f & User: Out-of-Province & \#Critical Comments & $88$ & -3.591 & 38 & -8.153 & -7.879 & -6.730 & -4.181 & -3.430 & -3.195 \\ 
 & & & & (1.536) & & (0.440) & (4.360) & (2.675) & (1.683) & (1.499) & (1.450)\\ 
Figure~1f & User: Oversea & \#Critical Comments & $96$ & -0.030 & 81 & 0.004 & 0.026 & -0.001 & -0.036 & -0.030 & -0.025\\ 
 & & & & (0.017) & & (0.013) & (0.024) & (0.016) & (0.016) & (0.017) &  (0.021)\\ 
\hline \\[-1.8ex]
Figure~2a &  & \%Replies & $96$ & 0.035 & 127 & 0.056 & 0.060 & 0.071 & 0.041 & 0.035 & 0.032\\ 
&  & & & (0.013) & & (0.021) & (0.029) & (0.028) & (0.015) & (0.013) & (0.012)\\ 
\hline \\[-1.8ex] 
\hline \\[-1.8ex] 
\end{tabular}}
\end{table}

\clearpage

\subsection{Stable post volume, topics, and censorship rate}

\begin{figure}[H]
    \centering
    \begin{subfigure}[b]{0.4\textwidth}
        \centering
        \includegraphics[width=\textwidth]{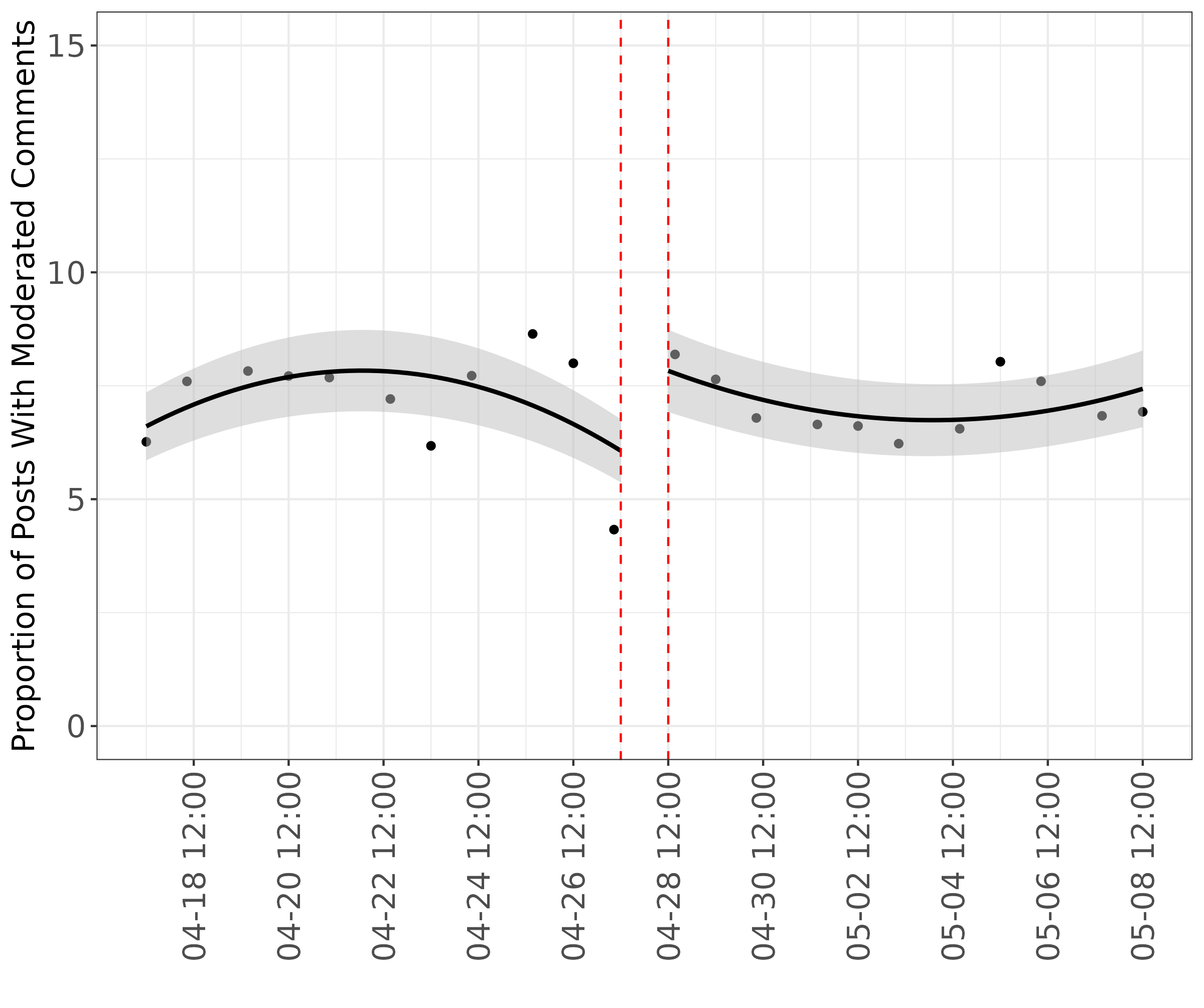}
        \caption{Number of Posts on Local Issues}
        \label{fig:pnas_stable_number_of_posts_bandwidth_best}
    \end{subfigure}
    \hspace{0.2cm} 
        \begin{subfigure}[b]{0.4\textwidth}
        \centering
        \includegraphics[width=\textwidth]{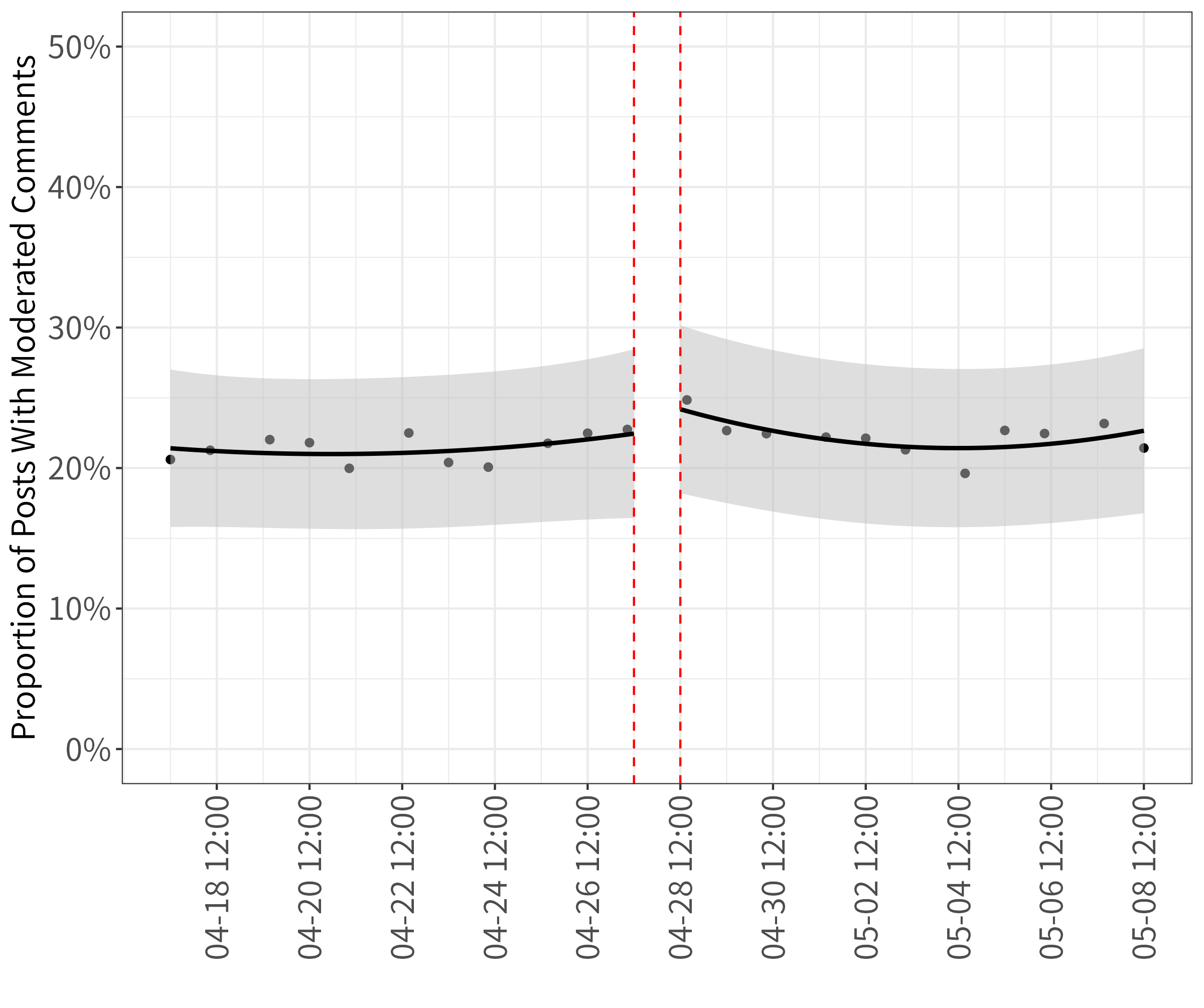}
        \caption{Proportion of Posts with Moderated Comments}
    \label{fig:pnas_stable_proportion_comment_moderated_bandwidth_best}
        \end{subfigure}
    \caption{Stable Volume, Posts Topics, and Censorship}
    \label{fig:stable_posts}
\caption*{\textbf{Notes:} Panel \ref{fig:pnas_stable_number_of_posts_bandwidth_best} displays the number of posts on local issues before and after the user location disclosure policy implementation, indicating stable or slightly increased posting volumes post-policy. Panel \ref{fig:pnas_stable_proportion_comment_moderated_bandwidth_best} presents the proportion of posts with moderated comments, which shows stable moderation levels before and after the policy implementation and ruling out intensified censorship as an explanation for the observed decline in comments. The red dashed lines in both panels mark the time of policy implementation.}
\end{figure}

We observed a decline in comments on local issues following the implementation of the user location disclosure policy. One possible explanation is that editors posted fewer local-topic posts, leaving users with less content to engage with. To evaluate this, Figure~\ref{fig:pnas_stable_number_of_posts_bandwidth_best} plots the number of posts on local topics before and after the policy change. The data show that local-topic posts remained stable—or even slightly increased—after implementation, ruling out reduced posting as the cause of declining engagement. 

We also assessed whether the drop in comments could reflect heightened censorship through selective comment release. In our data, we can observe whether a post’s comment section is subject to moderation—i.e., whether the account manager has enabled screening prior to public display—which offers a granular proxy for censorship. Figure~\ref{fig:pnas_stable_proportion_comment_moderated_bandwidth_best} shows that moderation levels remained stable after the policy change, suggesting that increased censorship is unlikely to account for the observed decline in comment volume.

\clearpage

\subsection{Robustness to logarithmic transformation}

\begin{figure}[H]
  \centering  
  \begin{subfigure}{0.32\textwidth}
    \centering
    \includegraphics[width=\textwidth]{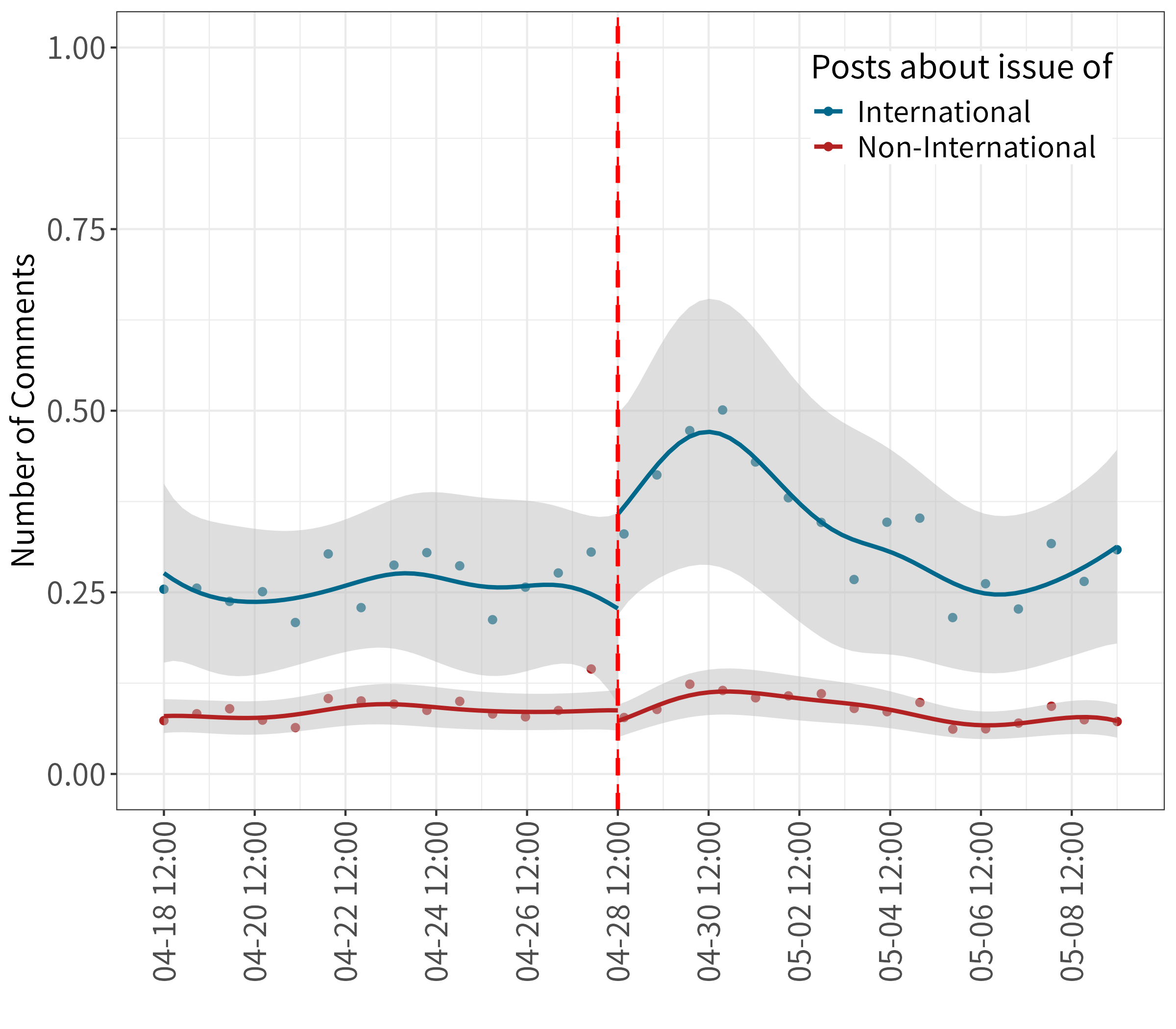}
    \caption{Posts on International Issues}
    \label{fig:pnas_volume_comments_oversea_origin_48h_bandwidth_best_log}
  \end{subfigure}
   \hspace{0.03\textwidth}
  \begin{subfigure}{0.32\textwidth}
    \centering
    \includegraphics[width=\textwidth]{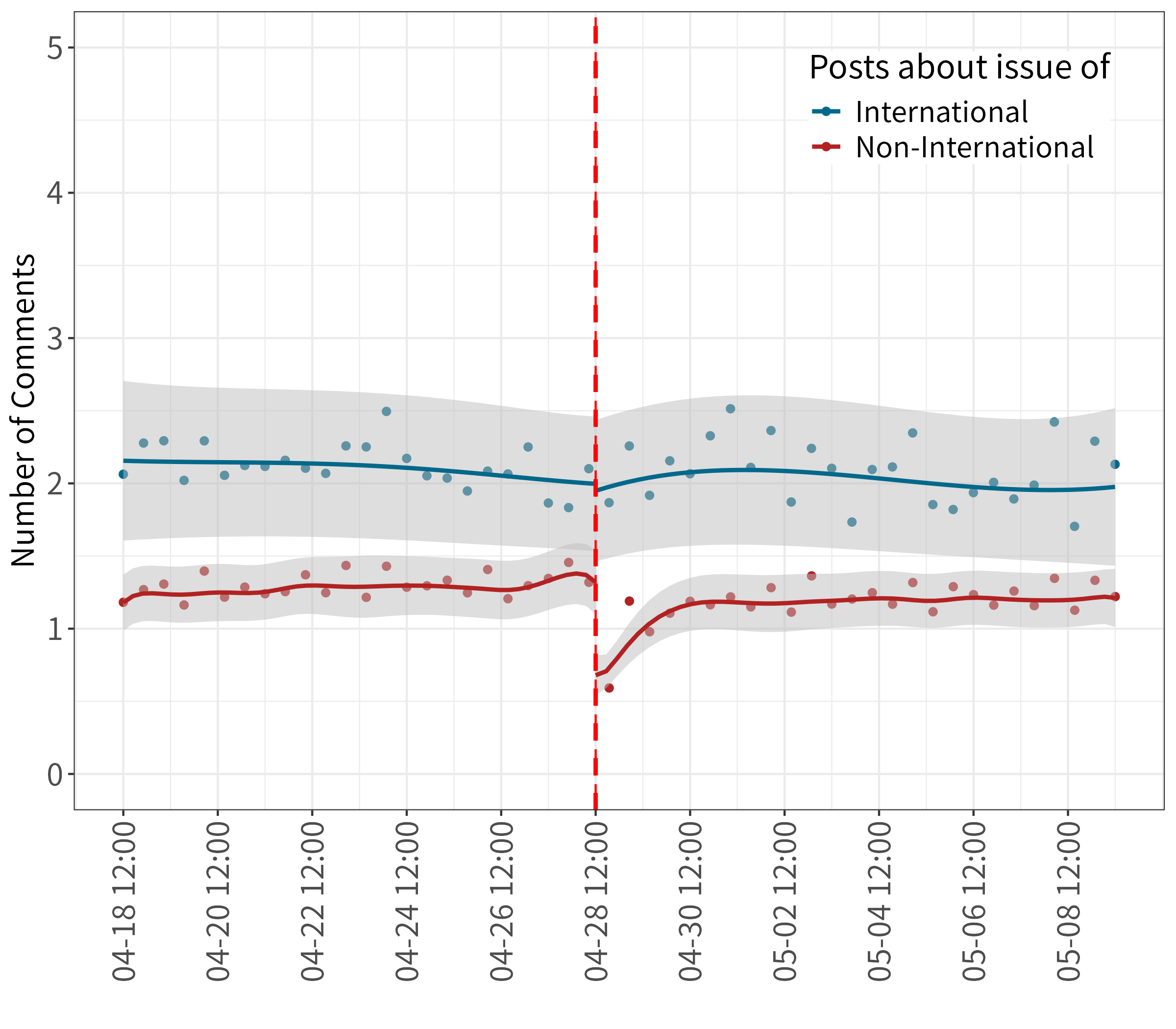}
    \caption{Posts on Domestic Issues}
    \label{fig:pnas_volume_comments_domestic_origin_48h_bandwidth_best_log}
  \end{subfigure}
  
  \vspace{0.8em} 
  
  \begin{subfigure}{0.32\textwidth}
    \centering
    \includegraphics[width=\textwidth]{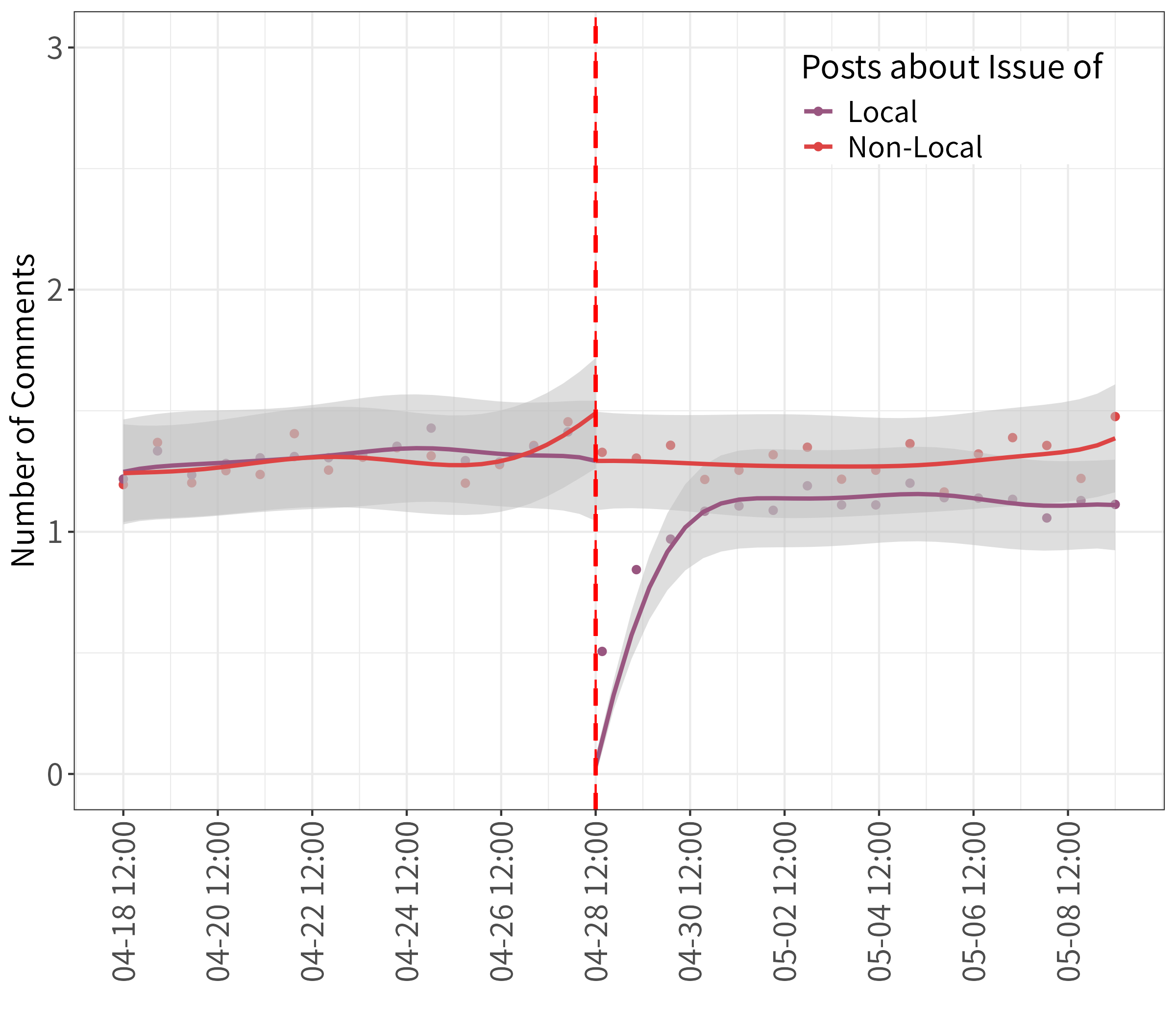}
    \caption{Posts on National and Local Issues}
    \label{fig:pnas_volume_comments_national_and_local_48h_bandwidth_best_log}
  \end{subfigure}
  \hfill
  \begin{subfigure}{0.32\textwidth}
    \centering
    \includegraphics[width=\textwidth]{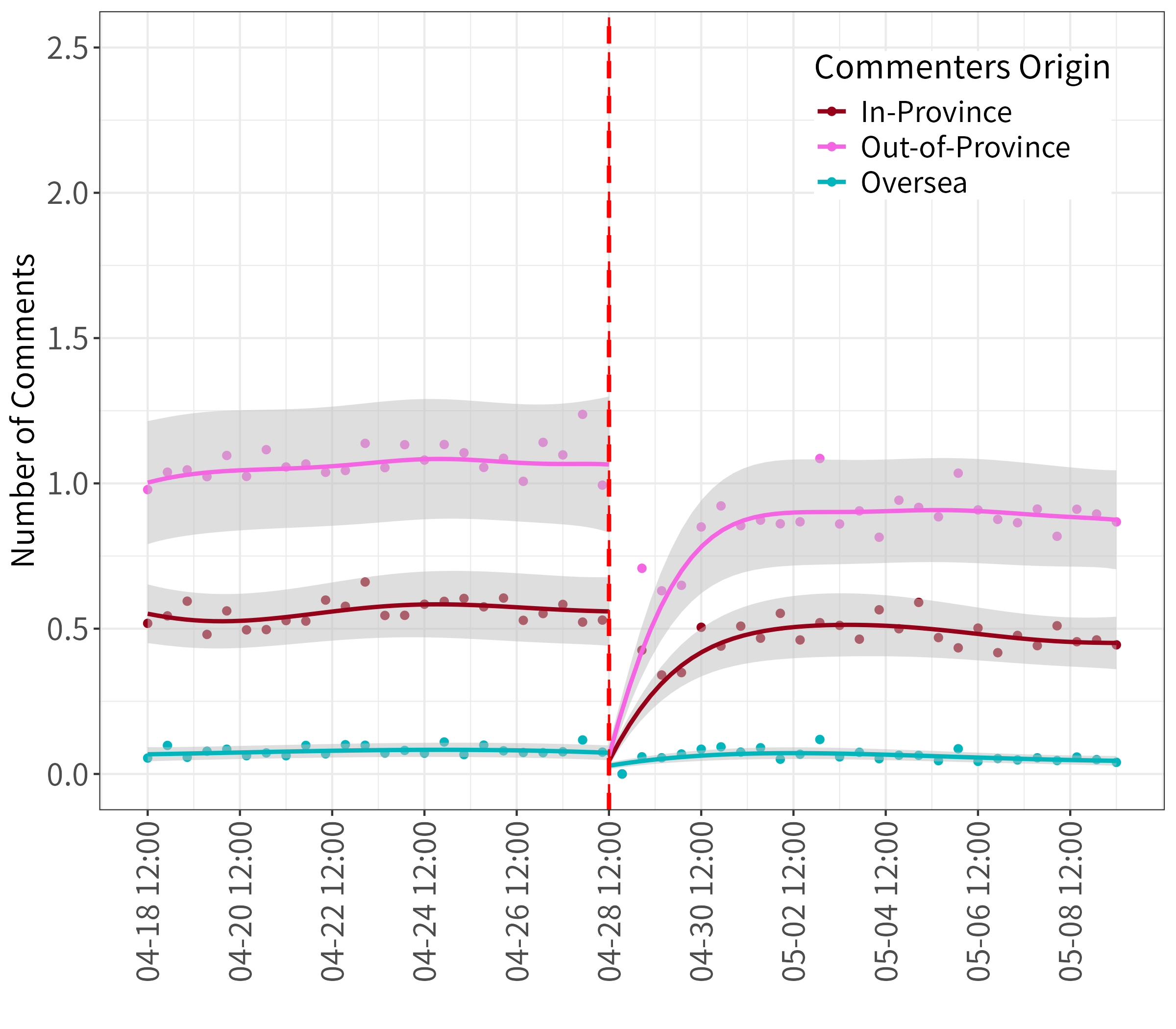}
    \caption{Decreased Comments}
    \label{fig:pnas_decreased_comments_local_by_origin_48h_bandwidth_best_log}
  \end{subfigure}
  \hfill
  \begin{subfigure}{0.32\textwidth}
    \centering
    \includegraphics[width=\textwidth]{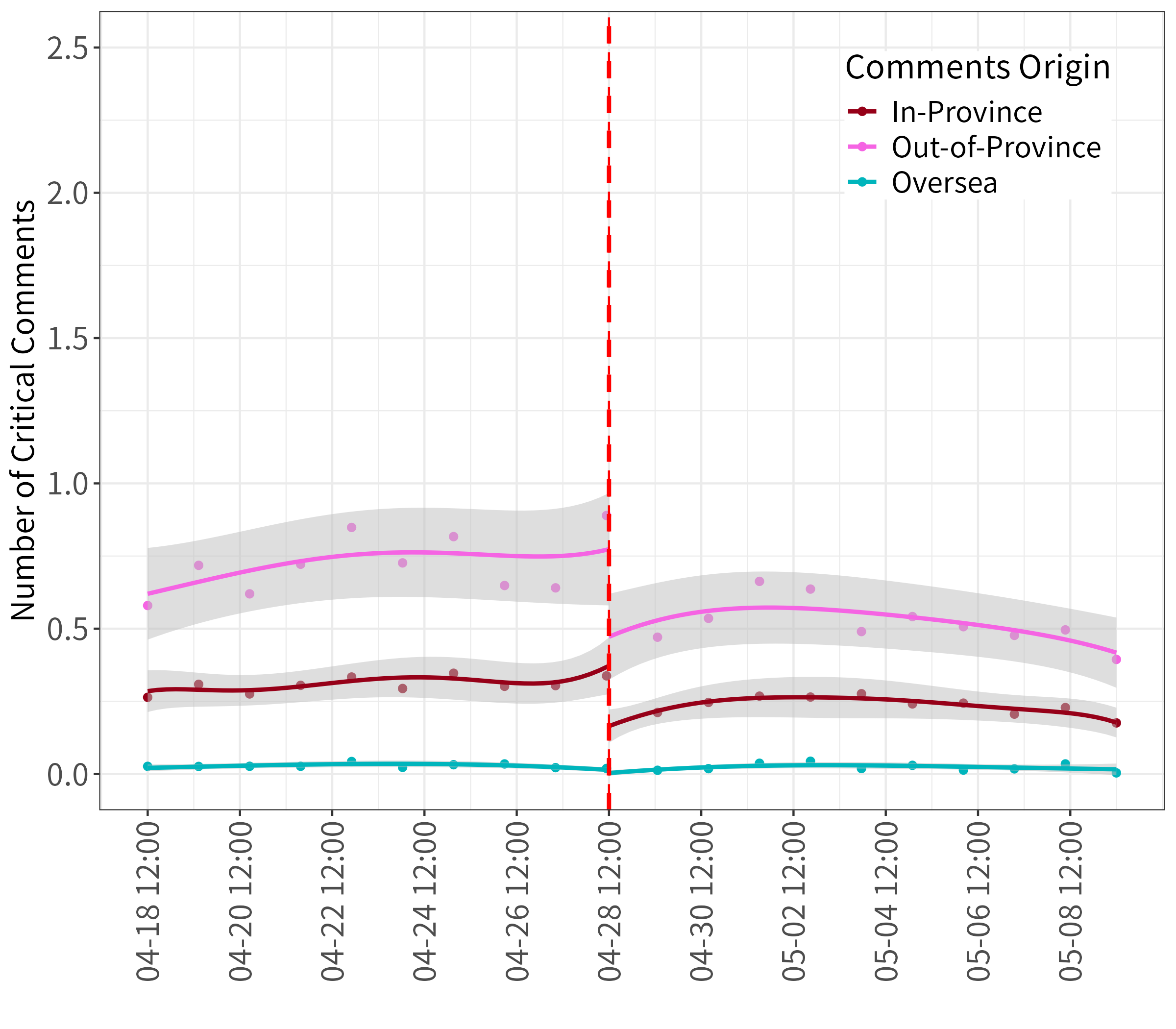}
    \caption{Decreased Critical Comments}
    \label{fig:pnas_number_critical_comments_local_48h_bandwidth_best_log}
  \end{subfigure}
  
  \caption{Robustness Check with Dependent Variables in Logarithmic Form}
  \label{fig:robustness_log_y_1}
  \caption*{\textbf{Notes:} Panels \ref{fig:pnas_volume_comments_oversea_origin_48h_bandwidth_best_log}-\ref{fig:pnas_number_critical_comments_local_48h_bandwidth_best_log} demonstrate robustness checks using dependent variables transformed into logarithmic form . Panel \ref{fig:pnas_volume_comments_oversea_origin_48h_bandwidth_best_log} shows comments on posts about international issues, while Panel \ref{fig:pnas_volume_comments_domestic_origin_48h_bandwidth_best_log} represents comments on posts about domestic issues, distinguishing between international and non-international posts. Panel \ref{fig:pnas_volume_comments_national_and_local_48h_bandwidth_best_log} focuses on posts about national and local issues. Panels \ref{fig:pnas_decreased_comments_local_by_origin_48h_bandwidth_best_log} and \ref{fig:pnas_number_critical_comments_local_48h_bandwidth_best_log} illustrate the robustness of the observed decline in overall and critical comments, respectively, categorized by commenters' geographic origins (in-province, out-of-province, and overseas). The red dashed lines mark the timing of the user location disclosure implementation, confirming that the main findings remain consistent when using logarithmically transformed dependent variables.}
\end{figure}

We conduct a series of robustness checks to validate our main findings. First, we assess whether transforming the dependent variables into log form affects our conclusions. Specifically, we re-estimate our models using $\log(Y + 1) $ as the outcome variable. As shown in Figure~\ref{fig:robustness_log_y_1}, the results remain consistent with our primary analysis, indicating that our conclusions are robust to this transformation.

\clearpage

\subsection{Robustness to bandwidth selection}

As shown in Table~\ref{table:treatment_effect}, our results are robust across a range of bandwidth choices. Below, we present interrupted time series plots using bandwidths of 8, 24, 72, 96, and 120 hours.

\begin{figure}[H]
  \centering
  
  \makebox[\textwidth][c]{%
    \begin{subfigure}{0.32\textwidth}
      \centering
      \includegraphics[width=\linewidth]{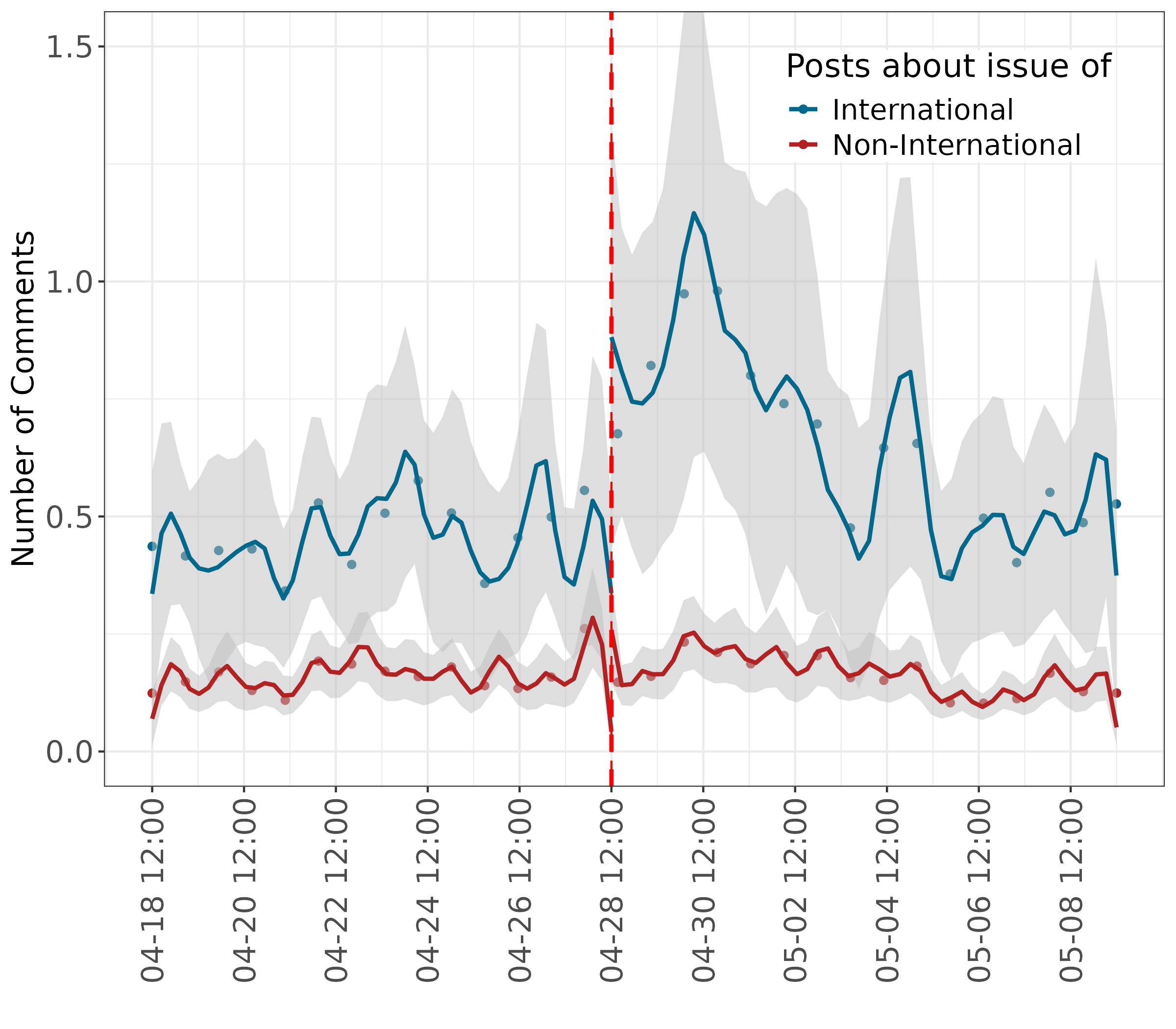}
      \caption{Bandwidth = 8 h}
      \label{fig:pnas_volume_comments_oversea_origin_48h_bandwidth_8h}
    \end{subfigure}
    \hspace{0.03\textwidth}
    \begin{subfigure}{0.32\textwidth}
      \centering
      \includegraphics[width=\linewidth]{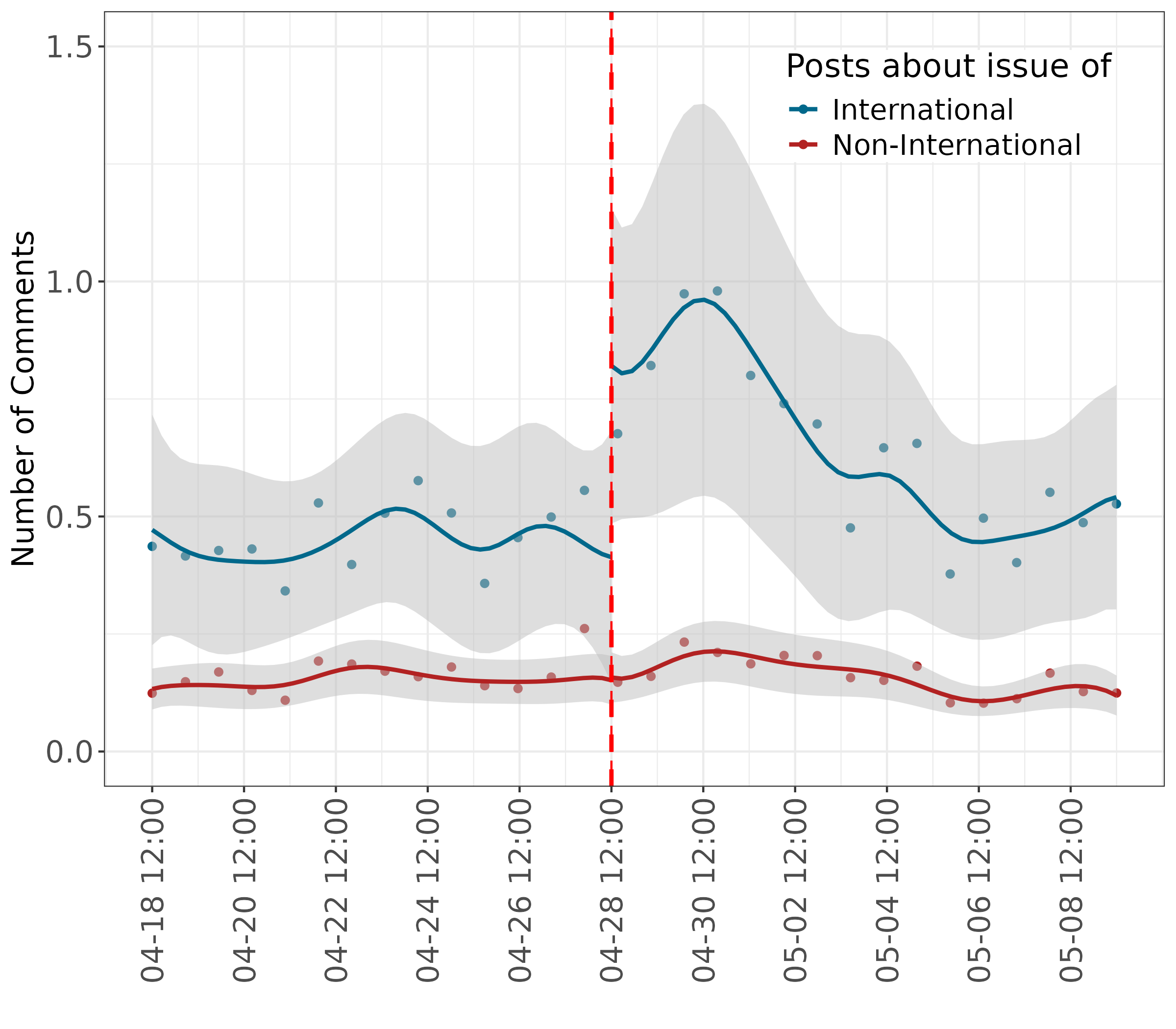}
      \caption{Bandwidth = 24 h}
      \label{fig:pnas_volume_comments_oversea_origin_48h_bandwidth_24h}
    \end{subfigure}
  }%
  
  \vspace{0.8em} 
  
  \makebox[\textwidth][c]{%
    \begin{subfigure}{0.32\textwidth}
      \centering
      \includegraphics[width=\linewidth]{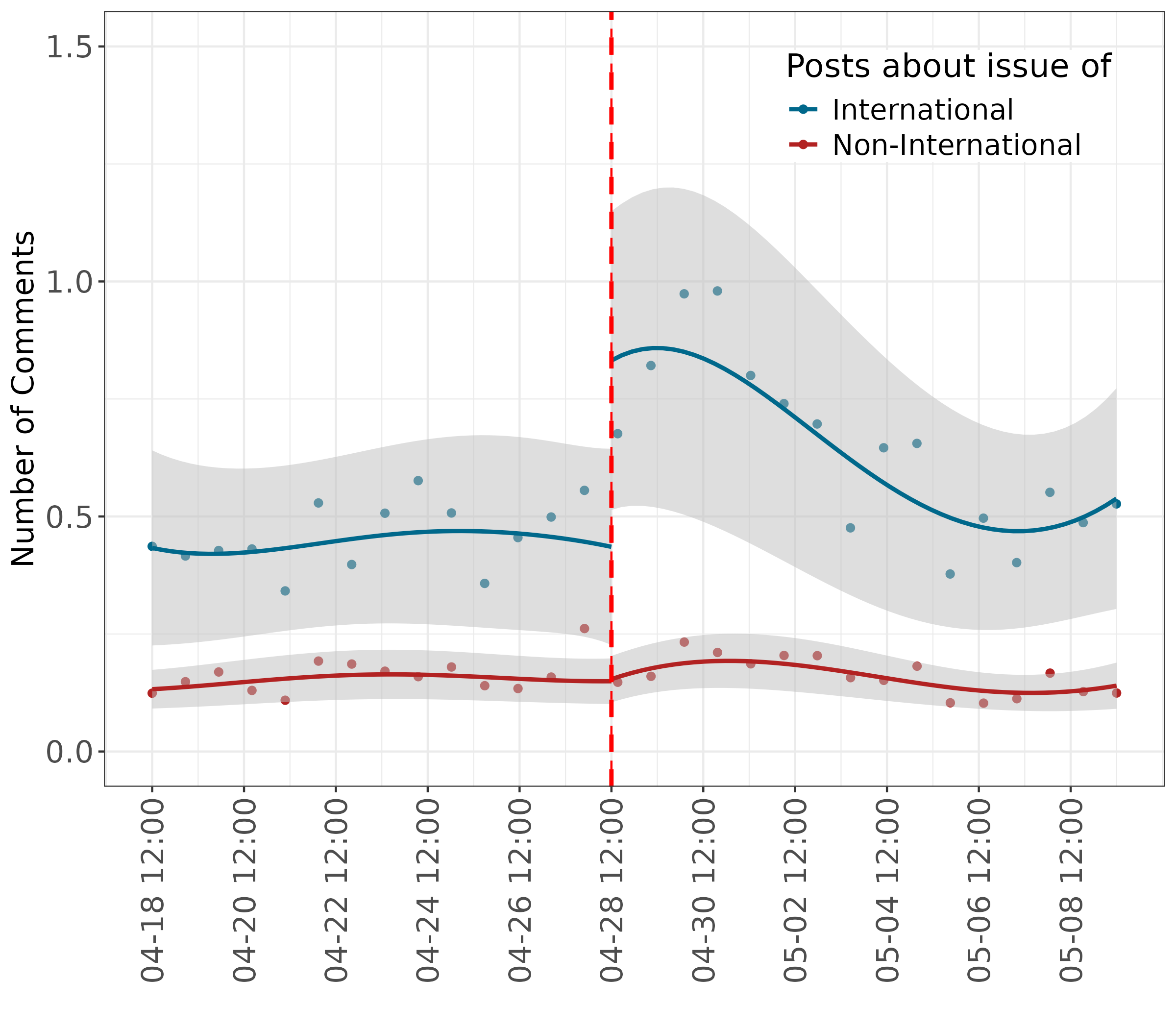}
      \caption{Bandwidth = 72 h}
      \label{fig:pnas_volume_comments_oversea_origin_48h_bandwidth_72h}
    \end{subfigure}
    \hspace{0.02\textwidth}
    \begin{subfigure}{0.32\textwidth}
      \centering
      \includegraphics[width=\linewidth]{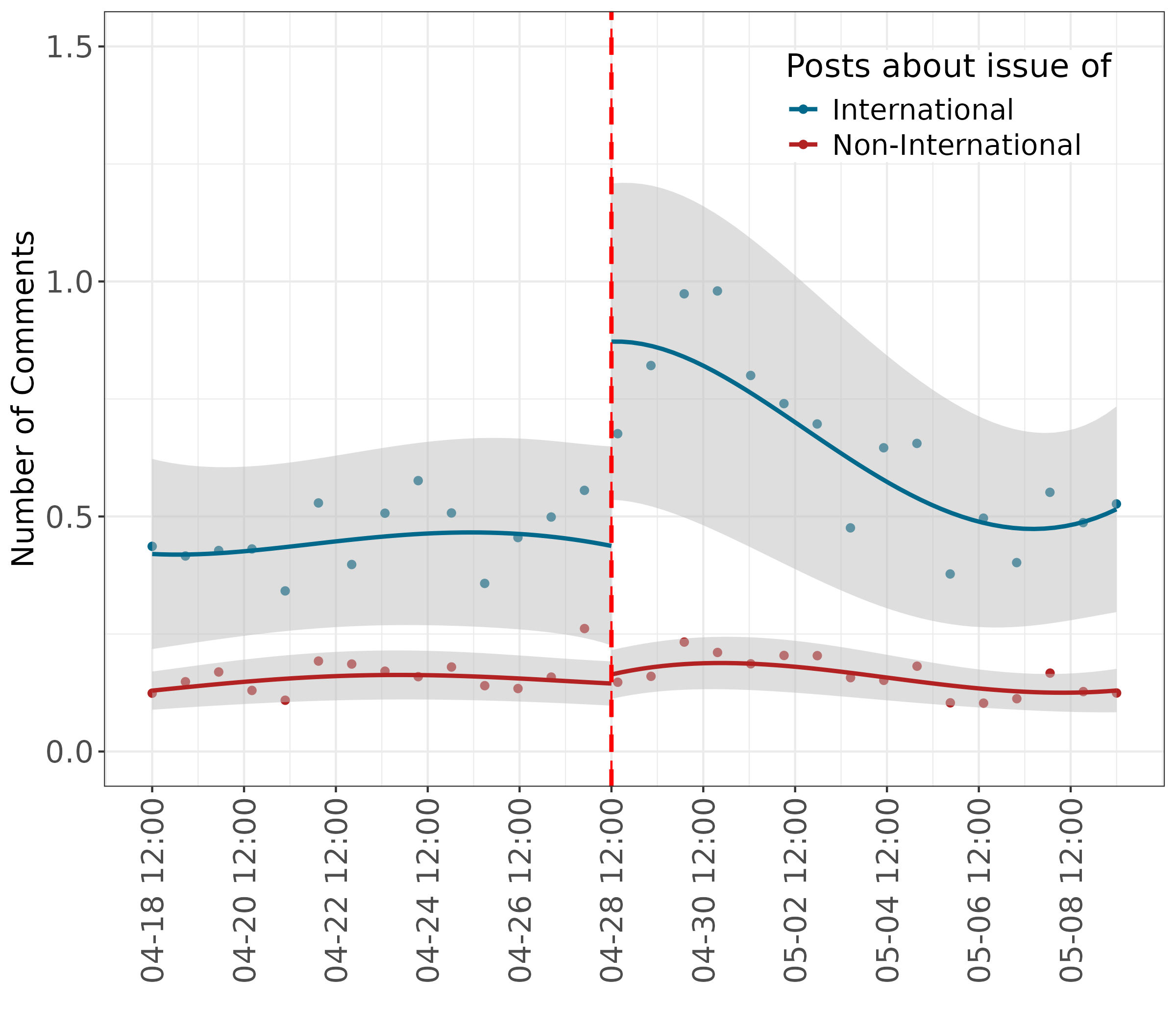}
      \caption{Bandwidth = 96 h}
      \label{fig:pnas_volume_comments_oversea_origin_48h_bandwidth_96h}
    \end{subfigure}
    \hspace{0.02\textwidth}
    \begin{subfigure}{0.32\textwidth}
      \centering
      \includegraphics[width=\linewidth]{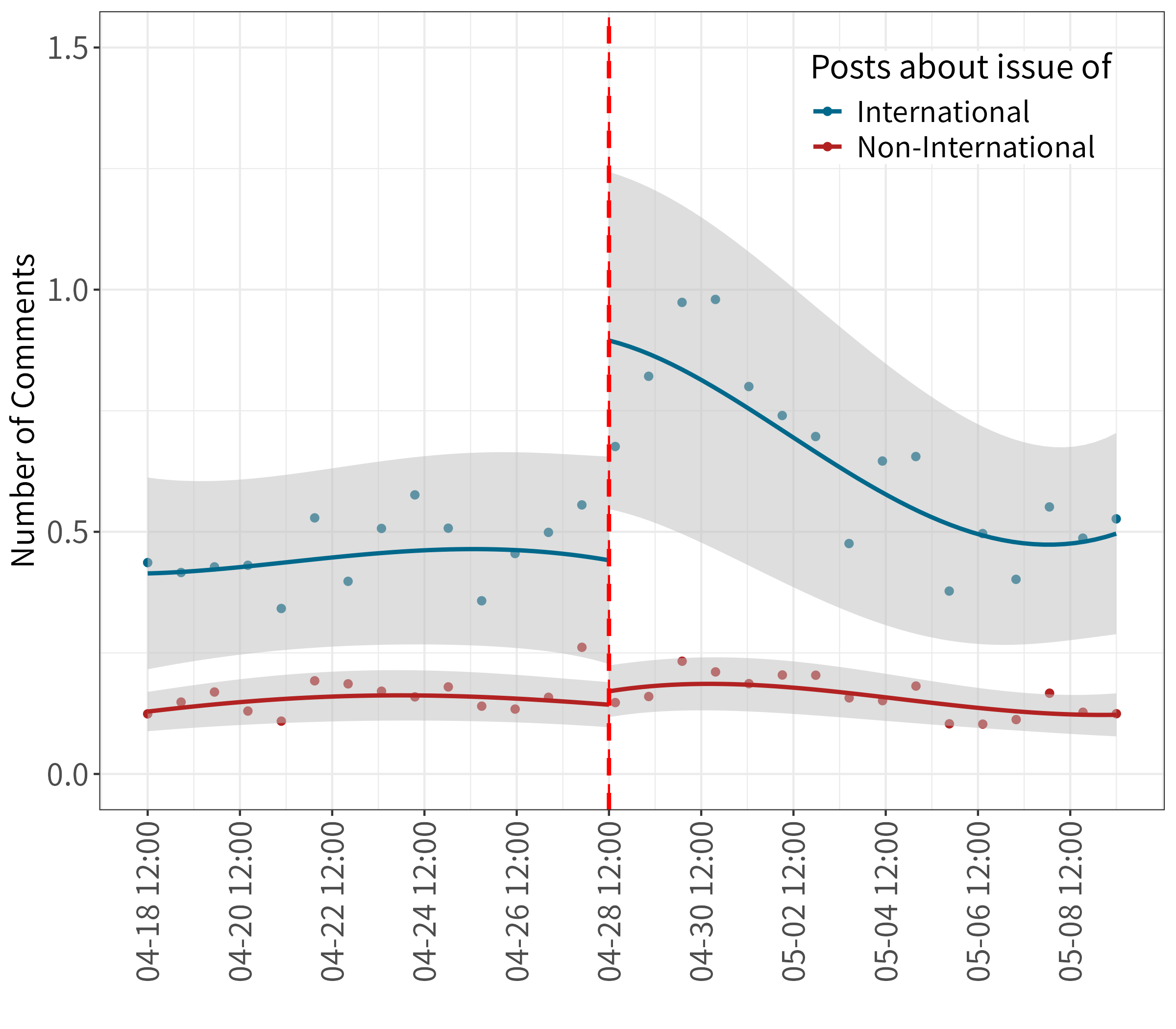}
      \caption{Bandwidth = 120 h}
      \label{fig:pnas_volume_comments_oversea_origin_48h_bandwidth_120h}
    \end{subfigure}
  }%
  
  \caption{Robustness Check with Different Bandwidths for Comments from Overseas Users}
  \label{fig:pnas_volume_comments_oversea_origin_48h_bandwidth}
    \caption*{\textbf{Notes:} Panels~\ref{fig:pnas_volume_comments_oversea_origin_48h_bandwidth_8h}--\ref{fig:pnas_volume_comments_oversea_origin_48h_bandwidth_120h} present robustness checks on the main finding regarding overseas users’ comments on international and non-international posts, using local smoothing with bandwidths of 8, 24, 72, 96, and 120 hours. Across all panels, we consistently observe a sharp increase in comments on international topics after the policy implementation, while comments on non-international topics remain stable. The red dashed lines mark the timing of the policy implementation.}
\end{figure}

To illustrate, we reanalyze the main result from Figure~1a, which shows a post-policy increase in comment volume from overseas users on international topics. Figure~\ref{fig:pnas_volume_comments_oversea_origin_48h_bandwidth} replicates this analysis using the five alternative bandwidths. With a narrower bandwidth (8 hours; Figure~\ref{fig:pnas_volume_comments_oversea_origin_48h_bandwidth_8h}), the smoothing line appears more variable, but the pattern remains: comments on international posts rise sharply, while those on non-international posts remain stable. This finding holds consistently with wider bandwidths of 24 hours (Figure~\ref{fig:pnas_volume_comments_oversea_origin_48h_bandwidth_24h}), 72 hours (Figure~\ref{fig:pnas_volume_comments_oversea_origin_48h_bandwidth_72h}), 96 hours (Figure~\ref{fig:pnas_volume_comments_oversea_origin_48h_bandwidth_96h}), and 120 hours (Figure~\ref{fig:pnas_volume_comments_oversea_origin_48h_bandwidth_120h}).

\clearpage

\begin{figure}[H]
  \centering
  
  \makebox[\textwidth][c]{%
    \begin{subfigure}{0.32\textwidth}
      \centering
      \includegraphics[width=\linewidth]{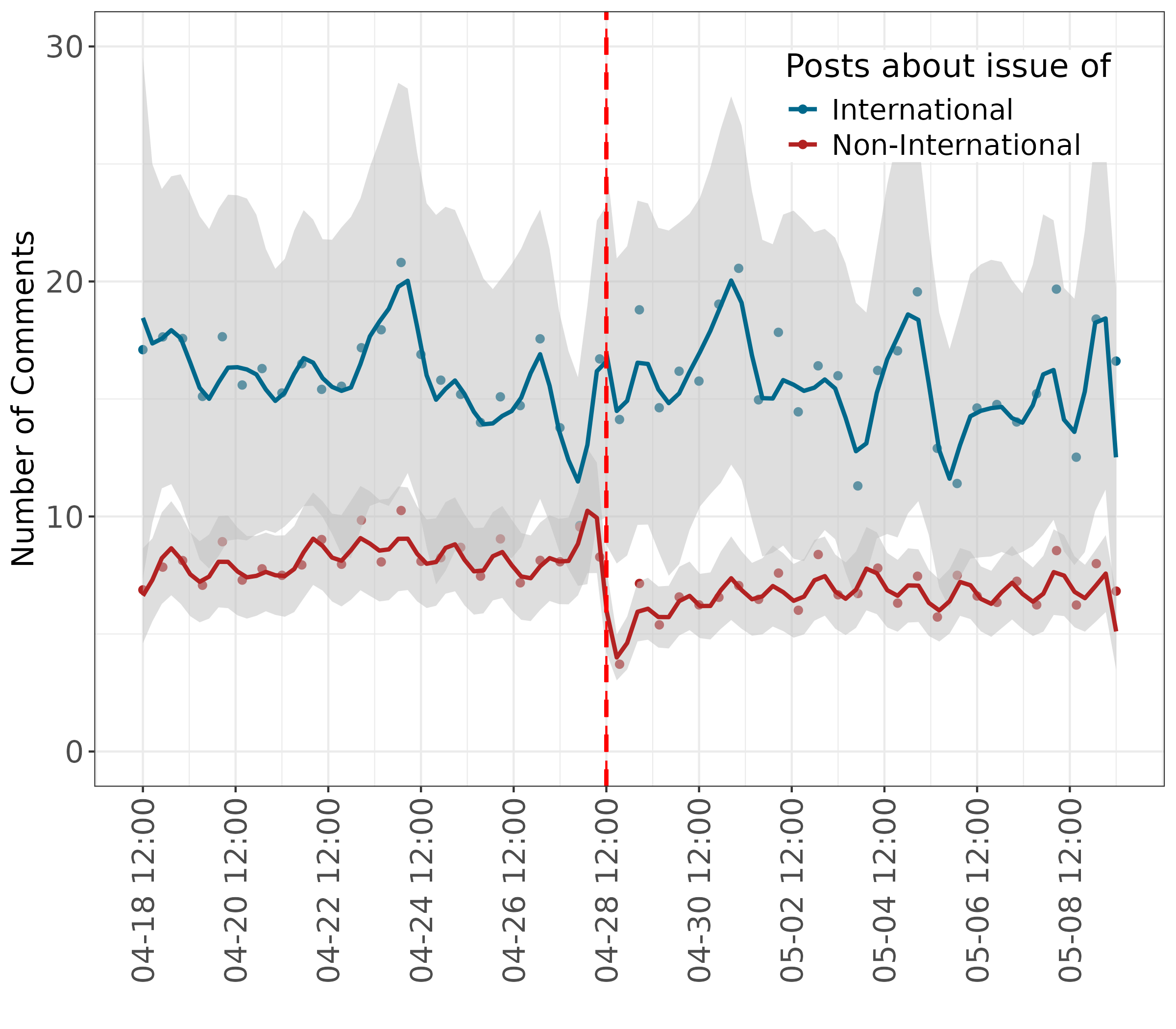}
      \caption{Bandwidth = 8 h}      \label{fig:pnas_volume_comments_domestic_origin_48h_bandwidth_8h}
    \end{subfigure}
    \hspace{0.03\textwidth}
    \begin{subfigure}{0.32\textwidth}
      \centering
      \includegraphics[width=\linewidth]{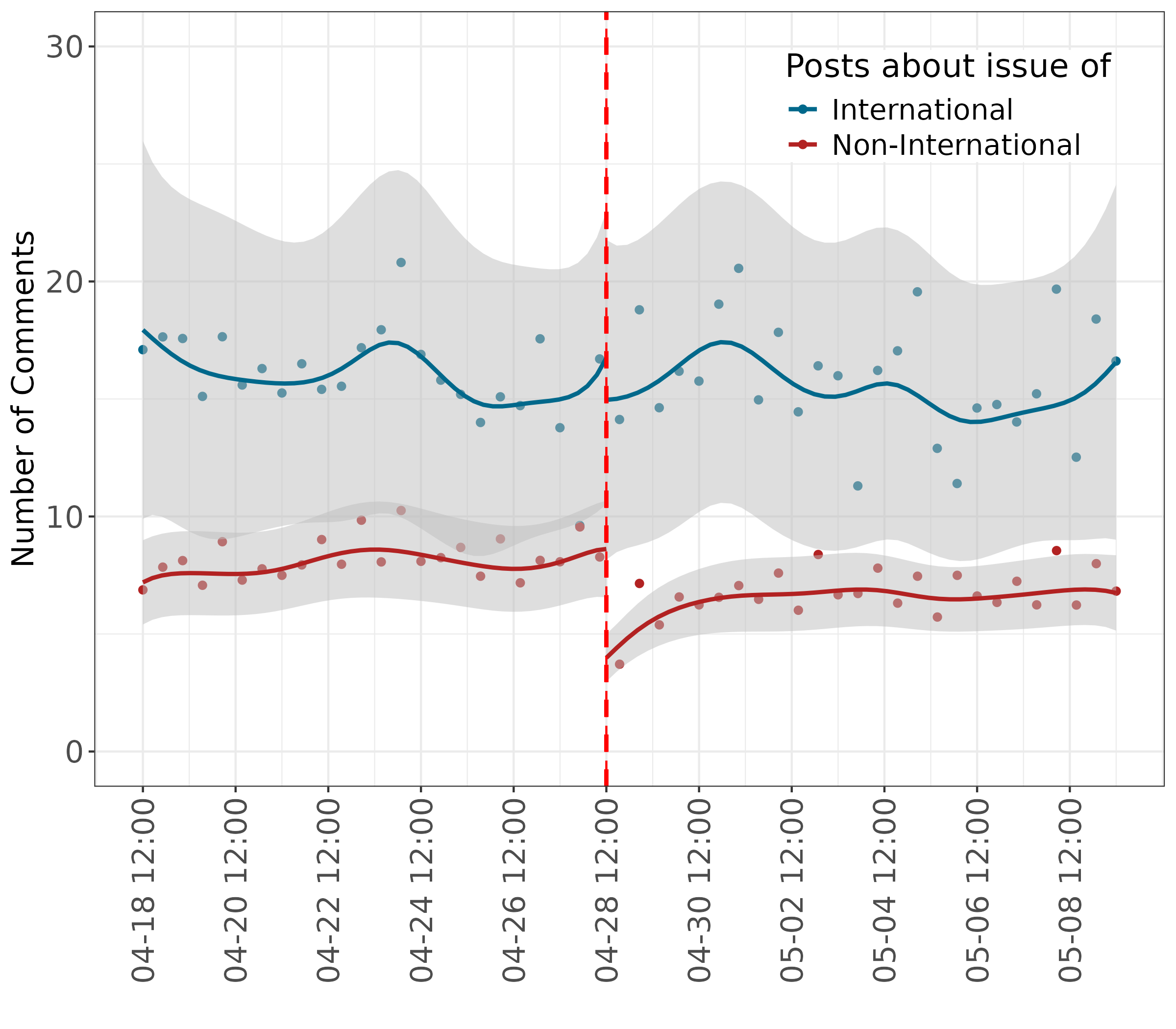}
      \caption{Bandwidth = 24 h}      \label{fig:pnas_volume_comments_domestic_origin_48h_bandwidth_24h}
    \end{subfigure}
  }%
  
  \vspace{0.8em} 
  
  \makebox[\textwidth][c]{%
    \begin{subfigure}{0.32\textwidth}
      \centering
      \includegraphics[width=\linewidth]{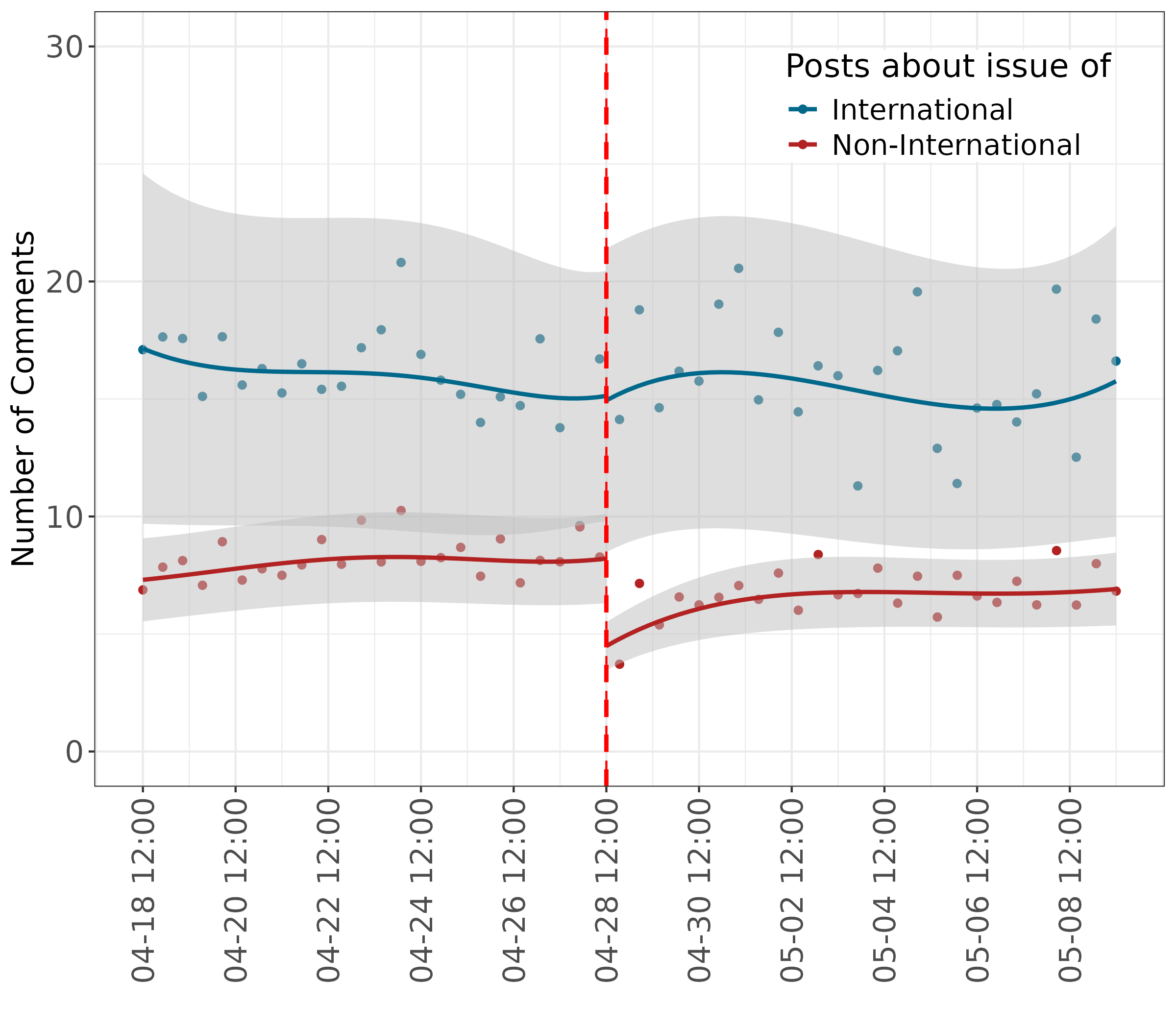}
      \caption{Bandwidth = 72 h}
      \label{fig:pnas_volume_comments_domestic_origin_48h_bandwidth_72h}
    \end{subfigure}
    \hspace{0.02\textwidth}
    \begin{subfigure}{0.32\textwidth}
      \centering
      \includegraphics[width=\linewidth]{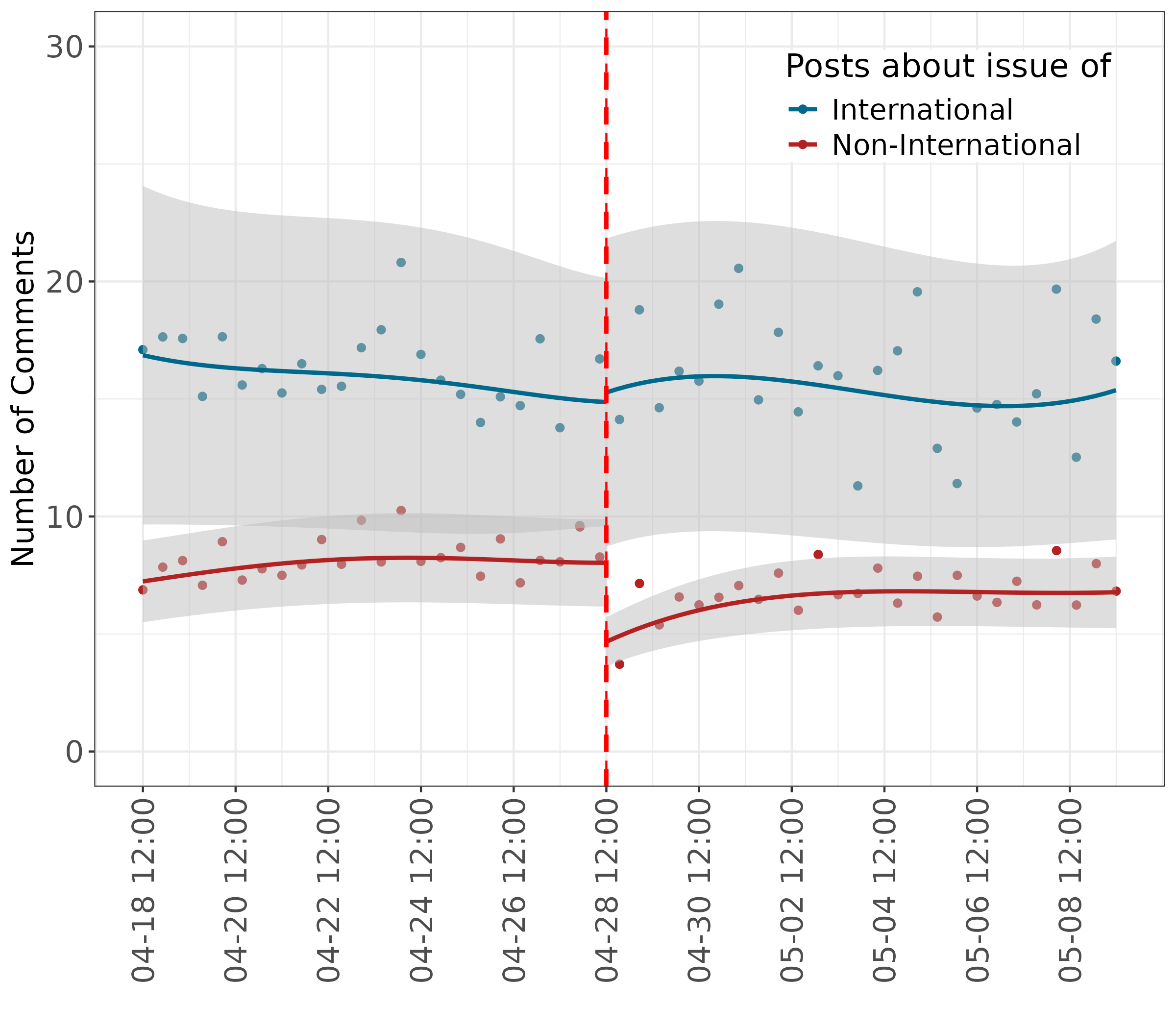}
      \caption{Bandwidth = 96 h}
      \label{fig:pnas_volume_comments_domestic_origin_48h_bandwidth_96h}
    \end{subfigure}
    \hspace{0.02\textwidth}
    \begin{subfigure}{0.32\textwidth}
      \centering
      \includegraphics[width=\linewidth]{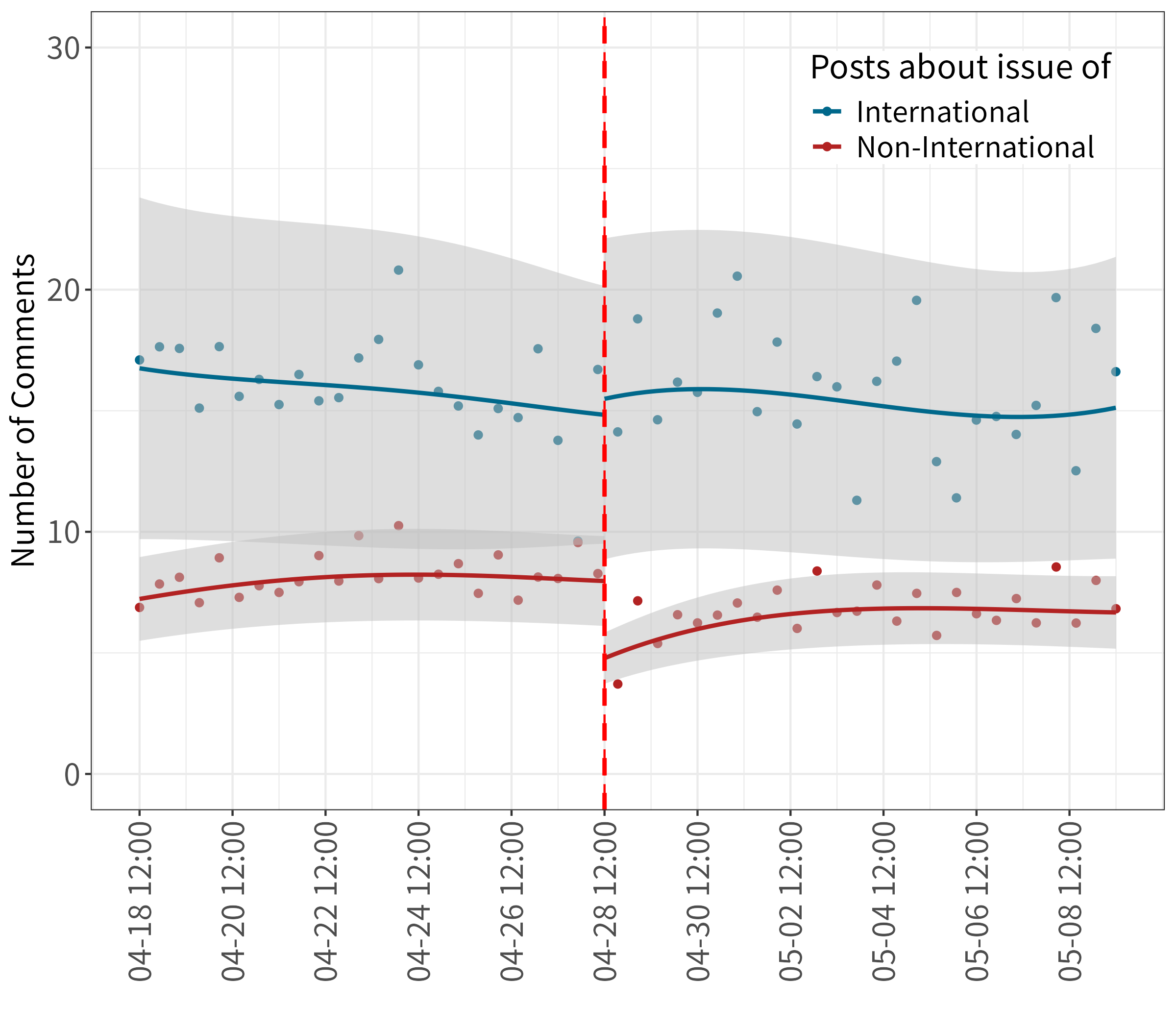}
      \caption{Bandwidth = 120 h}
      \label{fig:pnas_volume_comments_domestic_origin_48h_bandwidth_120h}
    \end{subfigure}
  }%
  
  \caption{Robustness Check with Different Bandwidths for Comments from Domestic Users}
  \label{fig:pnas_volume_comments_domestic_origin_48h_bandwidth}
    \caption*{\textbf{Notes:} Panels~\ref{fig:pnas_volume_comments_domestic_origin_48h_bandwidth_8h}--\ref{fig:pnas_volume_comments_domestic_origin_48h_bandwidth_120h} show robustness checks for domestic user comments on international and non-international posts using bandwidths of 8, 24, 72, 96, and 120 hours. Across all bandwidths, the decline in comments appears only on non-international topics after the policy implementation, while engagement on international topics remains stable. The red dashed lines mark the timing of the policy implementation.}
\end{figure}

Figure~\ref{fig:pnas_volume_comments_domestic_origin_48h_bandwidth} replicates Figure~1b using alternative bandwidths. Across all specifications, comments from domestic users consistently decline only on non-international posts, mirroring the pattern observed with the cross-validated bandwidth.

\clearpage

\begin{figure}[h]
  \centering
  
  \makebox[\textwidth][c]{%
    \begin{subfigure}{0.32\textwidth}
      \centering
      \includegraphics[width=\linewidth]{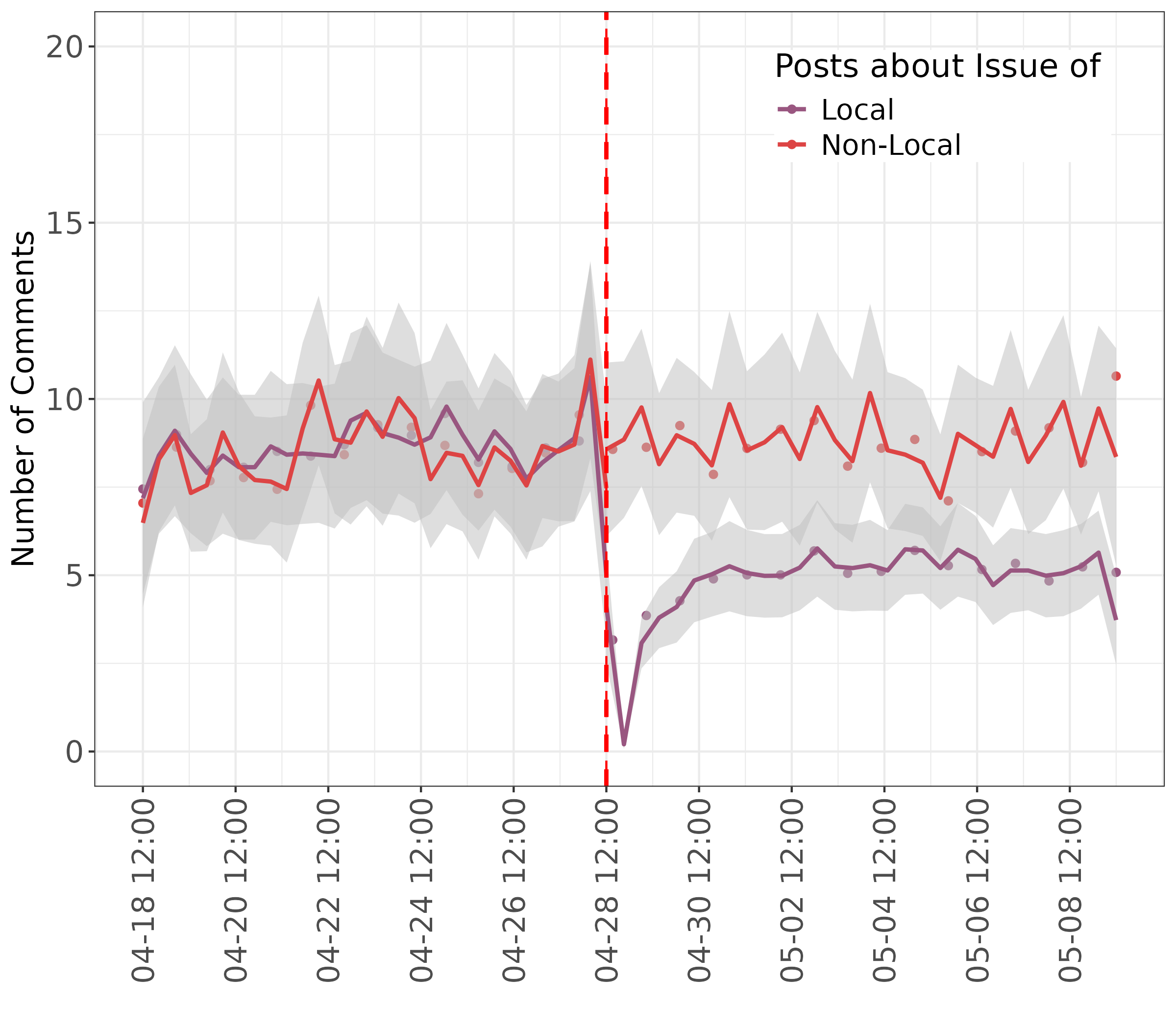}
      \caption{Bandwidth = 8 h}
      \label{fig:pnas_volume_comments_national_and_local_48h_bandwidth_8h}
    \end{subfigure}
    \hspace{0.03\textwidth}
    \begin{subfigure}{0.32\textwidth}
      \centering
      \includegraphics[width=\linewidth]{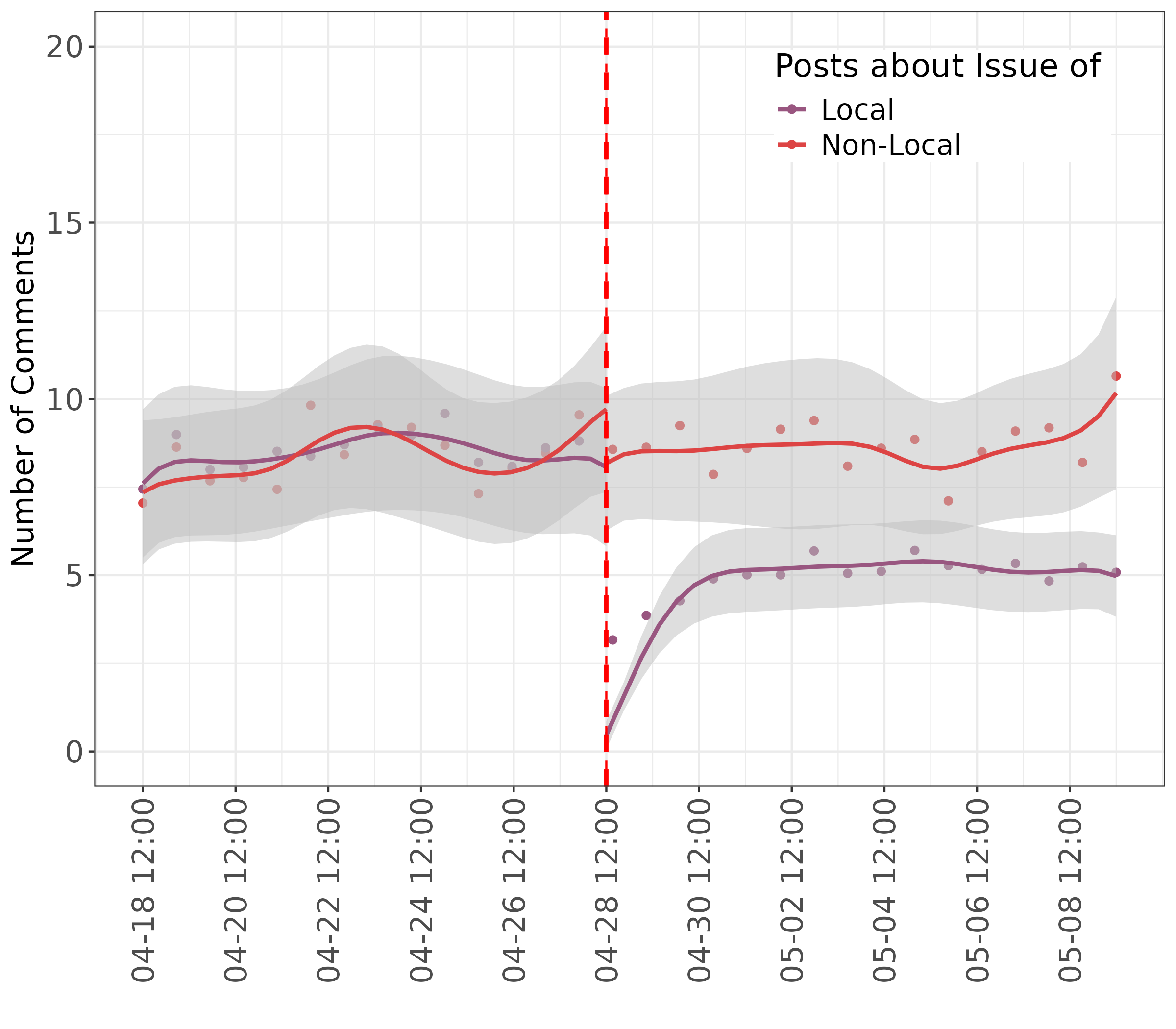}
      \caption{Bandwidth = 24 h}
      \label{fig:pnas_volume_comments_national_and_local_48h_bandwidth_24h}
    \end{subfigure}
  }%
  
  \vspace{0.8em} 
  
  \makebox[\textwidth][c]{%
    \begin{subfigure}{0.32\textwidth}
      \centering
      \includegraphics[width=\linewidth]{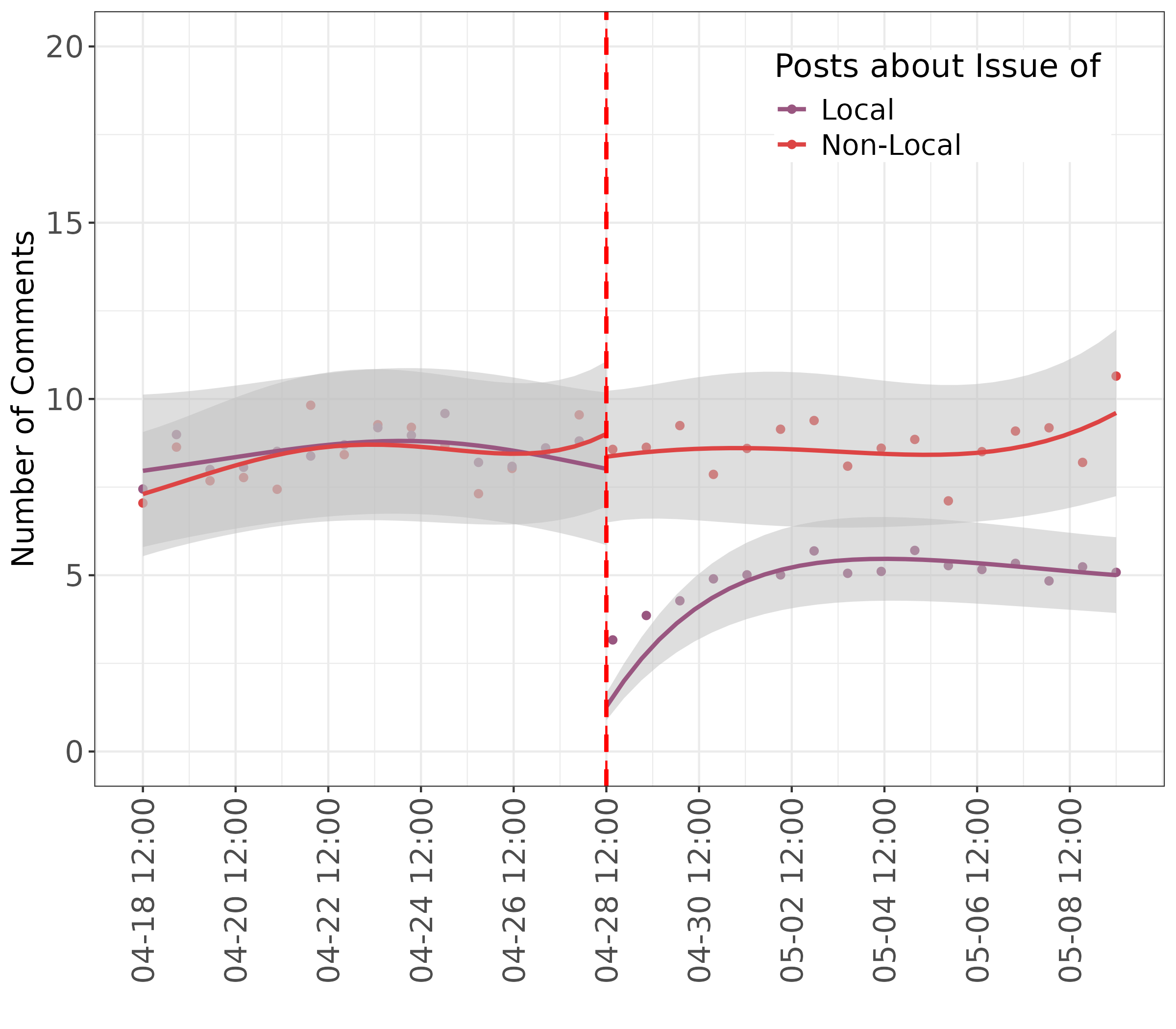}
      \caption{Bandwidth = 72 h}
      \label{fig:pnas_volume_comments_national_and_local_48h_bandwidth_72h}
    \end{subfigure}
    \hspace{0.02\textwidth}
    \begin{subfigure}{0.32\textwidth}
      \centering
      \includegraphics[width=\linewidth]{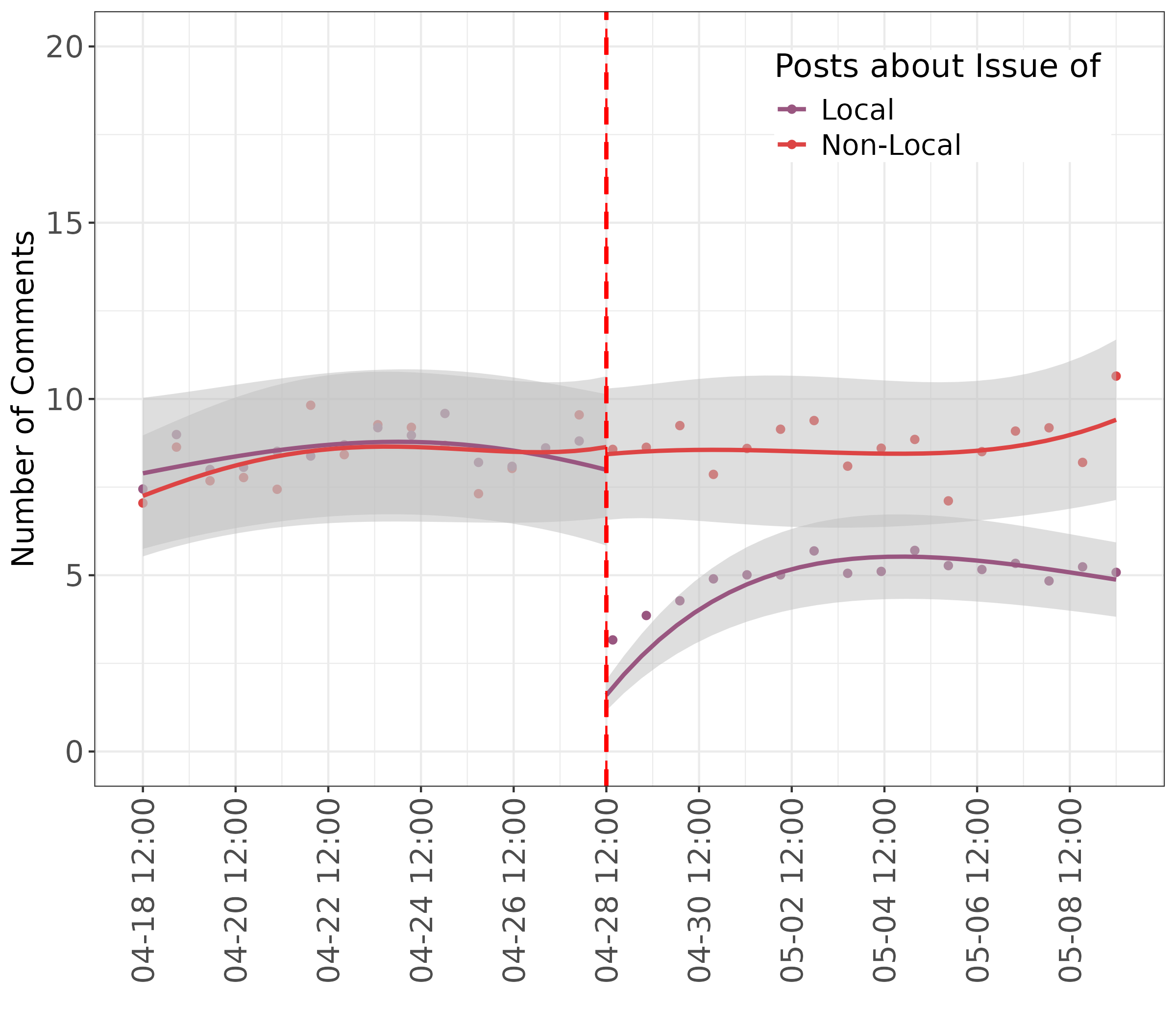}
      \caption{Bandwidth = 96 h}
      \label{fig:pnas_volume_comments_national_and_local_48h_bandwidth_96h}
    \end{subfigure}
    \hspace{0.02\textwidth}
    \begin{subfigure}{0.32\textwidth}
      \centering
      \includegraphics[width=\linewidth]{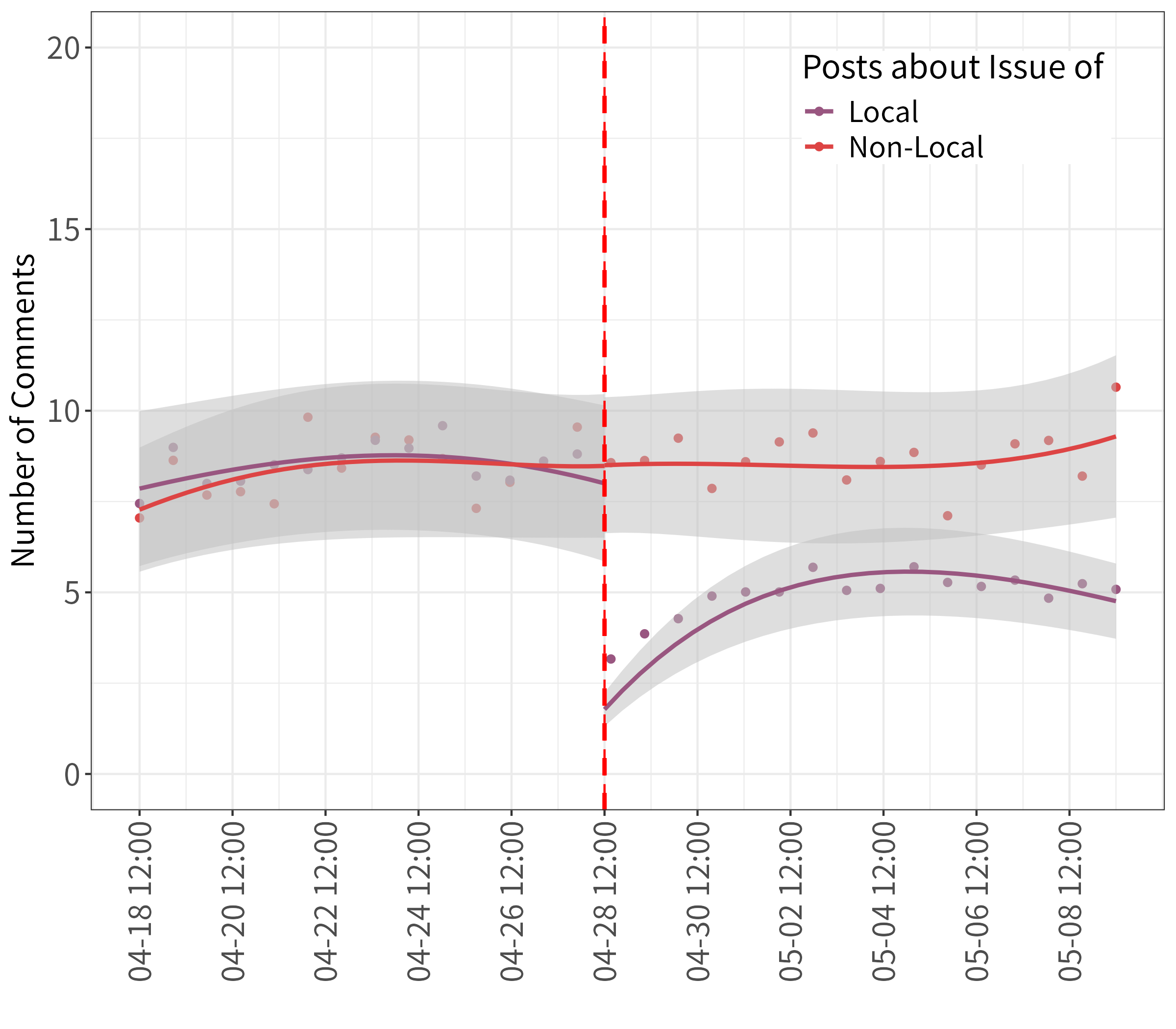}
      \caption{Bandwidth = 120 h}
      \label{fig:pnas_volume_comments_national_and_local_48h_bandwidth_120h}
    \end{subfigure}
  }%
  
  \caption{Zoom-In on Non-International Issues with Different Bandwidths}
  \label{fig:pnas_volume_comments_national_and_local_48h_bandwidth}
      \caption*{\textbf{Notes:} Panels \ref{fig:pnas_volume_comments_national_and_local_48h_bandwidth_8h}–\ref{fig:pnas_volume_comments_national_and_local_48h_bandwidth_120h} display comment volumes on non-international issues across bandwidths of 8, 24, 72, 96, and 120 hours. In every case, a clear decline in comments on local issues is evident following the user location disclosure, while comment volumes on non-local issues remain stable or slightly increase. These patterns underscore the robustness of our findings. The red dashed lines mark the timing of the policy implementation.}
\end{figure}

Figure~\ref{fig:pnas_volume_comments_national_and_local_48h_bandwidth} tests the robustness of our main finding in Figure~1c by examining comment dynamics on non-international posts under varying bandwidths (8, 24, 72, 96, and 120 hours). Across all bandwidth choices, the decline in comments on local issues after the user location disclosure policy remains consistent, confirming that the observed pattern is not sensitive to bandwidth selection and reinforcing the reliability of our results.

\clearpage

\begin{figure}[H]
  \centering  
  \makebox[\textwidth][c]{%
    \begin{subfigure}{0.32\textwidth}
      \centering
      \includegraphics[width=\linewidth]{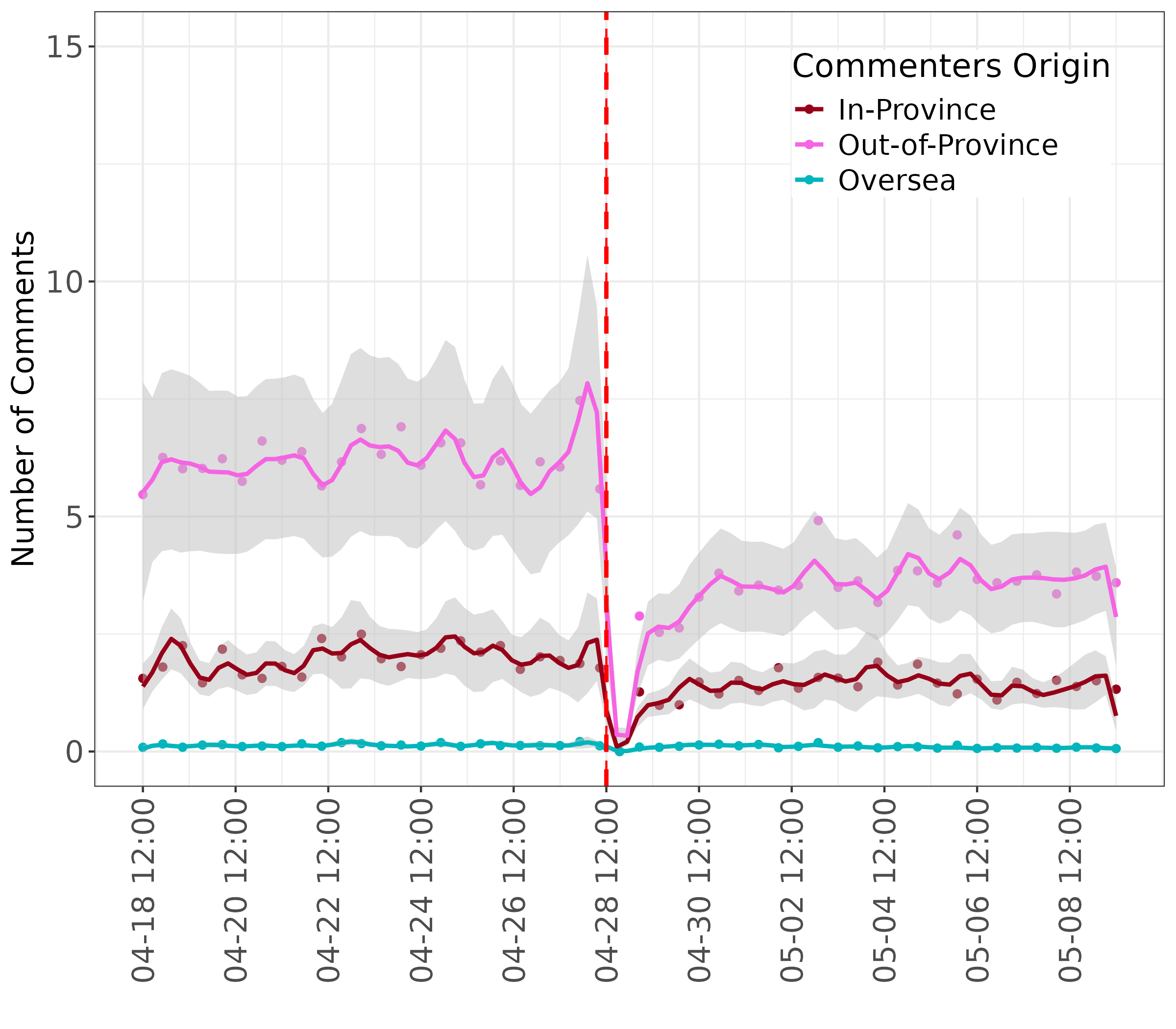}
      \caption{Bandwidth = 8 h}
      \label{fig:pnas_decreased_comments_local_by_origin_48h_bandwidth_8h}
    \end{subfigure}
    \hspace{0.03\textwidth}
    \begin{subfigure}{0.32\textwidth}
      \centering
      \includegraphics[width=\linewidth]{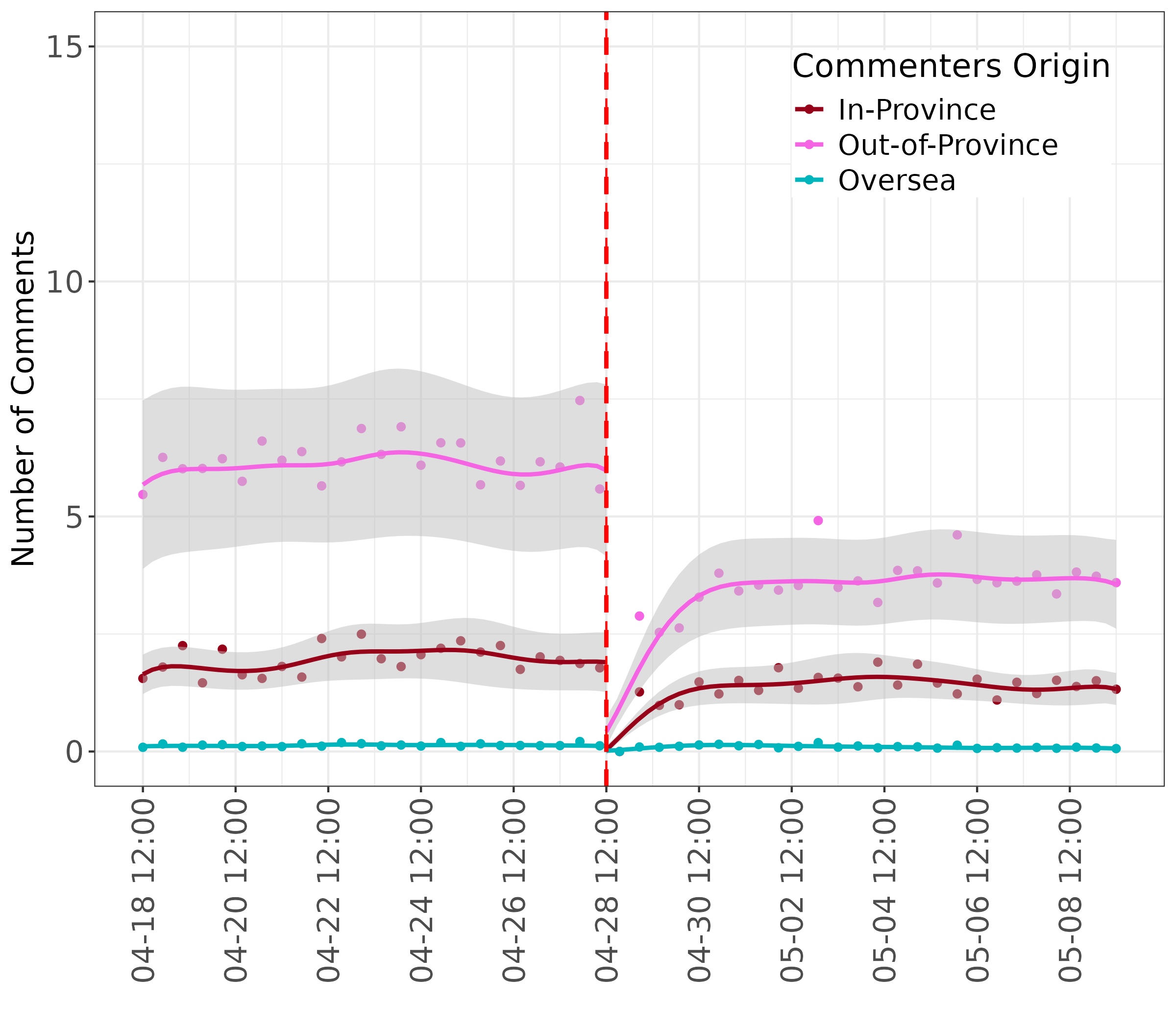}
      \caption{Bandwidth = 24 h}
      \label{fig:pnas_decreased_comments_local_by_origin_48h_bandwidth_24h}
    \end{subfigure}
  }%
  
  \vspace{0.8em} 
  
  \makebox[\textwidth][c]{%
    \begin{subfigure}{0.32\textwidth}
      \centering
      \includegraphics[width=\linewidth]{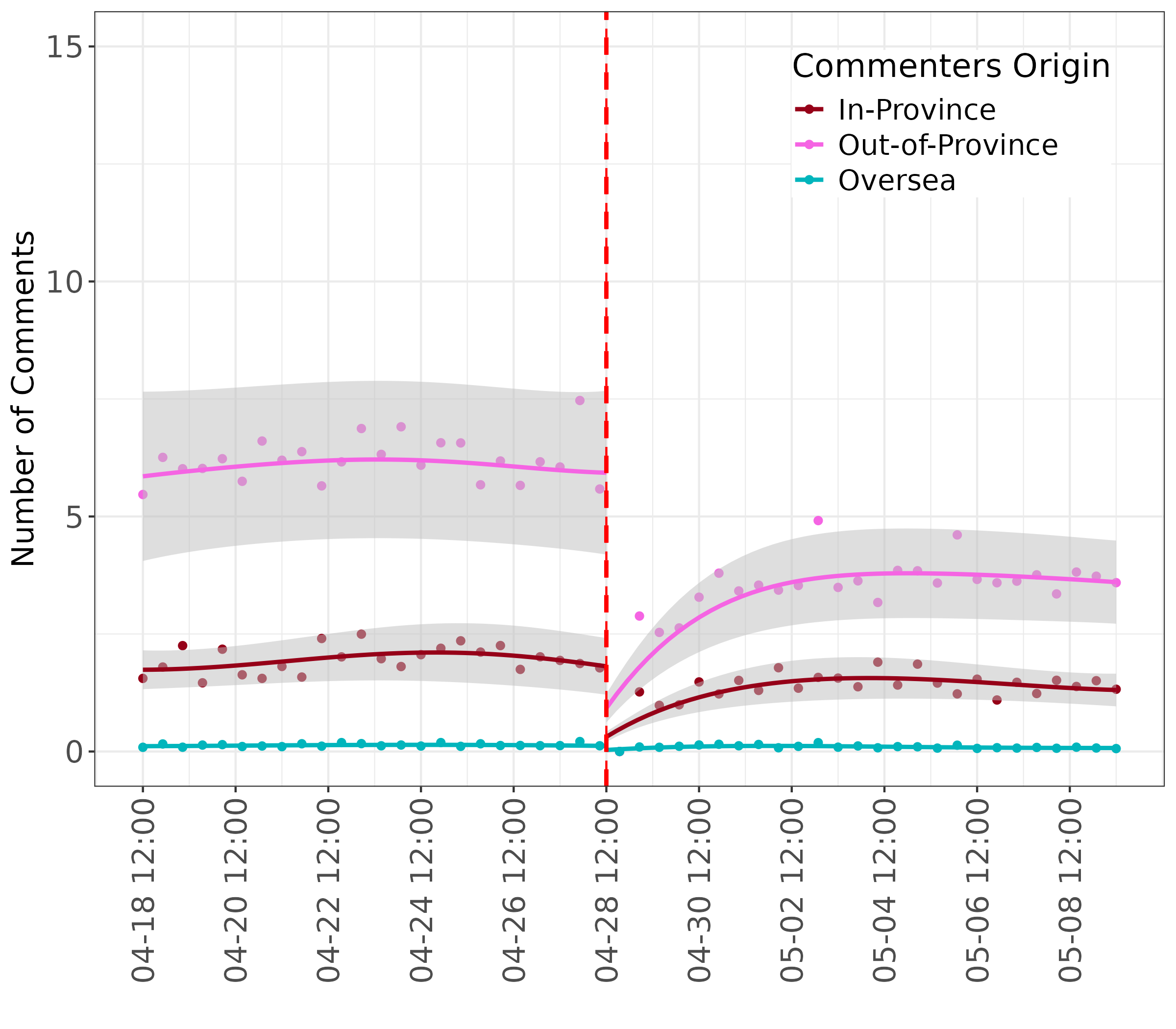}
      \caption{Bandwidth = 72 h}
      \label{fig:pnas_decreased_comments_local_by_origin_48h_bandwidth_72h}
    \end{subfigure}
    \hspace{0.02\textwidth}
    \begin{subfigure}{0.32\textwidth}
      \centering
      \includegraphics[width=\linewidth]{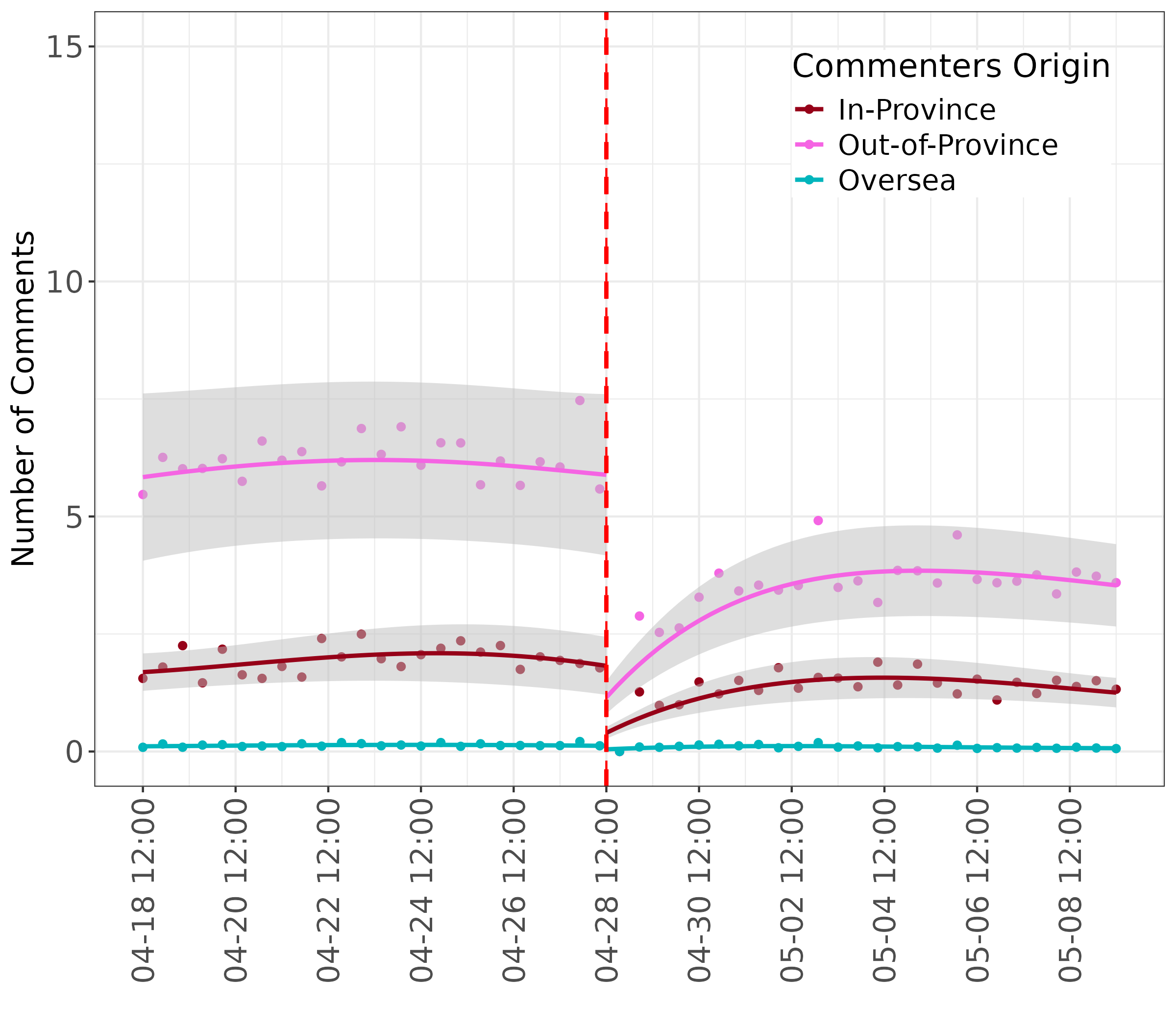}
      \caption{Bandwidth = 96 h}
      \label{fig:pnas_decreased_comments_local_by_origin_48h_bandwidth_96h}
    \end{subfigure}
    \hspace{0.02\textwidth}
    \begin{subfigure}{0.32\textwidth}
      \centering
      \includegraphics[width=\linewidth]{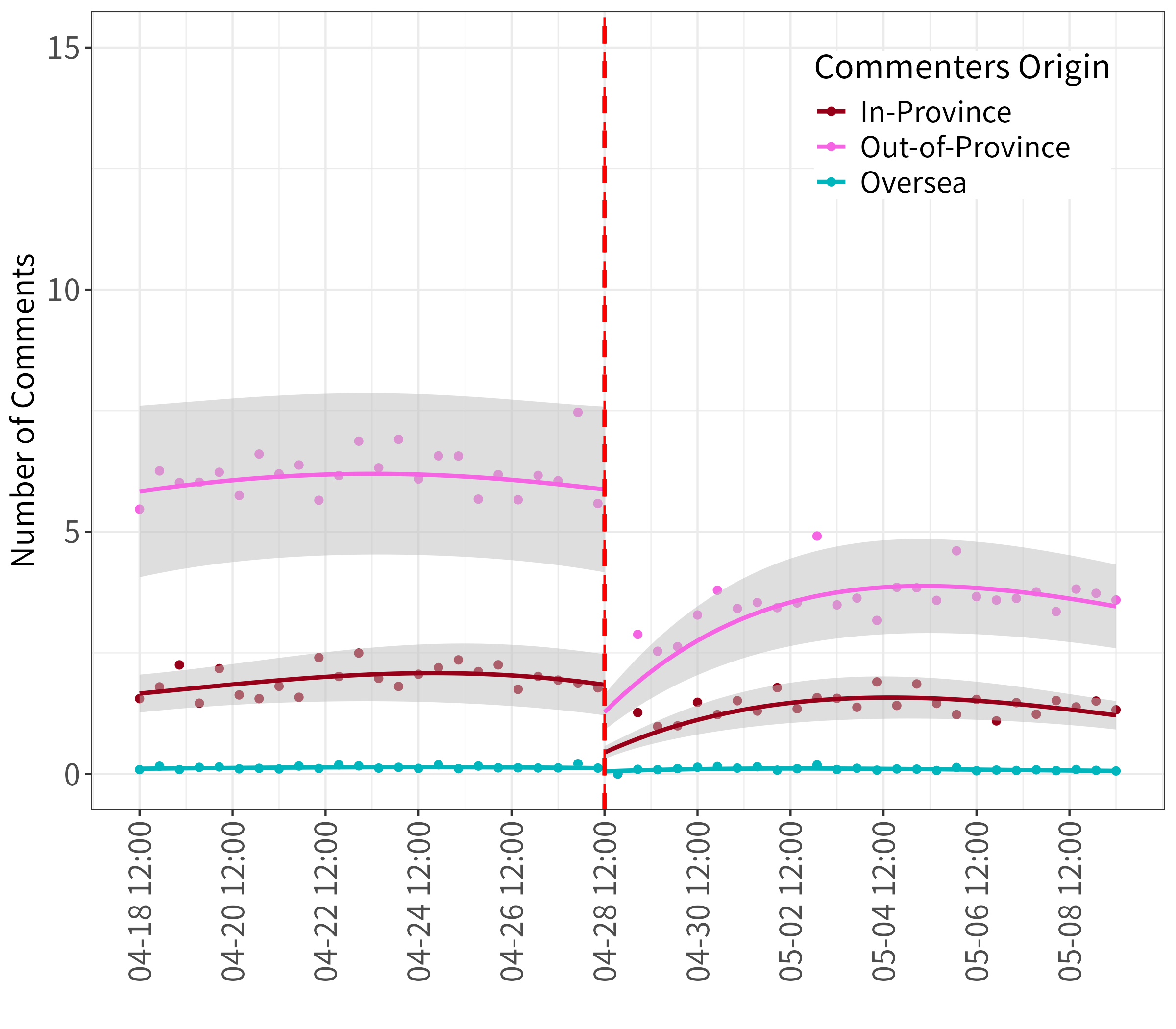}
      \caption{Bandwidth = 120 h}
      \label{fig:pnas_decreased_comments_local_by_origin_48h_bandwidth_120h}
    \end{subfigure}
  }%
  
  \caption{Decreased Comments on Local Issues with Different Bandwidths}
  \label{fig:pnas_decreased_comments_local_by_origin_48h_bandwidth}
  \caption*{\textbf{Notes:} Panels \ref{fig:pnas_decreased_comments_local_by_origin_48h_bandwidth_8h}–\ref{fig:pnas_decreased_comments_local_by_origin_48h_bandwidth_120h} present robustness checks on comment volume declines for local-topic posts, disaggregated by commenter origin (in-province, out-of-province, and overseas) across bandwidths of 8, 24, 72, 96, and 120 hours. Across all specifications, the most pronounced and consistent decline follows the policy implementation among out-of-province users. In-province and overseas comment levels remain stable or show only minor fluctuations. The red dashed lines mark the timing of the policy implementation.}
\end{figure}

Figure \ref{fig:pnas_decreased_comments_local_by_origin_48h_bandwidth} examines the decline in comments on local topics by commenter origin, replicating Figure~1e with varying bandwidths. Across all bandwidth choices, the estimated reduction in out-of-province comments remains robust, confirming that the user location disclosure policy disproportionately affected non-local engagement. This reinforces our conclusion that the policy significantly curtailed participation from geographically distant users.

\clearpage

\begin{figure}[H]
  \centering
  
  \makebox[\textwidth][c]{%
    \begin{subfigure}{0.32\textwidth}
      \centering
      \includegraphics[width=\linewidth]{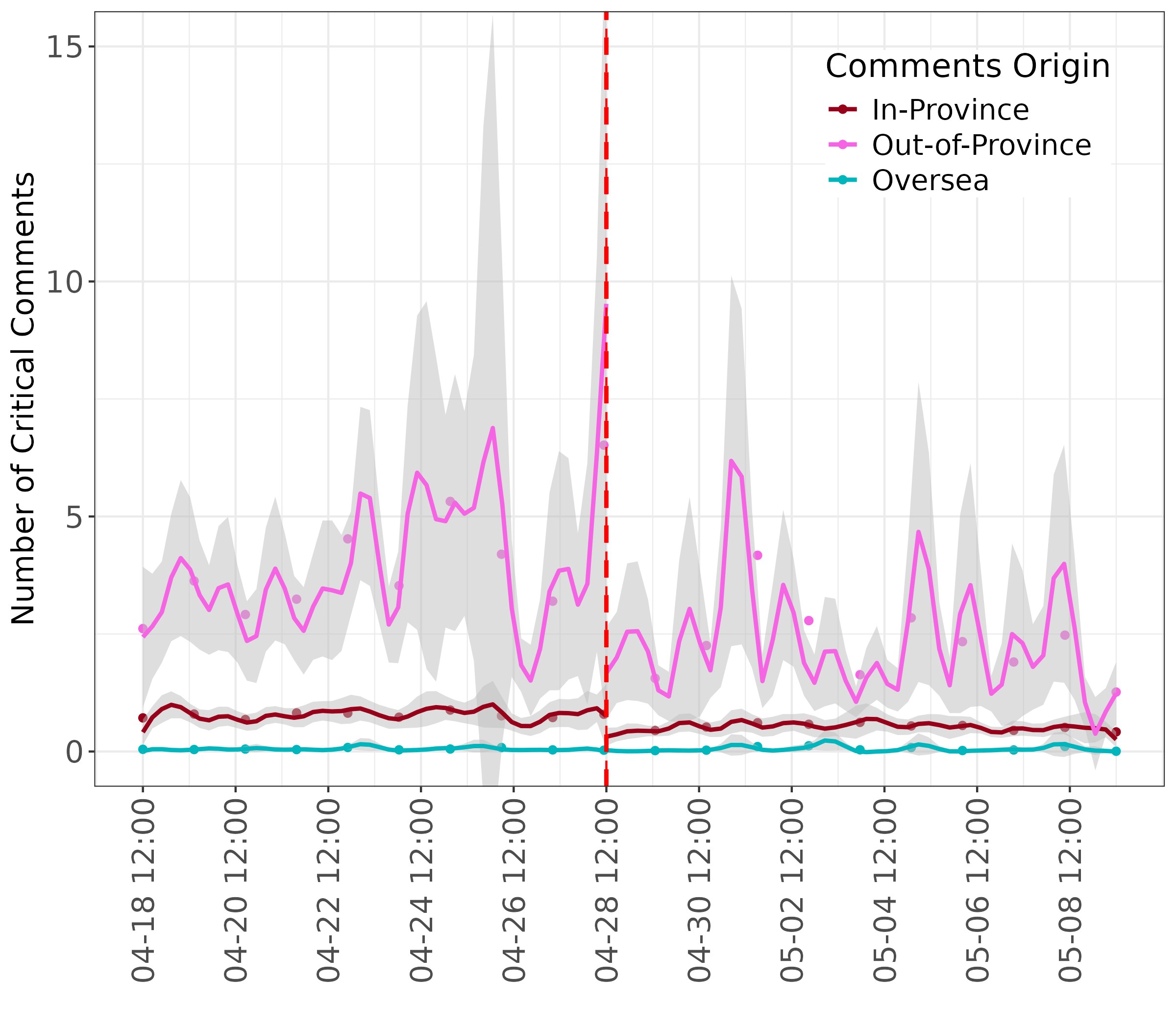}
      \caption{Bandwidth = 8 h}
      \label{fig:pnas_number_critical_comments_local_48h_bandwidth_8h}
    \end{subfigure}
    \hspace{0.03\textwidth}
    \begin{subfigure}{0.32\textwidth}
      \centering
      \includegraphics[width=\linewidth]{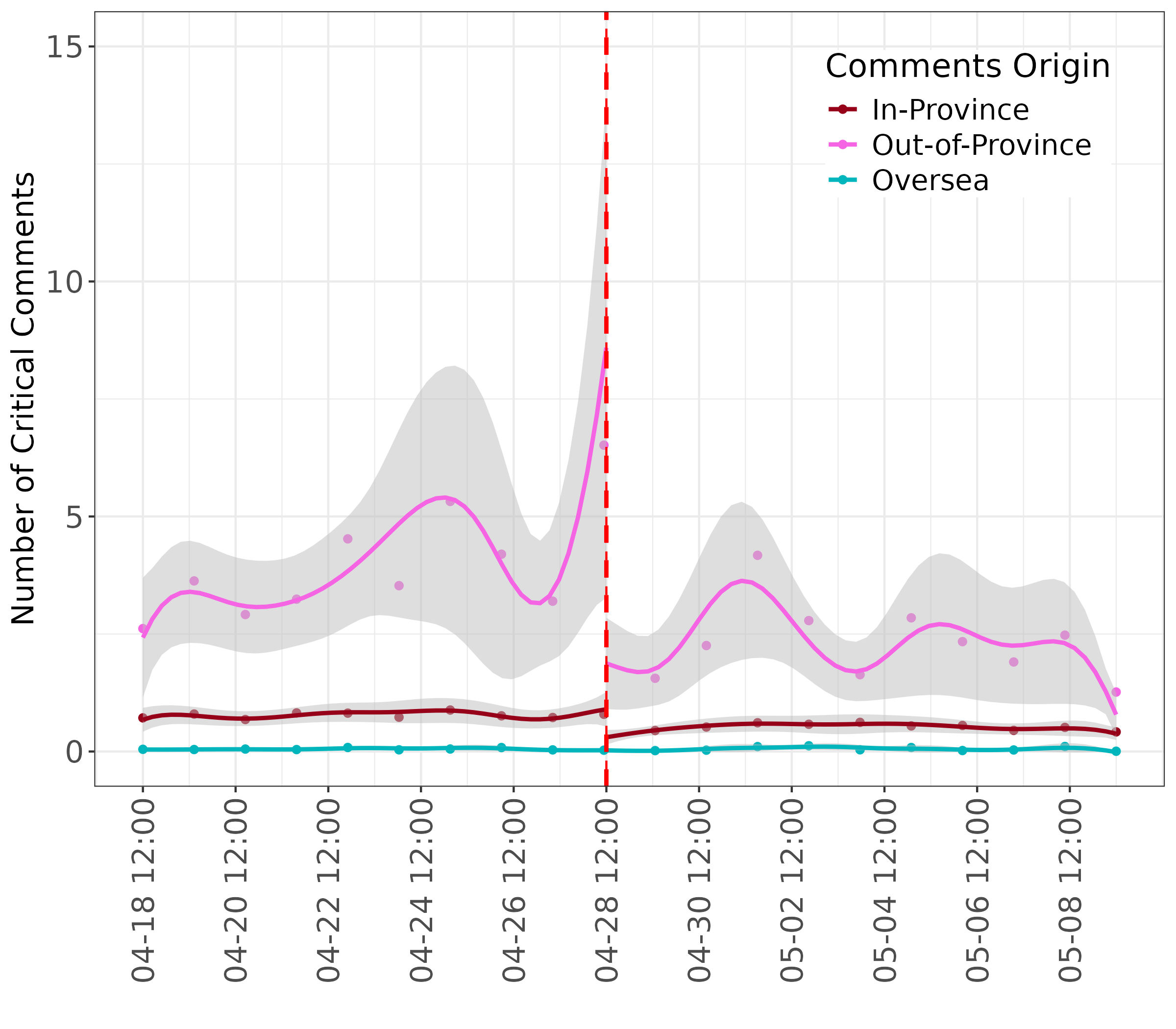}
      \caption{Bandwidth = 24 h}
      \label{fig:pnas_number_critical_comments_local_48h_bandwidth_24h}
    \end{subfigure}
  }%
  
  \vspace{0.8em} 
  
  \makebox[\textwidth][c]{%
    \begin{subfigure}{0.32\textwidth}
      \centering
      \includegraphics[width=\linewidth]{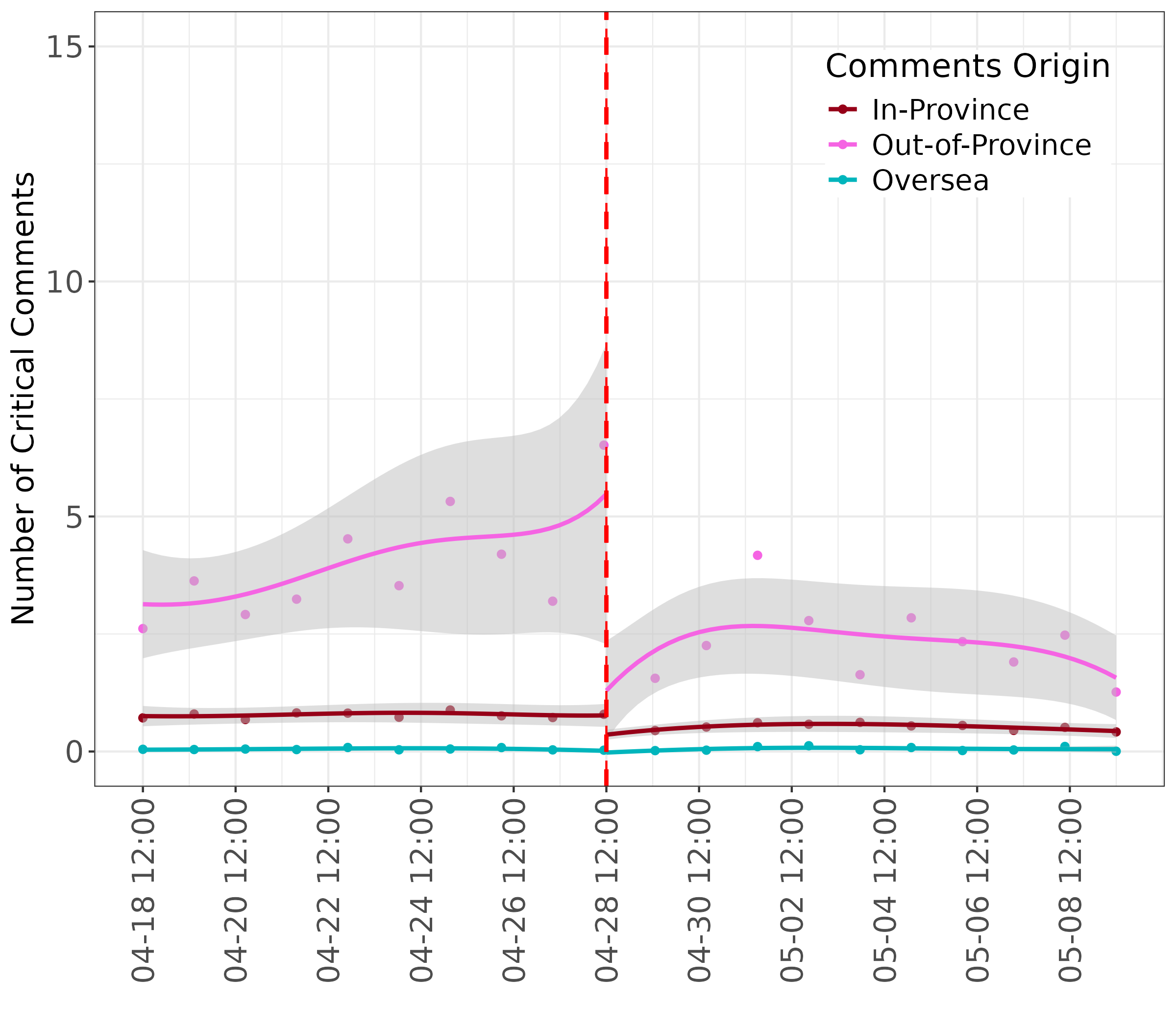}
      \caption{Bandwidth = 72 h}
      \label{fig:pnas_number_critical_comments_local_48h_bandwidth_72h}
    \end{subfigure}
    \hspace{0.02\textwidth}
    \begin{subfigure}{0.32\textwidth}
      \centering
      \includegraphics[width=\linewidth]{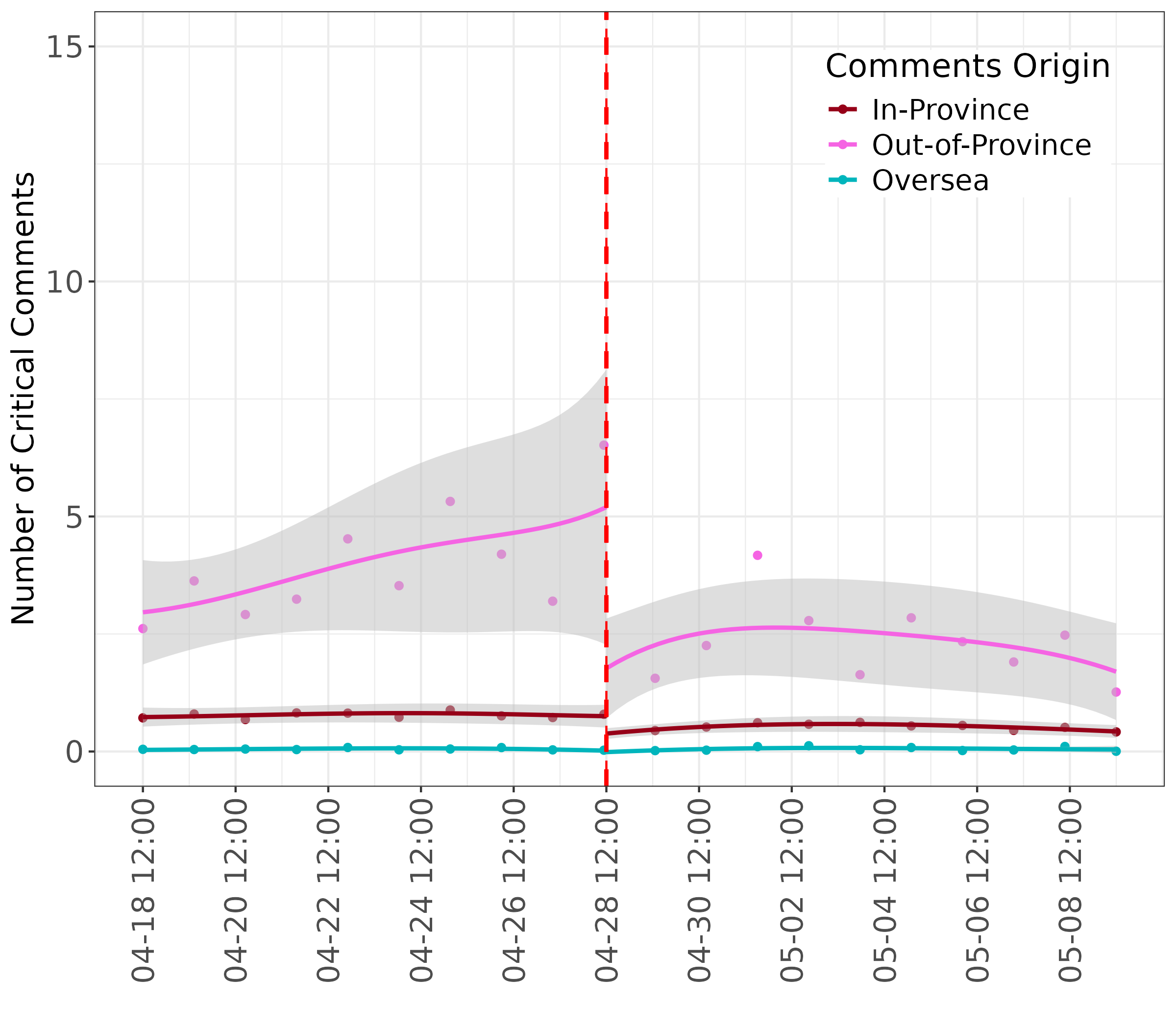}
      \caption{Bandwidth = 96 h}
      \label{fig:pnas_number_critical_comments_local_48h_bandwidth_96h}
    \end{subfigure}
    \hspace{0.02\textwidth}
    \begin{subfigure}{0.32\textwidth}
      \centering
      \includegraphics[width=\linewidth]{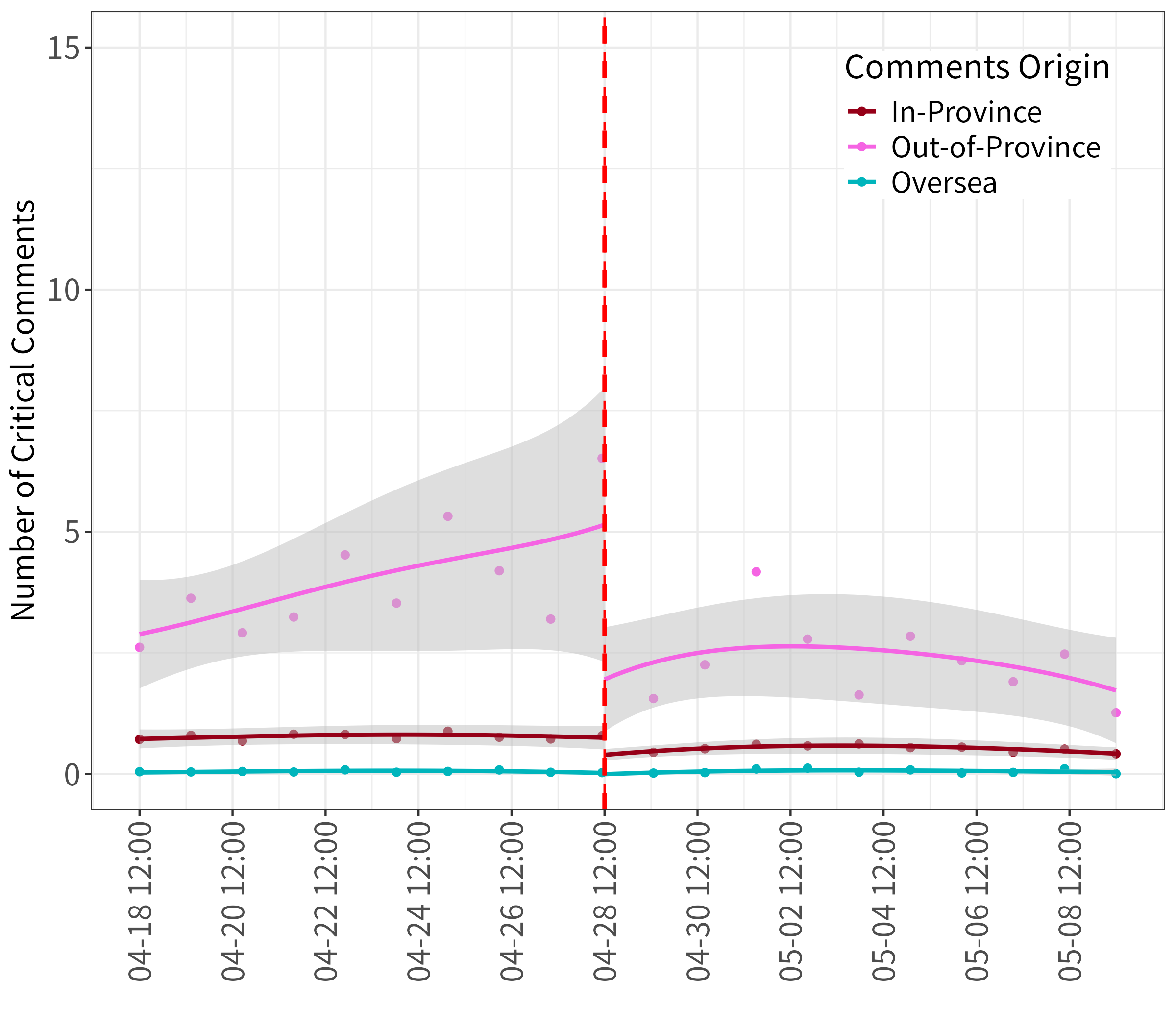}
      \caption{Bandwidth = 120 h}
      \label{fig:pnas_number_critical_comments_local_48h_bandwidth_120h}
    \end{subfigure}
  }%
  
  \caption{Decreased Critical Comments on Local Issues with Different Bandwidths}
  \label{fig:pnas_number_critical_comments_local_48h_bandwidth}
  \caption*{\textbf{Notes:} Panels \ref{fig:pnas_number_critical_comments_local_48h_bandwidth_8h} - \ref{fig:pnas_number_critical_comments_local_48h_bandwidth_120h} provide robustness checks examining the decrease in critical comments specifically on local issues, classified by commenters' geographic origin (in-province, out-of-province, and overseas) across different bandwidths (8, 24, 72, 96, and 120 hours). A consistent reduction in critical comments is primarily observed from out-of-province commenters after the implementation of the user location disclosure, whereas critical comments from in-province and overseas commenters remain relatively stable or show minor variations. The red dashed lines indicate the time when the user location disclosure was implemented.}
\end{figure}

Figure \ref{fig:pnas_number_critical_comments_local_48h_bandwidth} extends the analysis by specifically focusing on critical comments and their geographical origin (Figure~1f), further validating the robustness of the findings. The consistent decline observed in out-of-province critical comments highlights that the user location disclosure significantly affects negative or dissenting voices from geographically distant commenters, reinforcing the broader implications of the policy on public discourse.

\clearpage

\begin{figure}[H]
  \centering
  
  \makebox[\textwidth][c]{%
    \begin{subfigure}{0.32\textwidth}
      \centering
      \includegraphics[width=\linewidth]{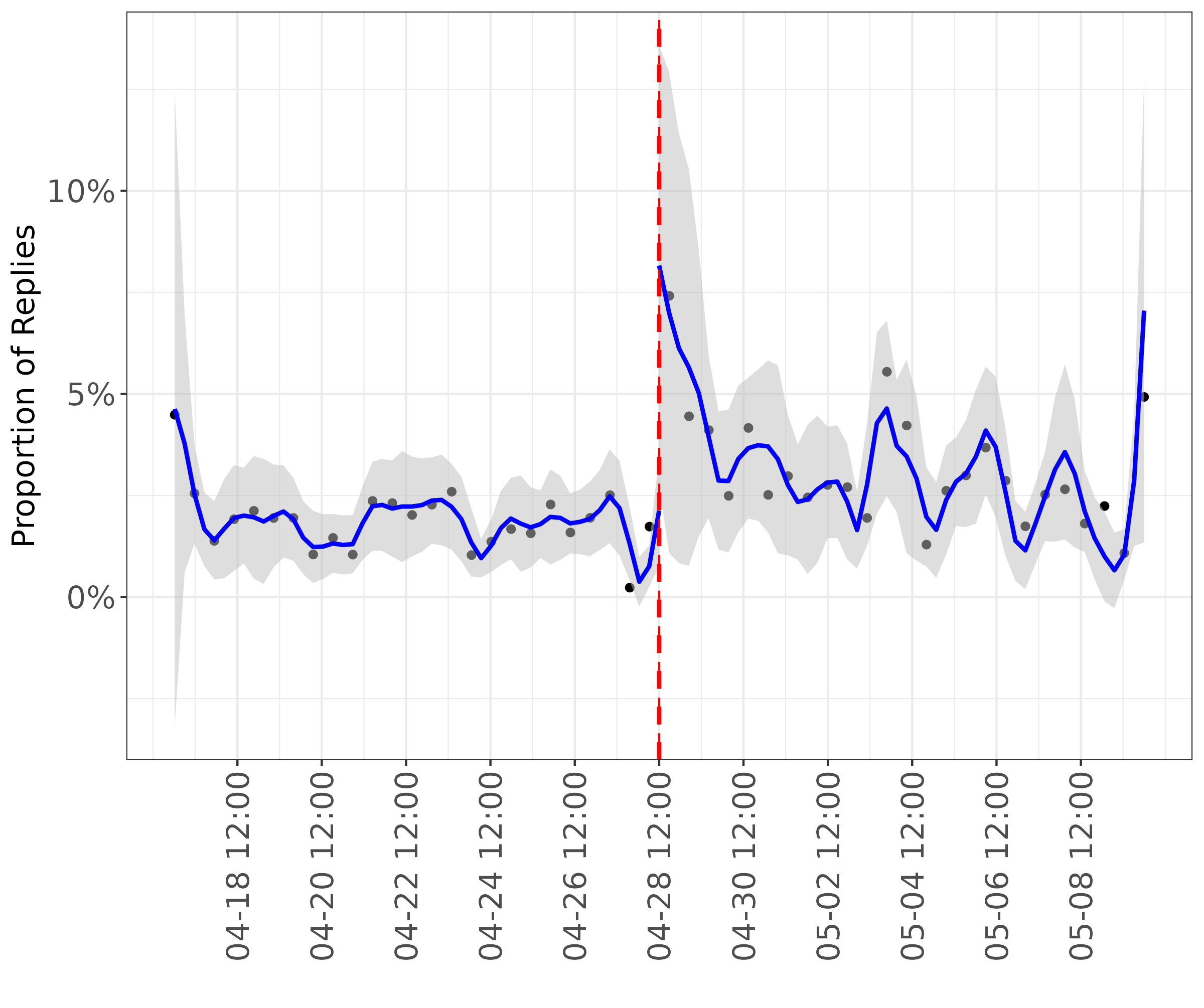}
      \caption{Bandwidth = 8 h}
      \label{fig:pnas_increased_discrimination_bandwidth_8h}
    \end{subfigure}
    \hspace{0.03\textwidth}
    \begin{subfigure}{0.32\textwidth}
      \centering
      \includegraphics[width=\linewidth]{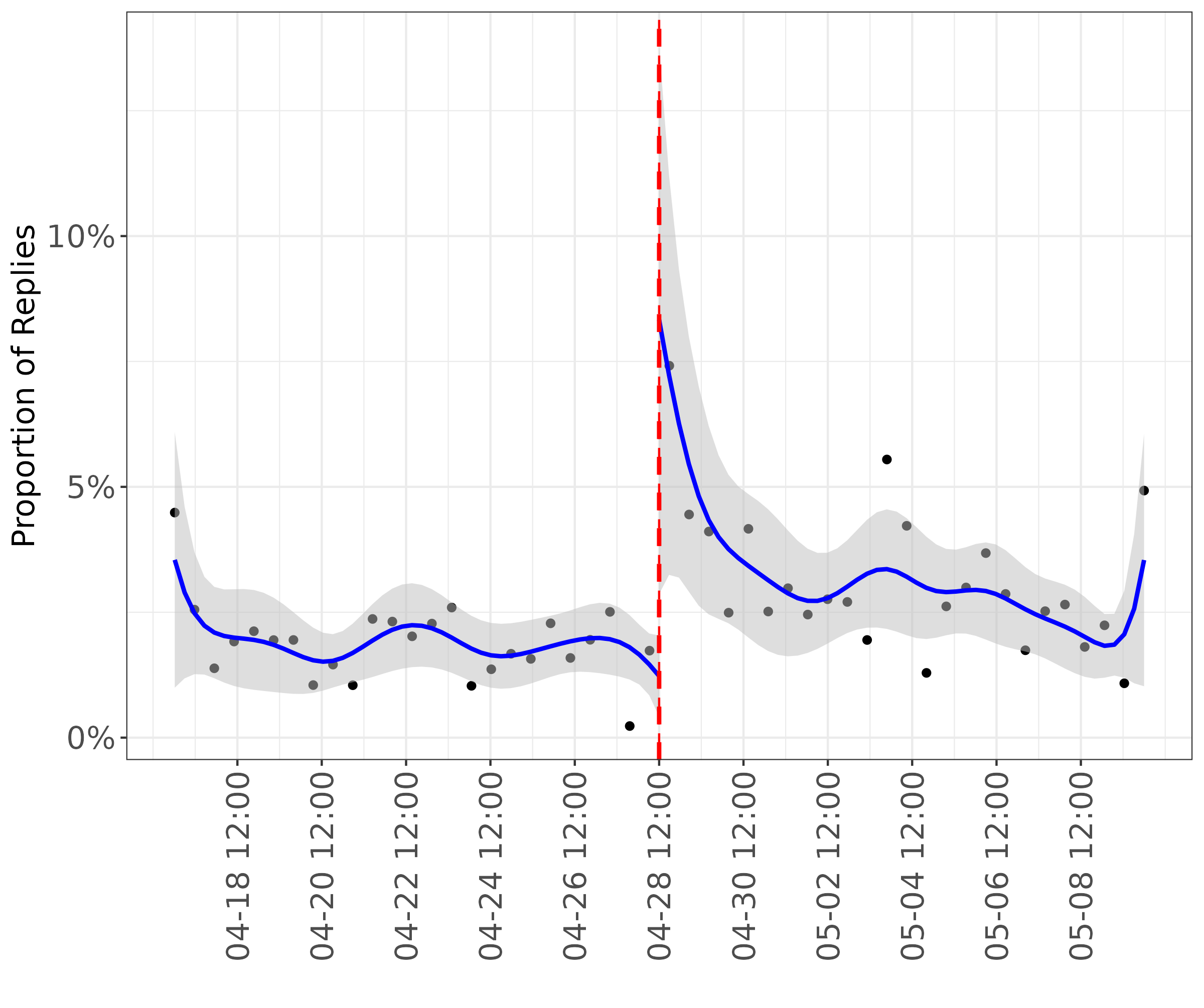}
      \caption{Bandwidth = 24 h}
      \label{fig:pnas_increased_discrimination_bandwidth_24h}
    \end{subfigure}
  }%
  
  \vspace{0.8em} 
  
  \makebox[\textwidth][c]{%
    \begin{subfigure}{0.32\textwidth}
      \centering
      \includegraphics[width=\linewidth]{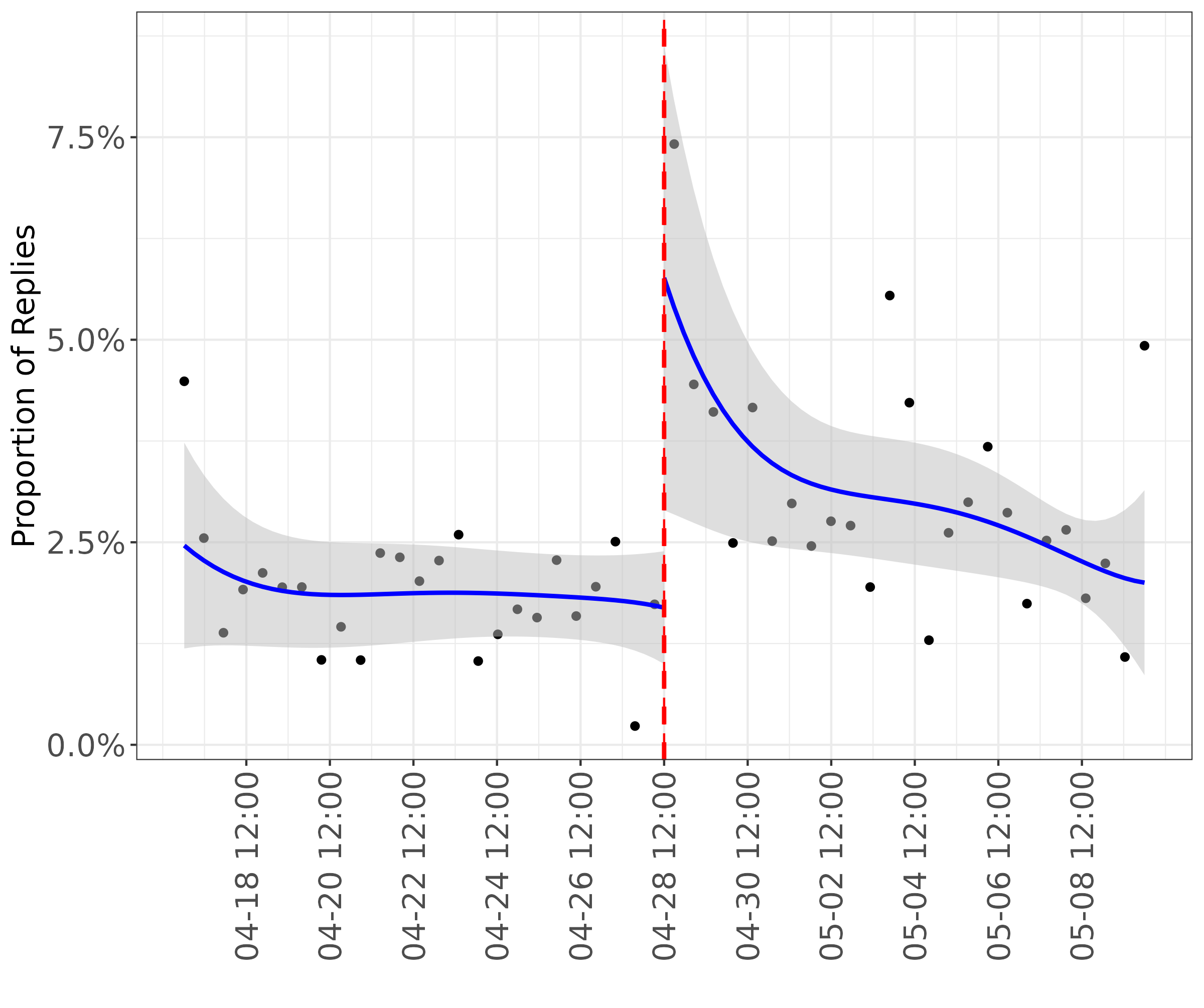}
      \caption{Bandwidth = 72 h}
      \label{fig:pnas_increased_discrimination_bandwidth_72h}
    \end{subfigure}
    \hspace{0.02\textwidth}
    \begin{subfigure}{0.32\textwidth}
      \centering
      \includegraphics[width=\linewidth]{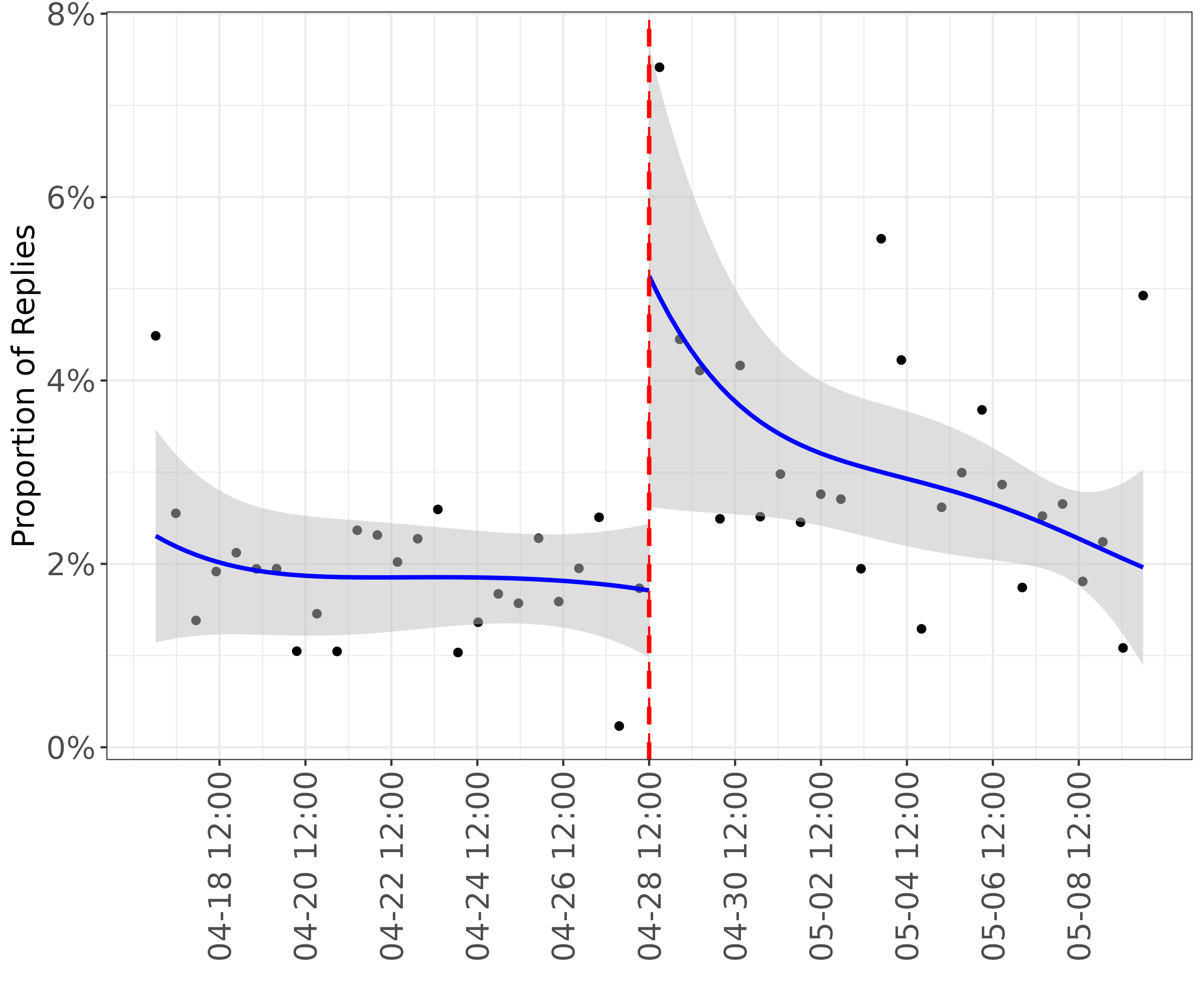}
      \caption{Bandwidth = 96 h}
      \label{fig:pnas_increased_discrimination_bandwidth_96h}
    \end{subfigure}
    \hspace{0.02\textwidth}
    \begin{subfigure}{0.32\textwidth}
      \centering
      \includegraphics[width=\linewidth]{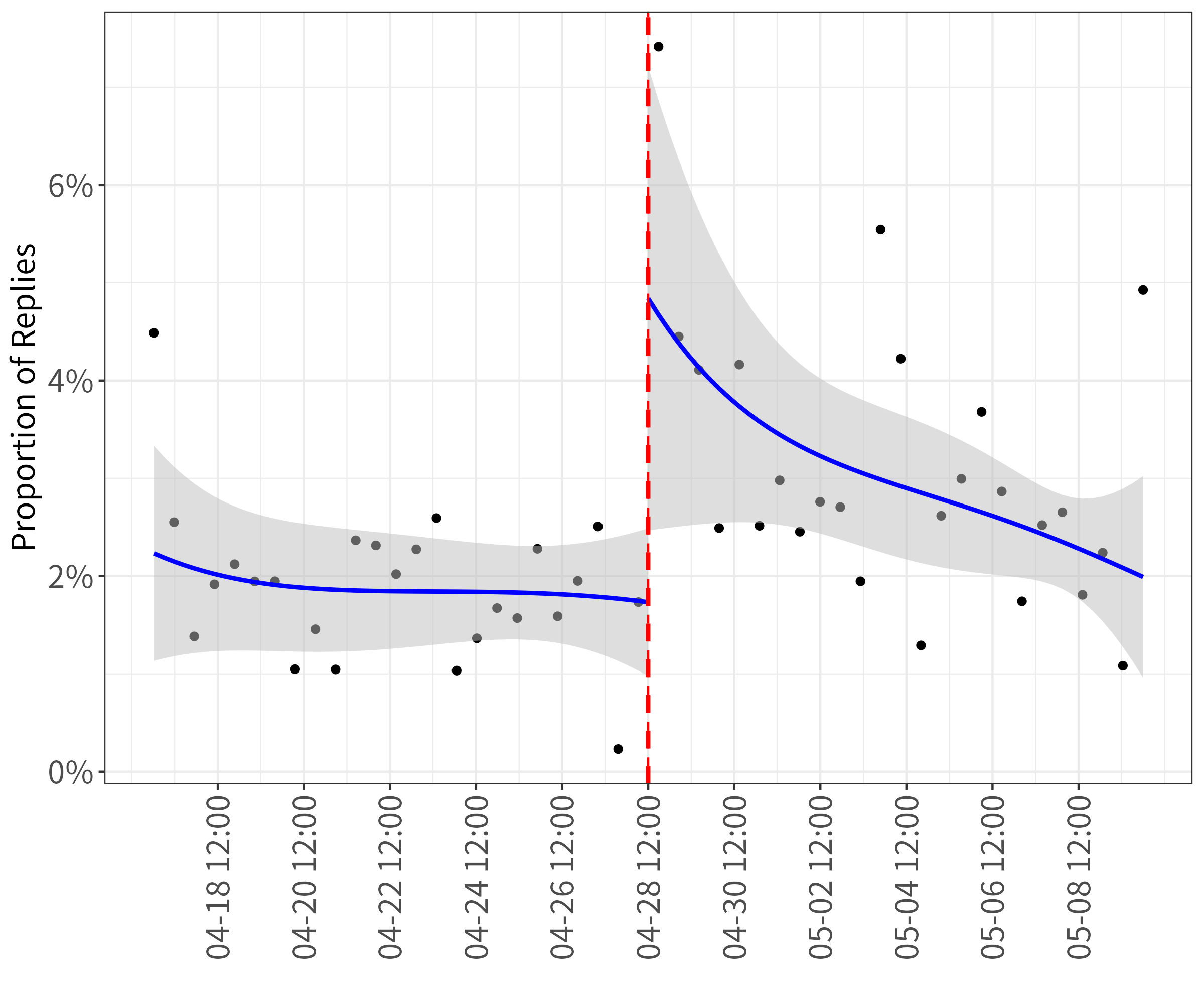}
      \caption{Bandwidth = 120 h}
      \label{fig:pnas_increased_discrimination_bandwidth_120h}
    \end{subfigure}
  }%
  
  \caption{Intensified Regional Discrimination with Different Bandwidths}
  \label{fig:pnas_increased_discrimination_bandwidth}
    \caption*{\textbf{Notes:} Panels \ref{fig:pnas_increased_discrimination_bandwidth_8h}–\ref{fig:pnas_increased_discrimination_bandwidth_120h} present robustness checks on the share of replies containing regional discrimination, using bandwidths of 8, 24, 72, 96, and 120 hours. Across all bandwidths, we consistently observe an increase in discriminatory replies following the implementation of the user location disclosure policy. These patterns shows the robustness of the finding that the policy intensified regional antagonism in online discourse. The red dashed lines indicate the time when the user location disclosure was implemented.}
\end{figure}

Figure~\ref{fig:pnas_increased_discrimination_bandwidth} replicates the analysis in Figure~2a using multiple bandwidths to assess the robustness of the observed rise in regional discrimination following the user location disclosure. Across all bandwidth choices, the increase in discriminatory replies remains robust, supporting our finding that the policy exacerbated regional antagonism in online interactions.

\clearpage

\subsection{Placebo analyses}

In addition to robustness checks using different bandwidths, we conduct placebo analyses to further validate our findings. Specifically, we shift the intervention date to one and two days prior to the actual implementation and re-estimate our models. These placebo tests assess whether the observed effects are truly caused by the user location disclosure policy or could be explained by unrelated temporal fluctuations. Null or insignificant effects in these placebo windows would strengthen the causal interpretation of our main results.

\begin{figure}[H]
    \centering
    \resizebox{\textwidth}{!}{
    \begin{subfigure}[b]{0.33\textwidth}
        \centering
        \includegraphics[width=\textwidth]{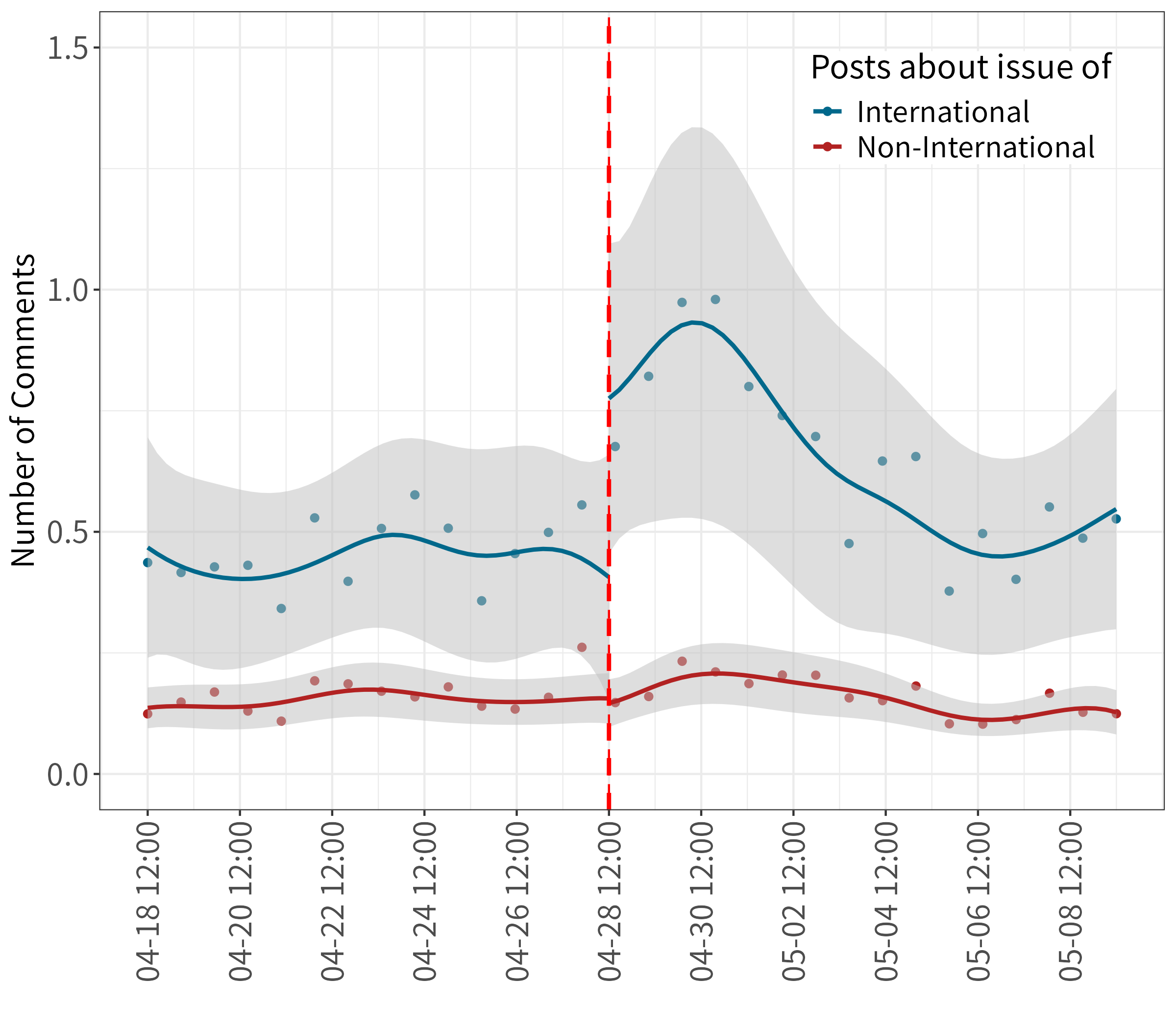}
        \caption{True Treatment Time}
        \label{fig:pnas_volume_comments_oversea_origin_48h_bandwidth_best_time_2022-04-28}
    \end{subfigure}
    \hspace{0.2cm} 
    \begin{subfigure}[b]{0.33\textwidth}
        \centering
        \includegraphics[width=\textwidth]{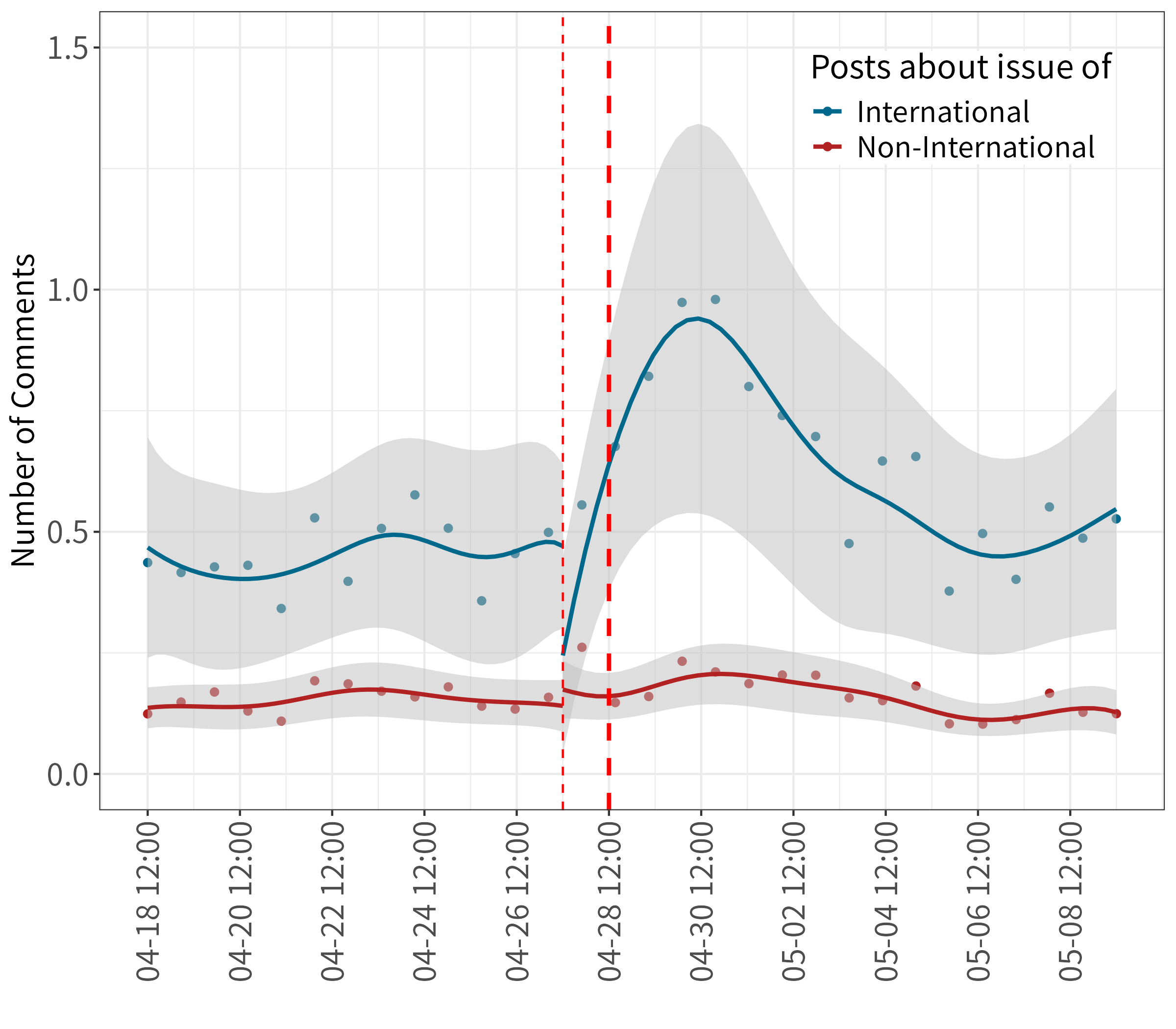}
        \caption{One Day Before True Treatment Time}
        \label{fig:pnas_volume_comments_oversea_origin_48h_bandwidth_best_time_2022-04-27}
        \end{subfigure}
    \hspace{0.2cm} 
        \begin{subfigure}[b]{0.33\textwidth}
        \centering
        \includegraphics[width=\textwidth]{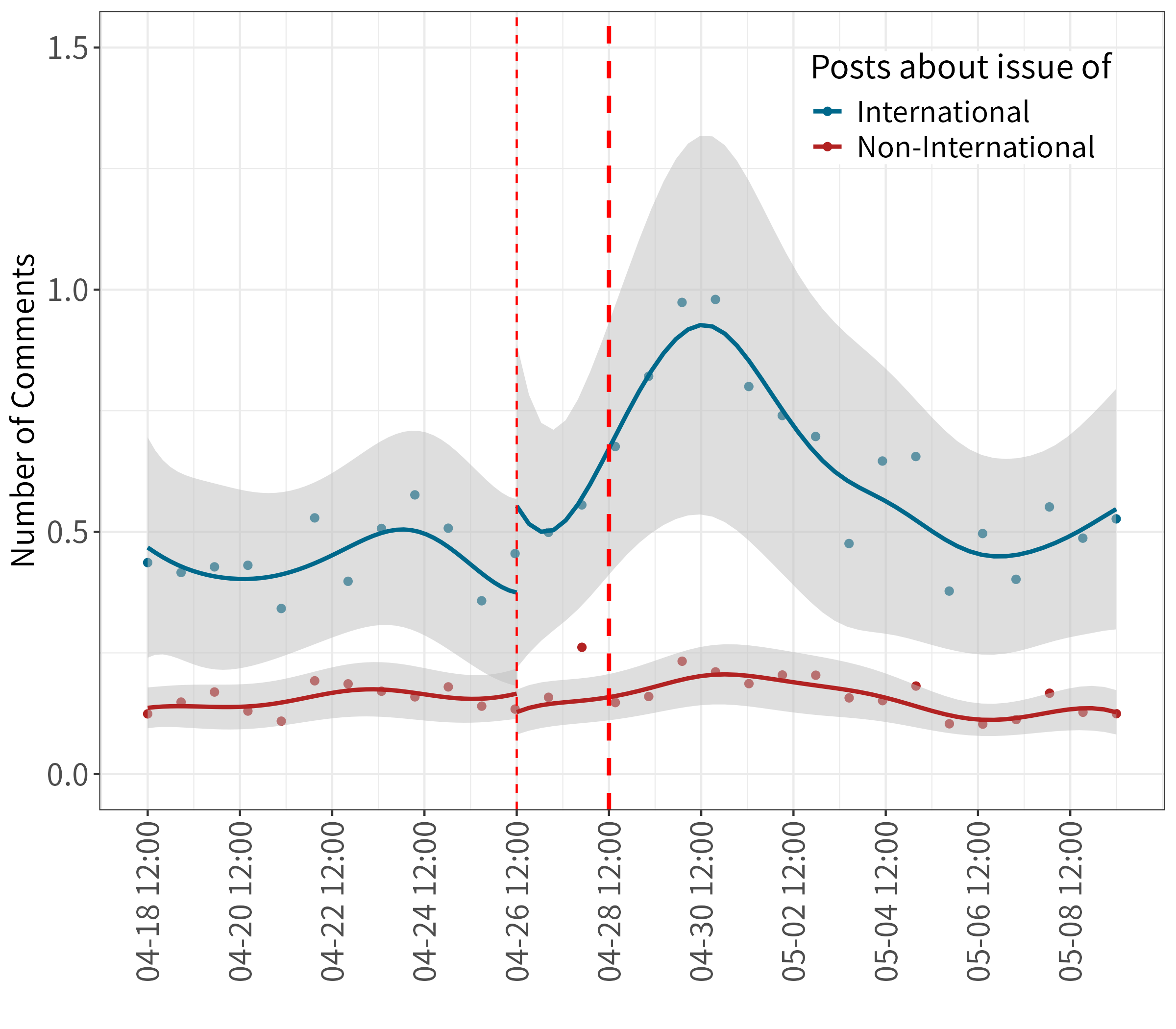}
        \caption{Two Days Before True Treatment Time}
    \label{fig:pnas_volume_comments_oversea_origin_48h_bandwidth_best_time_2022-04-26}
        \end{subfigure}}
    \caption{Comments from Oversea Users (Time Placebo)}
    \label{fig:placebo_time_2_1}
\end{figure}

Figure~\ref{fig:placebo_time_2_1} compares model estimates using the actual treatment time (solid red dashed line) with those using the placebo intervention times (dense dashed lines). The volume of comments from overseas users—on both international and non-international topics—shows no detectable change under the placebo conditions, suggesting that the actual implementation, not random timing, drove the observed shifts in engagement.

\begin{figure}[H]
    \centering
        \resizebox{\textwidth}{!}{
    \begin{subfigure}[b]{0.33\textwidth}
        \centering
        \includegraphics[width=\textwidth]{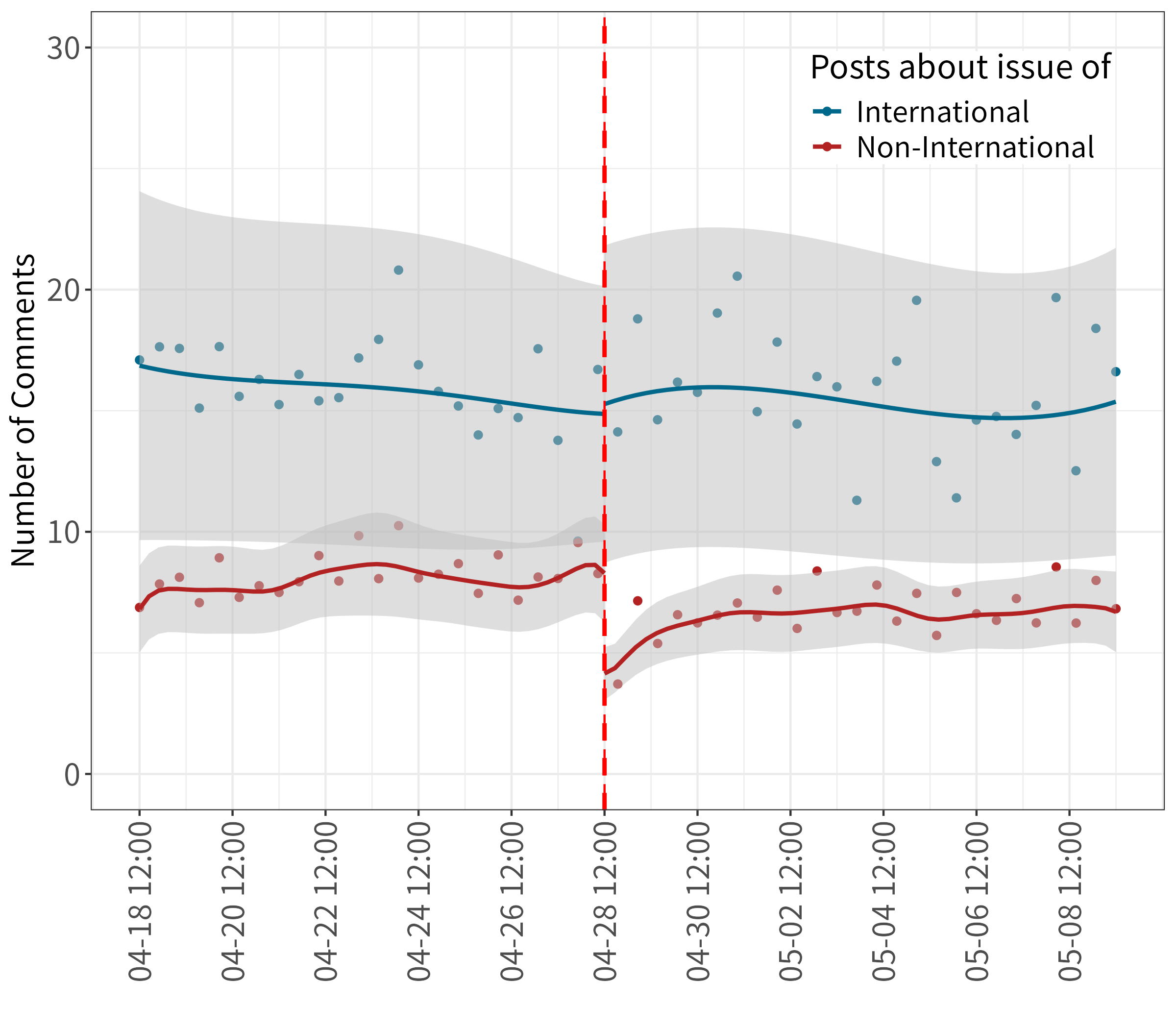}
        \caption{True Treatment Time}
        \label{fig:pnas_volume_comments_domestic_origin_48h_bandwidth_best_time_2022-04-28}
    \end{subfigure}
    \hspace{0.2cm} 
    \begin{subfigure}[b]{0.33\textwidth}
        \centering
        \includegraphics[width=\textwidth]{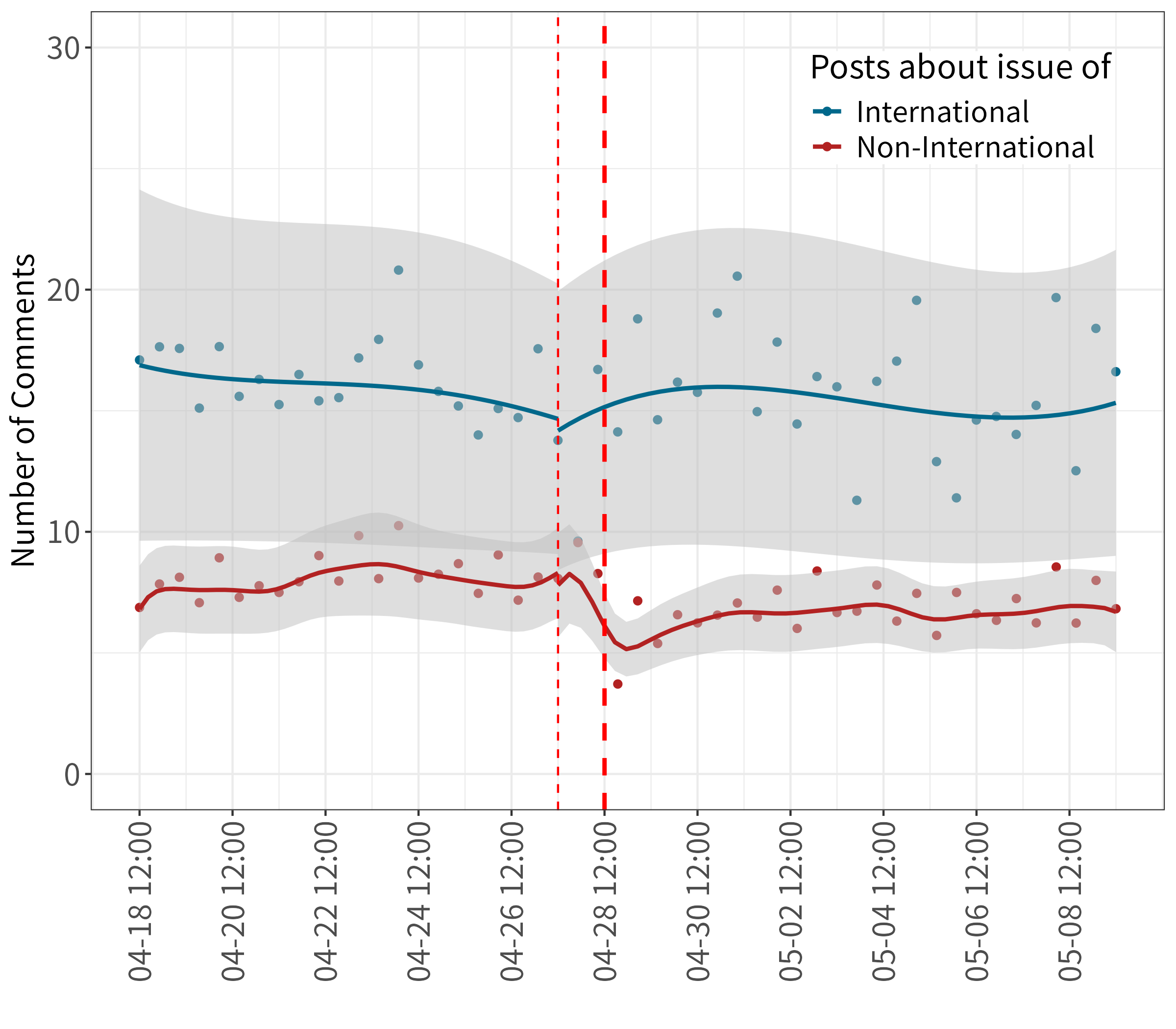}
        \caption{One Day Before True Treatment Time}
        \label{fig:pnas_volume_comments_domestic_origin_48h_bandwidth_best_time_2022-04-27}
        \end{subfigure}
    \hspace{0.2cm} 
        \begin{subfigure}[b]{0.33\textwidth}
        \centering
        \includegraphics[width=\textwidth]{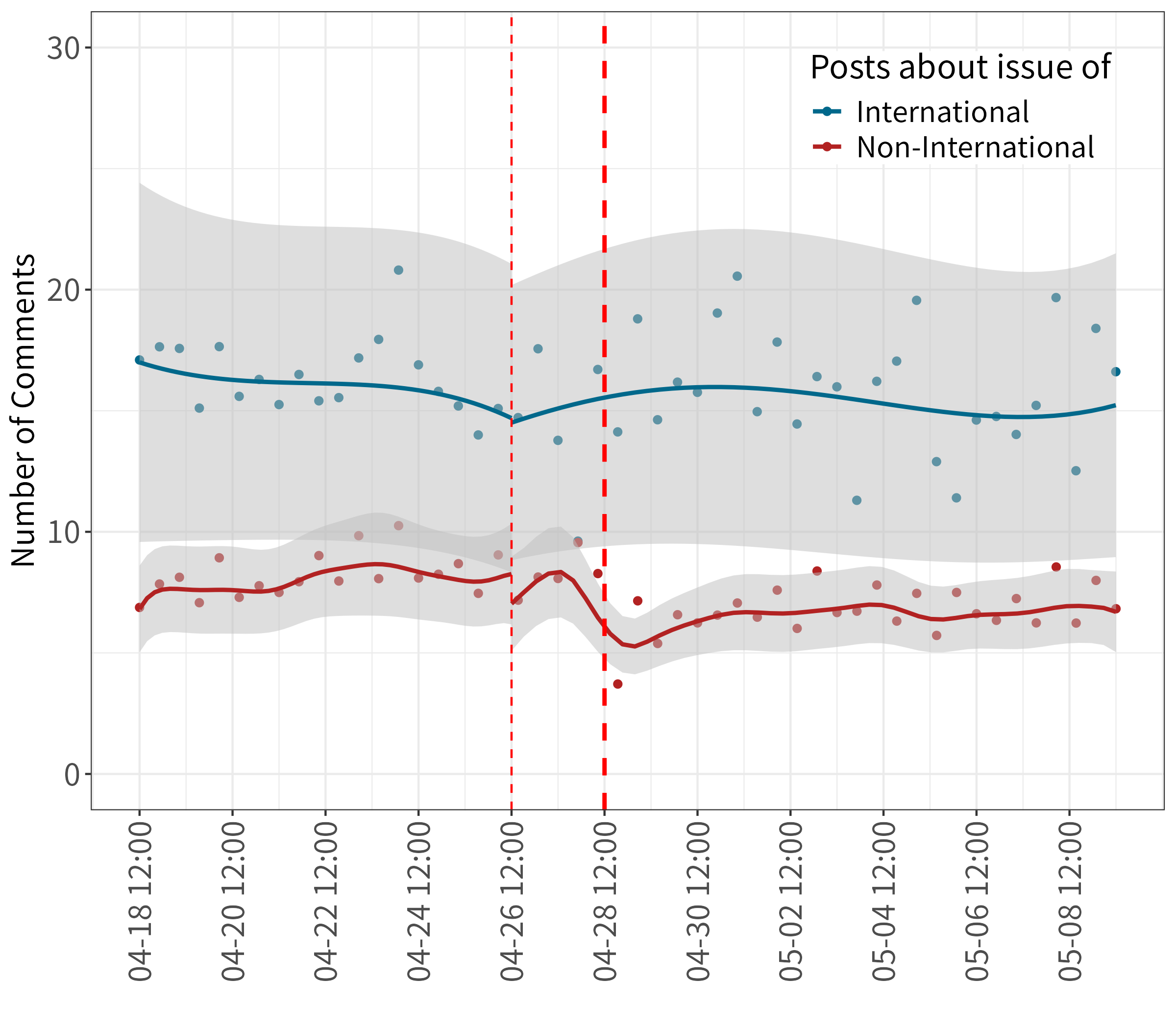}
        \caption{Two Days Before True Treatment Time}
    \label{fig:pnas_volume_comments_domestic_origin_48h_bandwidth_best_time_2022-04-26}
        \end{subfigure}}
    \caption{Comments from Domestic Users (Time Placebo)}
    \label{fig:placebo_time_2_2}
\end{figure}

Similar results are observed for comments from domestic users in Figure~\ref{fig:placebo_time_2_2}. The significant decline in comments on non-international topics—previously attributed to the policy—disappears when placebo treatment dates are used. This lack of effect under the shifted timelines further supports that the observed decrease in domestic engagement is causally linked to the actual implementation of the user location disclosure policy.

\begin{figure}[H]
    \centering
        \resizebox{\textwidth}{!}{
    \begin{subfigure}[b]{0.33\textwidth}
        \centering
        \includegraphics[width=\textwidth]{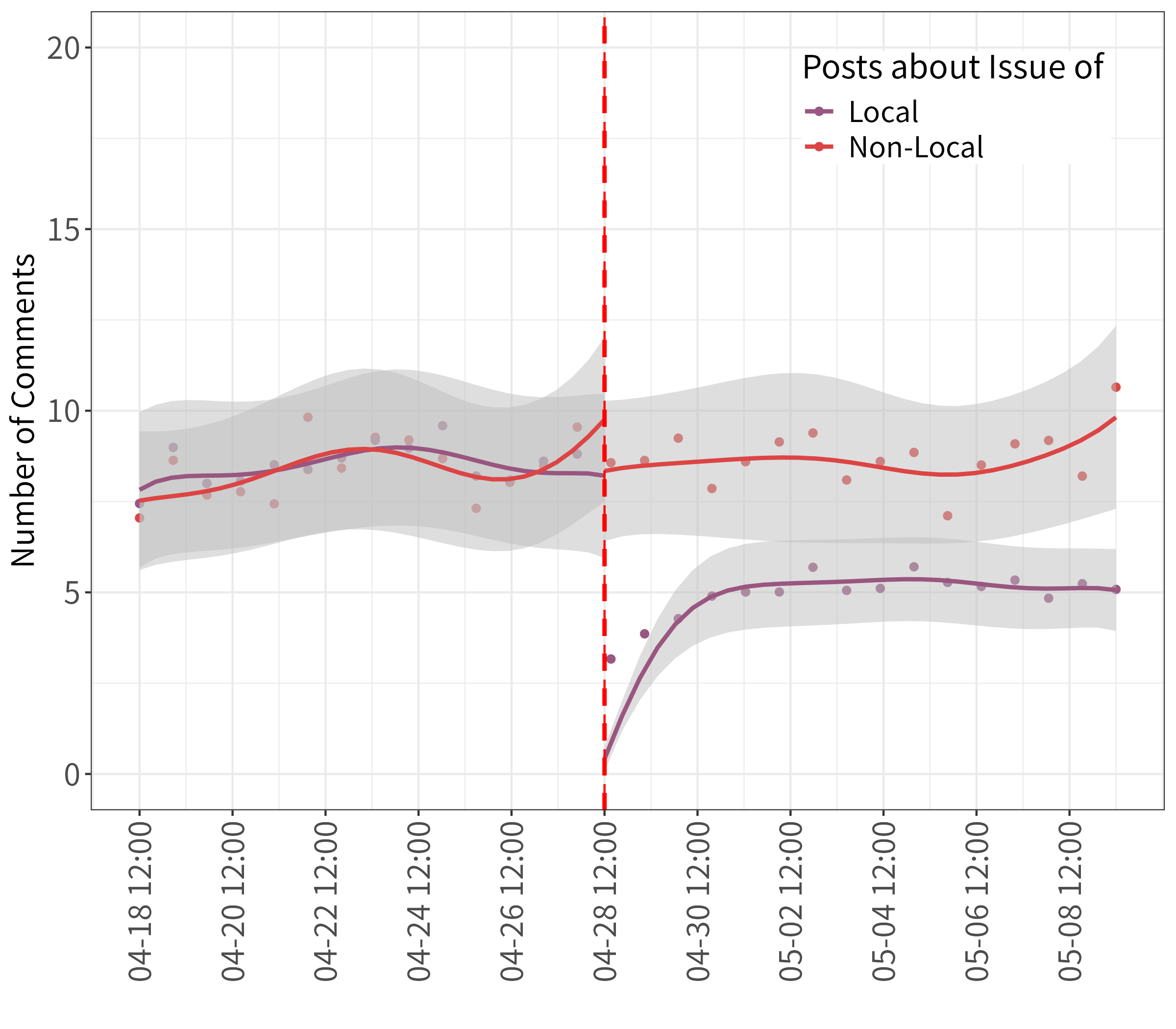}
        \caption{True Treatment Time}
        \label{fig:pnas_volume_comments_national_and_local_48h_bandwidth_best_time_2022-04-28}
    \end{subfigure}
    \hspace{0.2cm} 
    \begin{subfigure}[b]{0.33\textwidth}
        \centering
        \includegraphics[width=\textwidth]{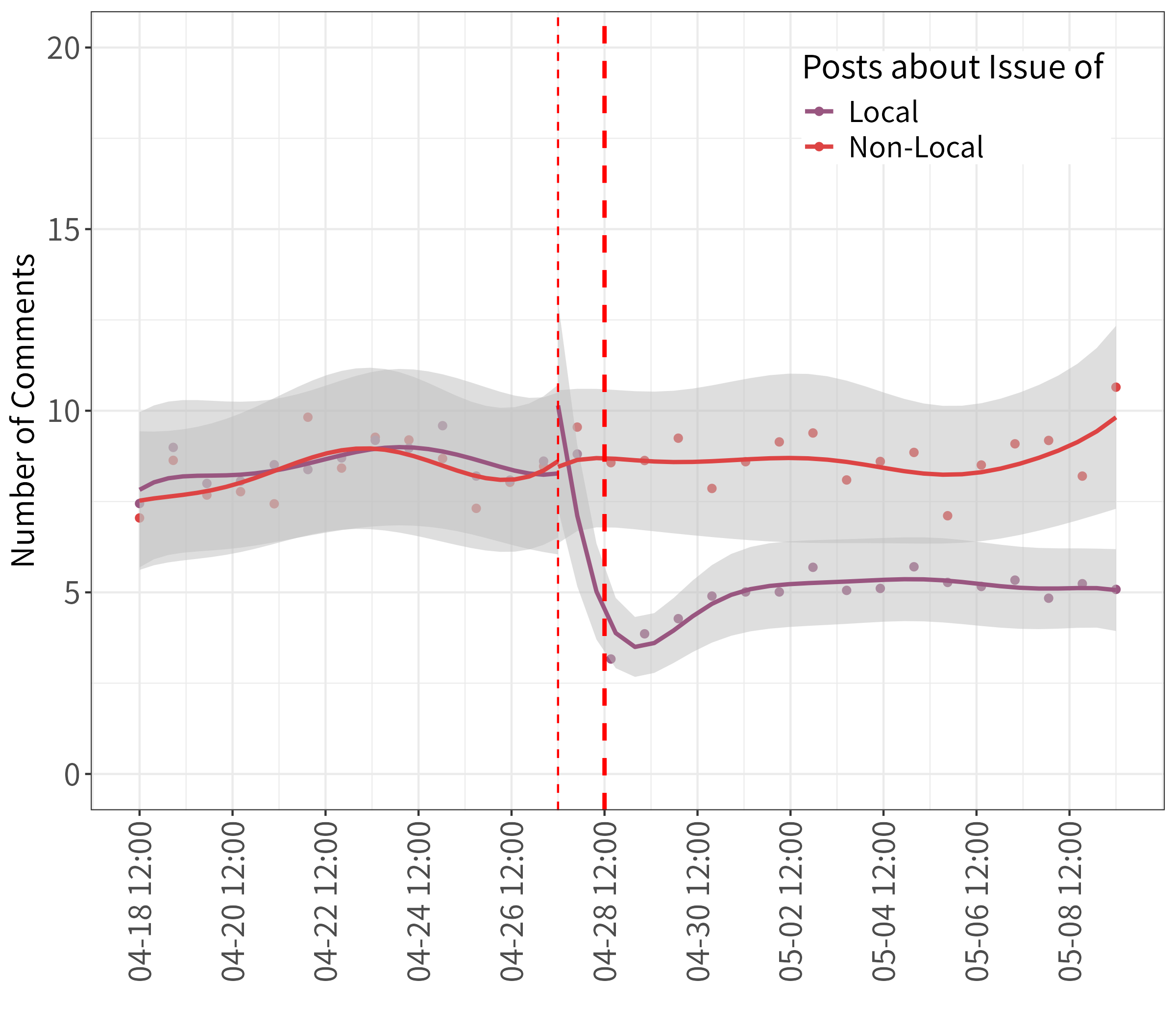}
        \caption{One Day Before True Treatment Time}
        \label{fig:pnas_volume_comments_national_and_local_48h_bandwidth_best_time_2022-04-27}
        \end{subfigure}
    \hspace{0.2cm} 
        \begin{subfigure}[b]{0.33\textwidth}
        \centering
        \includegraphics[width=\textwidth]{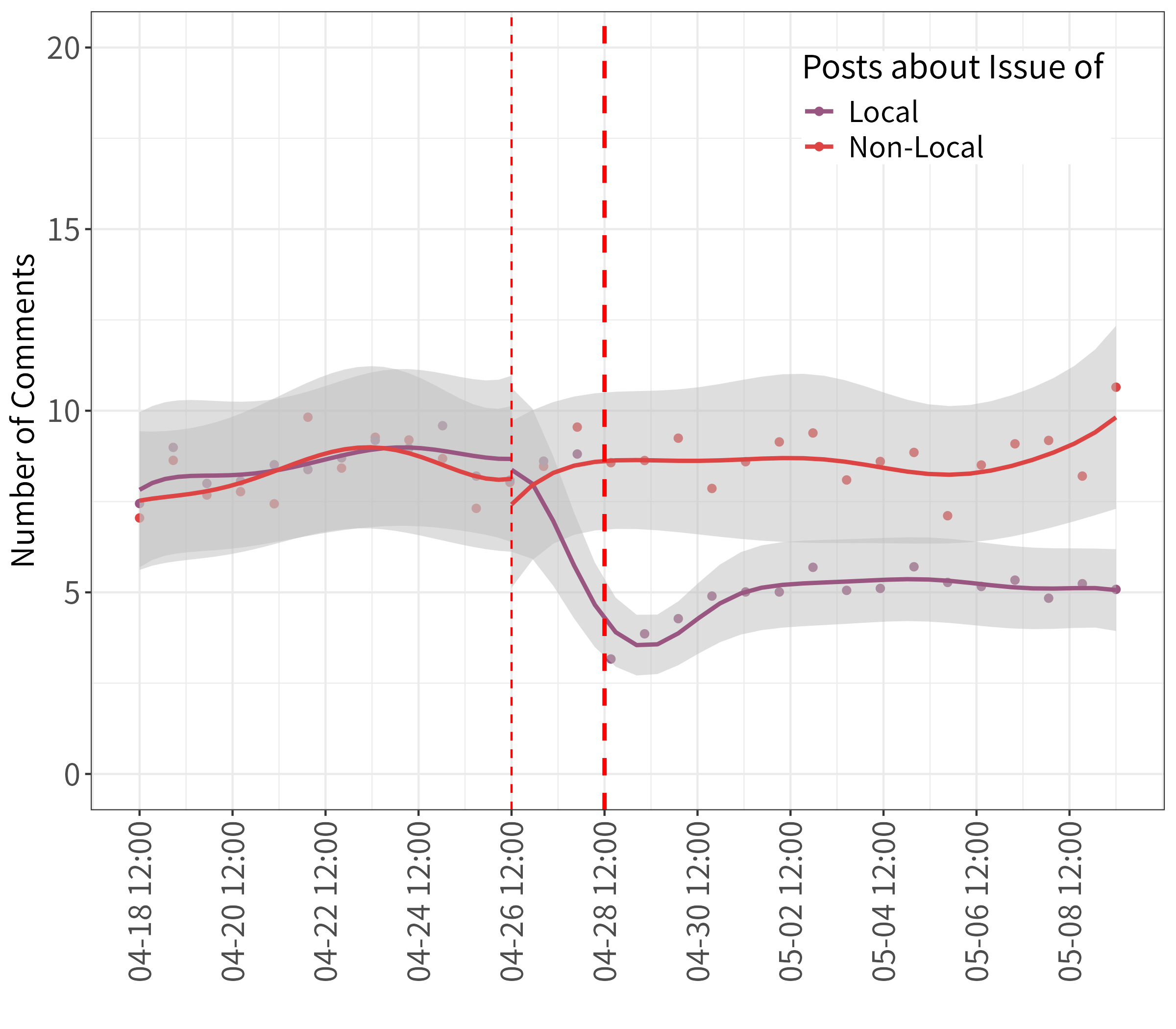}
        \caption{Two Days Before True Treatment Time}
    \label{fig:pnas_volume_comments_national_and_local_48h_bandwidth_best_time_2022-04-26}
        \end{subfigure}}
    \caption{Zoom in Non-International Issue (Time Placebo)}
    \label{fig:placebo_time_2_3}
\end{figure}

For all posts not related to international issues, Figure~\ref{fig:placebo_time_2_3} shows that placebo treatment dates—set one or two days prior to the actual policy implementation—yield no significant effects. Only when the true treatment time is used do we observe a significant decline in comments on posts discussing non-local issues. This pattern supports the validity of our findings and suggests that the observed effects are not driven by unrelated temporal fluctuations.

When further distinguishing commenters by geographic origin, we find that placebo treatment dates do not yield significant results, unlike the actual treatment date, which supports that the observed declines—particularly among out-of-province commenters—are specific to the true timing of the user location disclosure policy.

\begin{figure}[H]
    \centering
    \resizebox{\textwidth}{!}{
    \begin{subfigure}[b]{0.33\textwidth}
        \centering
        \includegraphics[width=\textwidth]{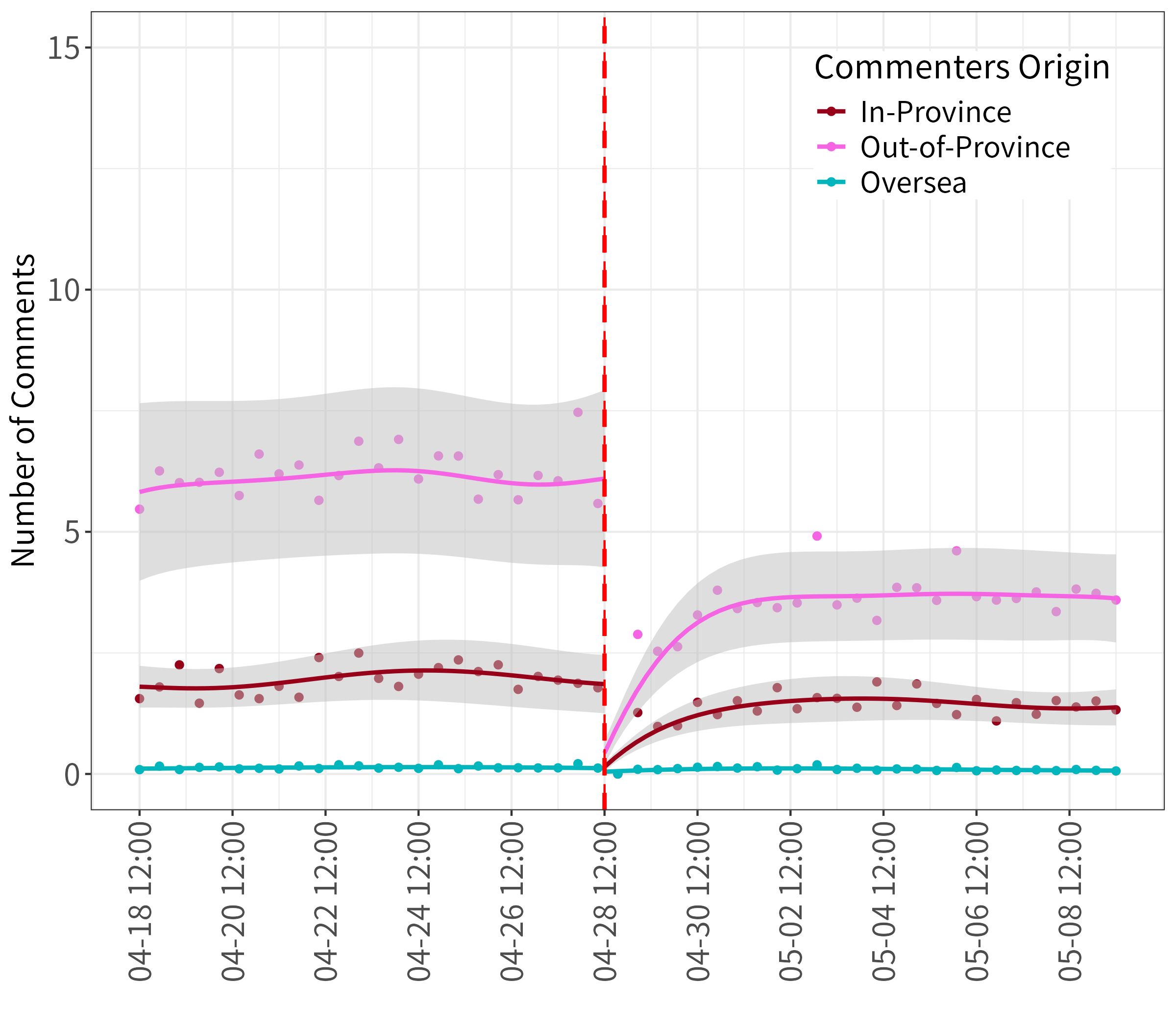}
        \caption{True Treatment Time}
        \label{fig:pnas_decreased_comments_local_by_origin_48h_bandwidth_best_time_2022-04-28}
    \end{subfigure}
    \hspace{0.2cm} 
    \begin{subfigure}[b]{0.33\textwidth}
        \centering
        \includegraphics[width=\textwidth]{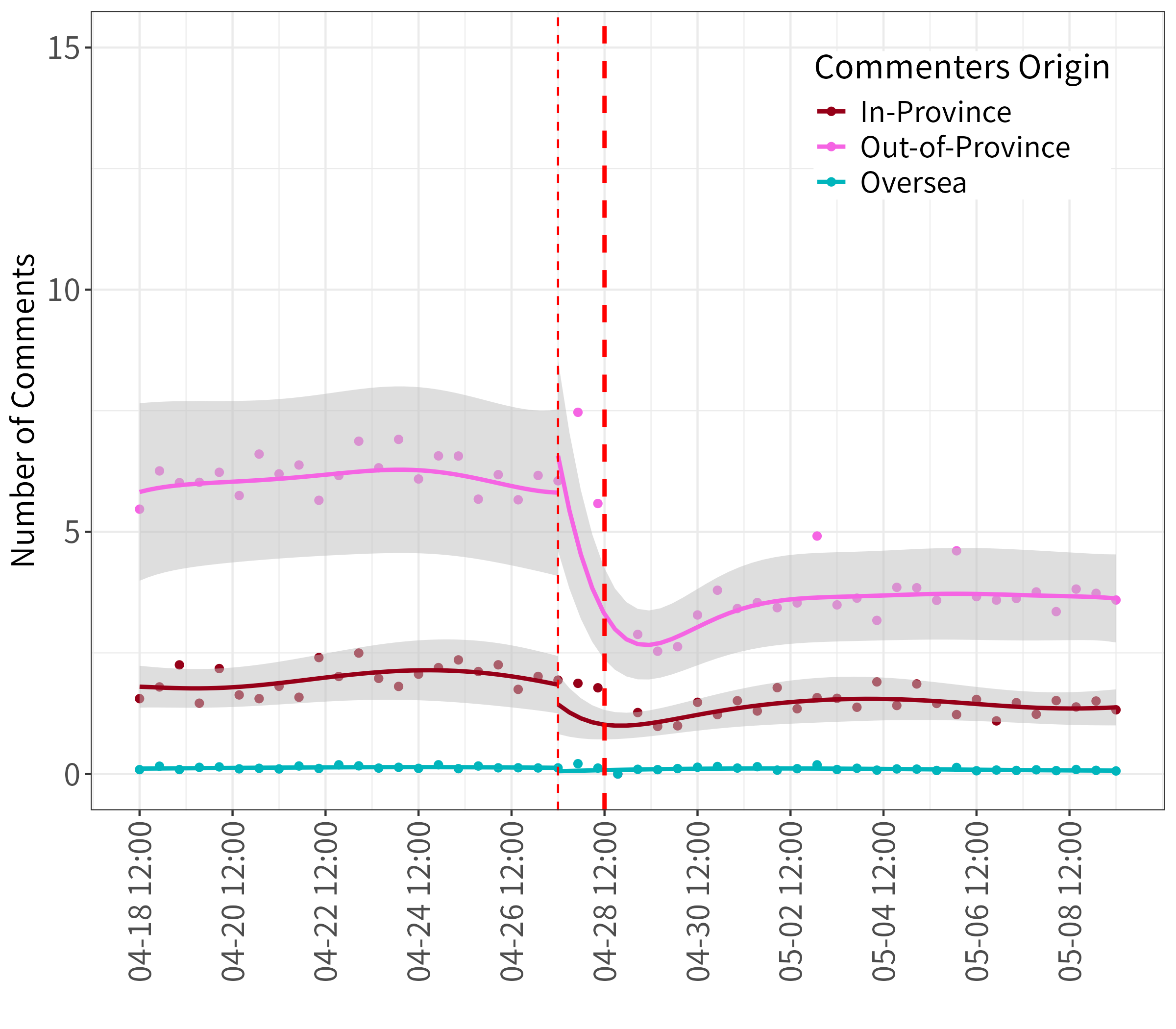}
        \caption{One Day Before True Treatment Time}
        \label{fig:pnas_decreased_comments_local_by_origin_48h_bandwidth_best_time_2022-04-27}
        \end{subfigure}
    \hspace{0.2cm} 
        \begin{subfigure}[b]{0.33\textwidth}
        \centering
        \includegraphics[width=\textwidth]{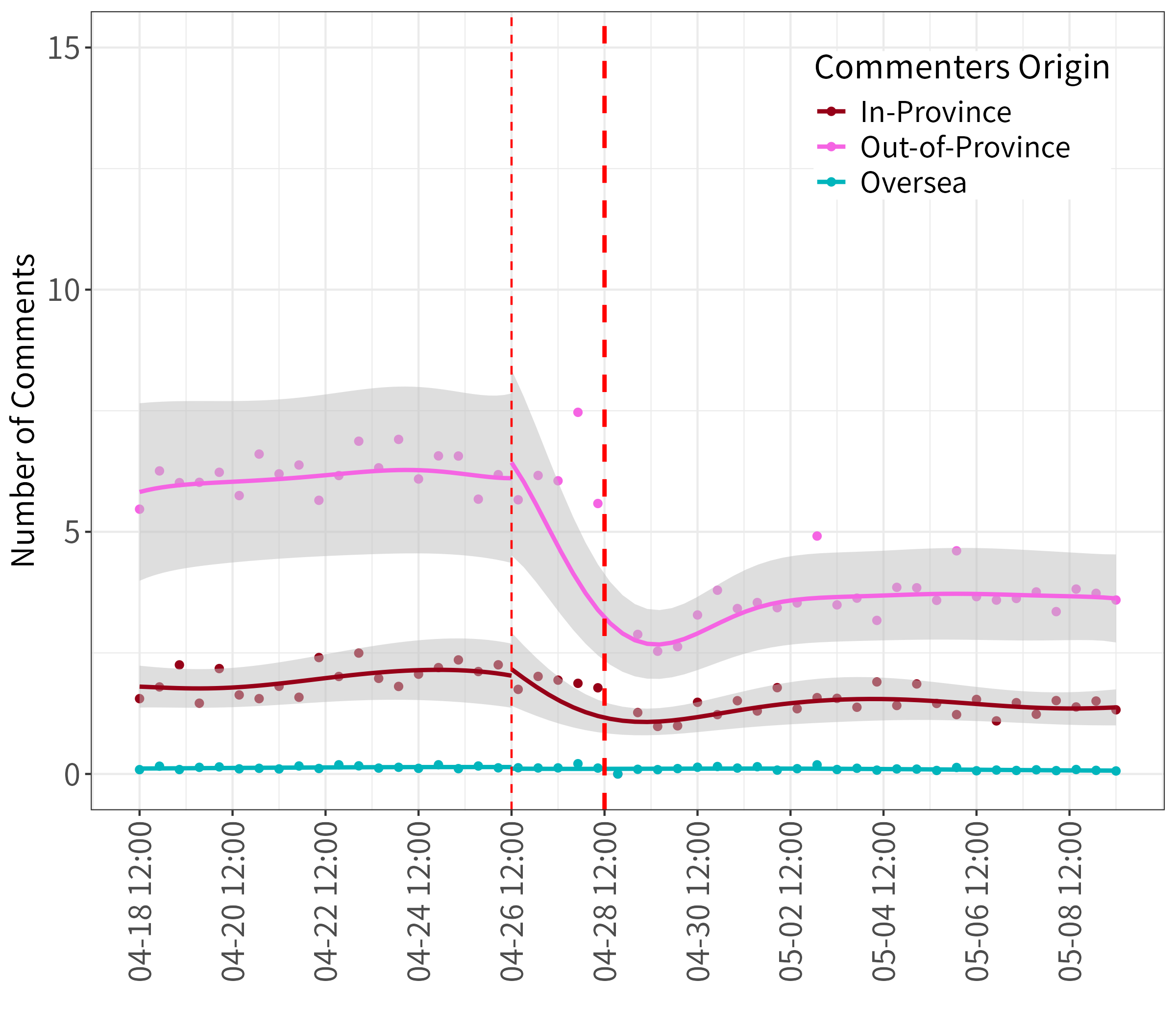}
        \caption{Two Days Before True Treatment Time}
    \label{fig:pnas_decreased_comments_local_by_origin_48h_bandwidth_best_time_2022-04-26}
        \end{subfigure}}
    \caption{Decreased Comments of Posts on Local Issues (Time Placebo)}
    \label{fig:placebo_time_3_2}
\end{figure}

So does the critical comments of posts on local issues.

\begin{figure}[H]
    \centering
    \resizebox{\textwidth}{!}{
    \begin{subfigure}[b]{0.33\textwidth}
        \centering
        \includegraphics[width=\textwidth]{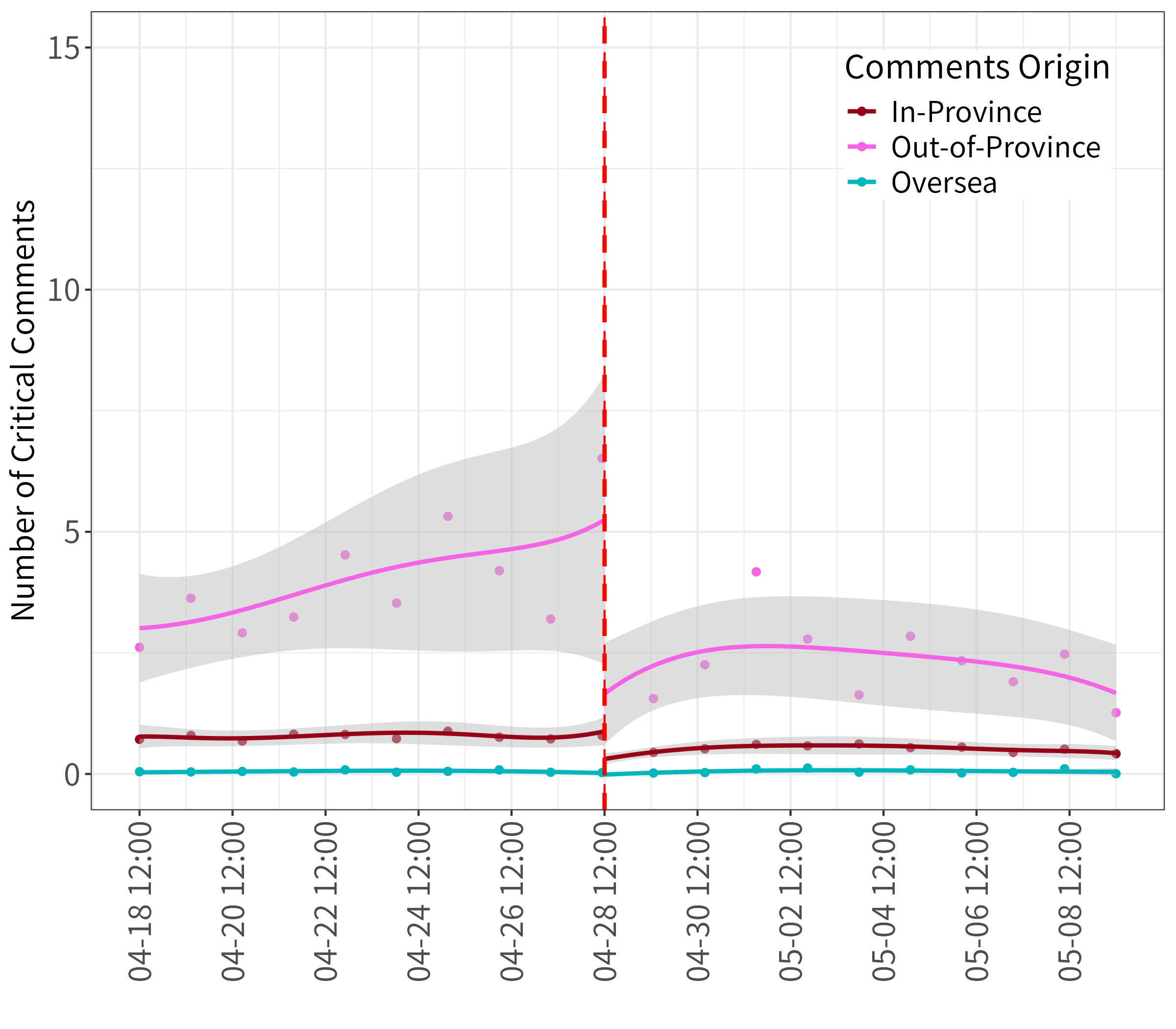}
        \caption{True Treatment Time}
        \label{fig:pnas_number_critical_comments_local_48h_bandwidth_best_time_2022-04-28}
    \end{subfigure}
    \hspace{0.2cm} 
    \begin{subfigure}[b]{0.33\textwidth}
        \centering
        \includegraphics[width=\textwidth]{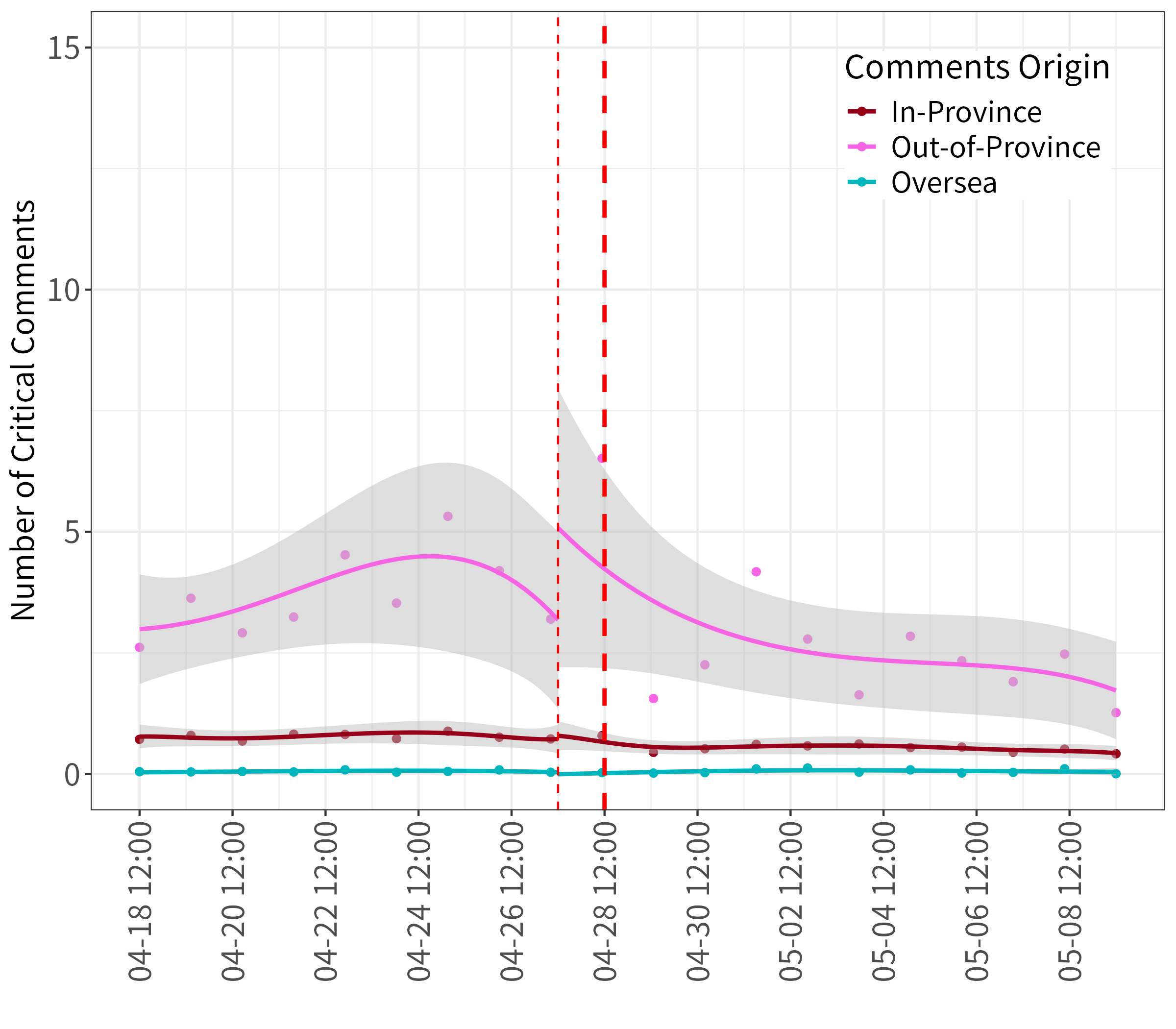}
        \caption{One Day Before True Treatment Time}
        \label{fig:pnas_number_critical_comments_local_48h_bandwidth_best_time_2022-04-27}
        \end{subfigure}
    \hspace{0.2cm} 
        \begin{subfigure}[b]{0.33\textwidth}
        \centering
        \includegraphics[width=\textwidth]{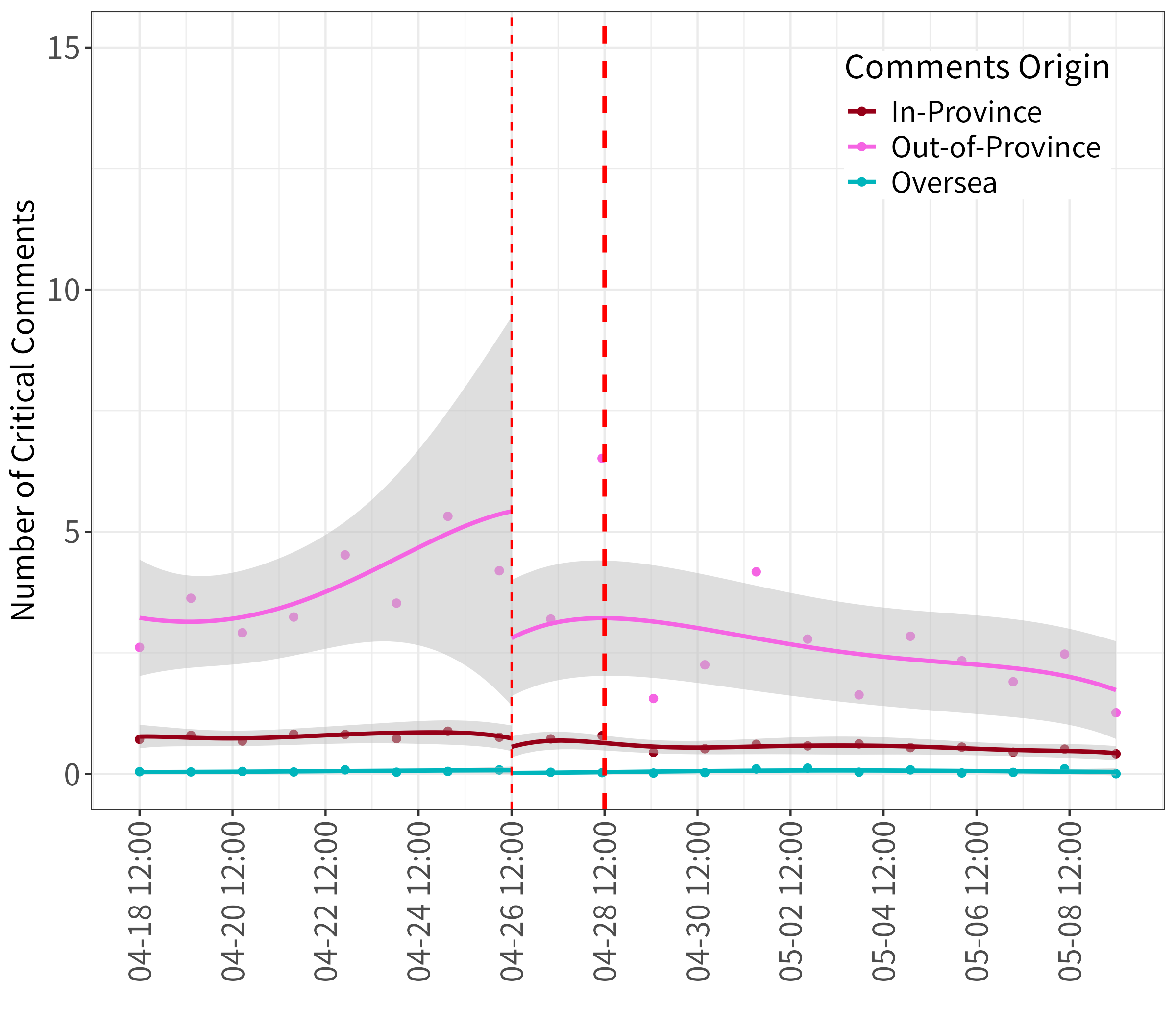}
        \caption{Two Days Before True Treatment Time}
    \label{fig:pnas_number_critical_comments_local_48h_bandwidth_best_time_2022-04-26}
        \end{subfigure}}
    \caption{Decreased Critical Comments of Posts on Local Issues (Time Placebo)}
    \label{fig:placebo_time_3_3}
\end{figure}

\begin{figure}[H]
    \centering
    \resizebox{\textwidth}{!}{
    \begin{subfigure}[b]{0.33\textwidth}
        \centering
        \includegraphics[width=\textwidth]{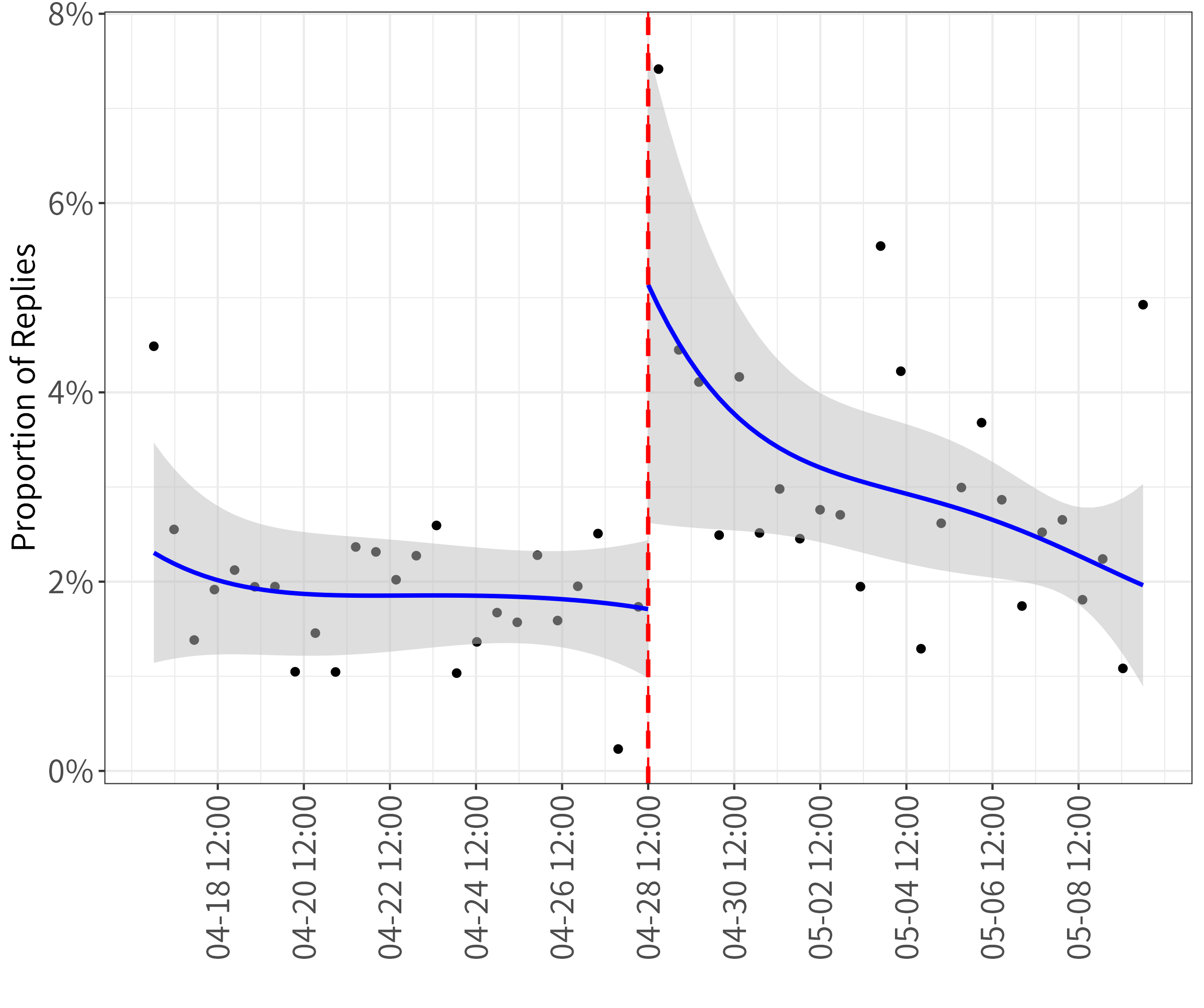}
        \caption{True Treatment Time}
        \label{fig:pnas_increased_discrimination_bandwidth_best_time_2022-04-28}
    \end{subfigure}
    \hspace{0.2cm} 
    \begin{subfigure}[b]{0.33\textwidth}
        \centering
        \includegraphics[width=\textwidth]{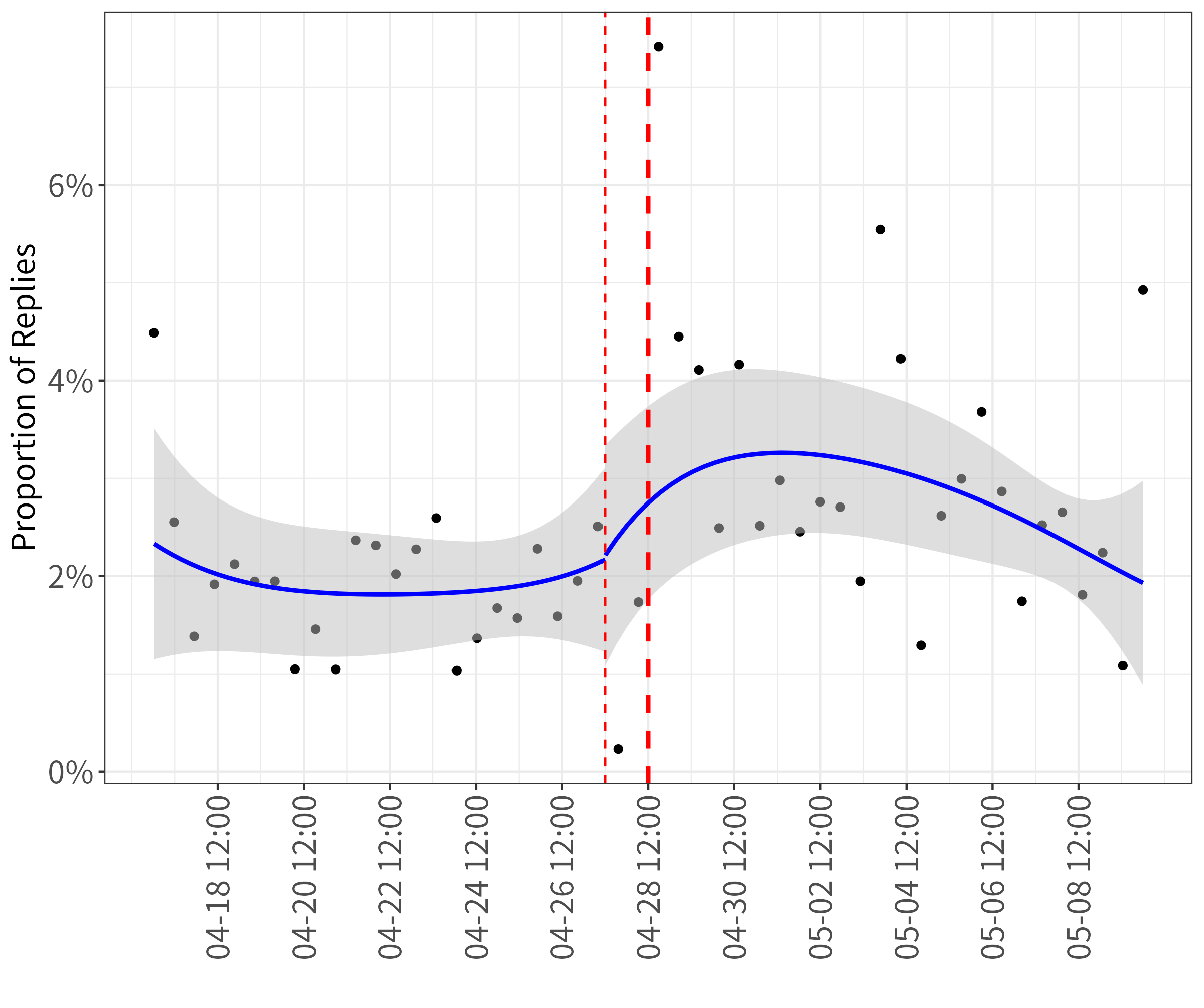}
        \caption{One Day Before True Treatment Time}
        \label{fig:pnas_increased_discrimination_bandwidth_best_time_2022-04-27}
        \end{subfigure}
    \hspace{0.2cm} 
        \begin{subfigure}[b]{0.33\textwidth}
        \centering
        \includegraphics[width=\textwidth]{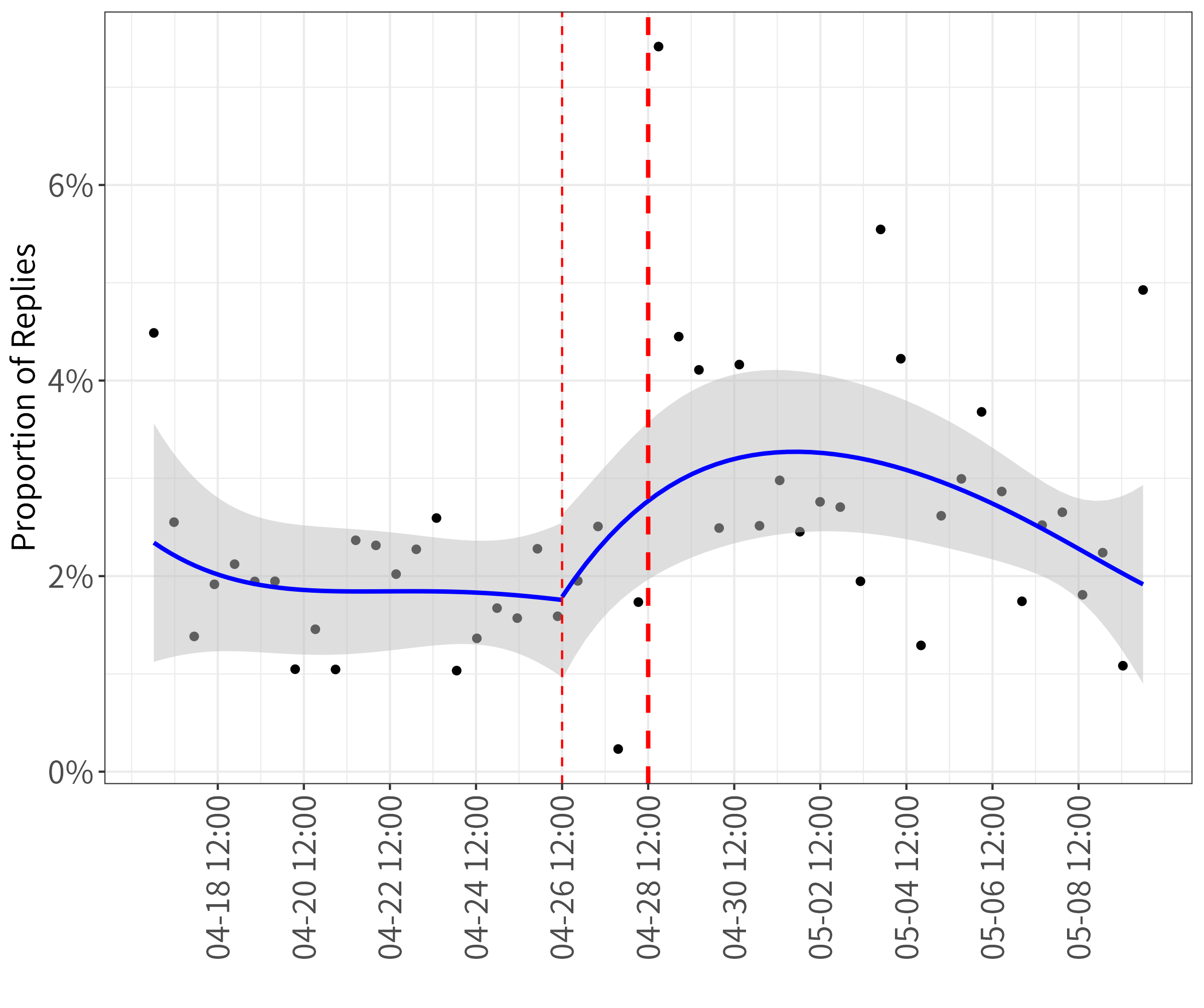}
        \caption{Two Days Before True Treatment Time}
    \label{fig:pnas_increased_discrimination_bandwidth_best_time_2022-04-26}
        \end{subfigure}}
    \caption{Proportion of Replies involving Regional Discrimination (Time Placebo)}
    \label{fig:placebo_time_4_1}
\end{figure}

In Figure \ref{fig:placebo_time_4_1}, we observe that only the true treatment time produces statistically significant results, while placebo dates do not.

Table \ref{table:treatment_effect_time_placebo} reports estimated treatment effects for the actual policy implementation time (April 28, 2022, at 12 p.m.) and two placebo dates set one and two days earlier. The placebo estimates are statistically insignificant across outcomes, in contrast to the significant effects observed at the true treatment time. This pattern supports that the documented effects are attributable to the user location disclosure and not to unrelated temporal fluctuations, bolstering the causal interpretation of our findings.

\begin{table}[!htbp] \centering 
  \caption{Placebo Estimates} 
  \label{table:treatment_effect_time_placebo} 
\resizebox{0.9\textwidth}{!}{
\begin{tabular}{@{\extracolsep{5pt}} clcccc} 
\\[-1.8ex]\hline 
\hline \\[-1.8ex] 
& & & \multicolumn{3}{c}{Coefficients for Actual and Placebo Times} \\
& & & \multicolumn{3}{c}{(Standard Error)} \\
\cmidrule{4-6}
Task & Group & Dependent Variable & Actual & Placebo (-1 Day) & Placebo (-2 Days) \\ 
\hline \\[-1.8ex] 
Figure~1a & Topic: International & \#Comments & 0.369 & -0.226 & 0.179 \\
 & & & (0.146) & (0.097) & (0.106) \\
Figure~1a & Topic: Non-International & \#Comments & -0.011 & 0.034 & -0.036 \\ 
 & & & (0.019) & (0.021) & (0.014) \\ 
Figure~1b & Topic: International & \#Comments & 0.407 & -0.487 & -0.186 \\ 
 & & & (1.363) & (1.072) & (0.990) \\ 
Figure~1b & Topic: Non-International & \#Comments & -4.110 & -0.504 & -1.149 \\ 
 & & & (0.736) & (0.738) & (0.502) \\ 
Figure~1c & Topic: Non-Local & \#Comments & -1.399 & -0.154 & -0.627 \\
 & & & (0.785) & (0.809) & (0.598) \\ 
Figure~1c & Topic: Local & \#Comments & -7.794 & 1.872 & -0.288 \\ 
& & & (1.077) & (0.847) & (0.654) \\
Figure~1e & User: In-Province & \#Comments & -1.710 & -0.406 & 0.137 \\ 
& & & (0.310) & (0.191) & (0.163) \\
Figure~1e & User: Out-of-Province & \#Comments & -5.665 & 0.761 & 0.324 \\ 
 & & & (0.870) & (0.546) & (0.480) \\ 
Figure~1e & User: Oversea & \#Comments & -0.073 & -0.070 & -0.032 \\
& & & (0.016) & (0.020) & (0.015) \\ 
Figure~1f & User: In-Province & \#Critical Comments & -0.572 & 0.070 & -0.178 \\ 
& & & (0.157) & (0.188) & (0.137) \\
Figure~1f & User: Out-of-Province & \#Critical Comments & -3.591 & 1.902 & -2.622 \\ 
& & & (1.536) & (1.339) & (1.956) \\ 
Figure~1f & User: Oversea & \#Critical Comments & -0.030 & -0.042 & -0.057 \\ 
& & & (0.017) & (0.024) & (0.035) \\ 
Figure~2a &  & \%Replies & 0.034 & 0.000 & 0.000 \\ 
 & & & (0.013) & (0.007) & (0.005) \\ 
\hline \\[-1.8ex] 
\end{tabular}}
\end{table}

\clearpage

\section{Other Findings}

\subsection{Null effects on post type and sentiment}

\begin{figure}[!ht]
    \centering
    \begin{subfigure}[b]{0.4\textwidth}
    \centering
    \includegraphics[width=\textwidth]{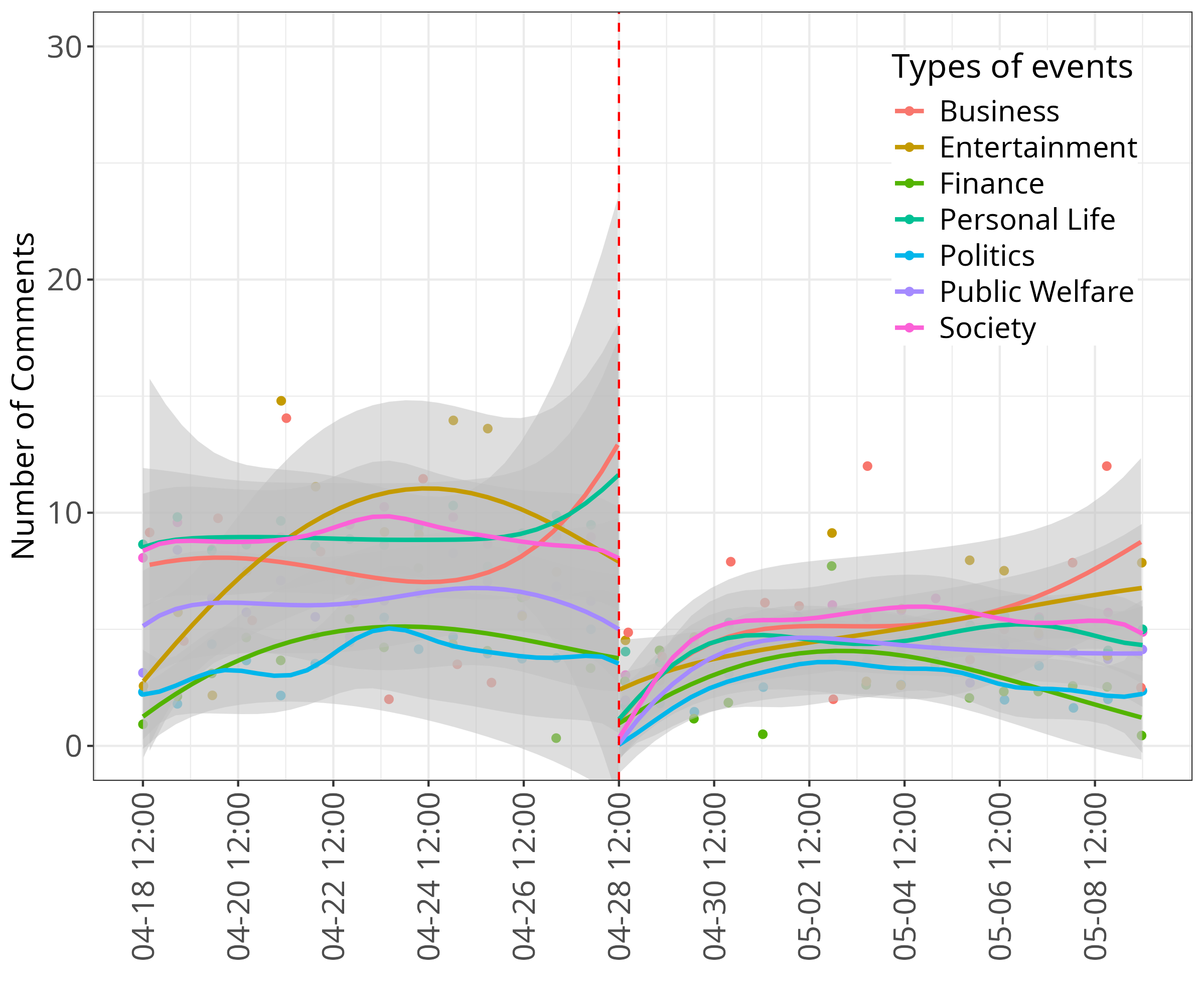}
    \caption{\#Comment of Posts on Different Topics}
    \label{fig:pnas_volume_comments_domestic_topic_48h_bandwidth_best}
    \end{subfigure}
    \hspace{0.2cm} 
    \begin{subfigure}[b]{0.4\textwidth}
        \centering
        \includegraphics[width=\textwidth]{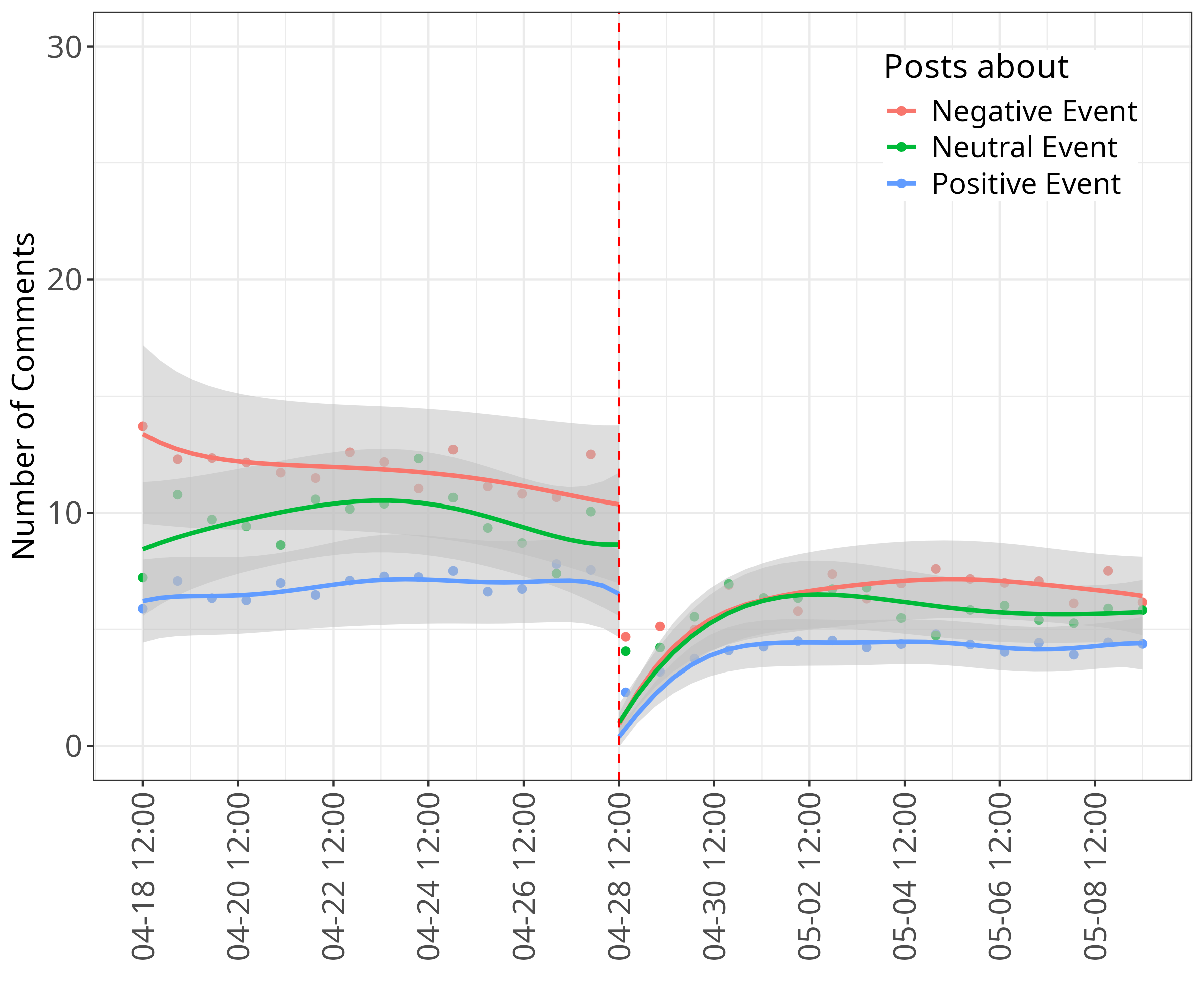}
        \caption{\#Comment of Posts on Different Event Sentiment}
        \label{fig:pnas_volume_comments_domestic_stance_48h_bandwidth_best}
    \end{subfigure}
    \caption{Decreased Comments under Posts featuring All Types of Events on domestic issues}
    \label{fig:null_effect_topic_stance}
\caption*{\textbf{Notes:} Figure~\ref{fig:pnas_volume_comments_domestic_topic_48h_bandwidth_best} plots the average number of comments on posts across seven event categories—Business, Entertainment, Finance, Personal Life, Politics, Public Welfare, and Society—between April 18 and May 8, 2022. The red dashed line indicates the implementation of the policy at noon on April 28. Comment trends across all categories decline almost simultaneously, suggesting that the drop in engagement is not driven by any specific topic. Figure~\ref{fig:pnas_volume_comments_domestic_stance_48h_bandwidth_best} repeats this analysis by sentiment, grouping posts as Negative (red), Neutral (green), or Positive (blue). All sentiment groups show a comparably sharp and parallel decline following the policy change, indicating that the decrease in interaction is unrelated to the emotional tone of the posts.}

\end{figure}

To assess whether the decline in user engagement is concentrated in specific types of content, we disaggregate the analysis by event topic and sentiment. Figure~\ref{fig:pnas_volume_comments_domestic_topic_48h_bandwidth_best} displays comment volumes across seven topic categories—Business, Entertainment, Finance, Personal Life, Politics, Public Welfare, and Society—classified using a BERT-based model trained on a hand-labeled corpus. 

Figure~\ref{fig:pnas_volume_comments_domestic_stance_48h_bandwidth_best} presents a parallel breakdown by sentiment, categorizing posts as Negative, Neutral, or Positive using the stance classifier described in Figure~\ref{fig:comment_sentiment_model}. In both cases, we plot locally smoothed comment trends. Across all topic and sentiment groups, comment volumes drop sharply—and in near parallel—following the implementation of the user location disclosure, indicating that the policy's effect is broad-based rather than limited to specific types of content.

\clearpage

\subsection{Effects on overseas users}

\begin{figure}[H]
    \centering
    \begin{subfigure}[b]{0.4\textwidth}
        \centering
        \includegraphics[width=\textwidth]{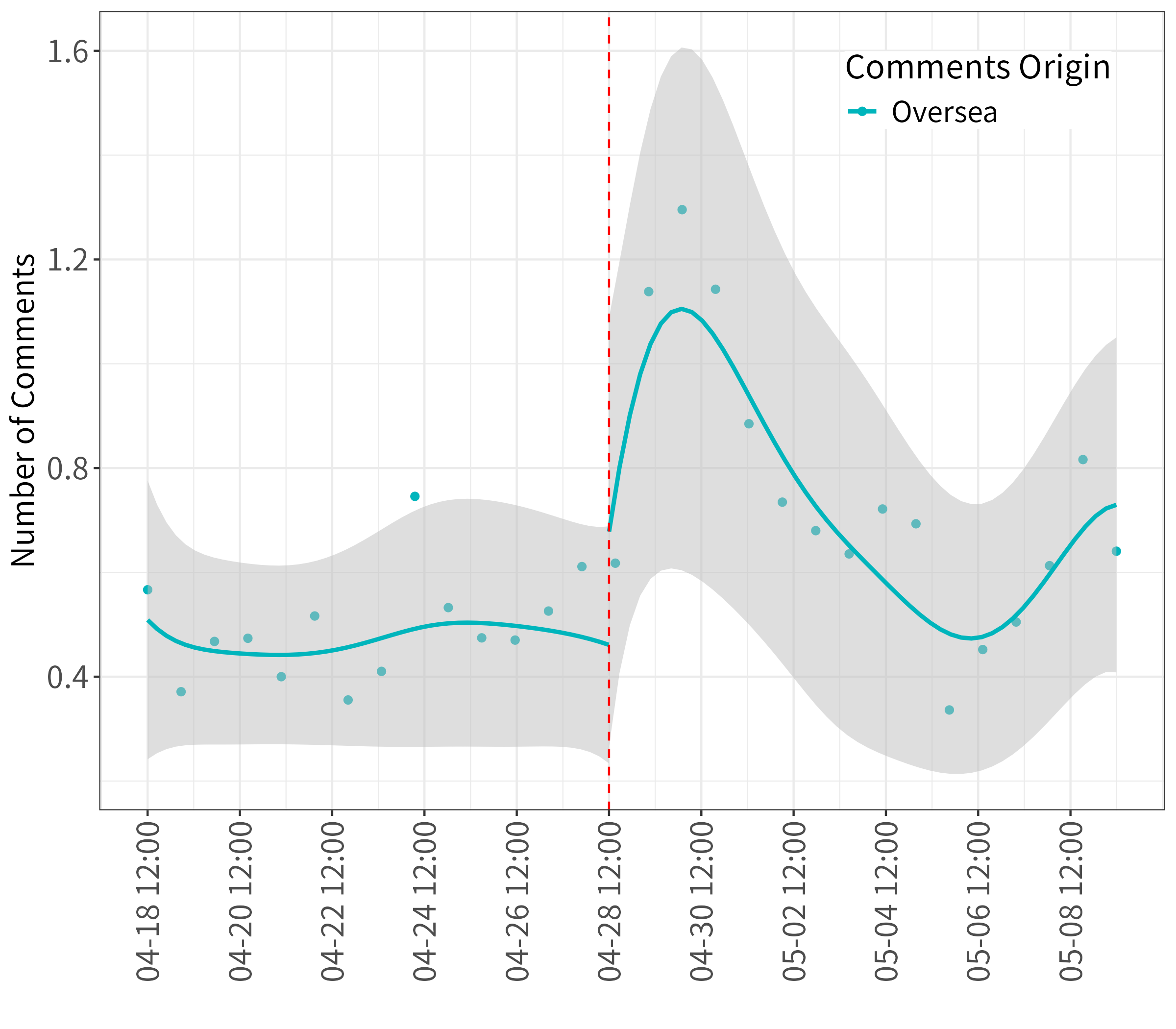}
        \caption{Increased Comments on Russian-Ukraine War}
    \label{fig:pnas_volume_comments_oversea_origin_ru_48h_bandwidth_best}
    \end{subfigure}
    \hspace{0.2cm} 
        \begin{subfigure}[b]{0.4\textwidth}
        \centering
        \includegraphics[width=\textwidth]{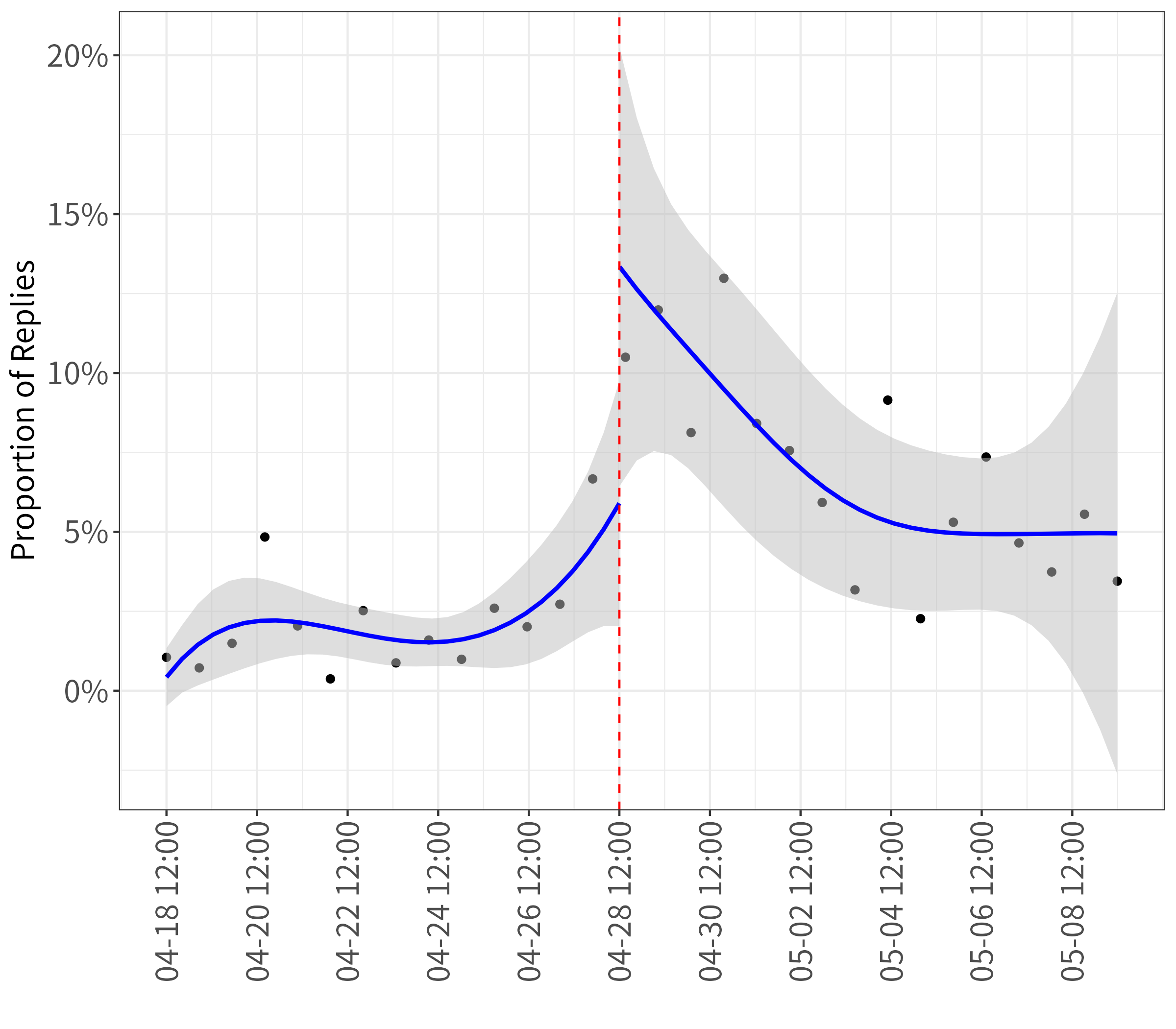}
        \caption{Increased Regional Discrimination}  \label{fig:pnas_increased_discrimination_bandwidth_best_international_issue}
        \end{subfigure}
    \caption{Oversea Users' Reaction on the Policy}
    \label{fig:oversea_user_reaction}
     \caption*{\textbf{Notes:} Panel~\ref{fig:pnas_volume_comments_oversea_origin_ru_48h_bandwidth_best} shows a marked increase in comments from overseas users on posts related to the Russia-Ukraine War following the implementation of the user location disclosure policy. Panel~\ref{fig:pnas_increased_discrimination_bandwidth_best_international_issue} reveals a simultaneous rise in regionally discriminatory replies targeting these users. Together, the panels indicate that although overseas engagement grew, it was met with heightened hostility from domestic commenters. The red dashed line denotes the timing of the policy rollout.}
\end{figure}

Instead of discouraging overseas users from commenting, the user location disclosure policy appeared to intensify their engagement on sensitive topics such as the Russia-Ukraine War. As shown in Figure~\ref{fig:pnas_volume_comments_oversea_origin_ru_48h_bandwidth_best}, overseas comment volume on this issue increased following the policy's implementation. At the same time, these users were more frequently targeted by domestic commenters. Figure~\ref{fig:pnas_increased_discrimination_bandwidth_best_international_issue} illustrates a rise in replies containing region-based discriminatory language, suggesting that geotags exposed overseas users to heightened hostility.

\clearpage

\subsection{Decompose regional discrimination}

\begin{figure}[H]
    \centering
    \begin{subfigure}[b]{0.43\textwidth}
        \centering
        \includegraphics[width=\textwidth]{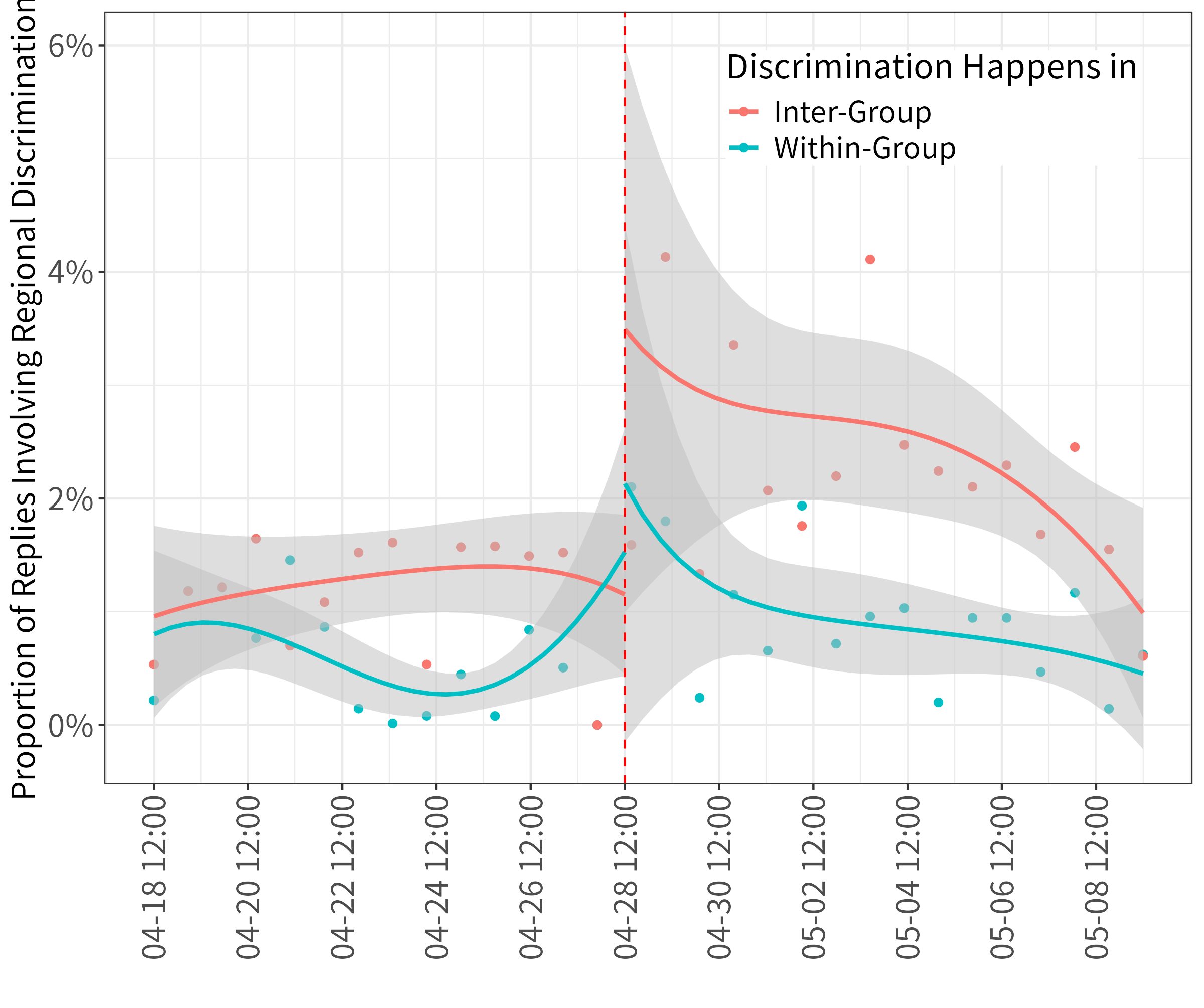}
        \caption{The Origin of Regional Discrimination}
        \label{fig:pnas_increased_discrimination_by_origin_bandwidth_best}
    \end{subfigure}
    \hspace{0.1cm} 
        \begin{subfigure}[b]{0.43\textwidth}
        \centering
        \includegraphics[width=\textwidth]{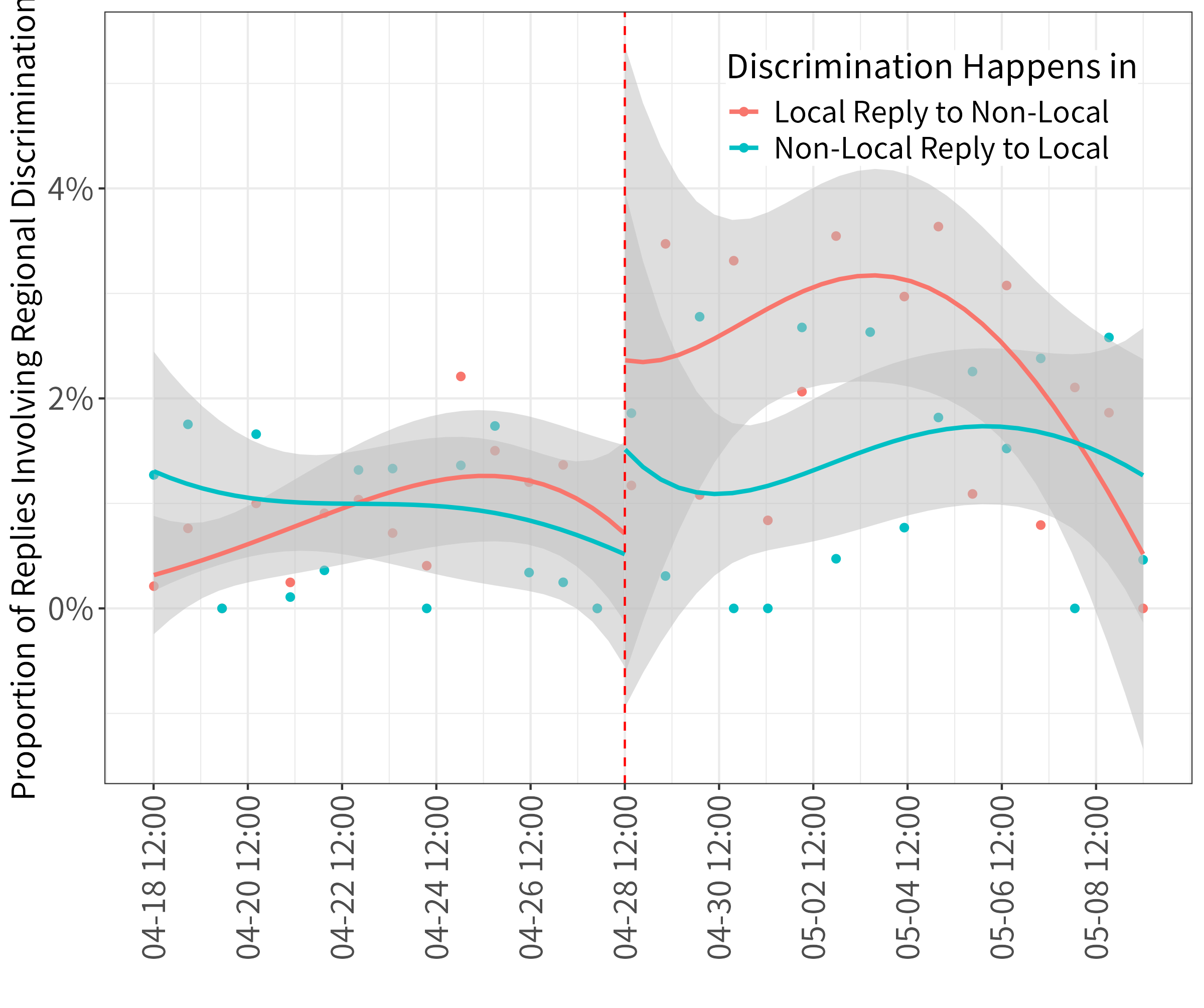}
        \caption{The Inter-Group Origin of Regional Discrimination}
    \label{fig:pnas_increased_discrimination_by_origin_inter_bandwidth_best}
        \end{subfigure}
    \caption{Decomposition of Increased Regional Discrimination}
    \label{fig:decomposed_regional_discrimination}
    \caption*{\textbf{Notes:} Panel \ref{fig:pnas_increased_discrimination_by_origin_bandwidth_best} differentiates between inter-group and within-group regional discrimination, showing a clear increase predominantly in inter-group interactions post-policy implementation. Panel \ref{fig:pnas_increased_discrimination_by_origin_inter_bandwidth_best} further specifies these inter-group interactions, illustrating that increased discrimination primarily involves local users targeting non-local users. The red dashed lines indicate the timing of the user location disclosure implementation, emphasizing the direct link between the policy and the observed rise in regional discrimination.}
\end{figure}

In Figure~2a, we documented an increase in regionally discriminatory replies after the implementation of the user location disclosure policy. To unpack this trend, Figure~\ref{fig:decomposed_regional_discrimination} disaggregates discriminatory replies by interaction type. Panel~\ref{fig:pnas_increased_discrimination_by_origin_bandwidth_best} shows that the rise is largely driven by inter-group interactions, while within-group discrimination remains stable. Panel~\ref{fig:pnas_increased_discrimination_by_origin_inter_bandwidth_best} further reveals that most inter-group discrimination comes from local users targeting non-local users, underscoring the role of geographic identity in shaping online hostility.

\subsection{Engagement analysis based on aggregate metrics}

\begin{figure}[H]
    \centering
    \begin{subfigure}[b]{0.33\textwidth}
        \centering
        \includegraphics[width=\textwidth]{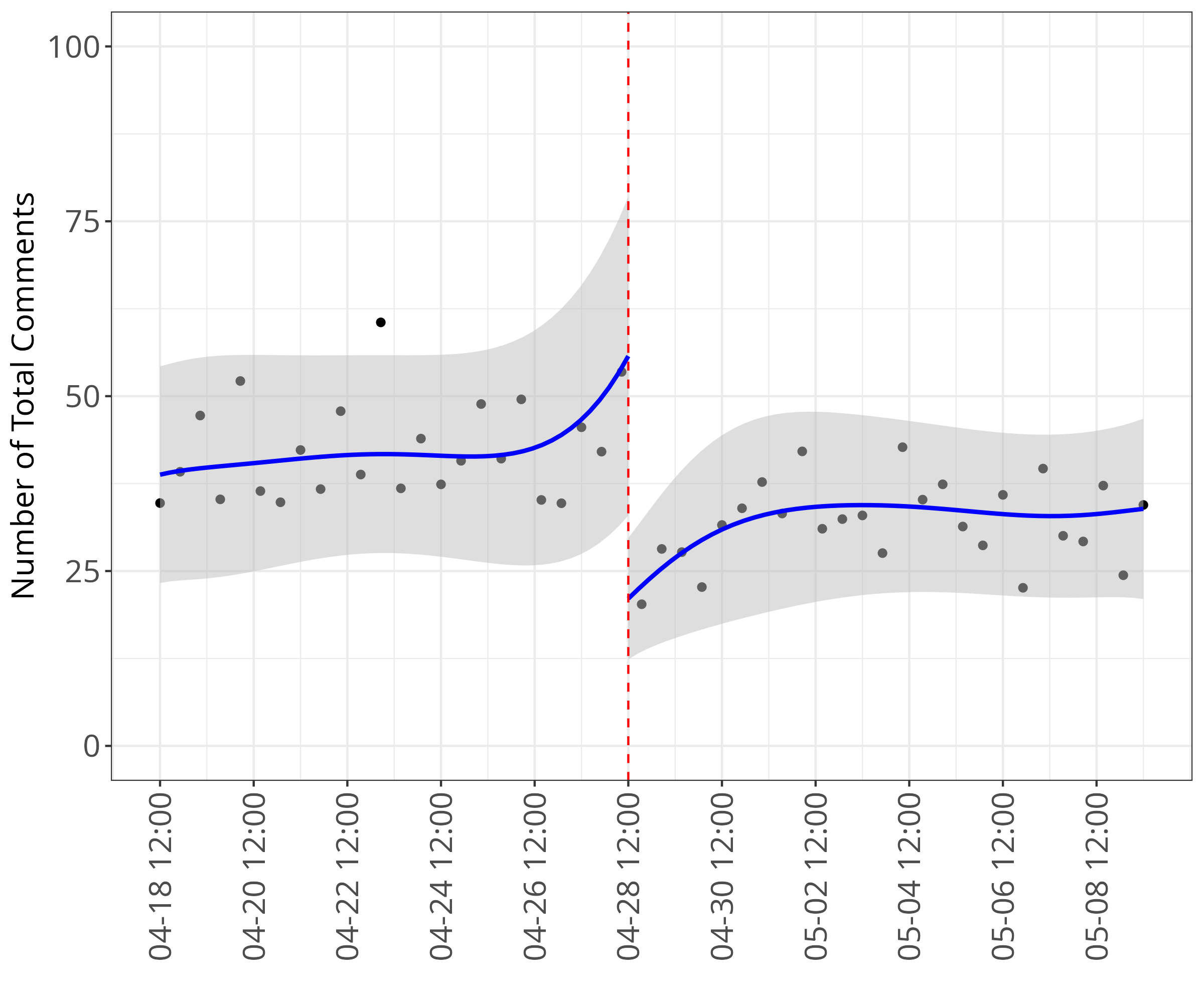}
        \caption{Total Number of Comments Received}
        \label{fig:pnas_total_comments_bandwidth_best}
    \end{subfigure}
    \hspace{0.2cm} 
    \begin{subfigure}[b]{0.33\textwidth}
        \centering
        \includegraphics[width=\textwidth]{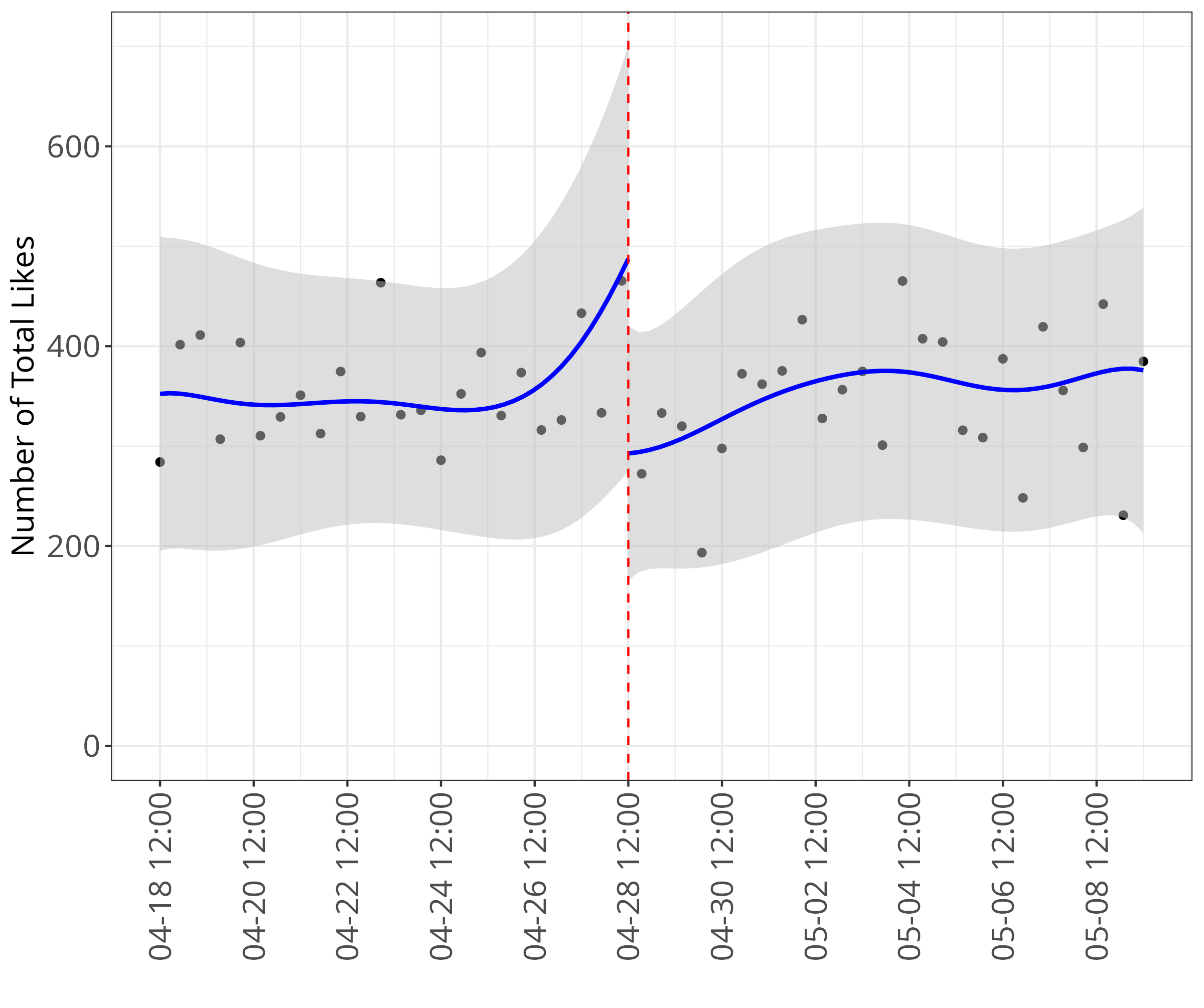}
        \caption{Total Number of Likes Received}
        \label{fig:pnas_total_likes_bandwidth_best}
        \end{subfigure}
    \hspace{0.2cm} 
        \begin{subfigure}[b]{0.33\textwidth}
        \centering
        \includegraphics[width=\textwidth]{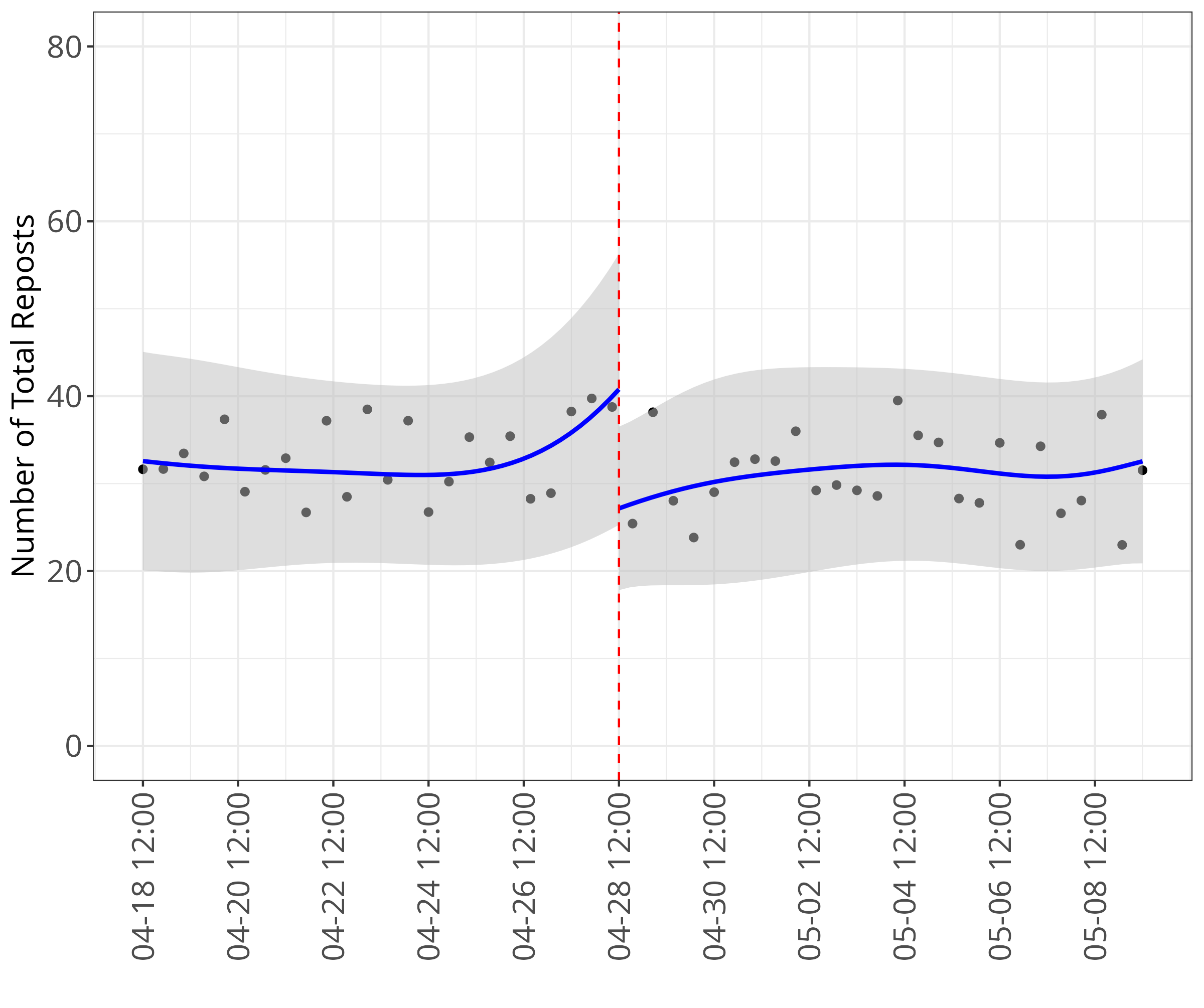}
        \caption{Total Number of Reposts Received}
    \label{fig:pnas_total_reposts_bandwidth_best}
        \end{subfigure}
    \caption{Differential Impact of user location disclosure on Aggregate Post Engagement Metrics}
    \label{fig:posts_total_metrics}
    \caption*{\textbf{Notes:} Panel~\ref{fig:pnas_total_comments_bandwidth_best} shows a statistically significant decline in the total number of comments after the user location disclosure was implemented. Panel~\ref{fig:pnas_total_likes_bandwidth_best} and panel~\ref{fig:pnas_total_reposts_bandwidth_best} display visible drops in the number of likes and reposts although the estimates are not statistically significant at the 5\% level. These results suggest that the policy primarily dampened user engagement through reduced commenting, but the effects on other engagemetn metrics are small.
}
\end{figure}

Figure~\ref{fig:posts_total_metrics} displays engagement trends based on aggregate metrics associated with each post concerning local issues—specifically, the total number of comments, likes, and reposts. Because each post was captured in multiple timestamped snapshots, we report values from the final snapshot to ensure data consistency. To mitigate the influence of extreme outliers, all values above the 99th percentile are winsorized to the 99th percentile.

\clearpage

\section{Case studies}
\subsection{Anecdotal evidence of intensified regional discrimination}

\begin{figure}[H]
    \centering
    \includegraphics[width=0.85\textwidth]{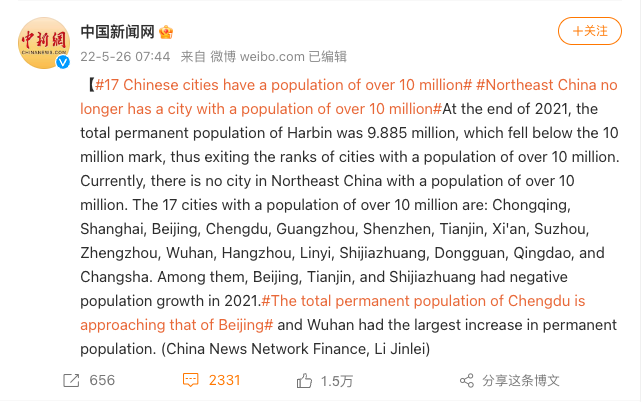}
    \caption{Anecdotal Evidence of Intensified Regional Discrimination (translated)}
    \label{fig:qualititive_post}
    \caption*{\textbf{Notes:} This figure offers anecdotal evidence of heightened regional discrimination following the implementation of the user location disclosure. The Weibo post, originally written in Chinese and presented here in translation, addresses population decline in major Chinese cities, including several in Northeast China. Its mention of specific provinces sparked regional tensions, as reflected in the discriminatory exchanges that followed in the comments.}
\end{figure}

To illustrate the rise in regional discrimination, we present a case study from the comments section following the implementation of the user location disclosure. In the post shown in Figure~\ref{fig:qualititive_post}, China News Network raised the issue of population decline in major Chinese cities, mentioning locations across several provinces—Heilongjiang, Chongqing, Shanghai, Beijing, and Sichuan—and referencing Northeast China in the hashtag (which includes Heilongjiang, Jilin, and Liaoning). Despite the neutral nature of the topic, the post sparked intense regional conflict in the comments.

Here we present two illustrative examples of comments on the post in Figure~\ref{fig:qualititive_post}.

\begin{figure}[h]
    \centering
    \begin{subfigure}[b]{0.49\textwidth}
        \centering
        \includegraphics[width=\textwidth]{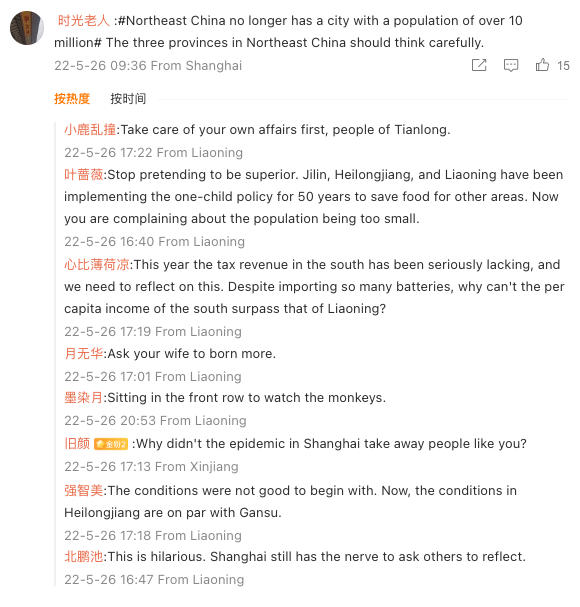}
        \caption{Example Comment 1}
        \label{fig:qualititive_comment_1}
    \end{subfigure}
    \hspace{0.01cm} 
        \begin{subfigure}[b]{0.49\textwidth}
        \centering
        \includegraphics[width=\textwidth]{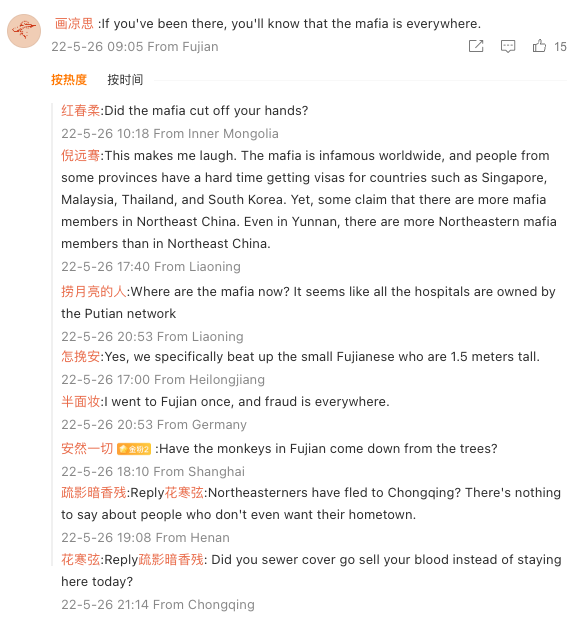}
        \caption{Example Comment 2}
    \label{fig:qualititive_comment_2}
        \end{subfigure}
    \caption{Anecdotal Evidence of Intensified Regional Discrimination: Comment Examples (translated)}
    \label{fig:division}
     \caption*{\textbf{Notes:} This figure presents two translated comment threads from the original Chinese Weibo post shown in Figure~\ref{fig:qualititive_post}, offering illustrative evidence of heightened regional discrimination following the user location disclosure. Usernames shown are pseudonyms and avatars have been replaced, while the comment content remains unchanged.}
\end{figure}

In example comment~\ref{fig:qualititive_comment_1}, a user from Shanghai urges the three northeastern provinces to think carefully.'' This remark triggers a wave of defensive and hostile replies from users in Liaoning, highlighting heightened regional tensions. One user retorts, ``Take care of your own affairs first, people of Tianlong,'' (天龙人, Tiān lóng rén, is a term borrowed from the Japanese manga *One Piece*, where it refers to an elitist group who see themselves as superior. On Chinese social media, it is used sarcastically to mock Shanghai residents' perceived sense of privilege). Another critiques historical policy: ``Jilin, Heilongjiang, and Liaoning implemented the one-child policy for 50 years to save food for other regions.'' Economic grievances surface as well: ``Despite importing so many batteries, why can't the south's per capita income surpass Liaoning's?'' The comment thread also features personal attacks and sarcasm, such as, ``Why didn't the epidemic in Shanghai take people like you?'' These replies underscore deep-rooted interregional resentment shaped by history, policy, and economic inequality.

\clearpage

In another example (Figure~\ref{fig:qualititive_comment_2}), a user from Fujian remarks that ``the mafia is everywhere,'' triggering a cascade of sarcastic and hostile replies rooted in regional stereotypes and animosities. A commenter from Inner Mongolia retorts mockingly,``Did the mafia cut off your hands?’’ while a user from Liaoning laughs at the idea that mafia activity is concentrated in the Northeast. A Heilongjiang user boasts about beating up people from Fujian, and another from Shanghai sneers, ``Have the monkeys in Fujian come down from the trees yet?’’ These exchanges underscore how the user location disclosure amplified entrenched regional prejudices, intensifying interprovincial conflict and misunderstanding.

\end{CJK}
\end{document}